\documentclass[showpacs,preprintnumbers,superscriptaddress,amsmath,amssymb,rmp]{revtex4}

\usepackage{amssymb}
\usepackage{amsfonts}

\usepackage{graphicx}
\usepackage{dcolumn}
\usepackage{bm}
\usepackage{multirow}

\newcommand{\ua}{\uparrow}
\newcommand{\da}{\downarrow}

\begin{document}

\title{Quantum Cloning Machines and the Applications}

\author{Heng Fan}
\email{hfan@iphy.ac.cn}
\affiliation{Beijing National Laboratory for Condensed Matter Physics,
Institute of Physics, Chinese Academy of Sciences, Beijing 100190, China}
\affiliation{Collaborative Innovation Center of Quantum Matter, Beijing 100190, China}

\author{Yi-Nan Wang}

\author{Li Jing}

\affiliation{School of Physics, Peking University, Beijing 100871, China}

\author{Jie-Dong Yue}
\affiliation{Beijing National Laboratory for Condensed Matter Physics,
Institute of Physics, Chinese Academy of Sciences, Beijing 100190, China}

\author{Han-Duo Shi}

\author{Yong-Liang Zhang}

\author{Liang-Zhu Mu}

\affiliation{School of Physics, Peking University, Beijing 100871, China}

\date{\today}

\begin{abstract}
No-cloning theorem is fundamental for quantum mechanics and for quantum information science that states an unknown quantum state cannot be cloned perfectly.
However, we can try to clone a quantum state approximately with the optimal fidelity, or instead, we can try to clone it perfectly with the largest probability.
Thus various quantum cloning machines have been designed for different quantum information protocols.
Specifically, quantum cloning machines can be designed to analyze the security of quantum key distribution protocols such as BB84 protocol, six-state protocol, B92 protocol and their
generalizations. Some well-known quantum cloning machines include universal quantum cloning machine, phase-covariant cloning machine,
the asymmetric quantum cloning machine and the probabilistic quantum cloning machine etc.
In the past years, much progress has been made in studying quantum cloning machines and their applications and implementations,
both theoretically and experimentally. In this review, we will give a complete description of those important developments about quantum cloning and some related topics. On the other hand, this review is self-consistent, and in particular,
we try to present some detailed formulations so that further study can be taken based on those results.
\end{abstract}



\maketitle


\tableofcontents

\section{Introduction}
In the past years, the study of quantum computation and quantum information has been attracting much
attention from various research communities.
Quantum information processing (QIP) is based on principles of quantum mechanics \cite{Nielsen2000}.
It promises algorithms which may surpass their classical counterparts.
 One of those algorithms is Shor algorithm \cite{Shor94} which can factorize large number exponentially
 faster than the existing classical algorithms do \cite{Ekert-JozsaRMP96}. In this sense,
 the RSA public key cryptosystem \cite{RSA78} widely used in modern financial systems and networks
  might be attacked easily if a quantum computer exists, since the security of RSA system
  is based on assumption that it is extremely difficult
 to factorize a large number. On the other hand, QIP provides an unconditional secure
 quantum cryptography based a principle of quantum mechanics, no-cloning theorem \cite{Wootters1982}, which means that an unknown quantum state
 cannot be cloned perfectly.

For comparison, in classical information science, we use bit which is either ``0'' or ``1'' to carry the information,
while for quantum information, a bit of quantum information which is named as ``qubit'' is encoded in a quantum state which may be
a superposition of states $|0\rangle $ and $|1\rangle $.
For example, a general qubit takes the form $\alpha |0\rangle +\beta |1\rangle $,
where parameters $\alpha $ and $\beta $ are complex numbers according to quantum mechanics and are normalized as
$|\alpha |^2+|\beta |^2=1$. So a qubit can collapse to either $|0\rangle $ or $|1\rangle $ with some probability
if a measurement is performed.
The classical information can be copied perfectly. We know that we can copy a file in a computer without
any principal restriction. On the contrary, QIP is based on principles of quantum mechanics,
which is linear and thus an arbitrary quantum state cannot be cloned perfectly since of the no-cloning theorem.
We use generally terminology ``clone'' instead of ``copy'' for reason of no-cloning theorem in \cite{Wootters1982}.
No-cloning, however, is not the end of the story.

It is prohibited to have a perfect quantum clone. It is still possible that we can copy a quantum state
approximately or probabilistically. There are various quantum protocols for QIP which may
use tool of quantum cloning for different goals. Thus various quantum cloning machines have been created
both theoretically and experimentally. The study of quantum cloning is of fundamental interest in QIP.
Additionally, the quantum cloning machines can also be applied directly in various
quantum key distribution (QKD) protocols.
The first quantum key
distribution protocol proposed by Bennett and Brassard in 1984 (BB84) uses four different qubits,
BB84 states \cite{Bennett1984},
to encode classical information in transmission.
Correspondingly, the phase-covariant quantum clone machine, which can copy optimally all qubits located
in the equator of the Bloch sphere, is proved to be optimal for cloning of states similar as BB84 states.
The BB84 protocol can be extended to six-state protocol, the corresponding cloning machine is the universal
quantum cloning machine which can copy optimally arbitrary qubits. Similarly the probabilistic quantum
cloning machine is for B92 QKD protocol \cite{PhysRevLett.68.3121}. Quantum cloning is also related with some
fundamentals in quantum information science, for example, the no-cloning theorem is closely related with no-signaling
theorem which means that superluminal communication is forbidden. We can also use quantum cloning machines
for estimating a quantum state or phase information of a quantum state.
So the study of quantum cloning is of interest for
reasons of both fundamental and practical applications.

The previous well-accepted reviews of quantum cloning can be found in \cite{RevModPhys.77.1225}, and also in \cite{Cerf2006455}.
Quantum cloning, as other topics of quantum information, developed very fast in the past years.
An up-to-date review is necessary.
In the present review, we plan to give a full description of  results about quantum cloning and
some closely related topics.
This review is self-consistent and some fundamental knowledge is also introduced. In particular,
a main characteristic of this review is that it contains a large number of
detailed formulations for the main review topics. It is thus easy for the beginners
to follow those calculations for
further study on those quantum cloning topics.

The review
is organized as follows: In the next part of this section, we will present in detail some fundamental
concepts of quantum computation and quantum information including the form of qubit represented in Bloch sphere,
the definition of entangled state, some principles of quantum mechanics used in the review.
Then we will present in detail the developments of quantum cloning. Here let us introduce briefly
some results contained in this review.
In Section II, we will review
several proofs of no-cloning theorem from different points of view, including a simple presentation,
no-cloning for mixed states, the relationship between no-cloning and no-signaling theorems for quantum states, no-cloning from information
theoretical viewpoints. In Section III, we will review the universal quantum cloning machine. We will present
the universal quantum cloning machine for qubit and qudit including both symmetric and asymmetric cases. We then will
present a unified quantum cloning machine which can be easily reduced to several universal cloning machines.
We will also show some schemes for cloning of mixed states.
We will show that the universal quantum cloning machine, by definition, can copy arbitrary input state,
is necessary for a six-state input which are used in QKD. Further, the universal cloning machine is
necessary for a four-state input which is the minimal input set. In Section V, the phase-covariant quantum cloning machines
will be presented. One important application of this cloning machine is to study the well-known BB84 quantum
cryptography. The phase-covariant quantum cloning machines include the cases of qubit and of higher dimension.
In particular, a unified phase-covariant quantum cloning machine will be presented which can be adjusted for an arbitrary subset of the
mutually unbiased bases. We can also show that the minimal input for phase-covariant quantum cloning is
a set of three states with equal
phase distances in the equator of Bloch sphere. The phase-covariant quantum cloning is actually state-dependent,
we thus will present some other cases of state-dependent quantum cloning.

In section IX, we present some detailed results of sequential quantum cloning. We expect that
further exploration is in order based on those results.

In order to have a full view of all developments in quantum cloning and some closely related topics, we
try to review some references briefly in one to two sentences. Those parts are generally named as
`other developments and related topics'. Our aim is
to cover as much as possible these
developments in quantum cloning, but we understand that some important references might still be missed in
this review.

\subsection{Quantum information, qubit and quantum entanglement}
\begin{figure}
\includegraphics[width=4cm]{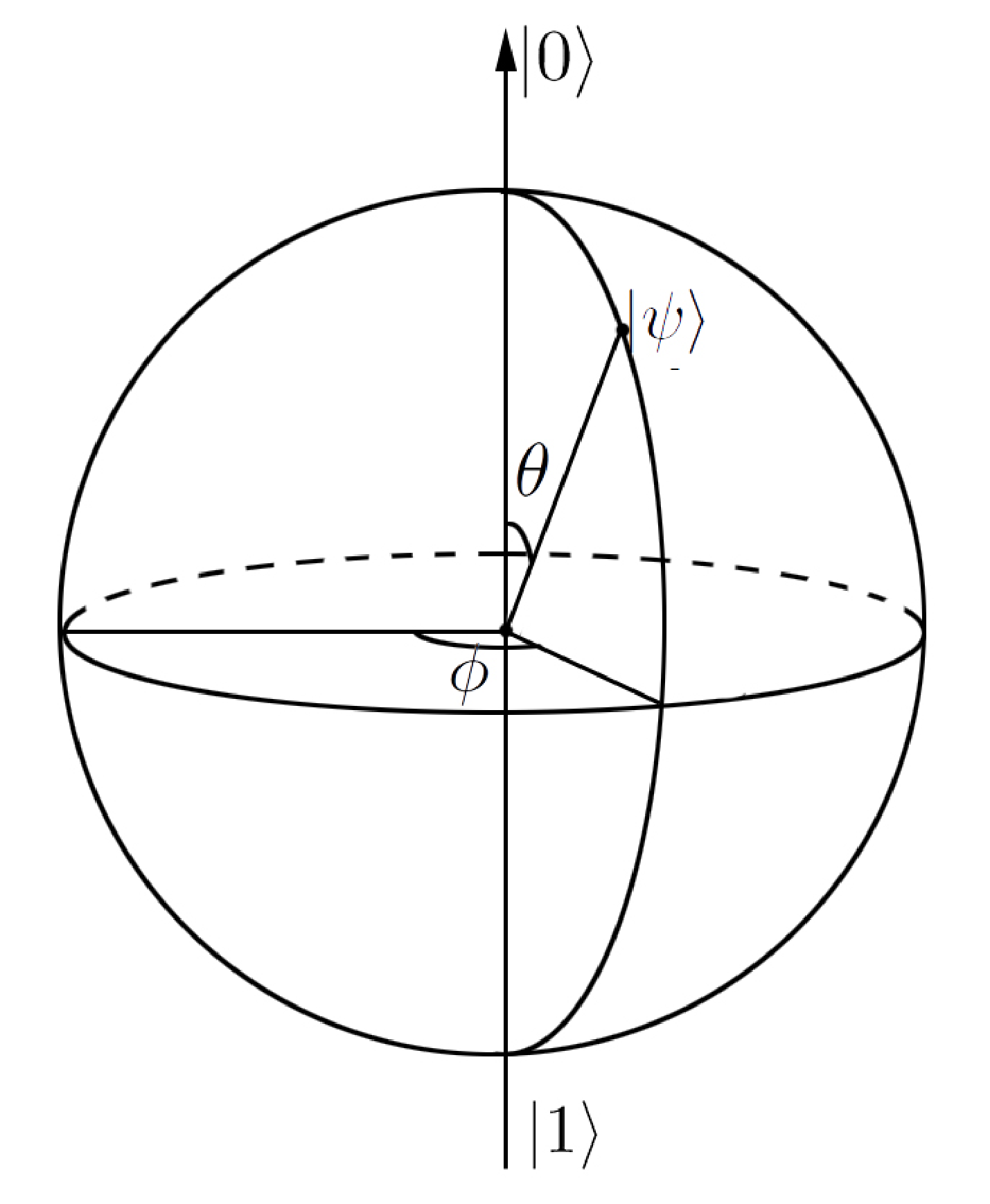}
\caption{(color online). A qubit in Bloch sphere, $|\psi \rangle
=\cos \frac {\theta }{2}|0\rangle +\sin \frac {\theta }{2}e^{i\phi }|1\rangle $, it contains
amplitude parameter $\theta $ and phase parameter $\phi $. }
\label{IBlochsphere}
\end{figure}

We have a quantum system constituted by two states $|0\rangle $ and $|1\rangle $. They are orthogonal,
\begin{eqnarray}
\langle 0|1\rangle =0.
\end{eqnarray}
Those two states can be energy levels of an atom, photon polarizations,
electron spins, Bose-Einstein condensate with two intrinsic freedoms or any physical material with quantum properties.
In this review, we also use some other standard notations $|0\rangle =|\ua \rangle ,|1\rangle =|\da\rangle $ and exchange
them without mentioning.
Simply, those two states can be represented as two vectors in linear algebra,
\begin{eqnarray}
|0\rangle =\left( \begin{array}{c}1\\ 0 \end{array}\right) ,~~~~ |1\rangle =\left( \begin{array}{c}0\\ 1 \end{array}\right).
\end{eqnarray}
Corresponding to bit in classical information science, a qubit in quantum information science is a superposition of two orthogonal states,
\begin{eqnarray}
|\psi \rangle =\alpha |0\rangle +\beta |1\rangle ,
\end{eqnarray}
where a normalization equation should be satisfied,
\begin{eqnarray}
|\alpha |^2+|\beta |^2=1.
\end{eqnarray}
 Here both $\alpha $ and $\beta $ are complex parameters which include amplitude and phase information, $\alpha =|\alpha |e^{i\phi _{\alpha }}$ and
$\beta =|\beta |e^{i\phi _{\beta }}$. So a qubit $|\psi \rangle $ is defined on a two-dimensional Hilbert space $\mathbb{C}^2$.
In quantum mechanics, a whole phase cannot be detected and thus can be omitted, only the relative phase of $\alpha $ and $\beta $ is important which
is $\phi =\phi _{\alpha }-\phi _{\beta }$. Now we can find that a qubit can be represented in another form,
\begin{eqnarray}
|\psi \rangle =\cos \frac {\theta }{2}|0\rangle +\sin \frac {\theta }{2}e^{i\phi }|1\rangle ,
\end{eqnarray}
where $\theta \in [0,\pi ], \phi \in [0, 2\pi\}$. It corresponds to a point in the Bloch sphere, see FIG. \ref{IBlochsphere}.

The two qubits in separable form can be written as,
\begin{eqnarray}
|\psi \rangle |\phi \rangle &=&(\alpha |0\rangle +\beta |1\rangle  )(\gamma |0\rangle +\delta |1\rangle  )
\nonumber \\
&=&\alpha \gamma |00\rangle +\alpha \delta |01\rangle +\beta \gamma |10\rangle +\alpha \delta |11\rangle .
\end{eqnarray}
If those two qubits are identical, one can find,
\begin{eqnarray}
|\psi \rangle ^{\otimes 2}&\equiv &(\alpha |0\rangle +\beta |1\rangle  )(\alpha |0\rangle +\beta |1\rangle  )
\nonumber \\
&=& \alpha ^2|00\rangle +\sqrt {2}\alpha \beta \frac {1}{\sqrt {2}}(|01\rangle +|10\rangle )
+\beta ^2|11\rangle .
\end{eqnarray}
For convenience, we write the second term as a normalized symmetric state $\frac {1}{\sqrt {2}}(|01\rangle +|10\rangle )$
which will be used later.

For two-qubit state, besides those separable state, we also have the entangled state, for example,
\begin{eqnarray}
|\Phi ^+\rangle \equiv \frac {1}{\sqrt {2}}(|00\rangle +|11\rangle ).
\label{IBell1}
\end{eqnarray}
This is state cannot be written as a product form like $|\psi \rangle |\phi \rangle $, so it is ``entangled''.
It is actually a maximally entangled state. In quantum information
science, quantum entanglement is the valuable resource which can be widely used in various tasks and protocols.
Complementary to entangled state $|\Phi ^+\rangle $, we have other three orthogonal and maximally entangled state
which constitute a complete basis for $\mathbb{C}^2 \otimes \mathbb{C}^2$. Those four states are Bell states,
here we list them all as follows,
\begin{eqnarray}
|\Phi ^+\rangle &\equiv & \frac {1}{\sqrt {2}}(|00\rangle +|11\rangle ),
\label{IBell0}
\\
|\Phi ^-\rangle &\equiv & \frac {1}{\sqrt {2}}(|00\rangle -|11\rangle ),
\label{IBell2}
\\
|\Psi ^+\rangle &\equiv & \frac {1}{\sqrt {2}}(|01\rangle +|10\rangle ),
\label{IBell3}
\\
|\Psi ^-\rangle &\equiv & \frac {1}{\sqrt {2}}(|01\rangle -|10\rangle ).
\label{IBell4}
\end{eqnarray}
Those four Bell states can be transformed to each other by local unitary transformations.

Consider three Pauli matrices defined as,
\begin{eqnarray}
\sigma _x=\left( \begin{array}{cc}0&1\\
1&0\end{array}\right) ,
~~~\sigma _y=\left( \begin{array}{cc}0&-i\\
i&0\end{array}\right) ,
~~~\sigma _z=\left( \begin{array}{cc}1&0\\
0&-1\end{array}\right).
\end{eqnarray}
Since $\sigma _y=i\sigma _x\sigma _z $, if the imaginary unit, ``$i$'', is the whole phase, we sometimes use $\sigma _x\sigma _z$
instead of $\sigma _y$. Bear in mind that we have $|0\rangle =\left( \begin{array}{c}1\\0 \end{array}\right) $, and $\langle 0|=(1,0)$, so in
linear algebra, we have the representation,
\begin{eqnarray}
|0\rangle \langle 0|=\left( \begin{array}{c}1\\0\end{array} \right)(1,0)=\left( \begin{array}{cc}
1&0\\0&0\end{array}\right).
\end{eqnarray}
Now three Pauli matrices have an operator representation,
\begin{eqnarray}
\sigma _x&=&|0\rangle \langle 1|+|1\rangle \langle 0|,\\
\sigma _y&=&-i|0\rangle \langle 1|+i|1\rangle \langle 0|,\\
\sigma _z&=&|0\rangle \langle 0|-|1\rangle \langle 1|.
\end{eqnarray}
In this review, we will not distinguish the matrix representation and the operator representation.
Acting Pauli matrices $\sigma _x$ and $\sigma _z$ on a qubit, we find,
\begin{eqnarray}
\sigma _x|0\rangle =|1\rangle ,~~\sigma _x|1\rangle =|0\rangle ,
\\
\sigma _z|0\rangle =|0\rangle ,~~\sigma _z|1\rangle =-|1\rangle ,
\end{eqnarray}
which are the bit flip action and phase flip action, respectively,  while $\sigma _y$ will cause both bit flip and phase flip for a qubit.
In this review, for convenience, we sometimes use notations $X\equiv \sigma _x, Z\equiv \sigma _z$ to represent the corresponding Pauli
matrices. Also, those Pauli matrices can also be defined in higher dimensional system,
while the same notations might be used if no confusion is caused.

For four Bell states, their relationship by local transformations can be as follows,
\begin{eqnarray}
|\Phi ^-\rangle &= &(I\otimes \sigma _z) |\Phi ^+\rangle ,\\
|\Psi ^+\rangle &= & (I\otimes \sigma _x) |\Phi ^+\rangle ,\\
|\Psi ^-\rangle &= & (I\otimes \sigma _x\sigma _z) |\Phi ^+\rangle ,
\end{eqnarray}
where $I$ is the identity in $\mathbb{C}^2$, the Pauli matrices are acting on the second qubit.

Here we have already used the tensor product. Consider two operators, $O_1=\left( \begin{array}{cc}A_1&B_1\\
C_1&D_1\end{array}\right) $, $O_2=\left( \begin{array}{cc}A_2&B_2\\
C_2&D_2\end{array}\right) $, the tensor product $O_1\otimes O_2$ is defined and calculated as follows,
\begin{eqnarray}
O_1\otimes O_2&=&\left( \begin{array}{cc}A_1O_2&B_1O_2\\
C_1O_2&D_1O_2\end{array}\right) \nonumber \\
&=&\left( \begin{array}{cccc}A_1A_2&A_1B_2&B_1A_2&B_1B_2\\
A_1C_2&A_1D_2&B_1C_2&B_1D_2\\
C_1A_2&C_1B_2&D_1A_2&D_1B_2\\
C_1C_2&C_1D_2&D_1C_2&D_1D_2\\\end{array}\right).
\end{eqnarray}
When we apply a tensor product of, $O_1\otimes O_2$, on a two-qubit quantum state, operator $O_1$
is acting on the first qubit, operator $O_2$ is acting on the second qubit.

We have already extended one qubit to two-qubit state. Similarly, multipartite qubit state
can be obtained. Also we may try to extend qubit from two-dimension to higher-dimensional system, generally
named as ``qutrit'' for dimension three and ``qudit'' for dimension $d$ in more general case,
the Hilbert space is extended from $\mathbb{C}^2$ to $\mathbb{C}^d$.
For example, we sometimes have ``qutrit'' when we consider a quantum state in three dimensional system.
For more general case, a qudit is also a superposed state,
\begin{eqnarray}
|\psi \rangle =\sum _{j=0}^{d-1}x_j|j\rangle ,
\end{eqnarray}
where $x_j, j=0,1,...,d-1$, are normalized complex parameters. Quantum entanglement can also be in higher dimensional,
multipartite systems.

A qubit $|\psi \rangle $ can be represented by its density matrix,
\begin{eqnarray}
|\psi \rangle \langle \psi |&=&(\alpha |0\rangle +\beta |1\rangle )(\alpha ^*\langle 0|+\beta ^*\langle 1|)
\nonumber \\
&=&\left( \begin{array}{cc}|\alpha |^2&\alpha \beta ^* \\
\alpha ^*\beta &|\beta |^2\end{array}\right) .
\end{eqnarray}
However, a general qubit may be not only the superposed state, which is actually the pure state, but also
a mixed state which is in a probabilistic form. It can only be represented by a density operator $\rho $,
\begin{eqnarray}
\rho =\sum _jp_j|\psi _j\rangle \langle \psi _j|,
\end{eqnarray}
where $p_j$ is the probabilistic distribution with $\sum p_j=1$.

A density matrix is positive semi-definite, and its trace equals to 1,
\begin{eqnarray}
\rho \ge 0, ~~~{\rm Tr}\rho =1.
\end{eqnarray}
The density matrices of a pure state and a mixed state can be easily distinguished by the following conditions,
\begin{eqnarray}
{\rm Tr}\rho ^2&=&1, ~~~{\rm pure ~state}; \\
{\rm Tr}\rho ^2&<&1, ~~~{\rm mixed ~state.}
\end{eqnarray}

For multipartite state, one part of the state
is the reduced density matrix obtained by tracing out other parts. For example, for two-qubit maximally entangled state
$|\Phi ^+_{AB}\rangle $ constituted by $A$ and $B$ parts, each qubit is a mixed state,
\begin{eqnarray}
\rho _A={\rm Tr}_B|\Phi ^+_{AB}\rangle \langle \Phi ^+_{AB}|=\frac {1}{2}I.
\end{eqnarray}
This case is actually a completely mixed state, here $I$ is the identity operator.
The identity operator can be written as any pure state and
its orthogonal state with equal probability,
\begin{eqnarray}
\rho _A=\frac {1}{2}I=\frac {1}{2}|\psi \rangle \langle \psi |+\frac {1}{2}|\psi ^{\perp}\rangle \langle \psi ^{\perp}|,
\end{eqnarray}
where if $|\psi \rangle =\alpha |0\rangle +\beta |1\rangle $, its orthogonal state can take the form,
\begin{eqnarray}
|\psi ^{\perp}\rangle =\beta ^*|0\rangle -\alpha ^*|1\rangle ,
\end{eqnarray}
where $*$ means the complex conjugation.

\subsection{Quantum gates}

In QIP, all operations should satisfy the laws of quantum mechanics such
as the generally used unitary transformation and quantum measurement. Similar as in classical computation,
all quantum computation can be effectively implemented by several fundamental gates. The single qubit rotation gate and controlled-NOT (CNOT) gate
constitute a complete set of fundamental gates for universal quantum computation \cite{elementarygates}. The single qubit
rotation gate is just a unitary transformation on a qubit,
$\hat {R}(\vartheta ,\phi )$, defined as
\begin{eqnarray}
\hat {R}(\vartheta ,\phi )|0\rangle &=&\cos \vartheta |0\rangle
+e^{i\phi }\sin \vartheta |1\rangle ,\nonumber \\
\hat {R}(\vartheta ,\phi )|1\rangle &=&
-e^{-i\phi }\sin \vartheta |0\rangle +\cos \vartheta |1\rangle ,
\label{singlegate}
\end{eqnarray}
where the phase parameter $\phi $ should also be controllable.
The CNOT gate is defined as a unitary transformation
on two-qubit system, one qubit is the controlled qubit and another qubit is the target qubit. For a CNOT gate, when
the controlled qubit is $|0\rangle $, the target qubit does not change; when the controlled qubit
is $|1\rangle $, the target qubit should be flipped. Explicitly it is defined as,

\begin{eqnarray}
&&CNOT: |0\rangle |0\rangle \rightarrow |0\rangle |0\rangle ;
\nonumber \\
&&CNOT: |0\rangle |1\rangle \rightarrow |0\rangle |1\rangle ;
\nonumber \\
&&CNOT: |1\rangle |0\rangle \rightarrow |1\rangle |1\rangle ;
\nonumber \\
&&CNOT: |1\rangle |1\rangle \rightarrow |1\rangle |0\rangle ,
\label{CNOT}
\end{eqnarray}
where the first qubit is the controlled qubit and the second qubit is the target qubit. By matrix representation, CNOT gate
takes the form,
\begin{eqnarray}
CNOT=\left( \begin{array}{cccc}
1&0&0&0\\
0&1&0&0\\
0&0&0&1\\
0&0&1&0\end{array}\right) .
\end{eqnarray}
Depending on physical systems, we can use different universal sets of quantum gates to realize the
universal quantum computation.

\newpage

\section{No-cloning theorem}

\subsection{A simple proof of no-cloning theorem}

For classical information, the possibility of cloning it is an essential feature. In classical systems, cloning, in other words, copying seems no problem. Information stored in computers can be easily made several copies as backup; the accurate semiconservative replication of DNA steadily passes gene information between generations. But for quantum systems, this is not the case. As proved by Wootters and Zurek \cite{Wootters1982}, deterministic cloning of pure states is not possible. After this seminal work, much interest has been shown in extending and generalizing the original no-cloning theorem\cite{PhysRevLett.76.2818,ISI:000253764500008,ISI:000266500900237,ISI:000277243800004,ISI:000280467400011}, which gives us new insight to boundaries of the classical and quantum. On the other hand, no-signaling, guaranteed by Einstein's theory of relativity, is also delicately preserved by no-cloning. This chapter will focus on these topics, hoping to give a thorough description of the no-cloning theorem.

As it is known, a single measurement on a quantum system will only reveal minor information about it, but as a result of which, the quantum system will collapse to an eigenstate of the measurement operator and all the other information about the original state becomes lost. Suppose there exists a cloning machine with a quantum operation U, which duplicates an arbitrary pure state
\begin{equation}
U(|\varphi\rangle\otimes|R\rangle\otimes|M\rangle) = |\varphi\rangle\otimes|\varphi\rangle\otimes|M(\varphi)\rangle
\end{equation}
here $|\varphi\rangle$ denotes an arbitrary pure state, $|R\rangle$ an initial blank state of the cloning machine, $|M\rangle$ the initial state of the auxiliary state(ancilla), and $M(\varphi)\rangle$ is the ancillary state after operation which depends on $|\varphi\rangle$.
With such machine, one can get any number of copies of the original quantum state, and then complete information of it can be determined. However, is it possible to really build such a machine? No-cloning theorem says no.

$\it{THEOREM}:$ No quantum operation exist which can perfectly and deterministically duplicate a pure state.

The proof can be in two methods.

(1). Using the linearity of quantum mechanics.
This proof is first proposed by Wootters and Zurek \cite{Wootters1982},
and also by Dieks \cite{dieks-no-cloning}. Suppose there exists a perfect cloning machine that can copy an arbitrary quantum state, that is, for any state $|\varphi\rangle$
$$ |\varphi\rangle|\Sigma\rangle|M\rangle \rightarrow |\varphi\rangle|\varphi\rangle|M(\varphi)\rangle $$
where $|\Sigma\rangle$ is a blank state, and $|M\rangle$ is the state of auxiliary system(ancilla).
Thus for state $|0\rangle$ and $|1\rangle$, we have
$$ |0\rangle|\Sigma\rangle|M\rangle \rightarrow |0\rangle|0\rangle|M(0)\rangle ,$$
$$ |1\rangle|\Sigma\rangle|M\rangle \rightarrow .
|1\rangle|1\rangle|M(1)\rangle $$
In this way, for the state $|\psi\rangle = \alpha|0\rangle+\beta|1\rangle$ $$ (\alpha|0\rangle+\beta|1\rangle)|\Sigma\rangle|M\rangle \rightarrow
\alpha|00\rangle|M(0)\rangle+\beta|11\rangle|M(1)\rangle $$
On the other hand, $|\psi\rangle$ itself is a pure state, so
$$ |\psi\rangle|\Sigma\rangle|M\rangle \rightarrow (\alpha^2|00\rangle+\alpha\beta|01\rangle+\alpha\beta|10\rangle+\beta^2|11\rangle)
|M(\psi)\rangle .$$
Obviously, the right hand sides of the two equations cannot be equal, as a result, the premise is false that such a perfect cloning machine exists, which concludes the proof. The linearity of quantum mechanics is also used to show that the superluminal is not possible \cite{dieks-no-cloning}.

(2). Using the properties of unitary operation.
This proof is first proposed by Yuen in \cite{yuen-no-cloning}, see also Sec. 9-4 of Peres's textbook\cite{Peres95}.
Consider the process of cloning machine as a unitary operator U, then for any two state $|\varphi\rangle$ and $|\psi\rangle$, since under unitary operation the inner product is preserved, we have
$$ \langle\psi|\varphi\rangle = \langle\psi|U^{\dagger}U|\varphi\rangle
= \langle\psi|\langle\psi|\varphi\rangle|\varphi\rangle
= \langle\psi|\varphi\rangle^2 $$
So $\langle\psi|\varphi\rangle$ is either 0 or 1. If the value is 0, it means the two states being copied should be orthogonal, while if 1, the two states are the same.

\subsection{No-broadcasting theorem}

Following the no cloning theorem for pure states, the impossibility of cloning a mixed state is later proved by Barnum \emph{et al.} \cite{PhysRevLett.76.2818}. In fact, rather than cloning, broadcasting, whose meaning will be presented later in this section, is prohibited by quantum mechanics. Correlations, as a fundamental theme of science, is also studied in quantum systems. An elegant no local broadcasting theorem for correlations in a multipartite state is proposed by Piani \emph{et al.} \cite{ISI:000253764500008}. With these two no-broadcasting theorems, it is natural to ask what is the relationship between them. Recently Luo \emph{et al.} have established the no-unilocal broadcasting theorem for quantum correlations, which proves to be the bridge between Barnum's and Piani's theorems and with it we are able to build the equivalence between them. The three theorems together would give us a unified picture of no-broadcasting in quantum systems.

We shall first elaborate on the original no-broadcasting theorem for non-commuting states
proposed by Barnum \emph{et al.}\cite{PhysRevLett.76.2818}. Suppose there are two parts A and B of a composite quantum system AB, A is prepared in one of the states $\{\sigma_i\}$, while B is prepared in the blank state $\tau$. If there exists a quantum operation $\mathcal{E}$ which can be performed on system AB, that is, $\sigma_k\otimes\tau\rightarrow\mathcal{E}(\sigma_k\otimes\tau)=\rho_k^{out}$ and the output state satisfies
\begin{equation*}
{\rm Tr}_a\rho_k^{out}=\sigma_k \ ~~~{\rm and} \ {\rm Tr}_b\rho_k^{out}=\sigma_k,\ \forall k,
\end{equation*}
we say $\mathcal{E}$ broadcasts the set of states $\{\sigma_i\}$. Here comes Barnum's theorem\cite{PhysRevLett.76.2818}.

\textit{Theorem 1.} A set of states $\{\sigma_i\}$ is broadcastable if and only if the states commute with each other.

Several kinds of proof for Theorem 1 have been found \cite{ISI:000251674300007,PhysRevLett.76.2818,PhysRevLett.100.210502,ISI:000079331000009},
one of them is provided as follows using the property of relative entropy \cite{PhysRevLett.100.210502}. We shall only prove Theorem 1 in the case that the set $\{\sigma_i\}$ only have two states $\sigma_1$ and $\sigma_2$, from which more complex cases can be easily extended to.

\textit{Proof for Theorem 1}

1) ``if'' part: since $\sigma_1$ and $\sigma_2$ commute, they can be expressed in the same orthonormal basis $\{|i\rangle\}$:
\begin{equation*}
\sigma_k = \sum_i \lambda_{k,i}|i\rangle\langle i|,\ k=1,2.
\end{equation*}

Because $\{|i\rangle\}$ is an orthonormal set, it can be cloned by an operator $\mathcal{E}$, so we get
\begin{equation*}
\rho_k^{out}=\mathcal{E}(\sigma_k\otimes\tau) = \sum_i \lambda_{k,i} |ii\rangle\langle ii|,\ k=1,2,
\end{equation*}
thus
\begin{equation*}
{\rm Tr}_a \rho_k^{out}=\sigma_k,\ ~~{\rm Tr}_b \rho_k^{out}=\sigma_k,\ k=1,2.
\end{equation*}
So we see $\sigma_1$ and $\sigma_2$ are broadcasted by $\mathcal{E}$.

2) ``only if'' part: first we shall introduce the concept of relative entropy. The relative entropy S of $\rho_1$ with respect to $\rho_2$ is defined as\cite{Umegaki1962}
\begin{equation*}
S(\rho_1|\rho_2)={\rm Tr}[\rho_1(\ln \rho_1-\ln \rho_2)].
\end{equation*}

When $ker(\rho_1)^{\perp}\bigcap ker(\rho_2)=0$, S is well-defined, otherwise S leads to $\infty$ \cite{RevModPhys.50.221}. We first consider the case $ker(\rho_2)\subseteq ker(\rho_1)$, then $S<\infty$. Denote $\rho _1^{in}=\sigma_1\otimes\tau$ and $\rho _2^{in}=\sigma_2\otimes\tau$, we get

\begin{align*}
S(\rho_1^{in}|\rho_2^{in})&={\rm Tr}[\sigma_1\otimes\tau(\log \sigma_1\oplus \log \tau-\log \sigma_2\oplus \log\tau]\\
&={\rm Tr}_a[\sigma_1(\log \sigma_1-\log \sigma_2)]{\rm Tr}_b\tau\\
&={\rm Tr}_a[\sigma_1(\log \sigma_1-\log \sigma_2)]\\
&=S(\sigma_1|\sigma_2).
\end{align*}
Quantum cloning process $\mathcal{E}$ corresponds to a unitary operator $U$ on input state and the ancillary state $\Sigma $ such that,
$\rho _k^{out}=U(\rho _k^{in}\otimes \Sigma )U^{\dagger }$.
Now, we have
\begin{equation*}
S(\sigma_1|\sigma_2)=S(\rho_1^{in}|\rho_2^{in})= S(\rho_1^{out}|\rho_2^{out}).
\end{equation*}
In general, we remark that for any quantum operation such as the cloning process $\mathcal{E}$, $S(\mathcal{E}(\rho_1)|\mathcal{E}(\rho_2))\le S(\rho_1|\rho_2)$,
see for example \cite{vedralRMP}. This is closely related with the monotonocity of relative entropy \cite{Lindblad1875},
\begin{equation*}
S(\rho_1^{ab}|\rho_2^{ab})\geq S(\rho_1^b|\rho_2^b),
\end{equation*}
where $\rho_1^b$ denotes the reduced density matrix of the composite system $\rho_1^{ab}$, and the equality holds if and only if the following condition is satisfied:
\begin{equation*}
\log \rho_1^{ab}-\log \rho_2^{ab}=I^a\otimes(\log \rho_1^b-\log \rho_2^b).
\end{equation*}
So we have
\begin{equation}
S(\rho_1^{out}|\rho_2^{out})\geq S(\rho_1^{k,out}|\rho_2^{k,out}),
\end{equation}
for k=a,b where $\rho_i^{k,out}$ denotes $Tr_{a(b)}\rho_i^{out}.$ The equality holds if and only if
\begin{align*}
\log \rho_1^{out}-\log \rho_2^{out}&=(\log \rho_1^{a,out}-\log \rho_2^{a,out})\otimes I^b \\
&= I^a\otimes(\log \rho_1^{b,out}-\log \rho_2^{b,out}).
\end{align*}
Under the broadcasting condition, we get
\begin{align*}
\log \rho_1^{out}-\log \rho_2^{out}&=(\log \sigma_1-\log \sigma_2)\otimes I^b \\
&= I^a\otimes(\log \sigma_1-\log \sigma_2).
\end{align*}
But the above equation holds only when $\sigma_1$ and $\sigma_2$ are diagonal or they can be diagonalized in the same basis, which means they commute.

For the case $S(\sigma_1|\sigma_2)=\infty $, we consider a mixed state $\sigma_{mix}=\lambda\sigma_1+(1-\lambda)\sigma_2$, where $0<\lambda<1$. If $\sigma_1$ and $\sigma_2$ can be broadcast, then so can be $\sigma_1$ and $\sigma_{mix}$, due to linearity of the operation. But $ker(\sigma_{mix})\subseteq ker(\sigma_1)$, thus $\sigma_1$ and $\sigma_{mix}$ commute, so $\sigma_1$ and $\sigma_2$ commute. Now we have finished the proof of Theorem 1.

We would like to comment that under a weak assumption, it is possible that broadcasting of some information
of quantum state is possible.
This one is the quantum state information broadcasting presented recently \cite{Horodecki-information-broadcasting}.

\subsection{No-broadcasting for correlations}
The quantum entanglement differs quantum world from classical world. Recently, it is also realized that quantum correlation,
which may be beyond the quantum entanglement, is also important for QIP.
Here we can first make a classification of states by correlation\cite{ISI:000253764500008}. For a bipartite state $\rho^{ab}$ shared by two parties A and B, it is called separable if it can be decomposed as
$$ \rho^{ab} = \Sigma_j p_j \rho_j^a \otimes \rho_j^b, $$
where $\{p_j\}$ denotes a probability distribution, $\{\rho_j^a\}$ and $\{\rho_j^b\}$ denote states of party a and b. Otherwise, the $\rho^{ab}$ is called entangled.

If $\rho^{ab}$ can be further decomposed as
$$ \rho^{ab} = \Sigma_i p_i |i\rangle\langle i| \otimes \rho_i^b, $$
with $\{p_i\}$ denoting a probability distribution, $\{|i\rangle\}$ a orthonormal set of party a and $\{\rho_i^b\}$ states of party b, we say it is classical-quantum.

If $\rho_i^b$ can also be represented in an orthonormal set $\{|j\rangle\}$, which makes
$$ \rho^{ab} = \Sigma_i p_{ij} |i\rangle\langle i| \otimes |j\rangle\langle j|, $$
where $\{p_{ij}\}$ represents a probability distribution for two variables, we say it is classical(or classical-classical).

As we know the correlation in $\rho^{ab}$ can be quantified by mutual information
\begin{equation*}
I(\rho^{ab}) = S(\rho^a) + S(\rho^b) - S(\rho^{ab}),
\end{equation*}
where S denotes the von Neumann entropy, that is, $S(\rho) = -{\rm Tr}(\rho \ln \rho)$.

We say the correlation in $\rho^{ab}$ is locally broadcast if there exists two quantum operations $\mathcal{E}^a:\mathcal{S}(H^a) \rightarrow \mathcal{S}(H^{a_1}\otimes H^{a_2})$ and $\mathcal{E}^b:\mathcal{S}(H^b) \rightarrow \mathcal{S}(H^{b_1}\otimes H^{b_2})$, here $\mathcal{S}(H)$ denotes the set of quantum states on Hilbert space H, such that $\rho^{ab}\rightarrow (\mathcal{E}^a\otimes\mathcal{E}^b)\rho^{ab}=\rho^{a_1a_2b_1b_2}$, and the amount of correlations in the two reduced states
$\rho^{a_1b_1} ={\rm Tr}_{a_2b_2}\rho^{a_1a_2b_1b_2}$ and $\rho^{a_2b_2} = {\rm Tr}_{a_1b_1}\rho^{a_1a_2b_1b_2}$
is identical to which of $\rho^{ab}$, that is
$$I(\rho^{a_1b_1}) = I(\rho^{a_2b_2}) = I(\rho^{ab}).$$
While suppose there is a quantum operation $\mathcal{E}^a$ performed on party a, and we get $(\mathcal{E}^a\otimes I^b)\rho^{ab}=\rho^{a_1a_2b}$, we say the correlation in $\rho^{ab}$ is locally broadcast by party a if
\begin{equation*}
I(\rho^{a_1b})=I(\rho^{a_2b})=I(\rho^{ab}),
\end{equation*}
where $\rho^{a_1b}={\rm Tr}_{a_2}\rho^{a_1a_2b}$ and $\rho^{a_2b}={\rm Tr}_{a_1}\rho^{a_1a_2b}$.

With the above definition, we can then state no-broadcasting theorems for correlation proposed by
Piani \emph{et al.} and Luo \emph {et al.}

\textit{Theorem 2:} The correlation in a bipartite state can be locally broadcast if and only if the state is classical.

\textit{Theorem 3:} The correlation in the bipartite state $\rho^{ab}$ can be locally broadcast by party a if and only if $\rho^{ab}$ is classical-quantum.

\subsection{A unified no-cloning theorem from information theoretical point of view}

Now we shall build equivalence among the theorems according to the method proposed by Luo \emph{et al.}\cite{ISI:000280467400011,ISI:000277243800004}, that is,

Theorem 1 $\Leftrightarrow$ Theorem 2 $\Leftrightarrow$ Theorem 3.

First we shall establish a lemma.

\textit{Lemma 1:} Any bipartite state can be decomposed as
$$\rho^{ab} = \sum_k X_k^a \otimes X_k^b,$$
where each $X_k^a$ is non-negative and $\{X_k^b\}$ forms a linearly independent set.

\textit{Proof:} Let $\{Y_j^a\}$ be a linearly independent set for party a, $\{Z_k^b\}$ a linearly independent set for party b. Then any bipartite state $\rho^{ab}$ can be decomposed in the basis $\{Y_j^a\otimes Z_k^b\}$, that is
\begin{equation*}
\rho^{ab} = \sum_{jk} \lambda_{jk} Y_j^a \otimes Z_k^b.
\end{equation*}
Obviously we can let $Z_k^a = \sum_j \lambda_{jk} Y_j^a$ and obtain
\begin{equation}\label{BipartiteDecomposition1}
\rho^{ab} = \sum_k Z_k^a \otimes Z_k^b.
\end{equation}

Notice that $Z_k^a$ need not to be non-negative, so we have not arrived at Lemma 1 yet. Starting from (\ref{BipartiteDecomposition1}), we take a fixed $|y\rangle\in H^b$ such that
$$c_1 = \langle y|Z_1^b|y\rangle \neq 0,$$
let $c_k = \langle y|Z_k^b|y \rangle$, we can write $\rho^{ab}$ as
\begin{align*}
\rho^{ab} &= \sum_k Z_k^a \otimes Z_k^b\\
&= \sum_k Z_k^a\otimes Z_k^b+\sum_{k\neq1}\frac{c_k}{c_1}Z_k^a\otimes Z_1^b - \sum_{k\neq 1}Z_k^a\otimes\frac{c_k}{c_1}Z_1^b\\
&= (Z_1^a+\sum_{k\neq 1}\frac{c_k}{c_1}Z_k^a)\otimes Z_1^b +\sum_{k\neq 1}Z_k^a\otimes(Z_k^b-\frac{c_k}{c_1}Z_1^b)\\
&= X_1^a\otimes Z_1^b+ \sum_{k\neq 1}Z_k^a \otimes \widetilde{Z_k^b},
\end{align*}
where $X_1^a = Z_1^a+\sum_{k\neq  1}\frac{c_k}{c_1}Z_k^a$ and $\widetilde{Z_k^b}=Z_k^b-\frac{c_k}{c_1}Z_1^b$

Because for any k,
$$ \langle y|\widetilde{Z_k^b}|y\rangle = \langle y|(Z_k^b-\frac{c_k}{c_1}Z_1^b)|y\rangle = c_k-\frac{c_k}{c_1}c_1 = 0, $$
together with the non-negative property of density operator $\rho^{ab}$,we have for any $|x\rangle \in H^a$
\begin{align*}
\langle x\otimes y|\rho^{ab}|x\otimes y\rangle
&= \langle x|X_1^a|x\rangle\langle y|Z_1^b|y\rangle + \sum_{k\neq 1}\langle x|Z_k^a|x\rangle\langle y|\widetilde{Z_k^b}|y\rangle\\
&= c_1\langle x|X_1^a|x\rangle\\
&\geq 0.
\end{align*}
Since $c_1\neq 0$, we see $X_1^a$ or $-X_1^a$ is non-negative depending on the sign of $c_1$.
Without loss of generality, we can always assume $X_1^a$ to be non-negative, because the negative sign can be absorbed by $Z_1^b$. Further we see the set $\{Z_1^b,\widetilde{Z_k^b}\}$ still forms a linearly independent set.

Now all the $Z_i^a(i\leq 2)$ and $Z_1^b$ remain unchanged, and replace $Z_1^a$ with $X_1^a$, $Z_j^b(j\geq 2)$ with $\widetilde{Z_j^b}$, we can find a $|\widetilde{y}\rangle\in H^b$ such that $\langle\widetilde{y}|\widetilde{Z_2^b}|\widetilde{y}\rangle \neq 0$, continue the above process, we would have got $X_2^a$ which is non-negative. Finally we can replace all the $Z_i^a$'s with $X_i^a$'s and thus the proof is completed.

Next we prove Theorem 1 $\Rightarrow$ Theorem 3.

\textit{Proof:} (``if'' part) Since $\rho^{ab}$ is a classical-quantum state, it can be rewritten as
\begin{equation*}
\rho^{ab}=\sum_i p_i|i\rangle\langle i|\otimes\rho_i^b,
\end{equation*}
where $\{|i\rangle\}$ is a linearly independent set, we can further assume it to be an orthonormal base, otherwise we may need to append some zero $p_i$. For any state $\sigma \in S(H^a)$, construct a quantum map $\mathcal{E}^a:S(H^a)\rightarrow S(H^{a_1})\otimes S(H^{a_2})$ such that
\begin{equation}\label{TrivialBroadcasting}
\mathcal{E}^a(\sigma) = \sum_{i}E_i\sigma E_i^{\dagger}
\end{equation}
where $E_i=|ii\rangle\langle i|$. Perform $\mathcal{E}$ on party a, then we have locally broadcast $\rho^a$, and of course the correlation in $\rho^{ab}$ is locally broadcast by party a as well.

(``only if'' part)
Suppose the correlation in $\rho^{ab}$ is locally broadcast by party a through the operator $\mathcal{E}^a:\mathcal{S}(H^a)\rightarrow\mathcal{S}(H^{a_1}\otimes H^{a_2})$, then
$$\rho^{a_1a_2b}=\mathcal{E}^a\otimes\mathcal{I}^b(\rho^{ab}),$$
where $\mathcal{I}^b$ is the identity operator on $H^b$. We have
$$I(\rho^{a_1b})=I(\rho^{a_2b})=I(\rho^{ab}).$$
Denote the operator $\mathcal{T}_{a_2}^{a_1a_2}:\mathcal{S}(H^{a_1}\otimes H^{a_2})\rightarrow \mathcal{S}(H^{a_1})$ as the partial tracing operator by tracing out $a_2$, thus
$$\rho^{a_1b}=(\mathcal{T}_{a_2}^{a_1a_2}\otimes\mathcal{I}^b)(\rho^{a_1a_2b})=
\mathcal{T}_{a_2}^{a_1a_2}\mathcal{E}^a\otimes\mathcal{I}^b(\rho^{ab}).$$
According to the condition
$$I(\rho^{a_1b})=I(\rho^{ab}),$$
and notice that $I(\rho^{ab})=S(\rho^{ab}|\rho^a\otimes\rho^b)$, where S is the relative entropy, $\rho^a$ and $\rho^b$ stand for reduced states, we have
\begin{equation}\label{RelativeEntropyEq}
S(\mathcal{T}_{a_2}^{a_1a_2}\mathcal{E}^a\otimes\mathcal{I}^b(\rho^{ab})|
\mathcal{T}_{a_2}^{a_1a_2}\mathcal{E}^a\otimes\mathcal{I}^b(\rho^a\otimes\rho^b))\\
=S(\rho^{ab}|\rho^a\otimes\rho^b).
\end{equation}
Now we shall introduce a theorem stating that\cite{Lindblad1875}
\begin{equation}\label{Monotonicity}
S(\rho|\sigma)\geq S(\mathcal{E}(\rho)|\mathcal{E}(\sigma))
\end{equation}
for any quantum state $\rho$ and $\sigma$, and any quantum map $\mathcal{E}:\mathcal{S}(H)\rightarrow\mathcal{S}(K)$.

The equality holds if and only if there exists an operator $\mathcal{F}:\mathcal{S}(K)\rightarrow\mathcal{S}(H)$ such that
$$\mathcal{F}\mathcal{E}(\rho)=\rho,\ \mathcal{F}\mathcal{E}(\sigma)=\sigma.$$
An explicit form of $\mathcal{F}$ is, see \cite{HaydenCMP},
\begin{equation}\label{MonoEqualExplicitForm}
\mathcal{F}(\tau)=\sigma^{1/2}\mathcal{E}^{\dagger}((\mathcal{E}(\sigma))^{-1/2}
\tau(\mathcal{E}(\sigma))^{-1/2})\sigma^{1/2},\ \tau\in\mathcal{S}(K).
\end{equation}
Apply (\ref{Monotonicity}) to (\ref{RelativeEntropyEq}), we know there exists an operator $\mathcal{F}^{a_1b}$ such that

$$\mathcal{F}^{a_1b}(\mathcal{T}_{a_2}^{a_1a_2}\mathcal{E}^a\otimes\mathcal{I}^b)(\rho^{ab})=\rho^{ab},$$
$$\mathcal{F}^{a_1b}(\mathcal{T}_{a_2}^{a_1a_2}\mathcal{E}^a\otimes\mathcal{I}^b)(\rho^a\otimes\rho^b)=\rho^a\otimes\rho^b.$$
Consider the explicit form of $\mathcal{F}^{a_1b}$ from (\ref{MonoEqualExplicitForm}) and the product structure of $\rho^a\otimes\rho^b$, we can express $\mathcal{F}^{a_1b}$ as $\mathcal{F}^{a_1}\otimes\mathcal{I}^b$, hence
$$(\mathcal{F}^{a_1}\otimes \mathcal{I}^b)(\mathcal{T}_{a_2}^{a_1a_2}\mathcal{E}^a\otimes\mathcal{I}^b)(\rho^{ab})=\rho^{ab}$$
Use Lemma 1, we obtain
\begin{equation*}
\rho^{ab}=\sum_i X_i^a\otimes X_i^b,
\end{equation*}
where $X_i^a$ is non-negative, $\{X_i^b\}$ constitutes a linearly independent set, thus
$$\sum_i\mathcal{F}^{a_1}\mathcal{T}_{a_2}^{a_1a_2}\mathcal{E}^a(X_i^a)\otimes X_i^b=\sum_i X_i^a\otimes X_i^b,$$
since $\{X_i^b\}$ is a linearly independent set, we have
$$\mathcal{F}^{a_1}\mathcal{T}_{a_2}^{a_1a_2}\mathcal{E}^a(X_k^a)=X_k^a,\ \forall k.$$
So $\{X_i^a\}$ is broadcastable, due to Theorem 1, $X_i^a$'s commute with each other, and hence can be diagonalized by the same basis $\{|i\rangle\}$,
now we obtain
\begin{equation*}
\rho^{ab} = \sum_i \lambda_i |i\rangle\langle i|\otimes Y_i^b,
\end{equation*}
it can be easily proved that $Y_i^b$ is non-negative, and hence $\rho^{ab}$ is indeed a classical-quantum state.

Now we prove Theorem 3 $\Rightarrow$ Theorem 2.

\textit{Proof:} We shall prove only the non-trivial part. Suppose the correlation in $\rho^{ab}$ can be locally broadcast by two operators respectively performed on party a and b:
\begin{align*}
\mathcal{E}^a:\mathcal{S}(H^a)\rightarrow\mathcal{S}(H^{a_1}\otimes H^{a_2}), \\
\mathcal{E}^b:\mathcal{S}(H^b)\rightarrow\mathcal{S}(H^{b_1}\otimes H^{b_2}),
\end{align*}
then we obtain
\begin{align*}
\rho^{a_1a_2b_1b_2}&=(\mathcal{E}^a\otimes\mathcal{E}^b)\rho^{ab}\\
&=(\mathcal{I}^{a_1a_2}\otimes\mathcal{E}^b)(\mathcal{E}^a\otimes\mathcal{I}^b)\rho^{ab}.
\end{align*}
So we have decomposed the operation $\mathcal{E}^a\otimes\mathcal{E}^b$ into two steps, each of which only deals with a single party. Through step one, we obtain
\begin{equation*}
S(\rho^{ab}|(\rho^a\otimes\rho^b))\geq S(\mathcal{E}^a(\rho^{ab})|\mathcal{E}^a(\rho^a\otimes\rho^b)),
\end{equation*}
that is, $I(\rho^{ab})\geq I(\rho^{a_1a_2b})$, since $I(\rho^{a_1a_2b})\geq I({\rm Tr}_{a_2}\rho^{a_1a_2b})=I(\rho^{a_1b})$, we have
$$I(\rho^{a_1b})\leq I(\rho^{a_1a_2b})\leq I(\rho^{ab}).$$
Similarly, through step two, we obtain
$$I(\rho^{a_1b_1})\leq I(\rho^{\rho^{a_1b_1b_2}})\leq I(\rho^{a_1b}).$$
With the condition $I(\rho^{a_1b_1})=I(\rho^{ab})$, we have $I(\rho^{a_1b})=I(\rho^{ab})$, which shows that the correlation in $\rho^{ab}$ is broadcast by party a, from Theorem 3, we know $\rho^{ab}$ is a classical-quantum state. Exchange a and b in the above discussion, it's obvious that $\rho^{ab}$ is also a quantum-classical state. So $\rho^{ab}$ is a classical state.

Next we prove Theorem 2 $\Rightarrow$ Theorem 1

\textit{Proof:} Again we shall only prove the non-trivial part. Suppose there exists a quantum operation $\mathcal{E}^b$ which can broadcast a set of states $\{\rho_i^b\}$. We can find a orthonormal set $\{|i\rangle\}$ and construct a composite system
\begin{equation*}
\rho^{ab}=\sum_i p_i |i\rangle\langle i|\otimes\rho_i^b,
\end{equation*}
where $\{p_i\}$ is a probability distribution. The party a can be easily broadcast by the operator $\mathcal{E}^a$ form (\ref{TrivialBroadcasting}), together with $\mathcal{E}^b$, $\rho^{ab}$ can be locally broadcast, so is the correlation. Thus from Theorem 2, $\rho^{ab}$ is a classical state, then $\rho_i^b$ commutes with each other.

From the above discussions, we have created a chain of equivalence among the three theorems: Theorem 1 $\Leftrightarrow$ Theorem 2 $\Leftrightarrow$ Theorem 3. This has provided us with a unified picture of the no-broadcasting theorem in quantum systems from the information theoretical point of view.

\subsection{No-cloning and no-signaling}

According to Einstein's relativity theory, superluminal signaling cannot be physically realized. Yet due to the non-local property of quantum entanglement, superluminal signaling is possible provided perfect cloning machine can be made. The scheme has been well-known since Herbert\cite{Herbert1982} first proposed his ``FLASH'' in 1982. The idea is as follows: suppose Alice and Bob, at an arbitrary distance, share a pair of entangled qubits in the state $|\psi\rangle = (1/\sqrt2)(|01\rangle-|10\rangle)$. Alice can measure her qubit by either $\sigma_x$ or $\sigma_z$. If the measurement is $\sigma_z$, Alice's qubit will collapse to the state $|0\rangle$ or $|1\rangle$, with probability 50\%. Respectively, this prepares Bob's qubit in the state $|1\rangle$ or $|0\rangle$. Without knowing the result of Alice's measurement, the density matrix of Bob's qubit is $\frac{1}{2}|0\rangle{\langle}0|+\frac{1}{2}|1\rangle{\langle}1| = \frac{1}{2}I$. On the other hand, if Alice's measurement is $\sigma_x$, Alice's qubit will collapse to the state $|\varphi_{x+}\rangle$ or $|\varphi_{x-}\rangle$, where $|\varphi_{x+}\rangle=1/\sqrt{2}(|0\rangle+|1\rangle)$, $|\varphi_{x-}\rangle=1/\sqrt{2}(|0\rangle-|1\rangle)$, being eigenvectors of $\sigma_x$. Thus Bob's qubit is prepared in the state $|\varphi_{x-}\rangle$ or $|\varphi_{x+}\rangle$ respectively, in this case the density matrix of Bob's qubit is still  $\frac{1}{2}|\varphi_{x+}\rangle\langle\varphi_{x+}|+\frac{1}{2}|\varphi_{x-}\rangle\langle\varphi_{x-}| = \frac{1}{2}I$. Obviously, Bob gets no information about which measurement is made by Alice.
While, if perfect cloning is allowed, the scenario will change. Bob can use the cloning machine to make arbitrarily many copies of his qubit, in which way he is able to determine the exact state of his qubit, that is, whether an eigenstate of $\sigma_z$ or $\sigma_x$. With this information, Bob knows the measurement Alice has taken.
Fortunately, since no-cloning theorem has been proved, the above superluminal signaling scheme cannot be realized, which leaves theory of relativity and quantum mechanics in coexistence.

Up to now, there are many cloning schemes found, naturally one may ask, whether it is possible by using imperfect cloning, to extract information about which measuring basis Alice has used. According to the property of quantum transformation, the answer is no. To see this, we may first consider a simple scheme, that is, Bob can use the universal quantum cloning machine (UQCM) proposed by Bu\v{z}ek and Hillery\cite{PhysRevA.54.1844} to process his qubit.
The UQCM transformation reads,
$$ |0\rangle|Q\rangle \rightarrow \sqrt{\frac{2}{3}}|00\rangle|\uparrow\rangle+\sqrt{\frac{1}{3}}|+\rangle|\downarrow\rangle, $$
$$ |1\rangle|Q\rangle \rightarrow \sqrt{\frac{2}{3}}|11\rangle|\downarrow\rangle+\sqrt{\frac{1}{3}}|+\rangle|\uparrow\rangle, $$
where $|Q\rangle$ is the original state of the copying-machine, $|+\rangle$ and $|-\rangle$ are two orthogonal states of the output, $|+\rangle = \frac{1}{\sqrt2}(|01\rangle+|10\rangle)$, $|-\rangle = \frac{1}{\sqrt2}(|01\rangle-|10\rangle)$, and $|\uparrow \rangle ,|\downarrow \rangle $
are the ancillary states which are orthogonal to each other.
If Alice chooses $\sigma_z$, the density matrix of Bob's qubit after the process is
\begin{align*}
\rho_b &= \frac{1}{2}(\frac{2}{3}|00\rangle\langle00|+\frac{1}{3}|+\rangle\langle+|)
+\frac{1}{2}(\frac{2}{3}|11\rangle\langle11|+\frac{1}{3}|+\rangle\langle+|) \\
&=
\frac{1}{3}(|00\rangle\langle00|+|11\rangle\langle11|+|+\rangle\langle+|).
\end{align*}
If Alice chooses $\sigma_x$, it can be easily verified that the density matrix will not change, thus no information can be gained by Bob.
In fact, Bruss \emph{et al.} have pointed out in \cite{PhysRevA.62.062302} that the density matrix of Bob's qubit will not change no matter what operation is taken on it, as long as the operation is linear and trace-preserving.
Suppose the original density matrix shared between Alice and Bob is $\rho^{ab}$, and Alice has done a measurement $\mathcal{A}_m$ on her qubit, Bob makes a transformation $\mathcal{B}$ on his, then the shared density matrix becomes $\mathcal{A}_m\otimes\mathcal{B}(\rho^{ab})$, here m specifies which measurement Alice has taken. In Bob's view, with the linear and trace-preserving property of $\mathcal{A}_m$, the density matrix of his qubit is
\begin{align*}
tr_a(\mathcal{A}_m\otimes\mathcal{B}(\rho^{ab})) &= \mathcal{B}tr_a(\mathcal{A}_m{\otimes}\mathcal{I}(\rho^{ab})) \\
&= \mathcal{B}tr_a(\rho^{ab}).
\end{align*}
Note that $tr$ and $Tr$ both denote trace similarly in this review.
Here we see that the density matrix of Bob's qubit has nothing to do with Alice's measurement $\mathcal{A}_m$, therefore no information is transferred to Bob.
Note that to get the above conclusion, we have only used the linear and trace-preserving property of $\mathcal{A}_m$. Since any quantum operator is linear and completely positive, no-signaling should always hold, thus providing a method to determine the fidelity limit of a cloning machine.

The situation might be more complicated when the no-signaling correlation is considered. It is found that, however,
no-signaling might be more non-local than that of quantum mechanics. Then it seems that besides of no-signaling, some
extra principle, like local orthogonality \cite{local-orthogonality}, should be satisfied such that the no-signaling non-locality might be realizable by quantum mechanics \cite{popescu-non-locality}.

Gisin studied the case of $1\rightarrow 2$ qubit UQCM in \cite{Gisin19981}. We continue the scheme that Alice and Bob share a pair of entangled states. Now Alice has done some measurement by $\sigma_x$ or $\sigma_z$, and thus Bob's state has been prepared in a respect mixed state. Let there be a UQCM, suppose the input density matrix is $|\varphi\rangle\langle \varphi|=\frac{1}{2}(I+\bm{m}\cdot\bm{\sigma})$, with $\bm{m}$ being the Bloch vector of $|\varphi\rangle$, then after cloning the reduced state on party a and b should read, $\rho^a=\rho^b=(1+\eta\bm{m}\cdot\bm{\sigma})$, yielding the fidelity to be $F=(1+\eta)/2$. According to the form of $\rho^a$ and $\rho^b$, the composite output state of the cloning machine should be
\begin{equation*}
\rho^{out}=\frac{1}{4}(I_4+\eta(\bm{m}\cdot\bm{\sigma}\otimes I+I\otimes\bm{m}\cdot\bm{\sigma})+\sum_{i,j=x,y,z}t_{ij}\sigma_i\otimes\sigma_j).
\end{equation*}
The universality of UQCM requires
\begin{equation}\label{NosignalCondition1}
\rho_{out}(U\bm{m})=U\otimes U\rho_{out}(\bm{m})U^{\dagger}\otimes U^{\dagger}.
\end{equation}
The no-signaling condition requires
\begin{equation}\label{NosignalCondition2}
\frac{1}{2}\rho^{out}(+x)+\frac{1}{2}\rho^{out}(-x)=\frac{1}{2}\rho^{out}(+z)+\frac{1}{2}\rho^{out}(-z),
\end{equation}
where $\rho^{out}(+z)$ represents the output state of the UQCM under the condition that Alice has take the measurement $\sigma_x$ and got result $+$.

Also we should notice that $\rho^{out}$ must be positive. Putting the positive condition together with (\ref{NosignalCondition1}) and (\ref{NosignalCondition2}), we shall get $\eta\leq\frac{2}{3}(F\leq\frac{5}{6})$. Although we have found an upper bound of F, the question remains whether it can be reached. But we know it can be, since a practical UQCM scheme with $F=\frac{5}{6}$ has been proposed\cite{PhysRevA.54.1844}.

Navez \emph{et al.} have derived the upper bound of fidelity for d-dimensional $1\rightarrow2$ UQCM using no-signaling condition\cite{PhysRevA.68.032313},
and the bound also has been proved to be tight. Simon \emph{et al.} have shown how no-signaling condition together with the static property of quantum mechanics can lead to properties of quantum dynamics \cite{PhysRevLett.87.170405}. By static properties we mean: 1) The states of quantum systems are described as vectors in Hilbert space. 2) The usual observables are represented by projections in Hilbert space and the probabilities for measurement are described by the usual trace rule. The two properties with no-signaling condition shall imply that any quantum map must be completely positive and linear, which is what we already have in mind. This may help to understand why bound derived by no-signaling condition is always tight. The experimental test of the no-signaling theorem is also performed in optical system \cite{experimentalnosignaling}. From no-signaling condition, the monogamy relation of violation of Bell inequalities can be derived, and this can be used to obtain the optimal fidelity for
asymmetric cloning \cite{bell-non-signaling}. And some general properties of no-signaling theorem
are presented in \cite{g-nonsignaling}. The relationship between optimal
cloning and no signaling is presented in \cite{Ghosh199917}. The no-signaling is shown
to be related with optimal state estimation \cite{Xiangbin-Hwang}. Also the no-signaling
is equivalent to the optimal condition in minimum-error quantum state
discrimination \cite{BaeHwang}, more results of those topics
can be found in \cite{Baepreprint0} for qubit case
and \cite{Baepreprint1} for the general case. The optimal cloning of arbitrary fidelity
by using no-signaling is studied in \cite{GedikPreprint}.

\subsection{No-cloning for unitary operators}

No-cloning is a fundamental theorem in quantum information science and quantum mechanics. It may be manifested
in various versions. Simply by calculation, and with the help of definition of CNOT gate, we may find the following relations,
\begin{eqnarray}
CNOT(\sigma _x\otimes I)CNOT&=&\sigma _x\otimes \sigma _x,
\nonumber \\
CNOT(\sigma _z\otimes I)CNOT&=&\sigma _z\otimes I,
\nonumber \\
CNOT(I\otimes \sigma _x)CNOT&=&I\otimes \sigma _x,
\nonumber \\
CNOT(I\otimes \sigma _z)CNOT&=&\sigma _z\otimes \sigma _z.
\label{opnoclone}
\end{eqnarray}
It implies that the bit flip operation is copied forwards (from first qubit to second qubit),
while the phase
flip operation is copied backwards. But we cannot copy simultaneously the
bit flip operation and phase flip operation. Those properties are important
for methods of quantum error correction and fault-tolerant quantum computation \cite{gottesman98}.
This is a kind of no-cloning theorem for
unitary operators. The quantum cloning of unitary operators is investigated in \cite{cloningunitary}.

\subsection{Other developments and related topics}

As a basic and fundamental theorem of quantum information and quantum mechanics,
no-cloning is related with many topics and has various applications.
In the following, we try to list some of those developments and closely related topics.

\begin{itemize}
\item
We may wonder what is the classical counterpart of no-cloning theorem in quantum world.
Some results are available.
Different from quantum case, classical broadcasting is possible with arbitrary high resolution \cite{broadcastingclassic}.
The difference between quantum copying and classical copying
is studied in \cite{Shen2011}, see also \cite{Fenyes2012}.
The classical no-cloning is also discussed \cite{classical-no-cloning}.

\item
The nonclassical correlations such as
entanglement are also fundamental phenomena in the quantum world.
They can be related with no-cloning theorem.
The no-cloning theorem for entangled states is shown in \cite{prl81.4264}.
No-cloning theorem prohibits perfect copying of nonorthogonal states, but as to orthogonal ones,
it says if we are allowed to use arbitrary unitary transformations, cloning of them can be deterministically done.
However, related with entanglement cloning, it is shown
that even orthogonal states in composite systems cannot be cloned \cite{Morprl1998},
related results are also available in \cite{prl77.3264,prl77.3265}.
It is also shown that no-cloning theorem is,
in principle, equivalent with no-increasing of entanglement \cite{ISI:000075192800004}.
By studying quantum correlation beyond quantum entanglement, the equivalence
between locally broadcastable and broadcastable is investigated in \cite{ISI:000291254100001},
see a review about quantumness of correlations \cite{RMP-discord}.
The combination of no-cloning, no-broadcasting and monogamy of entanglement can be found in \cite{ISI:000241723100033}.

\item
The no-cloning theorem can be described in other environments
and can be applied to other cases. Some of those results are the following.
The no-signaling principle and the state distinguishability is studied in \cite{Baepreprint2}.
The linear assignment maps for correlated system-environment
states is studied in \cite{ISI:000274001500049}, the connection between the violation of
the positivity of this linear assignments and the no-broadcasting theorem is found.
The transformations which preserve commutativity of quantum states are studied
in \cite{ISI:000268481100011}.
Related with quantum cloning, the quantum channels are studied in \cite{Bradler2011}.
No-cloning theorem means that two copies cannot be obtained out of a single copy,
and if we study the information content measured by Holevo quantity
of one copy and two copies,  a condition of states broadcasting can be obtained \cite{ISI:000237090400009}.
No-cloning can also be related with bounds on quantum capacity \cite{ISI:000244532300030}.
The no-cloning studied by wave-packet collapse of quantum measurement
is presented by Luo in \cite{Luo2010}.
It is also pointed out that no-cloning of non-orthogonal states does not necessarily
mean that inner product of quantum states should preserve \cite{ISI:000231391900002}.
We remark that no-cloning theorem is also pioneered in \cite{dieks-no-cloning,yuen-no-cloning},
interested readers could check them for reference, see also \cite{Wootters2009}.

\item
Since the first version of no-cloning theorem,
either inspiration is drawn from it, or generalizations are made, some similar ``no-'' theorems come up, which we shall list as below. 1) No-deleting. Being a reverse
process of quantum cloning, it is pointed out that it is also impossible to delete an unknown quantum state \cite{no-deleting}. 2) No-imprinting. See \cite{no-imprinting}, related results can also be found in \cite{ISI:000177872600044}. 3) No-stretching, which is a geometrical interpretation of
no-cloning theorem \cite{DAriano2009}.
4) No-splitting, which states that quantum information cannot be split into complementary parts  \cite{ISI:000235859600008}.
The impossibility of reversing or complementing an unknown
quantum state is a generalization of no-cloning theorem \cite{ISI:000232866900008}.

\item
Finally, let us remark some applications of no-cloning theorem.
We know that no-cloning theorem plays a key role in
quantum cryptography, which is close to practical industrialization.
Quantum key distribution (QKD), such as the BB84 protocol \cite{Bennett1984},
provides the unconditional security for secrete key sharing. The security of the quantum key distribution is based on no-cloning theorem since if we can copy perfectly the transferred state, we can always find its exact form by copying it to infinite copies so that its exact form can be found. For quantum cryptography protocol E91 \cite{PhysRevLett.67.661},
the security is based on the violation of Bell inequality \cite{Bell1964}.
The unified picture of no-broadcasting
theorem unifies those theorems together. This result is also shown in \cite{Acin2004}.
The study of quantum cryptography, on the other hand, suggests that the ultimate physical limits
of privacy might be possible under very weak assumptions \cite{Ekert-Renner-Privacy}.
One recent development may include that probabilistic super-replication of quantum information,
a different version of quantum cloning with limited aim, is possible \cite{qreplication-yao}.
Remarkably, these phenomena can be applied to achieve the ultimate limit of precision
in metrology provided by quantum mechanics,
which is the Heisenberg limit of quantum metrology.
 We would like to emphasize that the quantum cloning can be applied in quantum computation \cite{GalvaoHardy}.

\end{itemize}

\newpage

\section{Universal quantum cloning machines}

As we have shown in last section, there are various no-cloning theorems implied by the law of quantum mechanics.
They imply that one cannot clone an arbitrary qubit perfectly. On the other hand, the approximate quantum cloning
is not prohibited. So it is possible that one can get several copies that approximate the original state, with fidelity $F<1$.
Hence one naturally raises a question: can we achieve the same fidelity for any state on the Bloch sphere, for the qubit case, or more generally, for any state in a d-dimensional Hilbert space? And what is the best fidelity we can get?

A cloning machine that achieves equal fidelity for every state is called a universal quantum cloning machine (UQCM). This problem is equivalent to distribute information to different receivers, and it is natural to require the performance is the same for every input state, since we do not have any specific information about the input state ahead. According to no-cloning theorem, it is expected that the original input state will be destroyed and
become as one of the output copies. For the simplest case, one qubit is cloned to have two copies, those two copies can be
identical to each other, i.e., they are symmetric and of course they are different from the original input state.
On the other hand, those two copies can also be different, both of them are similar to the the original input state but with
different similarities, we mean that they are asymmetric. In this sense, there are symmetric and asymmetric UQCMs.

\subsection{Symmetric UQCM for qubit}

Consider a quantum cloning from 1 qubit to 2 qubits, a trivial scheme can be simply constructed as following:

(1), Measure the input state $|\vec{a}\rangle$ in a random base $|\vec{b}\rangle$. Here the vectors are on the Bloch sphere $S^2$.
The probability of obtaining result $|\pm\vec{b}\rangle$ is $p_\pm=\left(1\pm\vec{a}\cdot\vec{b}\right)/2$.

(2), Then duplicate the state $|\pm\vec{b}\rangle$ according to the measurement result. The fidelity is $F_+=|\langle\vec{a}|+\vec{b}\rangle|^2$ and $F_-=|\langle\vec{a}|-\vec{b}\rangle|^2$, respectively.

In an average sense, the fidelity is
\begin{equation}
F_{trivial}=\int_{S^2}p_+ F_++p_- F_-=\frac{1}{2}+\frac {1}{2}\int_{S^2}\left(\vec{a}\cdot\vec{b}\right)^2 \mathrm{d}\vec{b}=\frac{2}{3}.\label{Ftrivial}
\end{equation}

The problem is: can we design a better cloning machine? Bu\v{z}ek and Hillery \cite{PhysRevA.54.1844} proposed an optimal UQCM, namely, a unitary transformation on a larger Hilbert space:

\begin{align}
U|0\rangle_1|0\rangle_2|0\rangle_R&=\sqrt{\frac{2}{3}}|0\rangle_1|0\rangle_2|0\rangle_R+\sqrt{\frac{1}{6}}(|0\rangle_1|1\rangle_2+|1\rangle_1|0\rangle_2)|1\rangle_R,\\
U|1\rangle_1|0\rangle_2|0\rangle_R&=\sqrt{\frac{2}{3}}|1\rangle_1|1\rangle_2|1\rangle_R+\sqrt{\frac{1}{6}}(|0\rangle_1|1\rangle_2+|1\rangle_1|0\rangle_2)|0\rangle_R.
\end{align}
On the l.h.s of the equations, the first qubit 1 is the input state, the second is a blank state and
 the third with subindex $R$ is the ancillary state of the quantum cloning machine itself.
By a unitary transformation which is demanded by quantum mechanics, we find the output state on the r.h.s. of the
equations. We may find that the original qubit is destroyed and becomes as one of the output qubit in 1 while
the blank state is now changed as another copy in party 2, the ancillary state $R$ may or may not be changed
which will be traced out for the output. It is obvious that two output states are identical, so it is a symmetric quantum
cloning machine.

For an arbitrary normalized pure input state $|\psi\rangle=a|0\rangle+b|1\rangle$, since quantum mechanics is linear,
by applying $U$ on the state which is realized simply by following the above cloning transformation, we can find the copies.
After tracing out the ancillary state, the output density matrix take the form:
\begin{equation}
\rho_{out}=\frac{2}{3}|\psi\rangle\langle\psi|\otimes|\psi\rangle\langle\psi|+\frac{1}{6}(|\psi\rangle|\psi^\bot\rangle+|\psi^\bot\rangle|\psi\rangle)
(\langle\psi|\langle\psi^\bot|+\langle\psi^\bot|\langle\psi|).\label{1to2output}
\end{equation}
Here $|\psi^\bot\rangle=b^*|0\rangle-a^*|1\rangle$ is orthogonal to $|\psi\rangle$. We can further trace out one of the two states to get the single copy density matrix
\begin{equation}
\rho_1=\rho_2=\frac{2}{3}|\psi\rangle\langle\psi|+\frac{1}{6}I.
\end{equation}
Note this density matrix is of the form $\eta|\psi\rangle\langle\psi|+\frac{1-\eta}{d}I$ with $\eta$ called the ``shrinking factor''. This form is a linear combination of the original density matrix $|\psi\rangle\langle\psi|$ and the identity $I$ which corresponds to completely
mixed state and it is like a white noise.

In fact, in the original papers, the efficiency of the cloning machine is described by Hilbert-Schmidt norm
$d_{HS}^2={\rm Tr}[(\rho_{in}-\rho_{out})^\dagger(\rho_{in}-\rho_{out})]$, which also quantifies the
distance of two quantum states. The fidelity is a general accepted measure of merit of the quantum cloning \cite{ISI:000165200800033}.
We will generally use fidelity as the measure of the quality of the copies in this review.

We can obtain the single copy fidelity
\begin{equation}
F_1=\langle\psi|\rho_1|\psi\rangle=\frac{5}{6},
\end{equation}
 and the two copies fidelity (global fidelity),
\begin{equation}
F_2=^{\otimes2}\langle\psi|\rho_1|\psi\rangle^{\otimes2}=\frac{2}{3}.
\end{equation}
The single copy fidelity provides measure of similarity between state $\rho _1$ and the original input state.
If it is one, those two states are completely the same, while if it is zero, those two states
are orthogonal. One point may be noticed is that, the fidelity between a completely mixed state with $|\psi \rangle $ is
$1/2$. We know that a completely mixed state contains nothing about the input state, so fidelity $1/2$
should be the farthest distance between two quantum states.
Similarly, the global fidelity quantifies the similarity between the two-qubit output state with the ideal cloning case.
If it is one, we have two perfect copies.
We remark that the single copy fidelity does not depend on input state, so the quality of the copies
has the state-independent characteristic. In this sense, the corresponding cloning machine is ``universal''.
One may find that the above presented cloning machine achieves higher fidelity than the trivial one, and it is proved to be optimal
\cite{PhysRevLett.79.2153,Bruss1998,Gisin19981}).

Gisin and Massar \cite{PhysRevLett.79.2153} then generalize the cloning machine to $N\rightarrow M$ case, that is $M$ copies are
created from $N$ identical qubits. Their cloning machine is a transformation:
\begin{equation}
|N\psi\rangle|R\rangle \rightarrow\sum_{j=0}^{M-N}\alpha_j|(M-j)\psi,j\psi^\bot\rangle|R_j\rangle
\end{equation}
where
\begin{equation}
\alpha_j=\sqrt{\frac{N+1}{M+1}}\sqrt{\frac{(M-N)!(M-j)!}{(M-N-j)!M!}}
\end{equation} and$|(M-j)\psi,j\psi^\bot\rangle$ denote the normalized symmetric state with $M-j$ states $|\psi\rangle$ and $j$ states $|\psi^\bot\rangle$. Then the single copy fidelity is
\begin{equation}
F=\frac{M(N+1)+N}{M(N+2)}.\label{FNtoM}
\end{equation}

In \cite{PhysRevLett.79.2153} the optimality of this cloning machine is proved for cases $N=1,2,\dots,7$. The complete proof of the optimality is finished in \cite{PhysRevLett.81.2598} where the connection between optimal quantum cloning and optimal state estimation is established. The upper bound of $N$ to $M$ UQCM is found to be exactly equal to (\ref{FNtoM}), hence Gisin and Massar UQCM is optimal.

\subsection{Symmetric UQCM for qudit}

For further generalization, we may seek cloning machine for d-level systems. Bu\v{z}ek and Hillery proposed a 1 to 2 d-dimensional UQCM \cite{ISI:000084208300020}\cite{PhysRevLett.81.5003}: for a basis state $|i\rangle$, the transformation is
\begin{equation}
|i\rangle|0\rangle|R\rangle\rightarrow\frac{2}{\sqrt{2(d+1)}}|i\rangle|i\rangle|R_i\rangle+\frac{1}{\sqrt{2(d+1)}}\sum_{i\neq j}(|i\rangle|j\rangle+
|j\rangle|i\rangle)|R_j\rangle.
\end{equation}
Here $|R_i\rangle$ is a set of orthogonal normalized ancillary state. The resultant one copy fidelity is, $F=(d+3)/(2d+2)$.

Later, a general $N$ to $M$ UQCM is constructed in a concise way by Werner \cite{PhysRevA.58.1827}, see also \cite{zanardi} for related results. For N identical pure input state $|\psi\rangle$, the output density matrix is:
\begin{equation}
\rho_{out}=\frac{d[N]}{d[M]}s_M\left( (|\psi\rangle\langle\psi|)^{\otimes N}\otimes I^{\otimes(M-N)}\right)s_M
\label{Werner}
\end{equation}
where $d[N]=\tbinom{d+N-1}{N}$, (we also use notation $d[N]=C_{d+N-1}^N$), and $s_M$
is the projection onto the symmetric subspace of $\mathcal{H}^{\otimes M}$. As an example,
\begin{equation}
s_2=|00\rangle\langle00|+|11\rangle\langle11|+\frac{1}{2}(|01\rangle+|10\rangle)(\langle01|+\langle10|).
\end{equation}
If we insert this expression into formula (\ref{Werner}), we can get exactly the expression of output density matrix (\ref{1to2output}).
So this UQCM can recover the $N=1$, $M=2$, $d=2$ one.

For N to M case, the single copy fidelity is shown to be
\begin{align}
F_1&=\frac{d[N]}{d[M]}{\rm Tr}\left[\left(|\psi\rangle\langle\psi|\otimes I^{\otimes(M-1)}\right)s_M\left( (|\psi\rangle\langle\psi|)^{\otimes N}\otimes I^{\otimes(M-N)}\right)s_M\right]\nonumber\\
&=\frac{N(M+d)+M-N}{M(N+d)}.\label{UQCM1F}
\end{align}
In \cite{PhysRevA.58.1827}, this single copy fidelity is proved to be optimal under the restriction that the operation is a mapping into the symmetric Hilbert space. Generally, there might exist a cloning machine performing better without this constraint. Keyl and Werner studied the more general case and proved this cloning machine is indeed the unique optimal UQCM \cite{ISI:000081019000005}. As a special case, if we let $N=1$, $M=2$, the fidelity apparently reduces to the Bu\v{z}ek and Hillery 1 to 2 d-dimensional UQCM: $F=(d+3)/(2d+2)$. And if we take the $M\rightarrow\infty$ limit, the fidelity turns out to be $F=(N+1)/(N+d)$, this agrees with the state estimation result by Massar and Popescu \cite{Massar1995}.

We are also interested in the M copies fidelity (global fidelity), it can be found as follows \cite{PhysRevA.58.1827},
\begin{align}
F_M&=\frac{d[N]}{d[M]}tr\left[\left(|\psi\rangle\langle\psi|\right)^{\otimes M}s_M\left( (|\psi\rangle\langle\psi|)^{\otimes N}\otimes I^{\otimes(M-N)}\right)s_M\right]\nonumber\\&=\frac{d[N]}{d[M]}tr\left((|\psi\rangle\langle\psi|)^{\otimes M}\right)\nonumber\\&=\frac{d[N]}{d[M]}=\frac{M!(N+d-1)!}{N!(M+d-1)!}.
\label{UQCMMF}
\end{align}

The fidelity (\ref{UQCM1F}) quantifies the similarity between a single copy from the output state and one input state,
while the global fidelity (\ref{UQCMMF}) quantifies the similarity between the whole $M$ copies of the output and the
ideal $M$ copies of the input state $|\psi \rangle $. More generally, we may consider to choose arbitrary $L$ copies
from the output state and quantify how closeness of this state with $L$ ideal copies of the input state $|\psi \rangle $.
Recently, Wang \emph{et al.} \cite{PhysRevA.84.034302} proposed a more general definition ``$L$ copies fidelity'': $F_L=^{\otimes L}\langle\psi|\rho_{out,L}|\psi\rangle^{\otimes L}$ where $\rho_{out,L}$ is the $L$ copies output reduced density matrix. The expression is calculated as,
\begin{eqnarray}
F_L=\frac {(d+N-1)!(M-N)!(M-L)!}{(d+M-1)!M!N!}\times\sum _{m_1}\frac{(M-m_1+d-2)!(m_1!)^2}{(m_1-L)!(m_1-N)!(d-2)!(M-m_1)!}.
\end{eqnarray}
For $L=1$ and $L=M$, the expression will reduce to results presented above (\ref{UQCM1F},\ref{UQCMMF}).
For another special case $N=1$, expression of fidelity can be simplified by finding the explicit result of the summation,
it reads,
\begin{eqnarray}
F_L(N=1)=\frac {L!d![L(d+M)+M-L] }{(d+L)!M}.
\end{eqnarray}

Fan \emph{et al.} \cite{PhysRevA.64.064301} proposed another version of UQCM, written in more explicit form: let $\bm{n}=(n_1,\dots,n_d)$ denote a $d$-component vector. And $|\bm{n}\rangle=|n_1,\dots,n_d\rangle$ is a completely symmetric and normalized state with $n_i$ states in $|i\rangle$. These states is an orthogonal normalized basis of the symmetric Hilbert space $\mathcal{H}^{\otimes M}_+$. Then for an arbitrary input state $|\psi\rangle=\sum^d_{i=1}x_i|i\rangle$, the $N$-fold direct product $|\psi\rangle^{\otimes N}$ could be expanded as:
\begin{equation}
|\psi\rangle^{\otimes N}=\sum_{\bm{n}}^N\sqrt{\frac{N!}{n_1!\dots n_d!}}x_1^{n_1}\dots x_d^{n_d}|\bm{n}\rangle.
\label{expansion}
\end{equation}
The cloning transformation takes the form,
\begin{equation}
|\bm{n}\rangle|R\rangle\rightarrow\sum_{\bm{j}}^{M-N}\alpha_{\bm{n}\bm{j}}|\bm{n}+\bm{j}\rangle|R_{\bm{j}}\rangle\label{FanClone}
\end{equation}
The notation $\sum_{\bm{n}}^N$ means summation over all possible vectors $\bm{n}$ with $n_1+\dots +n_d=N$ and the $|R_{\bm{j}}\rangle$ is a set of orthogonal normalized ancillary states, as usual. The coefficients $\alpha_{\bm{n}\bm{j}}$ are:
\begin{equation}
\alpha_{\bm{n}\bm{j}}=\sqrt{\frac{(M-N)!(N+d-1)!}{(M+d-1)!}}\sqrt{\prod_{k=1}^d\frac{(n_k+j_k)!}{n_k!j_k!}}.
\label{Fancoefficients}
\end{equation}
This UQCM can achieve the same fidelities as the UQCM given by Werner \cite{PhysRevA.58.1827}. It is optimal.
Later Wang \emph{et al.} \cite{PhysRevA.84.034302} proved that
these two cloning machines are indeed equivalent by showing the output states
are the same. First, divide the symmetric state $|\vec {m}\rangle $ of $M$ qudits
into two parts with $N$ qudits and $M-N$ qudits, respectively,
\begin{eqnarray}
|\vec {m}\rangle =\frac {1}{\sqrt {C_M^N}}\sum _{\vec {k}}^{M-N}\prod _j\sqrt {\frac {m_j!}{(m_j-k_j)!k_j!}}|\vec {m}-\vec {k}\rangle
|\vec {k}\rangle .
\label{splitting}
\end{eqnarray}
The symmetry operator $s_M$ can be reformulated. After calculation, output density matrix in (\ref{Werner}) is shown to be:

\begin{eqnarray}
\rho_{out}=\frac{N!(M-N)!(N+d-1)!}{(M+d-1)!}\sum _{\vec {m},\vec {m}'}^M|\vec {m}\rangle \langle
\vec {m}'|
\times \left(
\sum _{\vec {k}}^{M-N}\prod _j\frac {x_j^{m_j-k_j}x^{*(m_j'-k_j)}\sqrt {m_j!m_j'!}}
{(m_j-k_j)!(m'_j-k_j)!k_j!}\right).
\end{eqnarray}

For the cloning machine (\ref{FanClone}), we can get the output density matrix after tracing out the ancillary state:
\begin{eqnarray}
{\rho '}^{out}=\frac{N!(M-N)!(N+d-1)!}{(M+d-1)!}\eta ^2\sum _{\vec {n},\vec {n}'}^N\sum _{\vec {k}}^{M-N}
|\vec {n}+\vec {k}\rangle \langle \vec {n}'+\vec {k}|
\times \left(
\prod _j\frac {x_j^{n_j}x_j^{*(n'_j)}\sqrt {(n_j+k_j)!(n_j'+k_j)!}}{n_j!n_j'!k_j!}\right).
\end{eqnarray}
These two expressions are apparently equivalent.

In \cite{PhysRevA.84.034302}, a unified form of the symmetric UQCM is presented, up to an unimportant overall normalization factor,
the transformation is,
\begin{equation}
|\psi\rangle^{\otimes N}|\Phi^+\rangle^{\otimes (M-N)}\rightarrow \left(s_M\otimes I^{\otimes (M-N)}\otimes I^{\otimes N}\right)|\psi\rangle^{\otimes N}|\Phi^+\rangle^{\otimes (M-N)}.\label{UnifiedUQCM}
\end{equation}
This cloning machine is realized by superposition of states in which some of the input states
 are permutated into one part of the maximally entangled states.
 Since $s_M=s_M(I^{\otimes N}\otimes s_{M-N})$, and the mapping of $s_{M-N}$ on the $M-N$ maximally entangled states is: $(s_{M-N}\otimes I^{M-N})|\Phi^+\rangle^{\otimes(M-N)}$, the cloning transformation may be rewritten as:
\begin{eqnarray}
&&\left( s_{M}\otimes I^{\otimes (M-N)}\right) |\vec {n}\rangle |\Phi ^+\rangle ^{\otimes (M-N)}
\nonumber \\
&&=\left( s_{M}\otimes I^{\otimes (M-N)}\right)
|\vec {n}\rangle \sum _{\vec {k}}^{M-N}|\vec {k}\rangle |\vec {k}\rangle
\nonumber \\
&&=\sum _{\vec {k}}^{M-N}\sqrt {\prod _j\frac {(n_j+k_j)!}{n_j!k_j!}}|\vec {n}+\vec {k}\rangle
|\vec {k}\rangle.
\end{eqnarray}
In fact this coincides with the UQCM (\ref{FanClone}).
Here the complicated coefficients (\ref{Fancoefficients})
proposed for optimal cloning machine can be naturally obtained.
Also it can be simply seen that the transformation (\ref{UnifiedUQCM}) is equivalent to the
construction (\ref{Werner}) if the ancillary states are traced out.
So the UQCM can be simply constructed, we can symmetrize the $N$ input pure states and halves of some maximally entangled states, while other halves of
the maximally entangled states are ancillary states. This simplify dramatically the construction of the
UQCM theoretically and its physical implementation becomes easier.
If maximally entangled states are available, the UQCM is to symmetrize the input
pure states with one sides of the maximally entangled states. Indeed, some experiments follow this scheme \cite{667603820020426}
which we will review in physical realization section.

\subsection{Asymmetric quantum cloning}

In the previous subsections we are considering symmetric cloning machines which provide identical output copies. However, naturally we may try to distribute information unequally among the copies. The 1 to 2 optimal asymmetric qubit cloner is found by Niu and Griffiths \cite{PhysRevA.58.4377}, Cerf \cite{PhysRevLett.84.4497} and Bu\v{z}ek, Hillery and Bednik \cite{ISI:000074966200008}. Their formalisms are slightly different, but they
lead to a same relation between A's fidelity $F_A$ and B's fidelity $F_B$:
\begin{equation}
\sqrt{(1-F_A)(1-F_B)}\geq F_A+F_B-\frac{3}{2}\label{AsyFrelation}
\end{equation}
So a tradeoff relation exists for the two fidelities, if one fidelity is large, correspondingly another fidelity
will become small. This will be presented further in the following.

The transformation can be written in the following form according to Bu\v{z}ek \emph{et al.}\cite{ISI:000074966200008}:
\begin{equation}
|\psi\rangle_A(a|\Phi^+\rangle_{BR}+b|0\rangle_B(|0\rangle_R+|1\rangle_R)/\sqrt{2})\rightarrow a|\psi\rangle_A|\Phi^+\rangle_{BR}+b|\psi\rangle_B|\Phi^+\rangle_{AR}\label{BuzekAsy}
\end{equation}
Here $R$ is an ancillary state, $|\Phi^+\rangle=(|00\rangle+|11\rangle)/\sqrt{2}$ is a Bell state. And the normalization condition of input state requires $a^2+ab+b^2=1$, which is an ellipse equation. The reduced density matrix of A and B are: $\rho_{A,B}=F_{A,B}|\psi\rangle\langle\psi|+(1-F_{A,B})|\psi^\bot\rangle\langle\psi^\bot|$, here
\begin{equation}
F_A=1-b^2/2, F_B=1-a^2/2\label{AsyF},
\end{equation}
which is just the fidelity of $A$ and $B$, respectively. It is easy to check (\ref{AsyF}) satisfy the inequality (\ref{AsyFrelation}). And as special cases, we can see if a=0, then $F_A=1$, $F_B=1/2$, hence the information all goes to $A$, and for $B$ it's all the same. If $a^2=b^2=1/3$, then it reduces to symmetric UQCM case, with fidelity $F_A=F_B=5/6$.

For completeness, here we would like to present a slightly different form for the asymmetric quantum cloning
which is named by Cerf as a Pauli channel\cite{PhysRevLett.84.4497}.
We start from qubit case. An arbitrary quantum pure state takes the form,
\begin{eqnarray}
|\psi \rangle =x_0|0\rangle +x_1|1\rangle , ~~
\sum _j|x_j|^2=1.
\end{eqnarray}
A maximally entangled state is written as
\begin{eqnarray}
|\Psi ^+\rangle =
\frac {1}{\sqrt{2}}(|00\rangle +|11\rangle ).
\end{eqnarray}
We can write the complete quantum state of three particles
as
\begin{eqnarray}
|\psi \rangle _A|\Psi ^+\rangle _{BC}
&=&\frac {1}{2}[
|\Psi ^+\rangle _{AB}|\psi \rangle _C
\nonumber \\
&&+(I\otimes X)
|\Psi ^+\rangle _{AB}X|\psi \rangle _C
\nonumber \\
&&+(I\otimes Z)
|\Psi ^+\rangle _{AB}Z|\psi \rangle _C
\nonumber \\
&&+(I\otimes XZ)
|\Psi ^+\rangle _{AB}XZ|\psi \rangle _C ],
\label{2level}
\end{eqnarray}
where $I$ is the identity, $X,Z$ are two Pauli matrices
and $XZ$ is another Pauli matrix up to a whole factor $i$.

Denote the unitary transformation
$U_{m,n}=X^mZ^n$, where
$m,n=0,1$,
and the relation (\ref{2level}) can be rewritten as
\begin{eqnarray}
|\psi \rangle _A|\Psi ^+\rangle _{BC}
&=&\frac {1}{2}\sum _{m,n}
(I\otimes U_{m,-n}\otimes
U_{m,n})
|\Psi ^+\rangle _{AB}|\psi \rangle _C .
\label{2levela}
\end{eqnarray}
Here we remark that $Z^{-1}=Z$ for 2-level
system. We write it in this form since this
relation can be generalized directly to the general
d-dimension system.

Now, suppose we do unitary transformation in the following form
\begin{eqnarray}
&&\sum _{\alpha ,\beta } a_{\alpha ,\beta}(U_{\alpha , \beta }
\otimes U_{\alpha ,-\beta }\otimes I) |\psi \rangle _A|\Psi
^+\rangle _{BC} \nonumber \\
&=&\frac {1}{2} \sum _{\alpha ,\beta ,m,n} (U_{\alpha , \beta
}\otimes U_{\alpha , -\beta }U_{m,-n}\otimes U_{m,n})|\Psi ^+\rangle
_{AB}|\psi \rangle _C
\nonumber \\
&=&\sum _{m,n}
b_{m,n}(I\otimes U_{m,-n}\otimes U_{m,n})
|\Psi ^+\rangle _{AB}|\psi \rangle _C ,
\label{unitary}
\end{eqnarray}

where we defined
\begin{eqnarray}
b_{m,n}=\frac {1}{2}\sum _{\alpha ,\beta }
(-1)^{\alpha n-\beta m}a_{\alpha ,\beta}.
\label{abrelation}
\end{eqnarray}
The amplitudes should be normalizaed $\sum _{\alpha , \beta
}|a_{\alpha ,\beta }|^2= \sum _{m,n}|b_{m,n}|^2=1$. This is actually
the asymmetric quantum cloning machine introduced by
Cerf\cite{PhysRevLett.84.4497}. We can find the quantum states of $A$ and $C$ now
take the form
\begin{eqnarray}
\rho _A&=&\sum _{\alpha ,\beta }|a_{\alpha ,\beta }|^2U_{\alpha ,\beta }
|\psi \rangle \langle \psi |U_{\alpha , \beta }^{\dagger } ,
\\
\rho _C&=&\sum _{m,n}|b_{m,n}|^2U_{m,n}
|\psi \rangle \langle \psi |U_{m,n}^{\dagger }.
\end{eqnarray}
The quantum state of $A$ is related with the quantum state $C$ by relationship
between $a_{\alpha ,\beta }$ and $b_{m,n}$.

The quantum state $\rho _A$ is the original quantum state
after the quantum cloning. The quantum state $\rho _C$ is the copy.

Now, let us see a special case,
\begin{eqnarray}
b_{0,0}=1, ~~~b_{0,1}=b_{1,0}=b_{1,1}=0.
\end{eqnarray}
Crrespondingly, we can choose
\begin{eqnarray}
a_{0,0}=a_{0,1}=a_{1,0}=a_{1,1}=\frac {1}{2}.
\end{eqnarray}
So, we know the quantum states of $A$ and $C$ have the form
\begin{eqnarray}
\rho _A=\frac {1}{2}I, ~~~\rho _C=|\psi \rangle \langle \psi |.
\end{eqnarray}
As a quantum cloning machine, this means the original quantum
state in $A$, $|\psi \rangle $, is completely destroyed,

These results can be generalized to d-dimension system directly.

The asymmetric cloning machine was generalized to $d$-dimensional case by Braunstein, Bu\v{z}ek and Hillery\cite{Braunstein2001}. The setup is almost the same with $|\Phi^+\rangle$ instead defined in higher-dimension, $|\Phi ^+\rangle =\frac{1}{\sqrt{d}}\sum_{j=1}^d|jj\rangle$,
and hence the normalization relation is $a^2+b^2+2ab/d=1$. The output reduced density matrices are written in the form with shrinking factor:
\begin{equation}
\rho_A=(1-b^2)|\psi\rangle\langle\psi|+b^2\frac{I}{d}, \rho_B=(1-a^2)|\psi\rangle\langle\psi|+a^2\frac{I}{d}.
\end{equation}
Hence the fidelities are:
\begin{equation}
F_A=1-b^2\frac{d-1}{d}, F_B=1-a^2\frac{d-1}{d}.
\end{equation}
If $a^2=b^2=d/(2d+2)$, it reduces to the symmetric 1 to 2 $d$-dimensional UQCM case, with fidelity $(d+3)/(2d+2)$. A trade-off relation between $F_A$ and $F_B$ can be found as follows \cite{ISI:000278182800016}:
\begin{equation}
\frac{(\sqrt{(d+1)F_A-1}+\sqrt{(d+1)F_B-1})^2}{2(d+1)}+\frac{(\sqrt{(d+1)F_A-1}-\sqrt{(d+1)F_B-1})^2}{2(d-1)}\leq 1
\end{equation}
Optimality is satisfied when the inequality is saturated. They also give a similar inequality for $1\rightarrow 1+1+1$ case.

Cerf obtained the same result in a different way, here we present
the d-dimensional case\cite{383952220000215,PhysRevLett.88.127902}. This result can be reformulated
for other cases such as for state-dependent case presented in next sections. The transformation is:
\begin{equation}
|\psi\rangle_A\rightarrow\sum_{m,n=0}^{d-1}a_{m,n}U_{m,n}|\psi\rangle_A|B_{m,-n}\rangle_{BR}
\end{equation}
Here $U_{m,n}$ is ``generalized Pauli matrix'':
\begin{equation}
U_{m,n}=\sum_{k=0}^{d-1}e^{\frac{2\pi kni}{d}}|k+m\rangle\langle k|
\end{equation}
and $|B_{m,n}\rangle$ is one of the generalized Bell basis:
\begin{equation}
|B_{m,n}\rangle=\frac{1}{\sqrt{d}}\sum_{k=0}^{d-1}e^{\frac{2\pi kni}{d}}|k\rangle|k+m\rangle.
\end{equation}
The resultant reduced density matrix
\begin{equation}
\rho_A=\sum_{m,n=0}^{d-1}|a_{m,n}|^2U_{m,n}|\psi\rangle\langle\psi|U_{m,n}^\dagger.
\end{equation}
Hence the fidelity $F_A=\sum_{n=0}^{d-1}|a_{0,n}|^2$. For $B$, we replace $a_{m,n}$ by its Fourier transform $b_{m,n}=\frac{1}{d}\sum_{m',n'=0}^{d-1}e^{2\pi(nm'-mn')i/d}a_{m',n'}$.

To clone all states equally well, the matrix $a$ can be written in the following form:
\begin{eqnarray}
a=\left(\begin{array}{cccc}v&x&\cdots&x\\x&y&\cdots&y\\ \vdots&\vdots&\ddots&\vdots\\x&y&\cdots&y\end{array}\right)
\end{eqnarray}
with normalization relation $v^2+2(d-1)x^2+(d-1)^2 y^2=1$. In this form, $F_A=v^2+(d-1)x^2$ and the expression of $b_{m,n}$ is just to replace $x$ by $x'=[v+(d-2)x+(1-d)y]/d$, $y$ by $y'=(v-2x+y)/d$, $v$ by $v'=[v+2(d-1)x+(d-1)^2 y]/d$. Optimal cloning requires $y=0$, and if we let $a=v-x,b=dx$, these coincide with the parameters $a$ and $b$ in the first formalism. When $v=v'=\sqrt{2/(d+1)}, x=x'=\sqrt{1/(2d+2)}$, it reduces to the symmetric case. These results generalized the qutrit cloning presented by Durt and Gisin \cite{688607420020710}.

The optimality of these cloning machines were also proved by Iblisdir \emph{et al.}\cite{PhysRevA.72.042328}, Fiur\'{a}\v{s}ek, Filip and Cerf \cite{ISI:000234463800006}and Iblisdir, Ac\'{i}n and Gisin\cite{Iblisdir2005}. They also generalize the 1 to 2 asymmetric cloning machine to more general cases. Here we use $N\rightarrow M_1+\dots+M_p$ to denote such a problem: construct an asymmetric cloning machine resulting fidelity $F_1$ for $M_1$ copies, $F_2$ for $M_2$ copies, \dots, $F_p$ for $M_p$ copies.

The $1\rightarrow1+1+1$ $d$-dimensional cloning machine was constructed as following:
\begin{align}
|\psi\rangle\rightarrow&\sqrt{\frac{d}{2d+2}}[\alpha|\psi\rangle_A(|\Phi^+\rangle_{BR}|\Phi^+\rangle_{CS}+|\Phi^+\rangle_{BS}|\Phi^+\rangle_{CR})+
\beta|\psi\rangle_B(|\Phi^+\rangle_{AR}|\Phi^+\rangle_{CS}+|\Phi^+\rangle_{AS}|\Phi^+\rangle_{CR})+\nonumber\\&
\gamma|\psi\rangle_C(|\Phi^+\rangle_{AR}|\Phi^+\rangle_{BS}+|\Phi^+\rangle_{AS}|\Phi^+\rangle_{BR})]\label{1to3}
\end{align}
where $A,B,C$ are three output states and $R,S$ are ancillary states. $|\Phi^+\rangle =1/\sqrt{d}\sum_{k=0}^{d-1}|kk\rangle$ as usual. For normalization purpose, $\alpha,\beta,\gamma$ obey $\alpha^2+\beta^2+\gamma^2+\frac{2}{d}(\alpha\beta+\beta\gamma+\alpha\gamma)=1$. The final one copy fidelities for $A,B,C$ are:
\begin{align}
F_A&=1-\frac{d-1}{d}\left(\beta^2+\gamma^2+\frac{2\beta\gamma}{d+1}\right)\nonumber\\
F_B&=1-\frac{d-1}{d}\left(\alpha^2+\gamma^2+\frac{2\alpha\gamma}{d+1}\right)\nonumber\\
F_C&=1-\frac{d-1}{d}\left(\alpha^2+\beta^2+\frac{2\alpha\beta}{d+1}\right).\label{1to3F}
\end{align}

In \cite{PhysRevA.72.042328}, the $1\rightarrow 1+n$ cloning machine was found. The Hilbert space $\mathcal{H}^{\otimes(n+1)}$ is decomposed into two symmetric subspace $\mathcal{H}^+_{n+1}\oplus\mathcal{H}^+_{n-1}$. Let $s_{n+1}$ and $s_{n-1}$ denote the projection operator, respectively, the transformation can be written as:
\begin{equation}
T:\rho\rightarrow(\alpha^* s_{n+1}+\beta^* s_{n-1})(\rho\otimes I^{\otimes n})(\alpha s_{n+1}+\beta s_{n-1}).
\end{equation}
It is a generalization of the construction of symmetric UQCM (\ref{Werner}). The resulting fidelity is $F_A=1-\frac{2}{3}y^2$ for the '1' side, and $F_B=\frac{1}{2}+\frac{1}{3n}(y^2+\sqrt{n(n+2)}xy)$. Here $x$ and $y$ satisfy $x^2+y^2=1$.
A more general case, $N\rightarrow M_A+M_B$, is studied with similar method in \cite{Iblisdir2005}.

In studying asymmetric quantum cloning, the region of possible output fidelities for
one to three cloning is studied in \cite{ISI:000278182800016}, the case of one to many case is
studied in \cite{Cwiklinski2012}, the general case is studied in \cite{KayDagPreprint}.

\subsection{A unified UQCM}

Recently, Wang \emph{et al.} proposed a unified way to construct general asymmetric UQCM \cite{PhysRevA.84.034302}. The essence is to replace the symmetric operator $s_M$ in construction (\ref{UnifiedUQCM}) by a linear combination of identity $I$ and many permutation operators. Take the 1 to 2 qubit cloning case as the simplest example, $s_2=\frac{1}{2}(I^{\otimes2}+\mathcal{P})$, where $\mathcal{P}=|00\rangle\langle00|+|11\rangle\langle11|+|01\rangle\langle10|+|10\rangle\langle01|$ is a permutation(swap) operator($\mathcal{P}|jl\rangle=|lj\rangle$). If $s_2$ is replaced by $\alpha I^{\otimes2}+\beta\mathcal{P}$, the output density matrix exactly coincides with the output density matrix in construction (\ref{BuzekAsy}): $|\psi\rangle_A\rightarrow a|\psi\rangle_A|\Phi^+\rangle_{BR}+b|\psi\rangle_B|\Phi^+\rangle_{AR}$, with $\alpha=\frac{\sqrt{3}}{2}a,\beta=\frac{\sqrt{3}}{2}b$.

In order to introduce this method, here we present two examples to show explicitly that it
can be applied straightforwardly for various occasions.

For $1\rightarrow 3$ asymmetric qubit cloning case, we replace the symmetry operator $s_3$ by
\begin{equation}
\alpha I+\beta\mathcal{P}_{12}+\gamma\mathcal{P}_{13}+\delta\mathcal{P}_{23}+\mu\mathcal{P}_{123}+\nu\mathcal{P}_{132}.\label{3AsyOp}
\end{equation}
Note $\mathcal{P}_{mn}$ is the operator that swap the $m$ qubit and the $n$ qubit, and $\mathcal{P}_{123}$ is a cyclic operator that move the first qubit to the second place, the second qubit to the third place, and the third qubit to the first place. $\mathcal{P}_{132}$ is its inverse transformation. In fact these six components in (\ref{3AsyOp}) are just the elements of permutation group $S_3$. The symmetry operator $s_3=\frac{1}{6}(I+\mathcal{P}_{12}+\mathcal{P}_{13}+\mathcal{P}_{23}+\mathcal{P}_{123}+\mathcal{P}_{132})$, is retrieved when $\alpha=\beta=\gamma=\delta=\mu=\nu=\frac{1}{6}$. The $1\rightarrow 3$ asymmetric qubit cloning can be obtained by replacing $s_3$ by
(\ref{3AsyOp}) and insert it to the cloning transformation (\ref{UnifiedUQCM}). Here we would like to remark that
the number of essential permutations for the specific $1\rightarrow 3$ case are actually three. There are only three independent
parameters corresponding to cases: the input state is in first, second, and third positions, respectively. This
will be shown explicitly later. Now, if we trace out the
ancillary states, it is equivalent to modify (\ref{Werner}).
The final density operator is:
\begin{align}
\rho&=\frac{1}{2}(\alpha I+\beta\mathcal{P}_{12}+\gamma\mathcal{P}_{13}+\delta\mathcal{P}_{23}+\mu\mathcal{P}_{123}+\nu\mathcal{P}_{132})(|\psi\rangle\langle\psi|\otimes I\otimes I)(\alpha I+\beta\mathcal{P}_{12}+\gamma\mathcal{P}_{13}+\delta\mathcal{P}_{23}+\mu\mathcal{P}_{123}+\nu\mathcal{P}_{132})\nonumber\\
&=\frac{1}{2}[\alpha|\psi\rangle\langle\psi|\otimes I\otimes I+\beta(|0\psi\rangle\langle\psi0|+|1\psi\rangle\langle\psi1|)\otimes I_3+\gamma(|0\psi\rangle\langle\psi0|+|1\psi\rangle\langle\psi1|)_{13}\otimes I_2\nonumber\\&+\delta|\psi\rangle\langle\psi|_1\otimes(|00\rangle\langle00|+|11\rangle\langle11|+|01\rangle\langle10|+|10\rangle\langle01|)+
\mu(|0\psi0\rangle\langle\psi00|+|1\psi0\rangle\langle\psi01|+|0\psi1\rangle\langle\psi10|+|1\psi1\rangle\langle\psi11|)\nonumber\\
&+\nu(|00\psi\rangle\langle\psi00|+|01\psi\rangle\langle\psi01|+|10\psi\rangle\langle\psi10|+|11\psi\rangle\langle\psi11|)](\alpha I+\beta\mathcal{P}_{12}+\gamma\mathcal{P}_{13}+\delta\mathcal{P}_{23}+\mu\mathcal{P}_{123}+\nu\mathcal{P}_{132})\nonumber\\
&=\frac{1}{2}[(\alpha^2+\delta^2)|\psi\rangle\langle\psi|\otimes I\otimes I+(\beta^2+\mu^2)I\otimes|\psi\rangle\langle\psi|\otimes I+(\gamma^2+\nu^2) I\otimes I\otimes|\psi\rangle\langle\psi|\nonumber\\
&+(\alpha\beta+\mu\delta)(|\psi0\rangle\langle0\psi|+|\psi1\rangle\langle1\psi|+|0\psi\rangle\langle\psi0|+|1\psi\rangle\langle\psi1|)_{12}\otimes I_3\nonumber\\&+(\alpha\gamma+\nu\delta)(|\psi0\rangle\langle0\psi|+|\psi1\rangle\langle1\psi|+|0\psi\rangle\langle\psi0|+|1\psi\rangle\langle\psi1|)_{13}\otimes I_2\nonumber\\&+(\beta\gamma+\mu\nu)(|00\psi\rangle\langle0\psi0|+|01\psi\rangle\langle1\psi0|+|10\psi\rangle\langle0\psi1|+|11\psi\rangle\langle1\psi1|+trans.)
\nonumber\\&+(\delta\beta+\alpha\mu)(|0\psi0\rangle\psi00|+|0\psi1\rangle\psi10|+|1\psi0\rangle\langle\psi01|+|1\psi1\rangle\langle\psi11|+trans.)\nonumber\\
&+\delta\gamma(|0\psi0\rangle\langle00\psi|+|0\psi1\rangle\langle10\psi|+|1\psi0\rangle\langle01\psi|+|1\psi1|\rangle\langle11\psi|+trans.)\nonumber\\&
+(\beta\nu+\gamma\mu)I_1\otimes(|\psi0\rangle\langle0\psi|+|\psi1\rangle\langle1\psi|+|0\psi\rangle\langle\psi0|+|1\psi\rangle\langle\psi1|)_{23}\nonumber\\&
+\alpha\nu(|00\psi\rangle\langle\psi00|+|01\psi\rangle\langle\psi01|+|10\psi\rangle\langle\psi10|+|11\psi\rangle\psi11|+trans.)\nonumber\\&
+2\alpha\delta|\psi\rangle\langle\psi|_1\otimes(|00\rangle\langle00|+|01\rangle\langle10|+|10\rangle\langle01|+|11\rangle\langle11|)_{23}\nonumber\\&
+2\beta\mu|\psi\rangle\langle\psi|_2\otimes(|00\rangle\langle00|+|01\rangle\langle10|+|10\rangle\langle01|+|11\rangle\langle11|)_{13}\nonumber\\&
+2\gamma\nu|\psi\rangle\langle\psi|_3\otimes(|00\rangle\langle00|+|01\rangle\langle10|+|10\rangle\langle01|+|11\rangle\langle11|)_{12}\label{U1to3DM}
\end{align}
Here \emph{trans.} denotes the transposition of previous terms, for example,
term $|00\psi \rangle \langle 0\psi 0|$ is followed by its transposition $|0\psi 0\rangle \langle 00\psi |$. Trace out the second and third qubits, we obtain the single qubit reduced density matrix,
\begin{align}
\rho_1&=[2(\alpha^2+\delta^2+\alpha\delta)+\alpha(2\beta+2\gamma+\mu+\nu)+\delta(2\mu+2\nu+\beta+\gamma)+\mu\nu+\beta\gamma]|\psi\rangle\langle\psi|\nonumber\\
&+(\beta^2+\mu^2+\gamma^2+\nu^2+\beta\mu+\beta\nu+\gamma\mu+\gamma\nu)I
\end{align}
Hence a normalization relation is easily obtained:
\begin{align}
&2(\alpha^2+\delta^2+\alpha\delta)+\alpha(2\beta+2\gamma+\mu+\nu)+\delta(2\mu+2\nu+\beta+\gamma)+\mu\nu+\beta\gamma+\nonumber\\
&2(\beta^2+\mu^2+\gamma^2+\nu^2+\beta\mu+\beta\nu+\gamma\mu+\gamma\nu)=1
\end{align}

Similarly we can find the reduced density matrices of the second and third copies, their fidelities are:
\begin{align}
F_1&=1-\frac{1}{2}[(\beta+\mu)^2+(\beta+\nu)^2+(\gamma+\mu)^2+(\gamma+\nu)^2]\nonumber\\
F_2&=1-\frac{1}{2}[(\alpha+\delta)^2+(\alpha+\nu)^2+(\gamma+\delta)^2+(\gamma+\nu)^2]\nonumber\\
F_3&=1-\frac{1}{2}[(\alpha+\delta)^2+(\alpha+\mu)^2+(\beta+\delta)^2+(\beta+\mu)^2]\label{U1to3F}
\end{align}
It reduces to the symmetric cloning case when $\alpha=\beta=\gamma=\delta=\mu=\nu=\frac{1}{6}$, and the fidelity is 7/9, which exactly coincide with the UQCM fidelity formula (\ref{UQCM1F}). To see its relation with the previous $1\rightarrow 3$ asymmetric UQCM (\ref{1to3}), we can explicitly compute out
the density matrix in (\ref{1to3}):
\begin{align}
\rho_{out}&=\frac{1}{12}\{4(\alpha'|\psi00\rangle+\beta'|0\psi0\rangle+\gamma'|00\psi\rangle)(\alpha'\langle\psi00|+\beta'\langle0\psi0|+\gamma'\langle00\psi|)
\nonumber\\&+2[\alpha'(|\psi01\rangle+\psi10\rangle)+\beta'(|0\psi1\rangle+|1\psi0\rangle)+\gamma'(|01\psi\rangle+|10\psi\rangle)]\nonumber\\&
[\alpha'(\langle\psi01|+\langle\psi10|)+\beta'(\langle0\psi1|+\langle1\psi0|)+\gamma'(\langle01\psi|+\langle10\psi|)]\nonumber\\&+
4(\alpha'|\psi11\rangle+\beta'|1\psi1\rangle+\gamma'|11\psi\rangle)(\alpha'\langle\psi11|+\beta'\langle1\psi1|+\gamma'\langle11\psi|)\}
\end{align}
For clarity purpose we replace the coefficients $\alpha, \beta, \gamma$ in (\ref{1to3}) with $\alpha', \beta', \gamma'$. Compare this expression with (\ref{U1to3DM}), we found if the following equations are satisfied, they are equivalent:
\begin{align}
&\frac{\alpha'^2}{3}=(\alpha+\delta)^2, \frac{\beta'^2}{3}=(\beta+\mu)^2, \frac{\gamma'^2}{3}=(\gamma+\nu)^2\nonumber\\
&\alpha'^2=12\alpha\delta,\beta'^2=12\beta\mu,\gamma'^2=12\gamma\nu.
\end{align}
This implies $\alpha=\delta=\alpha'/(2\sqrt{3}), \beta=\mu=\beta'/(2\sqrt{3}), \gamma=\nu=\gamma'/(2\sqrt{3})$. And in this case the fidelity expressions (\ref{U1to3F}) exactly coincide with the previous results (\ref{1to3F}). The cloning machine here has six parameters,
which indicates that it is a general form of asymmetric UQCM, and we do not need to consider the
specific input positions.

We can study the $2\rightarrow3$ case similarly. The resultant density matrix is:
\begin{align}
\rho&=\frac{1}{2}(\alpha I+\beta\mathcal{P}_{12}+\gamma\mathcal{P}_{13}+\delta\mathcal{P}_{23}+\mu\mathcal{P}_{123}+\nu\mathcal{P}_{132})(|\psi\psi\rangle\langle\psi\psi|\otimes I)(\alpha I+\beta\mathcal{P}_{12}+\gamma\mathcal{P}_{13}+\delta\mathcal{P}_{23}+\mu\mathcal{P}_{123}+\nu\mathcal{P}_{132})\nonumber\\
&=\frac{1}{2}[(\alpha+\beta)|\psi\psi\rangle\langle\psi\psi|\otimes I_3+(\gamma+\mu)(|0\psi\psi\rangle\langle\psi\psi0|+|1\psi\psi\rangle\langle\psi\psi1|)\nonumber\\
&+(\delta+\nu)(|\psi0\psi\rangle\langle\psi\psi0|+|\psi1\psi\rangle\langle\psi\psi1|)](\alpha I+\beta\mathcal{P}_{12}+\gamma\mathcal{P}_{13}+\delta\mathcal{P}_{23}+\mu\mathcal{P}_{123}+\nu\mathcal{P}_{132})\nonumber\\
&=\frac{1}{2}[(\alpha+\beta)^2|\psi\psi\rangle\langle\psi\psi|\otimes I_3+(\gamma+\mu)^2 I_1\otimes |\psi\psi\rangle\langle\psi\psi|+
(\delta+\nu)^2|\psi\rangle\langle\psi|\otimes I_2\otimes|\psi\rangle\langle\psi|\nonumber\\
&+(\alpha+\beta)(\gamma+\mu)(|0\psi\psi\rangle\langle\psi\psi0|+|1\psi\psi\rangle\langle\psi\psi1|+trans.)\nonumber\\
&+(\alpha+\beta)(\delta+\nu)(|\psi0\psi\rangle\langle\psi\psi0|+|\psi1\psi\rangle\langle\psi\psi1|+trans.)\nonumber\\
&+(\gamma+\mu)(\delta+\nu)(|\psi0\psi\rangle\langle0\psi\psi|+|\psi1\psi\rangle\langle1\psi\psi|+trans.)\label{U2to3DM}
\end{align}

We can see that there are only three independent parameters in the final expression: $\alpha+\beta$, $\gamma+\mu$, $\delta+\nu$, so we denote them by $A$, $B$ and $C$ respectively. We trace out the other two states to obtain the following one copy reduced density matrices:
\begin{align}
\rho_1&=(A^2+C^2+AB+AC+BC)|\psi\rangle\langle\psi|+\frac{B^2}{2}I\nonumber\\
\rho_2&=(A^2+B^2+AB+AC+BC)|\psi\rangle\langle\psi|+\frac{C^2}{2}I\nonumber\\
\rho_3&=(B^2+C^2+AB+AC+BC)|\psi\rangle\langle\psi|+\frac{A^2}{2}I
\end{align}

The coefficients apparently satisfy a normalization relation: $A^2+B^2+C^2+AB+BC+CA=1$. From the one copy reduced density matrices we simply read out the fidelities:
\begin{equation}
F_1=1-\frac{B^2}{2},F_2=1-\frac{C^2}{2},F_3=1-\frac{A^2}{2}
\end{equation}
For symmetric cloning case, we let $A=B=C=1/\sqrt{6}$, then the fidelity is $11/12$, which exactly coincide with the UQCM fidelity formula (\ref{UQCM1F}).

\subsection{Singlet monogamy and optimal cloning}
In quantum information science, entanglement is a resource for various QIP applications.
On the other hand, the entanglement cannot be shared freely among multi-parties.
For example, for a multipartite quantum state, one party cannot be maximally entangled
independently with other two parties simultaneously. It means that entanglement is
monogamous. There are some monogamy inequalities  for entanglement sharing \cite{ckw2000,Osborne2006,Ouyc2007,Ouyc2008}.

In this review, we consider the singlet monogamy in application of quantum cloning.
We know that singlet is a natural maximally entangled state, we use the name of
singlet monogamy to describe the restrictions on entanglement sharing.

Quantitatively, the amount of entanglement between $A$ and $B$ can be defined as
\begin{equation}
p_{A,B}=\max_{U,V}\langle \Phi^+|U\otimes V\rho_{A,B}U^\dagger\otimes V^\dagger|\Phi^+\rangle
\end{equation}
where $|\Phi^+\rangle =\sum_{i=0}^{d-1}|ii\rangle/\sqrt{d}$ is the d-dimensional maximally entangled state, which
is standard in this review. This quantity describes, under local unitary operations, the fidelity between
state $\rho _{A,B}$ and the maximally entangled state.
In \cite{Kay2009}, Kay, Kaszlikowski and Ramanathan discovered the relation between singlet monogamy and $1\rightarrow N$ optimal asymmetric UQCM. In their approach, a setup proposed by Fiur\'{a}\v{s}ek\cite{PhysRevA.64.062310} is used: $\Lambda_{out}(\psi_{in})$ is a reduced density matrix so that the efficiency of this cloning machine $F$ is measured by averaging $Tr[\sqrt{\sqrt{\rho_{in}}\Lambda_{out}(\psi_{in})\sqrt{\rho_{in}}}]^2$. Note this is a ``average'' definition of fidelity. In \cite{PhysRevA.64.062310} it is proved $F\leq d\lambda$ where $\lambda$ is the maximal eigenvalue of the matrix
\begin{equation}
R=\int\mathrm{d}\psi_{in}[|\psi_{in}\rangle\langle\psi_{in}|^T \otimes\Lambda_{out}(\psi_{in})].\label{monogamyR}
\end{equation}

For the specific $1\rightarrow N$ asymmetric cloning case, $\Lambda_{out}(\psi_{in})$ is defined to be
\begin{equation}
\Lambda_{out}(\psi_{in})=\sum_{i=1}^N \alpha_i I_1\otimes\dots\otimes I_{i-1}\otimes|\psi_{in}\rangle\langle\psi_{in}|_i\otimes I_{i+1}\otimes\dots\otimes I_N.
\label{singletout}
\end{equation}
Here $\alpha_i$ is a set of parameters to describe the required asymmetry of output states, which satisfies $\sum_{i=1}^N\alpha_i=1$. Rewriting $|\psi_{in}\rangle$ as $U|0\rangle$, where $U$ is a unitary operator in $d$-dimensional Hilbert space, then from (\ref{monogamyR}) we find
\begin{equation}
R=\int \mathrm{d} U\sum_{i=1}^N\alpha_i U^T\otimes U|00\rangle\langle00|_{0,i}U^*\otimes U^\dagger ,
\end{equation}
where the subscript $0$ denotes the port of state $|\psi_{in}\rangle\langle\psi_{in}|^T$ appeared in expression (\ref{monogamyR}) which is now
expressed as $U^T|0\rangle $. This equation is obtained by substituting Eq.(\ref{singletout}) into
Eq. (\ref{monogamyR}), so subscript $i$ corresponds to port of state $|\psi_{in}\rangle\langle\psi_{in}|$ appeared in Eq.(\ref{singletout}).
The form of state transposition denoted as $T$ is due to an identity ${\rm Tr}_1(|\psi _{in}\rangle \langle \psi _{in}|_1|\Phi ^+\rangle \langle \Phi ^+|_{0,1})=\frac {1}{d}|\psi _{in}\rangle \langle \psi _{in}|_0^T$.
After calculation it turns out to be
\begin{equation}
R=\frac{1}{d(d+1)}\sum_{i=1}^N\alpha_i(I+d|\Phi^+\rangle\langle\Phi^+|)_{0,i}.
\end{equation}
To find out the eigenvalue of this matrix, an ansatz is proposed:
\begin{equation}
|\Psi\rangle=\sum_{i=1}^N\beta_i|\Phi^+\rangle|\Phi\rangle_{1\dots(i-1)(i+1)\dots N}.
\end{equation}
$\beta_i$ is parameters satisfy normalization condition $(\sum_{i=1}^N\beta_i)^2+(d-1)\sum_{i=1}^N\beta_i^2=d$, and $|\Phi\rangle$ means a normalized superposition of all permutation of $|\Phi^+\rangle^{\otimes(N-1)/2}$ for odd $N$ and $|\Phi^+\rangle^{\otimes(N-2)/2}|0\rangle$ for even $N$. It satisfies
\begin{equation}
(|\Phi^+\rangle\langle\Phi^+|_{0,j}\otimes I)|\Phi\rangle_{0,i}|\Phi\rangle_{1\dots(i-1)(i+1)\dots N}=\gamma_{i,j}|\Phi^+\rangle_{0,j}|\Phi\rangle_{1\dots(j-1)(j+1)\dots N}.
\end{equation}
Here $\gamma_{i,j}=[1+\delta_{ij}(d-1)]/d$ and hence we know $|\Psi\rangle$ is a eigenvector of $R$ if $\alpha_i d\sum_{j=1}^N\gamma_{i,j}\beta_j=[d(d+1)-\lambda]\beta_i$ for every $i$. Combine this constraint with the expression of singlet monogamy of $|\Psi\rangle$: $p_{0,i}=(\sum_{j=1}^N\gamma_{i,j}\beta_j)^2$, as well as the normalization condition, we get the singlet monogamy relation for $1\rightarrow N$ asymmetric cloning machine:
\begin{equation}
\sum_{i=1}^N p_{0,i}\leq\frac{d-1}{d}+\frac{1}{N+d-1}(\sum_{i=1}^N\sqrt{p_{0,i}})^2.
\end{equation}

It is straightforward to find the one copy fidelity $F_i=(p_{0,i}d+1)/(d+1)$. For symmetric UQCM case, one let all $p_{0,i}$ to be equal to $(N+d-1)/dN$ and then the previous result $F=(2N+d-1)/[N(d+1)]$ is regenerated. In \cite{Kay2009} it is also shown that the previous $1\rightarrow 1+1+1$ asymmetric cloning and $1\rightarrow1+N$ asymmetric cloning cases are consistent with this approach.

The $1\rightarrow 4$ asymmetric cloning can be similarly studied \cite{Renxj2011}. The normalization condition in this specific case turns out to be:
\begin{equation}
\beta _{1}^{2}+\beta _{2}^{2}+\beta _{3}^{2}+\beta _{4}^{2}+\frac{2}{d}%
(\beta _{1}\beta _{2}+\beta _{1}\beta _{3}+\beta _{1}\beta _{4}+\beta
_{2}\beta _{3}+\beta _{2}\beta _{4}+\beta _{3}\beta _{4})=1
\end{equation}
and the fidelity of these four copies are:
\begin{eqnarray}
F_{1} &=&1-\frac{d-1}{d}[\beta _{2}^{2}+\beta _{3}^{2}+\beta _{4}^{2}+\frac{%
2(\beta _{2}\beta _{3}+\beta _{2}\beta _{4}+\beta _{3}\beta _{4})}{d+1}], \\
F_{2} &=&1-\frac{d-1}{d}[\beta _{1}^{2}+\beta _{3}^{2}+\beta _{4}^{2}+\frac{%
2(\beta _{1}\beta _{3}+\beta _{1}\beta _{4}+\beta _{3}\beta _{4})}{d+1}], \\
F_{3} &=&1-\frac{d-1}{d}[\beta _{1}^{2}+\beta _{2}^{2}+\beta _{4}^{2}+\frac{%
2(\beta _{1}\beta _{2}+\beta _{1}\beta _{4}+\beta _{2}\beta _{4})}{d+1}], \\
F_{4} &=&1-\frac{d-1}{d}[\beta _{1}^{2}+\beta _{2}^{2}+\beta _{3}^{2}+\frac{%
2(\beta _{1}\beta _{2}+\beta _{1}\beta _{3}+\beta _{2}\beta _{3})}{d+1}].
\end{eqnarray}
With general asymmetric quantum cloning machine available, we can expect that the corresponding
relationship between quantum cloning and entanglement monogamy can be constructed.

\subsection{Mixed-state quantum cloning}

In the previous discussions of quantum cloning, the input state is assumed to be pure. What if the input state is mixed state $\rho$? Sometimes since we only look at the resulted local one-copy reduced density matrix, this procedure is named ``broadcasting''\cite{PhysRevLett.76.2818,ISI:000231017700008},
as we already presented in this review. In \cite{PhysRevLett.76.2818}, Barnum \emph{et al.} proved the $1\rightarrow 2$ no cloning theorem can be extended to mixed state case, that is, for two non-commuting input density matrices, the cloning machine cannot copy both perfectly, as we have already
shown in previous sections. Then D'Ariano, Macchiavello and Perinotti studied the extended $N\rightarrow M$ case and constructed the optimal UQCM \cite{ISI:000231017700008}. They found a non-trivial result that the no-broadcasting theorem cannot be generalized to more than one input case. Specifically, for UQCM, it is even possible to purify the input states when $N\geq4$, this phenomenon is called ``superbroadcasting''. Note here UQCM does not mean constant fidelity reached for every mixed state, since the existence of such cloning machine ($1\rightarrow M$) was nullified by Chen and Chen \cite{ISI:000247624300060}.
For mixed state cloning, it seems reasonable to use the shrinking factor as the measure of merit for the quantum cloning machine. This is for cloning of mixed states
in symmetric subspace \cite{Fan-mixed}.
The property of ``universal'' for mixed cloning machine is in the sense that the shrinking factor of the single output is
independent of the input mixed state.

In this subsection, we try to review some explicit results of mixed state cloning studied in \cite{ISI:000248425500008,ISI:000249154900043,Yang2007}.
The UQCM for pure state (\ref{Werner},\ref{FanClone}) can be applied apparently to one input mixed state. But if we input the direct product of two identical $\rho$, direct application of (\ref{Werner}) cannot give the optimal output. This can be easily figured out if we consider the $2\rightarrow2$ case. The optimal transformation is just leaving it unchanged, but if we apply the symmetrization projection, since $\rho\otimes\rho$ contains an asymmetric part, this part is deleted so the final state changes. So we need to find out other ways to achieve maximal fidelity.

We suppose the input state is identical copies of $\rho=z_0|0\rangle\langle0|+z_1|0\rangle\langle1|+z_2|1\rangle\langle0|+z_3|1\rangle\langle1|$, and we use the notation $|m,n\rangle$ to denote the totally symmetric state with $m$ $|0\rangle $s and $n$ $|1\rangle $s. Additionally we introduce $|\widetilde{m,n}\rangle$ which is constructed by multiplying each components in $|m,n\rangle$ by a different power of $\omega=\exp[2\pi im!n!/(m+n)!]$ ranging from 0 to $(m+n)!/(m!n!)-1$. For example, $|\widetilde{2,1}\rangle=(|001\rangle+\omega |010\rangle+\omega ^2|100\rangle)/\sqrt{3}$, with $\omega =\exp[2\pi i/3]$. Obviously $|m,n\rangle$ is orthogonal to $|\widetilde{m,n}\rangle$.

Then the $2\rightarrow3$ transformation is written as:
\begin{align}
&|2,0\rangle|R\rangle\rightarrow\sqrt{\frac{3}{4}}|3,0\rangle|R_0\rangle+\sqrt{\frac{1}{4}}|2,1\rangle|R_1\rangle\nonumber\\
&|1,1\rangle|R\rangle\rightarrow\sqrt{\frac{1}{2}}|2,1\rangle|R_0\rangle+\sqrt{\frac{1}{2}}|1,2\rangle|R_1\rangle\nonumber\\
&|0,2\rangle|R\rangle\rightarrow\sqrt{\frac{1}{4}}|1,2\rangle|R_0\rangle+\sqrt{\frac{3}{4}}|0,3\rangle|R_1\rangle\nonumber\\
&|\widetilde{1,1}\rangle|R\rangle\rightarrow\sqrt{\frac{1}{2}}|\widetilde{2,1}\rangle|R_0\rangle+\sqrt{\frac{1}{2}}|\widetilde{1,2}\rangle|R_1\rangle.
\end{align}
It can be verified that the output single copy reduced density matrix is $\frac{5}{6}\rho+\frac{I}{12}$. The shrinking factor $5/6$,
apparently coincide with the maximal shrinking factor of $2\rightarrow3$ UQCM in pure state case. The more general $2\rightarrow M$ mixed state cloner is constructed in similar way:
\begin{align}
&|2,0\rangle|R\rangle\rightarrow\sum_{k=0}^{M-2}\alpha_{0k}|M-k,k\rangle|R_k\rangle\nonumber\\
&|1,1\rangle|R\rangle\rightarrow\sum_{k=0}^{M-2}\alpha_{1k}|M-k-1,k+1\rangle|R_k\rangle\nonumber\\
&|0,2\rangle|R\rangle\rightarrow\sum_{k=0}^{M-2}\alpha_{2k}|M-k-2,k+2\rangle|R_k\rangle\nonumber\\
&|\widetilde{1,1}\rangle|R\rangle\rightarrow\sum_{k=0}^{M-2}\alpha_{1k}|\widetilde{M-k-1,k+1}\rangle|R_k\rangle
\end{align}
where
\begin{equation}
\alpha _{jk}= \sqrt{\frac {6(M-2)!(M-j-k)!(j+k)!}{(2-j)!(M+1)!
(M-2-k)!j!k!}}.
\end{equation}
By calculations, it can also be shown that the shrinking factor leads to previous results(\ref{UQCM1F}) corresponding for
pure state case, hence it's optimal.

In \cite{ISI:000249154900043} this construction is generalized to $N\rightarrow M$ case:
\begin{equation}
|N-m,m\rangle|R\rangle\rightarrow\sum_{k=0}^{M-N}\beta_{mk}|M-m-k,m+k\rangle|R_{M-N-k,k}\rangle,
\end{equation}
the coefficients are:
\begin{equation}
\beta _{mk} =\sqrt{\frac{\left( M-N\right) !\left( N+1\right) !}{\left(
M+1\right) !}}\sqrt{\frac{\left( M-m-k\right) !}{\left( N-m\right) !\left(
M-N-k\right) !}}
\cdot \sqrt{\frac{\left( m+k\right) !}{m!k!}}.
\end{equation}

\subsection{Universal NOT gate}
Similar to the quantum cloning problem, one can ask if there is some transformation $U$ that convert an arbitrary state $|\psi\rangle=\alpha|0\rangle+\beta|1\rangle$ to its conjugate state $|\psi^\bot\rangle=\beta^*|0\rangle-\alpha^*|1\rangle$. For two states $|\psi_1\rangle=\alpha|0\rangle+\beta|1\rangle$ and $|\psi_2\rangle=\gamma|0\rangle+\delta|1\rangle$, we have $\langle\psi_2|\psi_1\rangle=\gamma^*\alpha+\delta^*\beta=\left(\langle\psi_2|U^\dagger U|\psi_1\rangle\right)^*$, hence $U$ is an anti-unitary operator. And it is not completely positive hence cannot be applied to a small system, as argued by Bu\v{z}ek, Hillery and Werner in \cite{PhysRevA.60.R2626}.

Then it is a question whether we can have a universal NOT gate approximately.
A general $N\rightarrow M$ universal NOT gate is constructed by using the universal cloning machine.
The final single copy output density matrix is $\rho_{out,1}=\frac{N}{N+2}|\psi^\bot\rangle\langle\psi^\bot|+\frac{1}{N+2}I$, regardless of $M$. In fact, this is exactly the reduced density matrix of the ancilla in the UQCM. This is an interesting result since it shows the ancilla has the ``anti-clone'' meaning. The optimality of this universal NOT gate is also proved \cite{Buzek2000}.
The universal NOT gate or anti-cloning is the same as the universal spin flip machine \cite{PhysRevLett.83.432}.
Related, it is found that a pair of antiparallel spins can contain more information than that of parallel spins.
The universal NOT gate is studied for continuous variable system in \cite{cerf-cv-not}.
The experimental implementation of universal NOT gate in optical system is reported in \cite{martininature}.
The universal controlled-NOT gate is studied in \cite{Siomau2010a}.

\subsection{Minimal input set, six-state cryptography and other results}

\begin{figure}
\includegraphics[height=5cm]{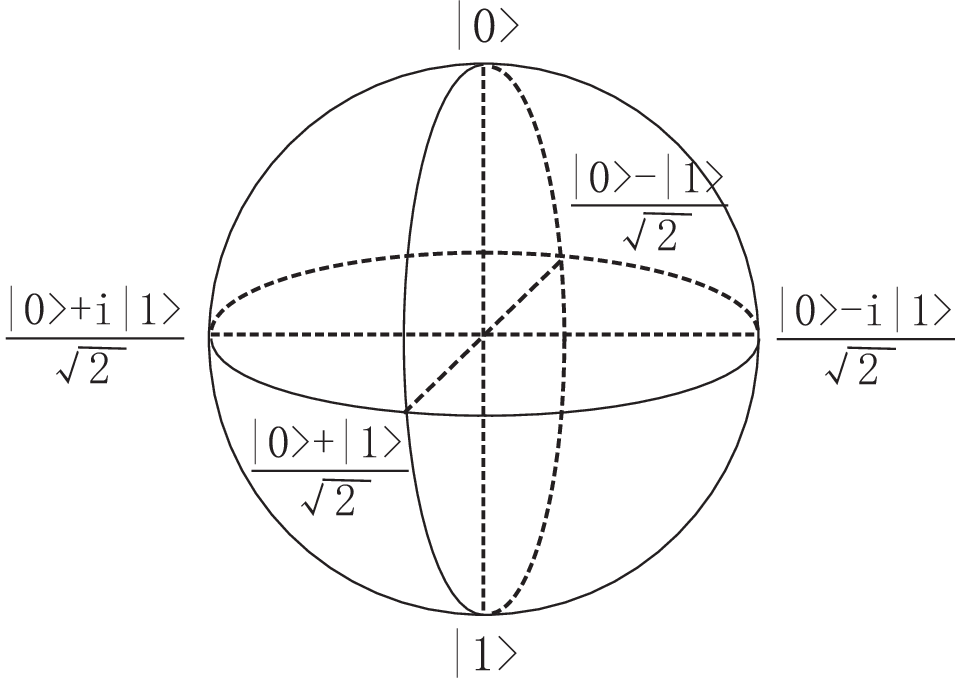}
\caption{The six states used in quantum key distribution, the optimal cloning machine to clone those
six states is a UQCM.}
\label{figure-6-state}
\end{figure}

Bruss showed that the $1\rightarrow2$ optimal cloning of the following
six states with equal fidelity for each state is equivalent to the qubit UQCM\cite{PhysRevLett.81.3018},
\begin{eqnarray}
&&\{ |0\rangle,|1\rangle\} ;\nonumber \\
&&\{ \frac {1}{\sqrt{2}}(|0\rangle+|1\rangle),\frac {1}{\sqrt{2}}(|0\rangle-|1\rangle)\} ;
\nonumber \\
&&\{ \frac {1}{\sqrt{2}}(|0\rangle+i|1\rangle),\frac {1}{\sqrt{2}}(|0\rangle-i|1\rangle)\} .
\label{6states}
\end{eqnarray}
These six states can be represented on Bloch sphere as FIG.\ref{figure-6-state}.

These six states are exactly the three basis used in the six-state QKD protocol,  and indeed the UQCM can be regarded as the optimal way to attack the quantum channel in this protocol \cite{Bechmann-Pasquinucci1999}.
This is an interesting phenomenon, it means that the optimal cloning machine for those six states can actually clone arbitrary qubits optimally.
On the other hand, it also means that we cannot clone six states better than a UQCM does.

\begin{figure}
\includegraphics[height=6cm]{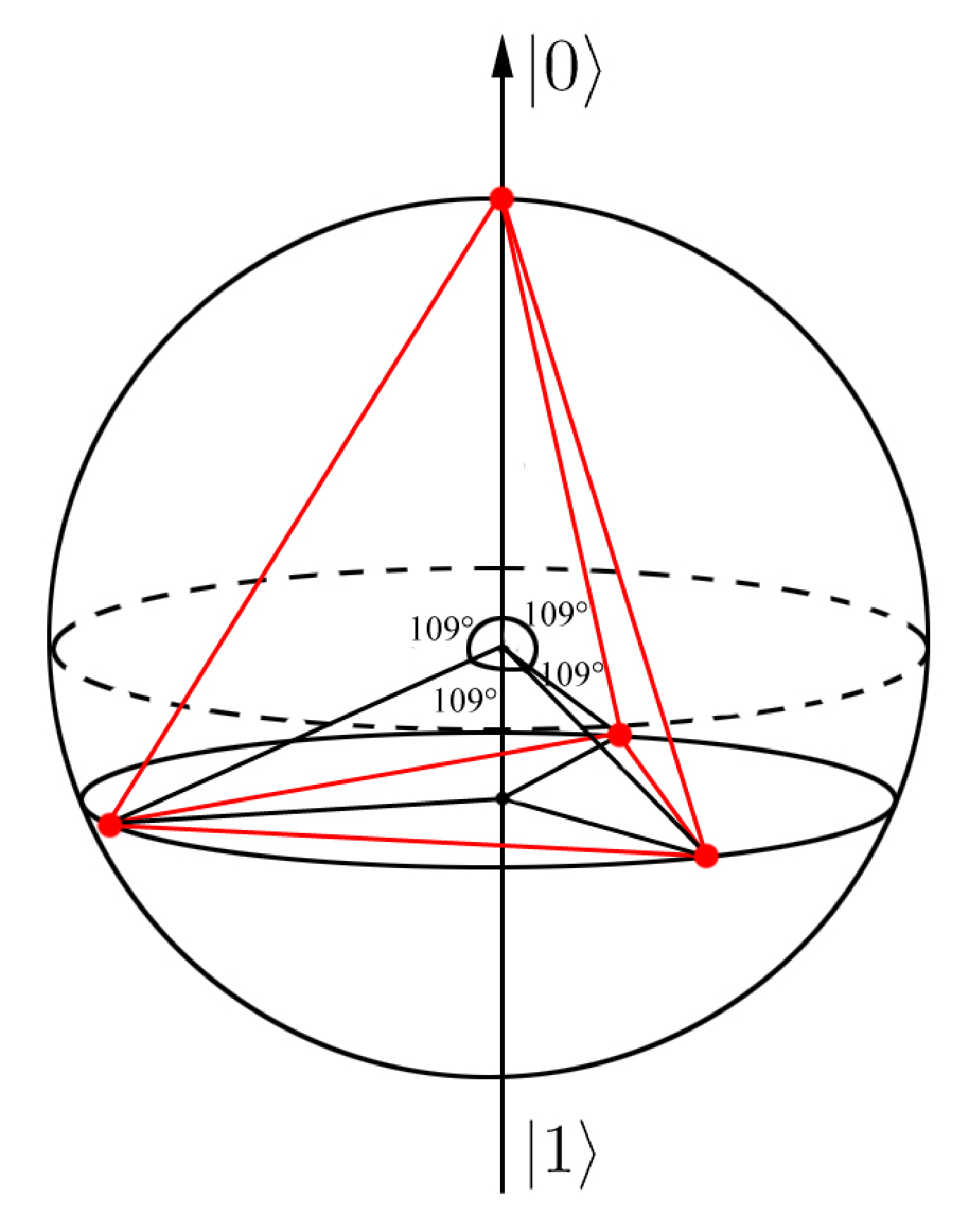}
\caption{The four states on the vertex of a tetrahedron, which determines a UQCM in $1\rightarrow2$ cloning case.}
\label{4states}
\end{figure}

A reverse question might be interesting: how many states are enough to define a UQCM? More explicitly, what is the minimal number of the states in the input set, such that the optimal cloning machine that achieves equal fidelity for them is equivalent to the UQCM? Jing \emph{et al.} \cite{Jing2013} solved this problem in the $1\rightarrow2$ cloning case. The minimal set turns out to be four states on the vertex of a tetrahedron:
\begin{eqnarray}
|\psi _1\rangle &=&|0\rangle ,
\nonumber \\
|\psi _2\rangle &=&\cos(\theta/2)|0\rangle+\sin(\theta/2)|1\rangle,
\nonumber \\
|\psi _3\rangle &=&\cos(\theta/2)|0\rangle+\sin(\theta/2)e^{2i\pi/3}|1\rangle,
\nonumber \\
|\psi _4\rangle &=&\cos(\theta/2)|0\rangle+\sin(\theta/2)e^{4i\pi/3}|1\rangle ,
\end{eqnarray}
where $\theta$ satisfies $\cos(\theta/2)=\frac{\sqrt{3}}{3}$, see FIG.\ref{4states}.

There is a similar phenomenon for states on the equator of the Bloch sphere, which will be demonstrated in the following section.

\subsection{Other developments and related topics}

The quantum cloning machine is originally proposed by considering arbitrary input states, thus it
in general has the universal property \cite{PhysRevLett.81.5003,PhysRevA.54.1844}.
For qubit case, by mapping several identical pure states into the cloning
of output states assisted by ancillary states, the general universal cloning machine
is realized \cite{PhysRevLett.79.2153}. The optimality of the fidelity is later
proved by considering that the case of infinite copies should be equivalent to
quantum state estimation \cite{PhysRevLett.81.2598}.
Along this line, the one to many universal cloning machine for higher-dimensional case is also studied in \cite{ShaomingFei2000}.
By using the symmetric projection on identical pure states and tensor product of the identities,
which are completely mixed states instead of the intuitively assumed blank states,  Werner proposed
the optimal universal cloning \cite{PhysRevA.58.1827}. The fidelity between all copies of the output
density operator with the ideal copies is used as the figure of merit for this cloning machine.
The optimal fidelity between single copy and a single input state is later obtained \cite{ISI:000081019000005}.
The cloning of higher dimensional state is also studied independently in \cite{zanardi}.
Equivalently but differently by explicit transformation method, the general many to many universal
quantum cloning of higher dimensional state is proposed in \cite{PhysRevA.64.064301}.
The combination of this universal cloning machine with the one by projection method \cite{PhysRevA.58.1827}
is proposed resulting in a unified universal cloning machine \cite{PhysRevA.84.034302}.
Also,the fidelities which range from cases of single copy to multiple copies are all obtained.
Let us remark that the well-accepted theory of fidelity for mixed states can be found in \cite{Josza1994}.

The topic of universal quantum cloning is well studied.
Next, we try to list those closely related developments in two directions.
The first direction is in general about the concepts extension from universal quantum cloning.
The second direction is about the realization of quantum cloning by various schemes
and with various noises.

Let us first list the topics which can be related with universal quantum cloning,
some of those topics may leads to potential applications.

\begin{itemize}

\item
State estimation.
The state estimation is corresponding to one to infinity quantum cloning, and it roughly describes how to find the exact form of an unknown state. Asymptotically, the quantum cloning machine corresponds to state estimation \cite{prl80.1571,PhysRevLett.81.2598,prl97.030402,Bruss1999}.
The state estimation can be understood like the asymmetric quantum cloning
which keeps one copy untouched and the rest infinite copies are used to estimate the form of the input state.
It can be expected that the precision of estimation and the fidelity of the remained copy has a tradeoff relation.
This relation means that the more information gained from the estimation, the larger disturbance is induced to
the remained copy. This phenomena is demonstrated by a tradeoff relation between the information gain and the disturbance on the estimated state \cite{prl86.1366}, also in \cite{ISI:000237147700049} and \cite{ISI:000254475300022}.

\item
Measurement.
The optimal minimal measurements of mixed states is studied in \cite{ISI:000081247000022}, which set the limits to optimal cloning of mixed states. Two incompatible observables cannot be measured simultaneously for a quantum system, the cloning schemes are studied for this task to accomplish it optimally \cite{measure-incompatible}. The trade-off relations between measurement accuracy of two or three non-commuting observables of a qubit system is studied in \cite{ISI:000252862000045}, this leads to the no-cloning inequality. The application of this method in quantum communication and the separability of quantum and classical information is studied in \cite{Ricci2005}.

\item
Quantum key distributions.
The security of QKD can be analyzed by using quantum cloning machine to attack
the protocol. This attack can be considered
as a simple quantum attack used by the eavesdropper, usually named as Eve.
She can use a quantum cloning machine to keep a copy of the state and send another
copy to the legitimated receiver. The strength of the attack can be adjusted
by using the asymmetric case of quantum cloning.
After the announcing of bases used by the sender and the receiver in the QKD protocol,
Eve can decode her copy, which in general is not perfect, to find the secrete key.
In extreme case, Eve may have a perfect copy but the state in the legitimated receiver side
will be random and thus this attack can be easily detected.
The universal quantum cloning machine is directly used to analyze the security of
six states QKD protocol \cite{PhysRevLett.81.3018}. It is shown that
if we want to clone those six states equally well, we cannot do better
than the universal quantum cloning machine. Interestingly, is is apparent that
those six states is only a subset of any arbitrary states. We can actually
go step further, one may find that the universal quantum cloning machine
can be determined completely by only four input states located on the vertices
of a tetrahedron inside the Bloch sphere \cite{Jing2013}.
The general $d$ dimension QKD by using
$d+1$ mutually unbiased states is investigated by universal cloning machine in \cite{PhysRevLett.88.127902},
see also a unified method in \cite{PhysRevA.85.012334}. One problem might be that what is the
minimal input set which can determine a universal cloning machine in higher dimensional case.
On the other hand, the figure of merit
of quantum cloning machine is by the fidelity of input and output states. By applying the quantum
cloning machine in QKD, the mutual information between pair of the sender and the receiver
in comparing with the pair between the eavesdropper and the sender are used.
The relationships between quantum cloning, eavesdropping of QKD and the Bell inequalities
are presented in \cite{GisinHuttner}.

\item
Cloning other than identical pure states: The universal cloning is in general
to study the quantum cloning of identical pure states. The aim can be extended to other
practical and interesting cases.
The problem of learning an unknown unitary transformation from a finite number of examples is related to, but different from cloning, which is studied in \cite{learning}. The cloning of a quantum measurement is studied in \cite{clonemeasure}.
The repeatable quantum channels with quantum memory is studied in \cite{Rybar2008}, this topic is something like quantum cloning of
quantum channels.
It is interestingly observed that a pair of qubits with anti-parallel spins may encode more quantum information \cite{PhysRevLett.83.432}, collective and local measurements of them is studied in \cite{massar-antispin}, the cloning of those kind of states is studied in \cite{cloning-orthogonalqubits}.
The upper bound of global fidelity for mixed state universal cloning and state-dependent cloning are obtained in \cite{ISI:000180804600031} and in \cite{ISI:000185716700021}. In relativistical quantum information, a trade-off relation is studied for universal cloning of qudit \cite{Jochym-OConnor2011}. The high fidelity copies from asymmetric cloning machine are studied in \cite{Siomau2010}.  The reverse of quantum cloning is also studied in photon stimulated emission scheme \cite{revsecloning} and in continuous variable system \cite{filipreverse}.
Several cases of qubit quantum cloning combinations are investigated in \cite{Wu2012}.

\item
Some applications of quantum cloning.
The superbroadcasting, which combines broadcasting and simultaneous purification of the local output states together, is investigated in \cite{ISI:000243894100042}.
The information transfer, and the information in practical cloning machine are presented in \cite{ISI:000087567900023} and \cite{ISI:000089688700023}. The information flux in many body system and in quantum cloning machine is studied in \cite{DiFranco2007}. The measurements on various subsystems of the cloning machine is studied in \cite{bruss-clone-measurement}.
The cloning is also related with optimizing the completely maps using semidefinite programming \cite{pra65.030302r}.
Numerical calculations are performed to study the relationships between fidelities of cloning machines and the entanglement \cite{Durt2011}. Related, the optimal realization of the transposition maps is studied in \cite{Buscemi2003374}.
The UQCM is also adopted to investigate the entanglement and the quantum coherence of the output field in the high-gain quantum injected parametric amplification
\cite{Caminati2006}.
The application of cloning machine to improve the detectors is in \cite{ISI:000085168800007}.
On the other hand, there are also some no-go theorems. Non-existence of a universal quantum machine to examine the precision of unknown quantum states, which is related to UQCM, is investigated \cite{ISI:000298380700008}.

\end{itemize}

Secondly, we present the results about the realization of universal quantum cloning.

\begin{itemize}

\item
The optimal quantum cloning model is only proposed by considering ideal condition,
but in order to make practical cloning, we have to consider effects such as noise.
The introduction of interference in UQCM is investigated,
it is shown that this interference does not diminish the optimal fidelity $5/6$ for
1 to 2 qubit symmetric cloning \cite{ISI:000260574100041}. If the ancillary state is not ideally initialed, its effect on the optimal UQCM is studied in \cite{Zhang2012}. The influence of temperature in quantum cloning is analyzed in \cite{pla373.821}.
The comparison of fidelities of quantum cloning expressed in theory and under experimental conditions is investigated in \cite{ISI:000221424900003}. The possibility to improve the fidelity of the UQCM in the photon stimulated emission scheme is studied in \cite{ISI:000170297300031}.

\item
The ultimate aim of QCM is to be physically implemented, and many proposals have been put up.
The spin networks is possible to realize the UQCM \cite{cloning-spinnetwork}.
The realization of UQCM is also proposed in optical system \cite{PhysRevLett.92.047902,filip1,filip2}.
The Hamiltonian realization of UQCM via adiabatic evolution is proposed,
the maximal eigenvalue of this Hamiltonian matrix is the fidelity \cite{ISI:000273639000005}. The proposals to implement cloning machines in separate cavities are in \cite{Fang2011}, by superconducting quantum-interference device qubits in a cavity is presented in \cite{ISI:000254541100254}. The scheme for implementing a UQCM in cavity QED with atoms is studied in \cite{ISI:000223715500003}, by ion trap technique is proposed in \cite{ISI:000230374500016}, by cavity-assisted atomic collisions is proposed in \cite{Zou2003}, via cavity-assisted interaction is studied in \cite{Fang2012}.
The scheme of quantum cloning of atomic state into two photonic states is presented in \cite{Song2008}.
The broadband photon cloning and the entanglement creation of atoms in waveguide is studied in \cite{ValenteLi}. As we can see, photonic system can play an important role in implementing QCM, a recent review about photonic quantum information processing can be found in \cite{REM-DeMartini,REM-Pan}.

\item
Some experiments have realized successfully the universal quantum cloning.
The universal cloning by entangled parametric amplification is studied in \cite{DeMartini2000}. The asymmetric UQCM is realized experimentally by partial teleportation \cite{Zhaozhi05}. The asymmetric quantum cloning machine is realized experimentally by polarization states of single photons \cite{Cernoch2009a}. The experimental quantum cloning can also be realized by using photon's orbital angular momentum \cite{nagali}. If both polarization and orbital angular momentum degrees of freedom of photons are used, the four-dimensional quantum state can be encoded. The experimental cloning of four-dimension state by this scheme is demonstrated in \cite{nagaliexperiment}. The general UQCM realized by projective operators and stochastic maps is investigated both theoretically and experimentally presented in \cite{stochastic}. By photon polarization in optics system, the universal quantum cloning and universal NOT gate is
implemented experimentally \cite{Sciarrino200434}.

\end{itemize}

\newpage

\section{Probabilistic quantum cloning}

Concerning about the B92 protocol which involves only two non-orthogonal states \cite{PhysRevLett.68.3121}, we
can try to clone it with the largest probability. That is, by measuring a
detector, we can make sure that the involved state is cloned perfectly or we
know that the cloning process fails. The aim of this quantum cloning
is to achieve the optimal probability.

\subsection{Probabilistic quantum cloning machine}

While the previous mentioned quantum cloning machines can always
succeed, on the same time, the copies cannot be perfect. Duan and
Guo \cite{Duan1998261,PhysRevLett.80.4999,ISI:000080184300011} proposed a different quantum cloning machine:
while the coping task can succeed with probability, but if it is
successful, we can always obtain perfect copies. This kind of
quantum cloning machine is called probabilistic quantum cloning
machine.

This quantum cloning machine is useful, in particular, in studying the B92 quantum
key distribution protocol \cite{PhysRevLett.68.3121}. In this QKD protocol,
only two non-orthogonal states are used for key distribution so the attack is
simply to use a specified quantum cloning machine to clone those two non-orthogonal states.
In fact, this is the simplest case for probabilistic quantum cloning machine which is used to
copy two linearly independent states $S=\{ |\Psi _0\rangle , |\Psi
_1\rangle \}$\cite{Duan1998261}. The cloning transformation can be proposed
as:
\begin{eqnarray}
U(|\Psi _0\rangle |\Sigma \rangle |m_p\rangle )= \sqrt {\eta
_0}|\Psi _0\rangle |\Psi _0\rangle |m_0\rangle +\sqrt {1-\eta
_0}|\Phi ^0_{ABP}\rangle , \nonumber \\
U(|\Psi _1\rangle |\Sigma \rangle |m_p\rangle )= \sqrt {\eta
_1}|\Psi _1\rangle |\Psi _1\rangle |m_1\rangle +\sqrt {1-\eta
_1}|\Phi ^1_{ABP}\rangle ,
\end{eqnarray}
where $|m_p\rangle ,|m_0\rangle \rangle ,|m_1\rangle $ are ancillary
states. The measurements are performed in these states. And the states $|\Phi ^0_{ABP}\rangle$ and $|\Phi ^1_{ABP}\rangle$
 are chosen so that the reduced state of $P$ is orthogonal to $|m_0\rangle$ and $|m_1\rangle$. When the
measurements are $|m_0\rangle $ or $|m_1\rangle $, we know the
states $S=\{ |\Psi _0\rangle , |\Psi _1\rangle \}$ are copied
perfectly. Otherwise, the cloning task fails. The probabilities of
success are $\eta _0$ and $\eta _1$ for states $|\Psi _0\rangle $
and $|\Psi _1\rangle $, respectively. If we let $\eta _0=\eta
_1=\eta $, we know that
\begin{eqnarray}
\eta \le \frac {1}{1+|\langle \Psi _0|\Psi _1\rangle |}.
\end{eqnarray}
This is also a no-cloning theorem: only orthogonal states can be
cloned perfectly. And the optimal probabilistic quantum cloning is
to let $\eta =1/(1+|\langle \Psi _0|\Psi _1\rangle |)$. It is also
related with the problem of how to distinguish non-orthogonal
quantum states.

The more complicated case is to copy a set of linearly independent
states $S=\{ |\Psi _0\rangle , |\Psi _1\rangle \,...,|\Psi _n\rangle
\} $. The form of the probabilistic cloning machine is:
\begin{equation}
U(|\Psi_i\rangle|\Sigma\rangle|P_0\rangle)=\sqrt{\gamma_i}|\Psi_i\rangle|\Psi_i\rangle|P_0\rangle+\sum_{j=1}^n c_{ij}|\Phi^j_{AB}\rangle|P_j\rangle.\label{DuanGuo}
\end{equation}
$P_0\dots P_n$ is a set of orthonormal ancilla states. Hence if the measurement result of the ancilla turns out to be $P_0$, we know the state $|\Psi_i\rangle$
is perfectly cloned, with the probability $\gamma_i$.
Taking the inner product of different $i$ and $j$ in (\ref{DuanGuo}), there's a matrix equation
\begin{equation}
X^{(1)}-\sqrt{\Gamma}X^{(2)}\sqrt{\Gamma}=CC^\dagger\label{DuanGuoCons}
\end{equation}
where $X^{(k)}_{ij}=\langle \Psi _i|\Psi _j\rangle^k$, $\Gamma=\Gamma^\dagger=diag\{\gamma_1\dots\gamma_n\}$, $C_{ij}=c_{ij}$. If the input states $|\Psi_i\rangle$s are not linearly independent, $X^{(1)}$ is not positive definite.
And for generic positive definite matrix $\Gamma$, the right-handed side of (\ref{DuanGuoCons}) is not positive semidefinite, hence the equation is not valid as the matrix $CC^\dagger$ is positive semidefinite.
Hence such probabilistic cloning machine only exists for linearly independent states(This result is also confirmed by Hardy and Song using the no-signaling argument\cite{Hardy1999331}).
They then found the existence is equivalent to the positive semidefiniteness of $X^{(1)}-\Gamma$. The result
is called the Duan-Guo bound to distinguish linearly independent quantum
states\cite{PhysRevLett.80.4999}. The $1\rightarrow M$ cloning machine is also easy to formulate, just by adding the number of copies at the right-handed side of (\ref{DuanGuo}).
 Later, Zhang \emph{et al.}\cite{ISI:000087567900029} constructed a network using universal quantum logic states realizing this cloning machine.

Later, Azuma \emph{et al.} \cite{PhysRevA.72.032335} studied the case with supplementary information, that is, the $|\Sigma\rangle$ at the left-handed side of (\ref{DuanGuo}) is state dependent.
Li and Qiu \cite{ISI:000244768000011} explored the case with two ancilla systems, but it is shown that the performance cannot be improved.

\subsection{A novel quantum cloning machine}

For probabilistic cloning machine, Pati \cite{PhysRevLett.83.2849} explored the possibility that the output state contains all possible copies of the original state.
That is, for a set of input states $|\Psi_1\rangle\dots|\Psi_n\rangle$, does there exist a transformation $U$ in the following form:
\begin{equation}
U(|\Psi_i\rangle|\Sigma\rangle|P_0\rangle)=\sum_{j=1}^M\sqrt{p^{(i)}_j}|\Psi_i\rangle^{\otimes(j+1)}|0\rangle^{\otimes(M-j)}
|P_j\rangle+\sum_{k=M+1}^{N'}\sqrt{f^{(i)}_k}|\Phi^k_{AB}\rangle|P_k\rangle.\label{Pati}
\end{equation}

$|P_1\rangle,\dots ,|P_{N'}\rangle$ is a set of orthonormal ancilla states, as usual. In fact, this can be regarded as a superposition of the $1\rightarrow 2,\cdots ,1\rightarrow (M+1)$ cloning machines.
From the unitarity constraint, we have
\begin{equation}
\langle \Psi_i|\Psi_j\rangle=\sum_{k=1}^{M}\sqrt{p^{(i)}_k}\langle \Psi_i|\Psi_j\rangle^{k+1}\sqrt{p^{(j)}_k}+\sum_{l=M+1}^{N'}\sqrt{f^{(i)}_l f^{(j)}_l}.
\end{equation}
This equation can be rewritten as a equation of matrices
\begin{equation}
X^{(1)}=\sum_{k=1}^M P_k X^{(k+1)} P_k+\sum_{l=M+1}^{N'}F^{(l)}.
\end{equation}
Here $P_k=P^\dagger_k=diag\{p^{(1)}_k,\dots,p^{(n)}_k\}$ ,$X^{(k)}_{ij}=\langle \Psi _i|\Psi _j\rangle^k$ as usual and $F^{(l)}_{ij}=\sqrt{f^{(i)}_l f^{(j)}_l}$.
From this relation, they proved if the states are linearly independent, then the equation can be satisfied with positive definite $P_k$s and $F_l$s.
It's also simple to see the transformation doesn't exist if the set of input state contains a state that is a superposition of other states says $|\Psi\rangle=\sum_j c_j|\Psi_j\rangle$,
 since we can simply add the $U(|\Psi\rangle|\Sigma\rangle|P_0\rangle)=\sum_j c_jU(|\Psi_j\rangle|\Sigma\rangle|P_0\rangle)$.
 And from the right-handed side of (\ref{Pati}) we can see that it is inconsistent.

Under this framework, the cloning machine of Duan and Guo can be viewed as a special case of $M=1$, see \cite{Duan1998261}.

Later, Qiu \cite{ISI:000175743800069} proposed a combination of Pati's probabilistic cloning machine and the approximate cloning machine in the usual sense, which is a more general framework.
The condition with supplementary information is also explored, that is, the $|\Sigma\rangle$ at the input side is state dependent.
It is found that the probability of success may increase\cite{ISI:000238380400032}.

\subsection{Probabilistic quantum anti-cloning and NOT gate}

Similar to the approximate universal NOT gate in the UQCM section of this review, we can also construct a probabilistic
quantum anti-cloning and NOT gate in the framework of probabilistic cloning.
Our aim for probabilistic quantum anti-cloning and NOT gate is that we keep the input state
unchanged, at the same time, we create additionally an anti-cloning state which corresponds
to the NOT gate. Actually, this task can only be fullfilled probabilistically.
A $1\rightarrow 2$ probabilistic quantum anti-cloning and NOT gate is proposed by Hardy and Song in \cite{Hardy1999331}:
\begin{equation}
U(|\Psi_i\rangle|\Sigma\rangle|P_0\rangle)=\sqrt{f}|\Psi_i\rangle|\Psi^\bot_i\rangle|P_0\rangle+\sum_{j=1}^n c_{ij}|\Phi^j_{AB}\rangle|P_j\rangle.\label{ProbNOT}
\end{equation}
By measuring the probe states $|P_0\rangle $ and $|P_j\rangle $ which are orthonormal, we know whether
the realization of anti-cloning and NOT gate is successful or not. On the other hand, we already
know that the perfect universal NOT gate is impossible \cite{PhysRevA.60.R2626},
the realization of only the NOT gate probabilistically seems an interesting question, $|\Psi _i\rangle \rightarrow |\Psi^\bot_i\rangle $.

The input states are $|\Psi_1\rangle,\dots,|\Psi_n\rangle$, as usual. Taking the inner product of different $i,j$, we get
\begin{equation}
X^{(1)}=fX'+CC^\dagger
\end{equation}
where $X'_{ij}=\langle\Psi_i|\Psi_j\rangle\langle\Psi^\bot_i|\Psi^\bot_j\rangle$ and other notation is same as above. If the input states are linearly independent, then the Gram matrix at the left-handed side of above equation is positive definite. Hence for a sufficiently small $f$, we can guarantee $CC^\dagger$ is also positive semidefinite. So such a cloning machine always exists. As a simple example, we consider the case $n=2, a_{11}=a_{21}=\sqrt{1-f}, a_{12}=a_{22}=0$, then (\ref{ProbNOT}) can be written as:
\begin{align}
|\Psi_1\rangle&\rightarrow \sqrt{f}|\Psi_1\rangle|\Psi^\bot_1\rangle|P_0\rangle+\sqrt{1-f}|\Phi_{AB}\rangle|P_1\rangle\nonumber\\
|\Psi_2\rangle&\rightarrow \sqrt{f}|\Psi_2\rangle|\Psi^\bot_2\rangle|P_0\rangle+\sqrt{1-f}|\Phi_{AB}\rangle|P_1\rangle.
\end{align}

In this case, we have a constraint of $f$:
\begin{equation}
f\leq\frac{1-|\langle\Psi_1|\Psi_2\rangle|}{1-|\langle\Psi_1|\Psi_2\rangle||\langle\Psi^\bot_1|\Psi^\bot_2\rangle|}=\frac{1}{1+|\langle\Psi_1|\Psi_2\rangle|}
\end{equation}
which is identical to the Duan and Guo case \cite{Duan1998261}. Li \emph{et al.} \cite{Li2007a} extended the above case to the case
where the output state contains all of $|\Psi\rangle|\Psi^{\bot}\rangle,|\Psi\rangle|\Psi^{\bot}\rangle^{\otimes2},
\dots,|\Psi\rangle|\Psi^{\bot}\rangle^{\otimes M}$.

\subsection{Other developments and related topics}

The probabilistic quantum cloning \cite{Duan1998261,PhysRevLett.80.4999} was initiated to study the attack on a
QKD protocol proposed by Bennett (B92) which can exploits any two nonorthogonal states for key distribution
\cite{PhysRevLett.68.3121}. Different from the approximate quantum cloning, the
probabilistic cloning aims to have perfect copies by sacrificing the success probability,
i.e., in case of failing, the state owned by eavesdropper is useless, but in case of success,
the eavesdropper will have perfect copies.
The aim of the probabilistic quantum cloning is equivalent to discriminate probabilistically
different quantum states such as the two nonorthogonal states in B92.
Then we next try to list three directions of research in studying probabilistic quantum cloning.

\begin{itemize}

\item
The probabilistic quantum cloning is equivalent the quantum states discrimination.
Along this line, the relation between the cloning machine
and states discrimination can be found, for example, in \cite{ISI:000230887300043,0305-4470-31-50-007,Zhang2010b}.
The minimum-error discrimination ambiguously of mixed states
is studied in \cite{ISI:000252862000060}. The optimal unambiguous discrimination
of two density matrices is studied in \cite{ISI:000231564200074}.
The optimal observables for minimum-error state discrimination
is studied in \cite{ISI:000282433200038}.
By homodyne detection, distinguishing two single-mode Gaussian states is studied in \cite{ISI:000228632100051}.
It is also shown that according to Wigner-Araki-Yanase theorem that the repeatability and distinguishability cannot be reached simultaneously \cite{ISI:000240238300141}. In order to distribute a quantum state to a coupled two subsystems,
the strength of interaction should be above a threshold \cite{ISI:000243166700173}.
The unambiguous discrimination of two squeezed states using probabilistic cloning is investigated in \cite{Mishra2012}.

\item

Probabilistic quantum cloning of various states and different methods have been presented.
Fiur\'{a}\v{s}ek \cite{PhysRevA.70.032308} used the technique described in the ``Singlet Monogamy'' subsection in the UQCM part to analyze the optimality of probabilistic cloning machine.
The study of probabilistic cloning of qubits with real parameters is shown in \cite{Zhang2010a}
The assisted probabilistic quantum cloning is proposed by Pati in \cite{ISI:000085336900037}.
The broadcasting of mixed state using probabilistic cloning machine is shown in \cite{ISI:000264951100014}.
The optimal probabilistic ancilla-free, which is economic, phase-covariant qudit telecloning machine
is presented in \cite{Wang2009a}.
The probabilistic cloning of three symmetric states \cite{Jimenez2010} and equidistant states \cite{Jimenez2010a}
are also studied.

\item
The implementation, both theoretically and probabilistically, of probabilistic quantum cloning is also an important subject.
The scheme to implement probabilistic cloning of qubits via twin photons is proposed in \cite{twin-photons}.
The scheme by GHZ states is proposed in \cite{ZhangLiWangGuo}.

Experimentally, the accuracy of quantum state estimation is studied \cite{ISI:000185192100032},
this accuracy is also compared with asymptotic lower bound obtained theoretically by Cram\'{e}r-Rao inequality.
The probabilistic quantum cloning experimentally realized in NMR system is reported in \cite{duprobabilistic}.
By generalizing the probabilistic cloning to state amplification,
the experimental heralded amplification of the photon polarized state and entanglement
distillation are reported in \cite{XiangGuoYong1} and \cite{XiangGuoYong2}.

\end{itemize}

\newpage

\section{Phase-covariant and state-dependent quantum cloning}

In last section, we studied the quantum cloning machines which are universal.
That is the case that the
input states are arbitrary or we know nothing about the input state. Practically, it is
possible that we already know partial information of the input state. The point is whether
this partial information is helpful or not for us to obtain a better fidelity in quantum cloning.
In this section, we will show that depending on specified input states, we can design
some quantum cloning machines which perform better for those restricted input states than that
the universal cloning machines.

On the other hand, one of the most important applications of quantum cloning is to analyze the security of
quantum key distribution protocols. In security analysis, the quantum states transfer through a quantum
channel. We suppose that this quantum channel is controlled by the eavesdropper who is generally named as Eve.
Eve can perform any operations which is allowed by quantum mechanics. One direct attack is the
``receive-measure-resend'' attack where ``measure'' can be supposed to be a quantum operation.
However, quantum mechanics states that non-orthogonal quantum states cannot be distinguished perfectly.
So the measured results will in general not be perfect and thus the obtained measurement result is not
the original sent state. This will induce inevitable errors which can be detected by public discussions
between the legitimated sender and receiver, Alice and Bob, in QKD.

Eve can choose freely her attack schemes. The quantum cloning machines provide a quantum scheme
of eavesdropping attack. We just assume that the Eve has
an appropriate quantum cloning machine. By quantum cloning, Eve can keep
one copy of the transferring state and send another copy to the legitimate receiver, Bob.
Now Eve and Bob both have copies of the sending state.
By this process, we can find how much information can be obtained by Eve, and on the
other hand, how much errors are induced by this attack. This provides a security analysis of
QKD. In this eavesdropping, Eve intends to get some information secretly between
Alice and Bob's communication and wish to make the least possibility to be detected.
So the optimal quantum cloning machine is required.
Based on different QKD protocols, various cloning machines are designed specially.
The universal quantum cloning machines studied in the previous sections themselves might be optimal.
But it may not be optimal for the quantum states involved in a special QKD protocol.
So the state-dependent quantum cloning machines are necessary.
In this section, we will give all the examples of state-dependent cloning.

\subsection{Quantum key distribution protocols}
In this subsection, we intend to refer some quantum key distribution protocols and show how the eavesdropper attacks them. Each protocol may lead to a special kind state-dependent cloning machine. Initial protocols are based simply on 2-dimension system and later they are generalized to higher-dimension.
Next, we present in detail the well-known BB84 protocol \cite{Bennett1984} and briefly its generalizations.
An earlier review of QKD is in \cite{QKDreview}.

\begin{enumerate}
\item BB84 protocol \cite{Bennett1984} uses two sets of orthogonal 2-level states and intersection angle in Bloch-sphere between different sets is $90^{\circ}$. They can be written as following, see FIG.\ref{figure-6-state},
\begin{eqnarray}
&&|0\rangle,|1\rangle,\nonumber\\
&&\frac{1}{\sqrt2}(|0\rangle+|1\rangle), \frac{1}{\sqrt2}(|0\rangle-|1\rangle).
\end{eqnarray}

Note that by operating a unitary transformation, characteristics of these states remain unchanged. Therefore, we may also use the following four states in BB84 protocol which are still two sets of orthogonal 2-level states with $90^{\circ}$ intersection angle, also see all those states in FIG.\ref{figure-6-state}.

\begin{eqnarray}
\frac{1}{\sqrt2}(|0\rangle\pm|1\rangle),\nonumber\\
\frac{1}{\sqrt2}(|0\rangle\pm\mathrm{i}|1\rangle).
\end{eqnarray}

In BB84 protocol,
Alice sends one of these four qubits to Bob through a certain quantum channel which is controlled by Eve.
After receiving the qubit, Bob measures the obtained qubit with one of the two bases randomly.
After Bob has finished his measurement, Alice would announce the bases of each qubits.
If their bases coincidence, Bob's measurement result is surely correct. Alice and Bob will share
a common secrete key. If sending basis and the measurement basis are different, they simply discord
this bit of information. Also they may use some qubits as the checking qubits
to find out the error rate introduced by the quantum channel.
They can suppose that all errors are caused by Eve's attack.

The eavesdropper, Eve, will capture the qubits in the quantum channel and clone them.
She remains one part to copies and still sends the other part to Bob in the quantum channel.
As soon as Alice broadcasts the bases, Eve measures her own qubits sequentially to
derive the information sent between Alice and Bob.

This BB84 protocol is proved to be unconditional secure and the security
is based on principles of quantum mechanics.
The security proofs of BB84 protocol are given by several
groups, for example Mayers \cite{Mayers}, Lo and Chau \cite{LoChau99}, Shor and Preskil \cite{Shor2000}.
We remark that Ekert proposed a QKD strategy based on the non-locality of quantum mechanics \cite{PhysRevLett.67.661} which
is the same of the BB84 protocol.

\item B92\cite{PhysRevLett.68.3121} is a protocol which uses any two non-orthogonal states.
Tamaki \emph{et al.} \cite{TKI03} provide the security proof of this protocol.
In this review, we suppose those two states take the form,
\begin{equation}
\frac{1}{\sqrt2}(|0\rangle+\mathrm{e}^{\mathrm{i}\phi_k}|1\rangle),k=1,2.
\end{equation}

\item BB84 protocol can be generalized to 6-state protocol\cite{PhysRevLett.81.3018}.
The six states involved in this protocol are BB84 states plus two more states as shown in Eq.(\ref{6states}).
Interestingly, the optimal cloning of those six states is
the universal quantum cloning machine as already
shown in previous sections.

\item For higher-dimensional case,
the QKD protocols can use $2$-basis or $d+1$-basis in a $d$-level system as studied by Cerf \emph{et al.} \cite{PhysRevLett.88.127902}.

\item In $d$-dimension, there are altogether $d+1$ mutually unbiased bases (MUB), provided $d$ is prime.
Any $(g+1)$-basis from those MUBs, $g=1,2,...,d$, can actually be used for QKD \cite{PhysRevA.85.012334}.
Here, we briefly give the definition of MUB,
$\{|i\rangle\}$ and $\{|\tilde{i}^{(k)}\rangle\}$
$(k=0,1, ...,d-1)$, they are expressed as:
\begin{equation}
|\tilde{i}^{(k)}\rangle=\frac{1}{\sqrt{d}}\sum_{j=0}^{d-1}\omega^{i(d-j)-ks_j}|j\rangle,
\end{equation}
with $s_j=j+...+(d-1)$ and $\omega=e^{i\frac{2\pi}{d}}$. These states
satisfy the condition, $|\langle\tilde{i}^{(k)}|\tilde{l}^{(j)}\rangle|=\delta_{kj}\delta _{il}
+\frac {1}{\sqrt {d}}(1-\delta _{kj})$. States in different set of bases are
mutually unbiased.

\item Basing on the characteristics of MUB, we can design a retrodiction protocol using method of mean king game.
This special protocol, different from BB84 or other QKD protocols,
shows that Bob has a 100\% successful measurement scheme in comparison with the $1/(g+1)$ successful measurement
in such as BB84 protocols. Here we remark that quantum memory is not available for Bob.
We will present a detailed analysis of this retrodiction protocol.
\end{enumerate}

\subsection{General state-dependent quantum cloning}
As to the above QKD protocols, universal quantum cloning machine is sure to work well, but not surely
to be the optimal one.
Thus if we need a higher quality of the output from the cloning machine,
a state-dependent cloning machine is needed. In fact, each protocol corresponds to a special kind of state-dependent cloning machine based on the given ensembles of states.

Let us firstly consider a general case based on two equatorial states.
Obviously, it is equivalent to the B92 protocol \cite{PhysRevLett.68.3121}.
To be non-trivial and satisfy the B92 protocol, these states are nonorthogonal.
The cloning machine is designed to clone only these two states optimally and equally well without considering other states on the Bloch sphere.
This problem is studied in \cite{Bruss1998}.

The quantum cloning machine takes a completely unknown 2-level state $|\psi\rangle$ and makes two output qubits.
Each output state is described by a reduced density matrix with the following form,
\begin{equation}
\rho=\eta|\psi\rangle\langle\psi|+(1-\eta)\frac{\mathbb{I}}{2}.
\end{equation}
Here, $\eta$ described the shrinking of the initial Bloch vector $\vec{s}$ corresponding to the density matrix $|\psi\rangle\langle\psi|$.
In other words, the output state is
\begin{equation}
\rho=\frac{\mathbb{I}+\eta\vec{s}\cdot\vec{\sigma}}{2}\nonumber ,
\end{equation}
with the input state being
\begin{equation}
|\psi\rangle\langle\psi|=\frac{\mathbb{I}+\vec{s}\cdot\vec{\sigma}}{2}.
\end{equation}

We assume that any quantum cloning machine satisfies the following reasonable conditions according
to requirement of all QKD protocols: First, $\rho_1=\rho_2$, which is called symmetry condition.
Second, $F=Tr(\rho_{\psi}\rho_1)=const.$, which is called isotropy condition
meaning that the fidelity between each output and the input does not depend on the specified form of the input.
Stronger condition $\vec{s_1}=\eta_{\psi}\vec{\psi}$ is required by orientation invariance of the Bloch vector. It is obvious that
when the last condition is satisfied, the second will be satisfied.

Next let us investigate the explicit form of the quantum cloning machine. Bru{\ss} \emph {et al.} \cite{Bruss1998}
make a general ansatz for the unitary transformation $U$ performed by the cloning machine.
They are,
\begin{eqnarray}
U|0\rangle|0\rangle|X\rangle=a|00\rangle|A\rangle+b_1|01\rangle|B_1\rangle+b_2|10\rangle|B_2\rangle+c|11\rangle|C\rangle,\\
U|1\rangle|0\rangle|X\rangle=\tilde{a}|11\rangle|\tilde{A}\rangle+\tilde{b_1}|10\rangle|\tilde{B_1}\rangle+\tilde{b_2}|01\rangle|\tilde{B_2}\rangle+\tilde{c}|00\rangle|\tilde{C}\rangle.
\end{eqnarray}
where $|X\rangle$ is an input ancilla. And $|A\rangle,|B_i\rangle,...$ denote output ancilla states. Ancilla states may have any dimension but are required to be normalized. There are several constraints for these parameters.
Thanks to the unitarity of the cloning transformation, the complex parameters $a,b_i,c...$ satisfy the normalization conditions:
\begin{eqnarray}
|a|^2+|b_1|^2+|b_2|^2+|c|^2=1,\nonumber\\
|\tilde{a}|^2+|\tilde{b_1}|^2+|\tilde{b_2}|^2+|\tilde{c}|^2=1,
\end{eqnarray}
and the orthogonality condition:
\begin{equation}
a^*\tilde{c}\langle A|\tilde{C}\rangle+b_2^*\tilde{b_1}\langle B_2|\tilde{B_1}\rangle+b_1^*\tilde{b_2}\langle B_1|\tilde{B_2}\rangle+c^*\tilde{a}\langle C|\tilde{A}\rangle=0.
\end{equation}
Assume that the cloning machine works in symmetric subspace, more relations are derived
\begin{eqnarray}
|b_1|=|b_2|,&&|\tilde{b_1}|=|\tilde{b_2}|,\nonumber\\
|\langle B_1|\tilde{B_2}\rangle|=|\langle B_2|\tilde{B_1}\rangle|,&&|\langle B_1|\tilde{B_1}\rangle|=|\langle B_2|\tilde{B_2}\rangle|,
\end{eqnarray}
and
\begin{eqnarray}
a b_1^*\langle B_1|A\rangle+c^*b_2\langle C|B_2\rangle=a b_2^*\langle B_2|A\rangle+c^*b_1\langle C|B_1\rangle,\nonumber\\
\tilde{b_1^*}a\langle \tilde{B_1}|A\rangle+\tilde{a^*}b_1\langle\tilde{A}|B_1\rangle=\tilde{b_2}a\langle\tilde{B_2}|A\rangle+\tilde{a}^*b_2\langle \tilde{A}|B_2\rangle,\nonumber\\
b_1^*\tilde{c}\langle B_1|\tilde{C}\rangle+c^*\tilde{b_1}\langle C|B_1\rangle=b_2^*\tilde{c}\langle B_2|\tilde{C}\rangle+c^*\tilde{b_2}\langle C|\tilde{B_2}\rangle.
\end{eqnarray}
Moreover, letting shrink factor remaining constant ratio within each direction in Bloch sphere, one has,
\begin{equation}
\frac{s_{1_x}}{s_{\psi_x}}=\frac{s_{1_y}}{s_{\psi_y}}=\frac{s_{1_z}}{s_{\psi_z}}=\eta_{\psi}.
\end{equation}
Applied in the transformation, we may derive further constraints:
\begin{eqnarray}
&&|a|^2-|c|^2=|\tilde{a}|^2-|\tilde{c}|^2\nonumber\\
&&|a|^2-|c|^2=Re[\tilde{b_1}^*a\langle\tilde{B_1}|A\rangle+\tilde{a}^*b_1\langle\tilde{A}|B_1\rangle],\nonumber\\
&&Im[\tilde{b_1}^*a\langle\tilde{B_1}|A\rangle+\tilde{a}^*b_1\langle\tilde{A}|B_1\rangle]=0,\nonumber\\
&&b_1^*\tilde{c}\langle B_1|\tilde{C}\rangle +c^*\tilde{b_1}\langle C|\tilde{B_1}\rangle=0,\nonumber\\
&&b_2^*a\langle B_2|A\rangle+c^*b_1\langle C|B_1\rangle=0,\nonumber\\
&&\tilde{b_2}^*\tilde{a}\langle \tilde{B_2}|\tilde{A}\rangle+\tilde{c}^*\tilde{b_1}\langle \tilde{C}|\tilde{B_1}\rangle=0,\nonumber\\
&&\tilde{c}^*a\langle\tilde{C}|A\rangle-\tilde{a}^*c\langle\tilde{A}|C\rangle=0,\nonumber \\
{\rm and~symmetrically,~}&& 1\leftrightarrow 2.
\end{eqnarray}
Here, notation $1\leftrightarrow2$ indicates the above constraints changing indices 1 with 2 according to the symmetry condition.

Our task is to maximize the shrinking factor $\eta$ with its explicit form taken as,
\begin{equation}
\eta=|a|^2-|c|^2.
\end{equation}
The fidelity which is defined as
\begin{equation}
F=Tr(\rho_1|\psi\rangle\langle\psi|)=\frac{1}{2}(1+\vec{s_1}\cdot\vec{s_{\psi}})
\end{equation}
is related to the shrinking factor as
\begin{equation}
F=\frac{1}{2}(1+\eta).
\end{equation}
Note that that this relationship between fidelity and shrinking factor holds only for pure states.
The study of mixed state has already been presented in the previous sections.
The above discussions are regardless of the specified QKD protocols.

\subsection{Quantum cloning of two non-orthogonal states}
Next, we will consider the situation of B92 protocol in which only two qubits are required to be cloned.
Now, we firstly prove that ancilla is necessary in our cloning machine.
Without ancilla, we could write down constraints as:
$|a|^2-|c|^2=|\tilde{a}|^2-|\tilde{c}|^2$,
$|a|^2-|c|^2=Re[\tilde{b_1}^*a+\tilde{a}^*b_1]$,
$b_2^*a+c^*b_1=0$,
and $\tilde{b_2}^*\tilde{a}+\tilde{c}^*\tilde{b_1}=0$.
Adding symmetric ansatz, we have
$|b_1|=|b_2|=|b|$
and $|\tilde{b_1}|=|\tilde{b_2}|=|\tilde{b}|$.

From these constraints we would have four possible results:
(a),$|a|=|c|$ and $|\tilde{a}|=|\tilde{c}|$,
(b), $|a|=|c|$ and $|\tilde{b}|=0$,
(c), $|b|=0$ and $|\tilde{a}|=|\tilde{c}|$,
and (d), $|b|=0$, and $|\tilde{b}|=0$.
For each case, we have $\eta=0$ which seems meaningless.
Consequently, it is impossible to generate a symmetric quantum cloning machine without ancilla.

In the following, we will explicitly give the form of the quantum cloning machine and the fidelity
in this case.
Assume two pure states in a two-dimensional Hilbert space with expressions:
\begin{eqnarray}
|a\rangle &=&\cos\theta|0\rangle+\sin\theta|1\rangle,\\
|b\rangle &=& \sin\theta|1\rangle+\cos\theta|0\rangle,
\end{eqnarray}
where $\theta$ varying from $0$ to $\pi/4$.
Define $S=\langle a|b\rangle=\sin2\theta$.
We may imagine that the fidelity only dependents on $S$ because we could transform every 2 states into the above form by only unitary operation without influence the fidelity.

Since there are too many constraints to give strict algebraic calculations,
we utilize the symmetry in the B92 protocol to simplify the calculations.
Performing an unitary operator $U$ on the input states, we define final states $|\alpha\rangle$ and $|\beta \rangle$ as
\begin{eqnarray}
|\alpha\rangle= U|a\rangle|0\rangle,
|\beta\rangle=U|b\rangle|0\rangle.
\end{eqnarray}
Since $U$ is an unitary transformation, we could derive
\begin{equation}
\langle\alpha|\beta\rangle=\langle a|b\rangle=\sin2\theta=S.
\end{equation}

Using global fidelity $F_g$ to evaluate the quantum cloning, which is defined as
\begin{equation}
F_g=\frac{1}{2}(|\langle\alpha|aa\rangle|^2+|\langle\beta|bb\rangle|^2)
\end{equation}
Certainly, optimal cloning machine needs
that both $|\alpha\rangle$ and $|\beta\rangle$ lying in the space spanned by vectors $|aa\rangle$ and $|bb\rangle$.
Without complicated calculations, we would obtain maximal global fidelity as
\begin{equation}
F_g=\frac{1}{4}(\sqrt{1+\sin^22\theta}\sqrt{1+\sin2\theta}+\cos2\theta\sqrt{1-\sin2\theta})^2.
\end{equation}

Additionally, we are also interested with the local fidelity of each output qubit with the input one, which is defined as
\begin{equation}
F_l=Tr[\rho_\alpha|a\rangle\langle a|].
\end{equation}
The explicit result is,
\begin{equation}
F_{l,1}=\frac{1}{2}[1+\frac{1-S^2}{\sqrt{1+S^2}}+\frac{S^2(1+S)}{1+S^2}].
\end{equation}
We may notice that it is larger than $5/6$. That is to say, for this protocol,
state-dependent cloning machine works better than UQCM as expected.
It is also noticed that the Bloch vector not only shrinks  but also makes a rotation with a state-dependent angle $\vartheta$:
\begin{equation}
\vartheta=\arccos[\frac{1}{|\vec{s}|}\frac{\cos2\theta}{\sqrt{1+\sin^22\theta}}]-2\theta.
\end{equation}
This is caused by that one constraint presented previously is released.

We should emphasize that this result is derived under the request of maximum global fidelity rather than maximum local fidelity.
When we only need a better state-dependent cloning machine locally, we may have different consequences.
 And the fidelity is given by:
\begin{eqnarray}
F_{l,3}=\frac{1}{2}+\frac{\sqrt2}{32S}(1+S)(3-3S+\sqrt{1-2S+9S^2})\times\sqrt{-1+2S+3S^2+(1-S)\sqrt{1-2S+9S^2}}.
\end{eqnarray}
Moreover, it could be tested that the minimum value $F_{l,3}\approx0.987$ is derived when $S=1/2$.
And when $S=0$ and $S=1$, one finds $F=1$ as expected.

In addition, we should note that different concerning in the eavesdropping would lead to variant results.
In B92 protocol, direct cloning is not the most advisable action for Eve if she wishes to be most surreptitious.
In fact, Eve's main purpose is not to clone the quantum information which is embodied in the two nonorthogonal quantum states,
but rather to optimize the trade-off between obtaining most classical information versus making the least disturbance on the original qubit\cite{PhysRevA.53.2038}.
We may name it the optimal eavesdropping which is different from optimal cloning. In \cite{Bruss1998}, fidelity for optimal eavesdropping is expressed as
\begin{equation}
F_{l,2}=\frac{1}{2}+\frac{\sqrt2}{4}\sqrt{(1-2S^2+2S^3+S^4)+(1-S^2)\sqrt{(1+S)(1-S+3S^2-S^3)}}.
\end{equation}
Note that, for all $S$, $F_{l,2}\geq F_{l,3}$.

Here we have a short summary, the general state-dependent cloning machine works better than UQCM when
applied to a certain number of states. We give the special case of two nonorthogonal pure states. It is obviously that,
if we know the ensemble of states used in one QKD protocol, state-dependent cloning machine can be designed accordingly.
Besides for QKD protocols, various quantum machines themselves are of fundamental interests.
As an extension of B92 protocol, Koashi and Imoto considered the quantum cryptography by two mixed states \cite{koashi-crypto-mixed}.

\subsection{Phase-covariant quantum cloning: economic quantum cloning for equatorial qubits}

In this subsection, we will discuss quantum cloning machine for BB84 states, which
is first studied in \cite{Bruss2000}.
For convenience, we will also refer those four states $\{ (|0\rangle \pm |1\rangle )/\sqrt {2}$,
$(|0\rangle \pm i|1\rangle )/\sqrt {2}\} $ as the BB84 states.
In fact, the cloning machine of BB84 states is proved to be able to copy
all equatorial states optimally.
It has a higher fidelity than that of the UQCM. Moreover, this kind of quantum cloning machine is able to
work without the help of the ancilla states. It is thus the economic quantum cloning.

\begin{figure}
\includegraphics[height=5cm]{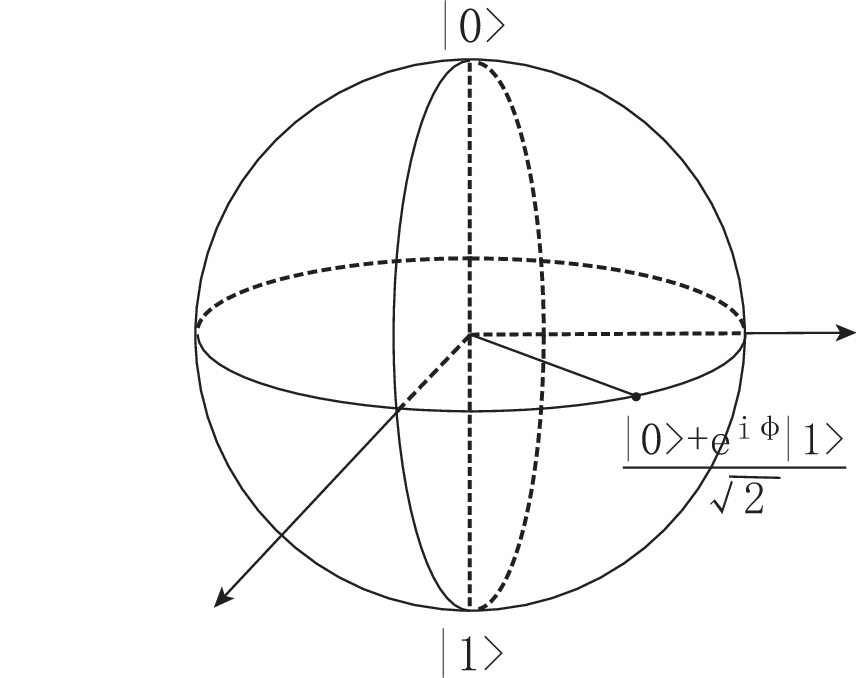}
\caption{The equatorial qubits are qubits which are located on the equtor of the
Bloch sphere. The optimal cloning machine for those states is the phase-covariant quantum
cloning machine. This machine is also optimal for cloning BB84 like states.}
\label{figure-equatorial}
\end{figure}

It is interesting to find that any quantum cloning machine that clones BB84 states equally well
will also clone equatorial states with the same fidelity. We know that the equatorial
qubits are located on the equator
of the Bloch sphere which take the form, $|\psi (\phi )\rangle =(|0\rangle +e^{i\phi }|1\rangle )/\sqrt {2}$
see FIG.\ref{figure-equatorial}.
Since each output qubit can be represented as the mixture of input state and the completely mixed state and
the corresponding fidelity does not depend on the
phase $\phi $, this kind of cloning machine is ``phase covariant''. It is named generally as the phase-covariant
quantum cloning machine.

Consider a completely positive map $T$ that could clone optimally the four states of BB84.
Perform $T$ on those states would lead to approximate result:
\begin{eqnarray}
T[|\pm x\rangle\langle\pm x|]=\eta |\pm x\rangle\langle\pm x|+(1-\eta)\frac{\mathbb{I}}{2},\\
T[|\pm y\rangle\langle\pm y|]=\eta |\pm y\rangle\langle\pm y|+(1-\eta)\frac{\mathbb{I}}{2}.
\end{eqnarray}

On the other hand, equatorial states could be written as
\begin{equation}
|\psi(\phi)\rangle\langle\psi(\phi)|=\frac{1}{2}(\mathbb{I}+\cos\phi \sigma_x+\sin\phi \sigma_y).
\label{xyequator}
\end{equation}
This is the qubits in $x-y$ equator. Similarly we have qubits in $x-z$ equator such as BB84 states
and in $y-z$ equator.
Perform linear operation $T$ on it and consider $T(\mathbb{I})=\mathbb{I}$, we derive
\begin{equation}
T[|\psi(\phi)\rangle\langle\psi(\phi)|]=\eta|\psi(\phi)\rangle\langle\psi(\phi)|+(1-\eta)\frac{\mathbb{I}}{2}
\label{scalar0}
\end{equation}
The shrinking factor $\eta$ remains unchanged.
Therefore, we could conclude that optimal cloning machine performed for
the BB84 states is equivalent to phase covariant cloning machine.

Next, we release the above constraint that the single output qubit takes
the scalar form (\ref{scalar0}). We only need that the fidelity does not
depend on the phase parameter $\phi $ in quantum cloning.
We fist consider the economic case, which is accomplished without ancilla.
Phase-covariant quantum cloning machine is presented in the following
as proposed by Niu and Griffiths \cite{PhysRevA.60.2764},
\begin{eqnarray}
&&|0\rangle|0\rangle\rightarrow|0\rangle|0\rangle,\nonumber \\
&&|1\rangle|0\rangle\rightarrow\cos\eta|1\rangle|0\rangle+\sin\eta|0\rangle|1\rangle,
\label{economic}
\end{eqnarray}
where $\eta\in[0,\pi/2]$ means the asymmetry between the two output states. And when $\eta=\pi/4$ the two output states are equivalent,
corresponding to the symmetric case.

For any equatorial state $|\psi(\phi)\rangle=\frac{1}{\sqrt2}(|0\rangle+e^{i\phi}|1\rangle)$ which is the input state,
we have
\begin{equation}
|\psi({\phi})\rangle|0\rangle\rightarrow\frac{1}{\sqrt2}(|00\rangle+\cos\eta e^{i\phi}|10\rangle+\sin\eta e^{i\phi}|01\rangle).
\end{equation}
So we could easily obtain the reduced matrix of each states,
\begin{eqnarray}
\rho_A=Tr_B(|\psi({\phi})\rangle\langle\psi({\phi})|)\nonumber\\
\rho_B=Tr_A(|\psi({\phi})\rangle\langle\psi({\phi})|).
\end{eqnarray}
Then, as to any equatorial state $|\psi(\phi)\rangle$, we have fidelity defined as $F=\langle\psi|\rho|\psi\rangle$:
\begin{eqnarray}
F_A=\frac{1}{2}(1+\cos\eta),\\
F_B=\frac{1}{2}(1+\sin\eta).
\end{eqnarray}
Obviously, fidelities are independent of $\phi$ as expected.
Particularly, for symmetric case $\eta=\pi/4$, the fidelity is
\begin{eqnarray}
F=1/2+1/\sqrt8\approx0.85355>\frac{5}{6}\approx 0.833333.
\label{phaseoptimal}
\end{eqnarray}
In other words, phase covariant cloning machine behaves better than UQCM in cloning equatorial states.
Phase-covariant quantum cloning machine can also be realized with ancillary states in a different form.
The related results of phase cloning can be found in \cite{PhysRevA.56.1173,Bruss2000,Acin2004a,PhysRevA.69.062316}. The experimental implementation
of this scheme is reported in optics system and nuclear magnetic resonance system \cite{Cernoch2006,PhysRevLett.94.040505}.

\subsection{One to many phase-covariant quantum cloning machine for equatorial qubits}

For quantum cloning, we are always interested in the case that multi-copies created from
some fewer identical input states. The simplest extension of $1\rightarrow 2$
is one to many quantum cloning,
i.e. $1\rightarrow M$
phase-covariant quantum cloning.
Based on the cloning transformations similar to the
UQCM \cite{PhysRevLett.79.2153}, for arbitrary equatorial qubits,
$|\Psi \rangle =(|\uparrow \rangle +e^{i\phi }|\downarrow
\rangle )/\sqrt{2}$, it is assumed that the cloning transformations take the
following form \cite{PhysRevA.65.012304},
\begin{eqnarray}
U_{1,M}|\uparrow \rangle \otimes R&=&
\sum _{j=0}^{M-1}\alpha _j|(M-j)\uparrow , j\downarrow \rangle \otimes R_j,
\nonumber \\
U_{1,M}|\downarrow \rangle \otimes R&=&
\sum _{j=0}^{M-1}\alpha _{M-1-j}
|(M-1-j)\uparrow , (j+1)\downarrow \rangle \otimes R_j,
\label{1m}
\end{eqnarray}
where $R$ denotes the initial state of the copy machine and $M-1$ blank
copies, $R_j$ are orthogonal normalized states of the ancillary (ancilla),
and $|(M-j)\psi, j)\psi _{\perp}\rangle $ denotes the symmetric and
normalized state with $M-j$ qubits in state $\psi $ and $j$ qubits
in state $\psi _{\perp}$. We already know the result of
universal case: For arbitrary input state, the case
$\alpha _j=\sqrt{\frac {2(M-j)}{M(M+1)}}$ is the optimal
$1\rightarrow M$ universal quantum cloning \cite{PhysRevLett.79.2153}.

Next we consider the case that the input states being
restricted to the equatorial qubits. It is assumed that
phase-covariant transformations satisfy some properties:
it possesses the orientation invariance of the Bloch vector
and that the output states are in symmetric subspace which
naturally ensure that we have identical copies.
The unitarity and the normalization is satisfied by $\sum _{j=0}^{M-1}\alpha _j^2=1$.
We now wish that the optimal phase-covariant cloning machine can be achieved.
Let us see fidelity which is found to take the form,
\begin{eqnarray}
F=\frac {1}{2}[1+\eta (1,M)],
\end{eqnarray}
where
\begin{eqnarray}
\eta (1,M)=\sum _{j=0}^{M-1}\alpha _j\alpha _{M-1-j}
\frac {C_{M-1}^j}{\sqrt{C_M^jC_M^{j+1}}}.
\end{eqnarray}
From this result, it is straightforward to examine two special cases, $M=2,3$.
For $M=2$, we have
$\alpha _0^2+\alpha _1^2=1$ and
$\eta (1,M)=\sqrt{2}\alpha _0\alpha _1$. In case
$\alpha _0=\alpha _1=1/\sqrt{2}$, the fidelity achieves the maximum.
For $M=3$, we have $\alpha _0^2+\alpha _1^2+\alpha _2^2=1$,
and
\begin{eqnarray}
\eta (1,3)=\frac {2}{3}\alpha _1^2
+\frac {2}{\sqrt{3}}\alpha _0\alpha _2.
\end{eqnarray}
For $\alpha _0=\alpha _2=0, \alpha _1=1$, we have
$\eta (1,3)=\frac {2}{3}$, which is the optimal value
and it reproduces the case of
quantum triplicator for $x-y$ equatorial qubits as presented below,
\begin{eqnarray}
|\ua \rangle \rightarrow \frac {1}{\sqrt {3}}(|\ua \ua \da\rangle +|\ua \da \ua \rangle
+|\da \ua \ua \rangle ),\nonumber \\
|\da \rangle \rightarrow \frac {1}{\sqrt {3}}(|\da \da \ua\rangle +|\da \ua \da \rangle
+|\ua \da \da \rangle ).\nonumber \\
\end{eqnarray}
Note that the fidelity of this quantum triplicator is $5/6$ which is the same
as the $1\rightarrow 2$ UQCM.

We next review the result of
1 to $M$ phase-covariant quantum cloning transformations.
When $M$ is even, we suppose
$\alpha _j=\sqrt{2}/2, j=M/2-1, M/2$
and $\alpha _j=0$, otherwise.
When $M$ is odd, we can suppose
$\alpha _j=1, j=(M-1)/2 $
and $\alpha _j=0$, otherwise.
The corresponding fidelities are
$F=\frac {1}{2}+\frac {\sqrt{M(M+2)}}{4M}$ for M is even,
and $F=\frac {1}{2}+\frac {(M+1)}{4M}$ for M is odd.
The explicit cloning transformations have already been presented in
(\ref{1m}).

The above fidelities for $M=2, 3$ cases can be found easily being optimal.
We next prove that
for general $M$, the fidelities shown above achieve the maximum as well \cite{PhysRevA.65.012304}.
As we just reviewed, the method introduced in \cite{PhysRevLett.79.2153} for UQCM can also be applied
in this phase-covariant case.
Here, we try to present a more general formula
by considering $N$ to
$M$ cloning transformation. This formula incorporates the coefficients
in the unitary transformation to the un-normalized ancillary states.
We expect that this formula can be used to study the general optimal $N\rightarrow M$
phase-covariant quantum cloning which is still an open question.
We then will reduce from the general formula to the simple
case $N=1$ to reach our conclusion.

By expansion, the $N$ identical
input states for equatorial qubits can be written as,
\begin{eqnarray}
|\Psi \rangle ^{\otimes N}=\sum _{j=0}^Ne^{ij\phi }\sqrt{C_N^j}
|(N-j)\uparrow ,j\downarrow \rangle .
\label{phaseinput}
\end{eqnarray}
The most general $N$ to $M$ quantum cloning machine for equatorial qubits is expressed as
\begin{eqnarray}
|(N-j)\uparrow ,j\downarrow \rangle \otimes R
\rightarrow
\sum _{k=0}^M|(M-k)\uparrow ,k\downarrow \rangle \otimes |R_{jk}\rangle ,
\label{phasecloning}
\end{eqnarray}
where $R$ still denotes the $M-N$ blank copies and the initial
state of the cloning machine, and $|R_{jk}\rangle $ are unnormalized final states of
the ancilla. By using the unitarity condition, we know that the ancillary states
should satisfy the following condition,
\begin{eqnarray}
\sum _{k=0}^M\langle R_{j'k}|R_{jk}\rangle =\delta _{jj'}.
\end{eqnarray}
Substitute the input state (\ref{phaseinput}) into the cloning transformation
(\ref{phasecloning}), we obtain the whole output state with ancillary state,
\begin{eqnarray}
|\Psi \rangle ^{\otimes N}\rightarrow \sum _{j=0}^Ne^{ij\phi }\sqrt{C_N^j}\sum _{k=0}^M|(M-k)\uparrow ,k\downarrow \rangle \otimes |R_{jk}\rangle .
\end{eqnarray}
By tracing out the ancillary state, the output state of $M$-qubit takes the form,
\begin{eqnarray}
\rho ^{out}=\sum _{j',k',j,k}e^{i(j-j')\phi }\sqrt {C_N^jC_N^{j'}}\langle R_{j'k'}|R_{jk}\rangle
 |(M-k)\uparrow ,k\downarrow \rangle \langle (M-k')\uparrow ,k'\downarrow |.
\end{eqnarray}
The one-qubit reduced density operators are the same which is ensured by
the symmetric space representation.
The fidelity between input and output of one-qubit can then be calculated as
\begin{eqnarray}
F=\langle \Psi |\rho ^{out}_{red.}|\Psi \rangle
=\sum _{j',k',j,k}\langle R_{j'k'}|R_{jk} \rangle A_{j'k'jk},
\end{eqnarray}
where $\rho ^{out}_{red.}$ is the density operator of each output qubit by
taking partial trace over $M-1$ output qubits with only one qubit left.
We impose the condition that the output density operator has
the property of Bloch vector invariance, and also we next consider
the simple case $N=1$,
\begin{eqnarray}
A_{j'k'jk}&=&\frac {1}{4}\{ \delta _{j'j}\delta _{k'k} +(1-\delta
_{j'j})[\delta _{k',(k+1)}\frac {\sqrt{(M-k)(k+1)}}{M} \nonumber \\
&& +\delta _{k,(k'+1)}\frac {\sqrt{(M-k')(k'+1)}}{M}]\},
\label{amatrix}
\end{eqnarray}
where $j,j'=0, 1$ for case $N=1$. The optimal fidelity of this cloning machine
for equatorial qubits corresponds to the maximal eigenvalue $\lambda
_{max}$ of matrix $A$ by $F=2\lambda _{max}$ \cite{PhysRevLett.79.2153}. The
matrix $A$ (\ref{amatrix}) is a block diagonal matrix with block $B$
given by,
\begin{eqnarray}
B=\frac {1}{4}\left( \begin{array}{cc}
1 & \frac {\sqrt {(M-k)(k+1)}}{M}\\
\frac {\sqrt {(M-k)(k+1)}}{M} & 1\end{array}\right) .
\end{eqnarray}
Thus we now can confirm that the optimal fidelities
of 1 to $M$ cloning machine for equatorial qubits takes the form,
\begin{eqnarray}
F=2\lambda _{max}=\left\{ \begin{array}{l}
\frac {1}{2}+\frac {\sqrt{M(M+2)}}{4M}, ~~~~{\rm M~is~even},\\
\frac {1}{2}+\frac {(M+1)}{4M}, ~~~~~~~~~{\rm M~is~odd}.\end{array}
\right.
\end{eqnarray}
Explicitly, the corresponding $1\rightarrow M$ optimal phase-covariant quantum
cloning can be written as:
\begin{enumerate}
\item M is even, suppose $M=2L$, we have
\begin{eqnarray}
|\uparrow \rangle &\rightarrow &\frac {\sqrt {2}}{2}|(L+1)\uparrow ,(L-1)\downarrow \rangle \otimes R_0+
+\frac {\sqrt {2}}{2}|L\uparrow ,L\downarrow \rangle \otimes R_1,
\nonumber \\
|\downarrow \rangle &\rightarrow &\frac {\sqrt {2}}{2}|L\uparrow ,L\downarrow \rangle \otimes R_0
+\frac {\sqrt {2}}{2}|(L-1)\uparrow ,(L+1)\downarrow \rangle \otimes R_1.
\label{withancilla}
\end{eqnarray}
\item M is odd, suppose $M=2L+1$, we have
\begin{eqnarray}
|\uparrow \rangle &\rightarrow &|(L+1)\uparrow ,L\downarrow \rangle ,
\nonumber \\
|\downarrow \rangle &\rightarrow &|L\uparrow ,(L+1)\downarrow \rangle .
\label{moddcase}
\end{eqnarray}
\end{enumerate}
Note that those transformations (\ref{withancilla}) have ancillary states $R_0, R_1$.
The simplest economic case without these ancillary states has been presented in (\ref{economic}).
The general economic case equivalent with Eq.(\ref{withancilla}) can be written as,
\begin{eqnarray}
|\uparrow \rangle &\rightarrow &|(L+1)\uparrow ,(L-1)\downarrow \rangle
\nonumber \\
|\downarrow \rangle &\rightarrow &|L\uparrow ,L\downarrow \rangle .
\label{generaleconomic}
\end{eqnarray}

The optimal phase-covariant quantum cloning for the general $N\rightarrow M$ case
still seems elusive, some related results and the phase-cloning of
qutrits can be found in \cite{PhysRevA.67.042306}.
The one to three
phase-covariant quantum cloning is realized in optics system \cite{1to3phasecloning}.

\subsection{Phase quantum cloning: comparison between economic and non-economic}

It seems that phase quantum cloning with input
$|\psi \rangle =\frac {1}{\sqrt{2}}(|0\rangle +e^{i\phi }
|1\rangle )$ can be realized by both economic and non-economic transformations
with completely the same optimal fidelity.
We suppose that qubit implemented by quantum device is precious,
so we should prefer to economic phase cloning.

On the other hand, there exist some subtle differences between those two
cases which are not generally noticed. For convenience, let us present
explicitly those transformations.
From the general results in Eq.(\ref{withancilla}),
the optimal phase-covariant cloning transformation takes the form,
\begin{eqnarray}
|0\rangle \rightarrow
\frac {1}{\sqrt{2}}|00\rangle |0\rangle _a
+{\frac 12}\left( |01\rangle +
|10\rangle \right) |1\rangle _a,
\nonumber \\
|1\rangle  \rightarrow
\frac {1}{\sqrt{2}}|11\rangle |1\rangle _a
+{\frac 12}\left( |01\rangle +
|10\rangle \right) |0\rangle _a,
\label{non-economic}
\end{eqnarray}
where the subindex $a$ denotes the ancillary state.
With the help of Eq.(\ref{generaleconomic}), the economic
phase-covariant cloning takes the following form, which is also presented
in Eq.(\ref{economic}) and here we choose asymmetric parameter $\eta =\pi /4$,
 \begin{eqnarray}
&&|0\rangle\rightarrow|0\rangle|0\rangle,\nonumber \\
&&|1\rangle\rightarrow \frac{1}{\sqrt {2}}(|1\rangle|0\rangle+|0\rangle|1\rangle),
\label{economic0}
\end{eqnarray}

We already know that the fidelities of both economic and non-economic are the same
and optimal, see Eq.(\ref{phaseoptimal}),
\begin{eqnarray}
F_{optimal}=\frac {1}{2}+\sqrt{\frac {1}{8}}.
\end{eqnarray}
The single qubit reduced density matrix of output from (\ref{non-economic}) can be
calculated as,
\begin{eqnarray}
\rho _{red.}=\frac {1}{\sqrt{2}}|\psi \rangle \langle \psi |+
\left( \frac {1}{2}-\sqrt{\frac {1}{8}}\right) \mathbb{I}
=\left( \begin{array}{ll}\frac {1}{2}& \frac {1}{\sqrt {8}}e^{-i\phi }\\
\frac {1}{\sqrt {8}}e^{i\phi }&\frac {1}{2}
\end{array}\right)
\label{scalar}
\end{eqnarray}
It takes the scalar form, i.e., the single output can be written as a mixture of
input qubit and a completely mixed state $\mathbb{I}/2$.

In comparison, the single qubit reduced density matrix of output from economic case is,
\begin{eqnarray}
\rho ^{eco.}_{red.}
=\left( \begin{array}{ll}\frac {3}{4}& \frac {1}{\sqrt {8}}e^{-i\phi }\\
\frac {1}{\sqrt {8}}e^{i\phi }&\frac {1}{4}
\end{array}\right).
\end{eqnarray}
This form does not satisfy the scalar form.
It also means relation $T(\mathbb{I})=\mathbb{I}$ is not satisfied.

In eavesdropping of well known BB84 QKD,
because all four states $|0\rangle ,|1\rangle , 1/\sqrt{2}(|0\rangle
+|1\rangle ), 1/\sqrt{2}(|0\rangle -|1\rangle )$ can be described by
$|\Psi \rangle =\cos \theta |0\rangle +\sin \theta |1\rangle $. So,
instead of the UQCM, we should at least use the cloning machine for
equatorial qubits in eavesdropping. Actually in individual attack,
we can not do better than the cloning machine for equatorial
qubits \cite{Bruss2000}. The cloning machine presented in
equations (\ref{economic0},\ref{non-economic}) can be used in analyzing the eavesdropping of other
two mutually unbiased bases $1/\sqrt{2}(|0\rangle -|1\rangle ),
1/\sqrt{2}(|0\rangle +|1\rangle ), 1/\sqrt{2}(|0\rangle +i|1\rangle
), 1/\sqrt{2}(|0\rangle -i|1\rangle )$ which belong to
$|\psi \rangle =(|0\rangle +e^{i\phi }|1\rangle )/\sqrt {2}$.

\subsection{Phase-covariant quantum cloning for qudits}

The phase quantum cloning can be applied to higher dimensional system.
For qutrit case, the optimal fidelity
was obtained by D'Ariano \emph {et al.}\cite{PhysRevA.64.042308} and Cerf \emph{et al.}\cite{688607420020710};
\begin{eqnarray}
F=\frac {5+\sqrt{17}}{12}, ~~~{\rm for}~d=3.
\label{level3}
\end{eqnarray}
In this review, we consider the general case in $d$-dimension \cite{PhysRevA.67.022317}.

The input state is restricted to have the sample amplitude parameter but have
arbitrary phases
\begin{eqnarray}
|\psi \rangle =\frac {1}{\sqrt{d}}\sum _{j=0}^{d-1} e^{i\phi
_j}|j\rangle , \label{dphase-input}
\end{eqnarray}
where phases $\phi _j\in [0,2\pi ), j=0, \cdots, d-1$.
A whole phase is not important, so we can assume $\phi _0=0$.
For comparing the input and the single qudit output, here we write
the density operator of input as
$\rho ^{(in)}={\frac {1}{d}}\sum _{jk}e^{i(\phi _j-\phi _k)}|j\rangle \langle k|$. Our aim
is to find the optimal quantum cloning transformations so that each output qudit
is close to this input density operator.

Considering the symmetries, we can propose the following simple transformations,
\begin{eqnarray}
U|j\rangle |Q\rangle =\alpha |jj\rangle |R_j\rangle +\frac {\beta
}{\sqrt{2(d-1)}}\sum _{l\not =j}^{d-1} (|jl\rangle +|lj\rangle
)|R_l\rangle , \label{dphaseclone}
\end{eqnarray}
where $\alpha ,\beta $ are real numbers, and $\alpha ^2+\beta ^2=1$.
Actually letting $\alpha , \beta $ to be complex numbers does not
improve the fidelity. $|R_j\rangle $ are orthonormal ancillary states.

Substituting the input state (\ref{dphase-input}) into the cloning
transformation and tracing out the ancillary states, the output state
takes the form
\begin{eqnarray}
\rho ^{(out)}&=&\frac {\alpha ^2}{d}\sum _j|jj\rangle \langle jj|
+\frac {\alpha \beta }{d\sqrt {2(d-1)}}\sum _{j\not =l} e^{i(\phi
_j-\phi _l)}\left[ |jj\rangle (\langle jl| \right. \nonumber \\
&&\left. +\langle |lj|) +(|jl\rangle +|lj\rangle )\langle ll|\right]
\nonumber \\
&+&\frac {\beta ^2}{2d(d-1)}\sum _{jj'}\sum _{l\not =j,j'} e^{i(\phi
_j-\phi _{j'})}(|jl\rangle +|lj\rangle ) (\langle lj'|+\langle
j'l|).
\end{eqnarray}
Then, we can obtain the single qudit
reduced density matrix of output
\begin{eqnarray}
\rho ^{(out)}_{red.}&=&{\frac {1}{d}}\sum _{j}|j\rangle \langle j|
\nonumber \\
&&+\left( \frac {\alpha \beta }{d}\sqrt{\frac {2}{d-1}} +\frac
{\beta ^2(d-2)}{2d(d-1)}\right) \sum _{j\not =k} e^{i(\phi _j-\phi
_k)}|j\rangle \langle k|.
\end{eqnarray}
The fidelity can be calculated as
\begin{eqnarray}
F={\frac {1}{d}}+\alpha \beta \frac {\sqrt{2(d-1)}}{d}
+\beta ^2\frac {d-2}{2d}.
\end{eqnarray}
Now, we need to optimize the fidelity under the restriction
$\alpha ^2+\beta ^2=1$. We can find the optimal fidelity
of 1 to 2 phase-covariant quantum cloning machine can
be written as
\begin{eqnarray}
F_{optimal}=\frac {1}{d}+\frac {1}{4d}(d-2+\sqrt{d^2+4d-4}).
\label{dfidelity}
\end{eqnarray}
The optimal fidelity is achieved when $\alpha ,\beta $ take
the following values,
\begin{eqnarray}
\alpha =\left( {\frac {1}{2}}-\frac {d-2}{2\sqrt{d^2+4d-4}}\right)
^{\frac {1}{2}},\nonumber \\
\beta =\left( {\frac {1}{2}}+\frac {d-2}{2\sqrt{d^2+4d-4}}\right)
^{\frac {1}{2}}.
\label{ab}
\end{eqnarray}
In case $d=2,3$, this results reduce to previous known results
(\ref{phaseoptimal},\ref{level3}), respectively.
As expected, this optimal fidelity of phase-covariant
quantum cloning machine is higher than the
corresponding optimal fidelity of UQCM,
\begin{eqnarray}
F_{optimal}>
F_{universal}=(d+3)/2(d+1).
\end{eqnarray}
These are the optimal phase-covariant quantum  cloning
machine for qudits (\ref{dphaseclone}, \ref{ab}) and
the optimal fidelity (\ref{dfidelity}).

\subsection{Symmetry condition and minimal sets in determining quantum cloning machines}

In this subsection, we will mainly discuss how symmetry condition determines the form of quantum cloning machine.
We will also consider the minimal sets
in determining those quantum cloning machines.

As we shown, the number of BB84 states is four. The optimal cloning of those four states
actually can clone optimally arbitrary corresponding equatorial qubits. This means that
BB84 states are enough in determining the phase-covariant quantum cloning machine.
We would come up with a question that whether they are the minimal input sets necessarily for the
phase-covariant cloning.
It is revealed that the set of BB84 states is not the minimal input set.
The minimal set which determines the phase cloning machine is supposed to possess the
 highest symmetry in Bloch sphere with the number three. Here, we give a brief proof.

Consider three input states $\frac{1}{\sqrt2}(|0\rangle+e^{i\phi}|1\rangle)$ where $\phi=0,2\pi/3,4\pi/3$
which are finite numbers. We suppose that the quantum
cloning machine works in symmetric subspace and is economic.
The most general form can be written as,
\begin{eqnarray}
|0\rangle\rightarrow a|00\rangle+b|01\rangle+c|10\rangle+d|11\rangle ,\nonumber \\
|1\rangle\rightarrow e|00\rangle+f|01\rangle+g|10\rangle+h|11\rangle ,
\end{eqnarray}
where $a$ to $h$ are complex numbers which satisfy constrains $|a|^2+|b|^2+|c|^2+|d|^2=1$,$|e|^2+|f|^2+|g|^2+|h|^2=1$ and $ae^*+bf^*+cg^*+dh^*=0$
due to orthogonal and normalizing conditions. Because the machine works in symmetric subspace, we have $b=c$ and $f=g$.
It is easily calculated that the fidelity for arbitrary input equatorial state is,
\begin{equation}
F_A(\phi)=\frac{1}{2}+\frac{1}{2}Re[ac^*e^{(i\phi)}+ag^*+ec^*e^{(2i\phi)}+eg^*e^{(i\phi)}+bd^*e^{(i\phi)}+fh^*e^{(i\phi)}+fd^*e^{(2i\phi)}+bh^*].
\end{equation}
Simplify the expression by utilizing constrains above, we find
\begin{equation}
F_A(\phi)=\lambda_1\cos(2\phi+\psi_1)+\lambda_2\cos(\phi+\psi_2)+\lambda_3,
\end{equation}
where $\lambda_i,i=1,2,3$, are real numbers. Explicit expressions
of these parameters are: $\lambda_1=\frac{1}{2}|ec^*+fd^*|$, $\psi_1=arg(ec^*+fd^*)$, $\psi_2=\frac{1}{2}|ac^*+eg^*+bd^*+fh^*|$, $\psi_2=arg(ac^*+eg^*+bd^*+fh^*)$, and $\lambda_3=\frac{1}{2}+\frac{1}{2}Re(ag^*+bh^*)$.

Additionally, we let the cloning fidelities for those three states being the same: $F(0)=F(2\pi/3)=F(4\pi/3)$.
Thus, we will obtain two more constraints: $\lambda_1\sin\psi_1=\lambda_2\sin\psi_2$, $\lambda_1\cos\psi_1+\lambda_2\cos\psi_2=0$.
With the help of some algebraic inequalities, one would find that $F$ reaches its maximum value if and only if $\lambda_1=\lambda_2=0$.
Now we are ready to find a simple form of the fidelity for the three input states,
\begin{equation}
F_A=\lambda_3.
\end{equation}
Remarkably, this result demonstrates that for any $\phi $, $F_A$ is independent of the phase parameter $\phi$.
This cloning machine becomes the standard phase-covariant quantum cloning machine.
Note that three states constituting the minimal input set have been studied from the viewpoint of
designing quantum measurement technique for optimal quantum information detection \cite{PeresWootters}.

We have just shown that the phase-covariant quantum cloning machine can be determined completely
 by a minimal input set with only three symmetric states.
Here we would like to remark two points: (i) We know that the phase cloning may take two different forms
with or without the ancillary states. The minimal input set is studied for the economic case in the above,
we would like to point out that the cloning fidelity cannot be increased with the help of the ancillary states.
Thus, the conclusion that this minimal input set can determine the phase-covariant quantum cloning are for both economic and non-economic cases.
(ii) We have just reviewed the 1 to 2 cloning. If we would like to clone equally well these three states
presented above to $M$ copies, the $1\rightarrow M$ phase-covariant quantum cloning machine is the optimal one.
So this minimal input set can also determine completely the $1\rightarrow M$ phase-covariant cloning machine.
Those two conclusions can be obtained by similar investigations as just reviewed.

We may find that the minimal input set contains three states which have a geometric symmetry
in two dimension Hilbert space.
It seems not obvious what kind of symmetry should be possessed for the
minimal input set for phase-covariant quantum cloning in higher dimensional system.
This is an open question and might be explored further.
Not only for the case of phase-covariant quantum cloning, the UQCM has similar question.
We already know that the minimal input set for UQCM of two dimension is constituted
by four states forming a tetrahedron on Bloch sphere, see Fig.(\ref{4states}).
It is not clear what is the minimal input set of UQCM in higher dimensional system.

We know that the fidelity of $1\rightarrow \infty $ phase quantum cloning
is corresponding to quantum phase estimation \cite{prl80.1571}.
The result, that the quantum phase cloning of states with arbitrary phase
is equivalent to the cloning of a finite ensemble including only three special states,
may shed light on the quantum state estimation of some fixed ensembles.
Besides the case that input states are restricted to the equator of the Bloch sphere,
there are some other cases where the input states may be located on a belt of the
Bloch sphere \cite{yuzw}, or with other distributions. It will interesting to study
the minimal input sets for those quantum cloning machines.

\subsection{Quantum cloning machines of arbitrary set of MUBs}

Here, we discussed the higher dimension quantum cloning of the mutual unbiased basis(MUB).
It is known that a Hilbert space of $d$ dimension contains $d+1$ sets of MUB, provided $d$ is prime.
In this review, when MUBs are used, we will restrict our attentions on case $d$ is prime.
We can design QKD protocols by using arbitrary sets of MUBs.
For example, in 2 dimensional system, we have two well accepted QKD protocols,
six-state protocol means 3 sets of MUBs and BB84 protocol means 2 sets.
In higher dimension, we can also propose corresponding cloning machines for those sets of MUBs.

Let us first present some characteristics of MUBs.
In a system of dimension $d$, there are $d+1$ MUBs \cite{roychowdhuryMUB}, namely $\{|i\rangle\}$ and $\{|\tilde{i}^{(k)}\rangle\}$
$(k=0,1, ...,d-1)$, are expressed as,
\begin{equation}
|\tilde{i}^{(k)}\rangle=\frac{1}{\sqrt{d}}\sum_{j=0}^{d-1}\omega^{i(d-j)-ks_j}|j\rangle,
\label{mubphase}
\end{equation}
with $s_j=j+...+(d-1)$ and $\omega=e^{i\frac{2\pi}{d}}$.
Any states in the same set are orthogonal  $\langle\tilde{i}^{(k)}|\tilde{l}^{(k)}\rangle=\delta_{il}$,
and any states in different sets satisfying $|\langle\tilde{i}^{(k)}|\tilde{l}^{(j)}\rangle|=\frac{1}{\sqrt{d}}$, $k\not =j$,
that is their overlaps are the same.
Define the generalized Pauli matrices $\sigma_x$ and $\sigma_z$ as,
$\sigma_|j\rangle=|j+1\rangle$ and $\sigma_z|j\rangle=\omega^j|j\rangle$. Note that, as usual, we omit module $d$ in equations.
Then there are $d^2-1$ independent Pauli matrices $U_{mn}=(\sigma _x)^m(\sigma _z)^n$ and
$U_{mn}|j\rangle=\omega^{jn}|j+m\rangle$. Those MUBs are eigenvectors of operators $\sigma _z, \sigma _x(\sigma _z)^k$,
$k=0,1,...,d-1$,
\begin{eqnarray}
\sigma _x(\sigma _z)^k|\tilde{i}^{(k)}\rangle =\omega ^i|\tilde{i}^{(k)}\rangle .
\end{eqnarray}
The result of MUBs can also be found in \cite{Wootters1989}.

A straightforward generalization of BB84 states in $d$-dimension is two sets of
bases from those $d+1$ MUBs, and the generalization of six-state protocol is to use all $d+1$ mutually
unbiased bases \cite{PhysRevLett.88.127902}.
Suppose two MUBs are $\{|k\rangle\},k=0,1,2,...,d-1$ and its dual under a Fourier transformation,
\begin{equation}
|\bar{l}\rangle=\frac{1}{\sqrt{d}}\sum_{k=0}^{d-1}e^{2\pi i(kl/d)}|k\rangle,
\end{equation}
where l=0,1,2,...,d-1. We follow the standard QKD,
Alice initially sends the state $|\psi\rangle$, Eve can use her quantum
clone machine to copy the state and the transferring state is disturbed which
is later still sent to Bob. Eve has a non-perfect copy of the sending state and
the ancillary state of her quantum cloning machine. The whole system is written as,
\begin{equation}
|\psi\rangle_{A}\rightarrow\sum_{m,n=0}^{d-1}a_{m,n}U_{m,n}|\psi\rangle_B|B_{m,-n}\rangle_{E,E'},
\label{whole48}
\end{equation}
where A,B,E, and E' represent Alice's qudit, Bob's clone, Eve's clone, and the cloning machine.
Obviously, parameters $a_{m,n}$ satisfy $\sum_{m,n=0}^{d-1}|a_{m,n}|^2=1$.
As we already know, $|B_{m,-n}\rangle_{EE'}$ stands for d-dimensional Bell states which is the maximally entangled states of two qubits with explicit form:
\begin{equation}
|B_{m,n}\rangle_{EE'}=\frac{1}{\sqrt{d}}\sum_{k=0}^{N-1}e^{2\pi i(kn/d)}|k\rangle_E|k+m\rangle_{E'},
\end{equation}
where $m,n=0,1,...,d-1$.
Note that the operators $U_{m,n}$ can be expressed as,
\begin{equation}
U_{m,n}=\sum_{k=0}^{d-1}e^{2\pi i(kn/d)}|k+m\rangle\langle k|.
\end{equation}
They actually form a group of qudit error operations where $m$ represents the shift errors and $n$ is related with
the phase errors. Trace off the joint states within Eve, Bob's clone will be a mixed state, it is the same
as the state $|\psi \rangle $ passing through a quantum channel which will cause decoherence,
\begin{equation}
\rho_B=\sum_{m,n=0}^{d-1}|a_{m,n}|^2U_{m,n}|\psi\rangle\langle\psi|U_{m,n}^{\dagger}.
\label{Bobdensity}
\end{equation}
Therefore, when Alice sends states $|k\rangle$, Bob's fidelity is
\begin{equation}
F=\langle k|\rho_B|k\rangle=\sum_{n=0}^{d-1}|a_{0,n}|^2.
\end{equation}
Also, when Alice sends states $|\bar{l}\rangle$, Bob's fidelity is
\begin{equation}
\bar{F}=\langle\bar{l}|\rho_b|\bar{l}\rangle=\sum_{m=0}^{d-1}|a_{m,0}|^2.
\end{equation}
Consider the requirement that the cloner works equally well with these states, we must choose the amplitude matrix as the following form,
\begin{equation}
(a_{m,n})=
\begin{pmatrix}
v & x & \cdots & x \\
x & y & \cdots & y \\
\vdots & \vdots & \ddots & \vdots \\
x & y & \cdots & y
\end{pmatrix}
\label{ampmatrix}
\end{equation}
where $x$, $y$ and $v$ are real number satisfying $v^2+2(d-1)x^2+(d-1)^2y^2=1$.
In this way, we find Bob's fidelity is
\begin{equation}
F=v^2+(d-1)x^2.
\end{equation}

Next let us consider the state of Eve's side. Eve performs the unitary transformation
on both the transferring state $|\psi \rangle $ and a maximally entangled state,
\begin{equation}
U=\sum_{m,n=0}^{d-1}a_{mn}(U_{mn}\otimes U_{m,-n}\times\mathbb{I}),
\end{equation}
as shown in Eq.(\ref{whole48}). We can find that this transformation can be rewritten
in a different form as follows,
\begin{eqnarray}
U|\psi\rangle_{A}|\Phi _{0,0}\rangle _{E,E'}&=&
\sum_{m,n=0}^{d-1}a_{m,n}U_{m,n}|\psi\rangle_B|B_{m,-n}\rangle_{E,E'}
\nonumber \\
&=&\sum _{mn}b_{m,n}|\Phi _{-m,n}\rangle _{BE'}\otimes U_{m,n}|\psi \rangle _E.
\label{wholeAB}
\end{eqnarray}
So coefficients $a_{m,n}$ are related with another set of coefficients $b_{m,n}$ as follows,
\begin{equation}
b_{m,n}=\frac{1}{d}\sum_{m',n'=0}^{d-1}e^{2\pi i(nm'-mn')/d}a_{m',n'}.
\label{fourier}
\end{equation}
By using coefficients $b_{m,n}$ as shown in Eq.(\ref{wholeAB}),
the density operator of Eve takes a simple form as
\begin{equation}
\rho_E=\sum_{m,n=0}^{d-1}|b_{m,n}|^2U_{m,n}|\psi\rangle\langle\psi|U_{m,n}^{\dagger}.
\end{equation}
This density operator is similar as Bob's density operator in (\ref{Bobdensity})
except that coefficients $b_{m,n}$ are used.
Further, we find that the fidelity for Eve can be expressed as,
\begin{equation}
F_E=v'^2+(d-1)x'^2
\end{equation}
where $v'$, $x'$ and $y'$ corresponds to parameters of coefficients $b_{m,n}$,
with similar structure as matrix in (\ref{ampmatrix}),
which are related with $a_{m,n}$ by the Fourier transformations (\ref{fourier}).
Explicitly, those parameters can be written as,
\begin{eqnarray}
x'&=&[v+(d-2)x+(1-d)y]/d,\nonumber\\
y'&=&(v-2x+y)/d,\nonumber\\
v'&=&[v+2(d-1)x+(d-1)^2y]/d.
\end{eqnarray}

Now, both fidelities of Eve and Bob are represented by the same parameters $v,x,y$.
Our purpose is to maximize Eve's fidelity $F_E$ under a given value of Bob's fidelity $F$.
The trade-off relation can then be found as,
\begin{equation}
F_E=\frac{F}{d}+\frac{(d-1)(1-F)}{d}+\frac{2}{d}\sqrt{(d-1)F(1-F)}.
\end{equation}

Next, we consider another protocol using all available $d+1$ bases.
Similarly, by considering that the same fidelity is necessary for all used bases
since they are applied randomly, we derive that amplitude matrix presented in Eq.(\ref{ampmatrix})
must satisfy $x=y$.
Hence, Bob's fidelity is
\begin{equation}
F=v^2+(d-1)x^2=1-d(d-1)x^2,
\end{equation}
and Eve's fidelity is,
\begin{equation}
F_E=v'^2+(d-1)x'^2=1-d(d-1)x'^2,
\end{equation}
where $v'$ and $x'$ are expressed as
\begin{eqnarray}
x'&=&(v-x)/d\nonumber\\
v'&=&[v+(d^2-1)x]/d.
\end{eqnarray}
The relations induce the trade-off between two fidelities of Bob and Eve.

\begin{figure}
\includegraphics[height=7cm]{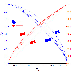}
\caption{The fidelities of Bob and Eve for dimension $d=5$. The number of
mutually unbiased bases runs from 2 to 6. These
results are presented in \cite{PhysRevA.85.012334}.}
\label{Figure-xiong}
\end{figure}

For higher dimension case, we may have more choices for QKD. Besides by using only two bases
or all $d+1$ bases, we may choose any sets of mutually unbiased bases.
Then corresponding cloning machines are necessary in analyzing the security.
Those general QKD protocols are studied recently in \cite{PhysRevA.85.012334}.
By using the same arguments about the symmetry, we can find,
\begin{eqnarray}
\rho_B&=&\sum_{m,n=1}^{d-1}|a_{mn}|^2|i+m\rangle\langle i+m|,\\
\tilde{\rho}_B^{(k)}&=&\sum_{m,n=0}^{d-1}|a_{mn}|^2(U_{mn}|\tilde{i}^{(k)}\rangle_B)(_B\langle\tilde{i}^{(k)} U_{mn}^{\dagger}),\\
\rho_E&=&\sum_{m,n=1}^{d-1}|b_{mn}|^2|i+m\rangle\langle i+m|,\\
\tilde{\rho}_E^{(k)}&=&\sum_{m,n=0}^{d-1}|b_{mn}|^2(U_{mn}|\tilde{i}^{(k)}\rangle_E)(_E\langle\tilde{i}^{(k)} U_{mn}^{\dagger}),
\end{eqnarray}
where $k=0,1,...,g-1$. Therefore, one may easily derive the fidelities,
\begin{eqnarray}
F_B&=&\sum_n|a_{0n}|^2,\\
\tilde{F}_B^{(k)}&=&\sum_m|a_{m,km}|^2,\\
F_E&=&\frac{1}{d}\sum_m|\sum_na_{mn}|^2,\\
\tilde{F}_E^{(k)}&=&\frac{1}{d}\sum_n|\sum_ma_{m,n+km}|^2,\\
\end{eqnarray}
where $k=0,1,...,g-1$.
Assuming that Eve's attack is balanced, or we say she induces an equal probability of error for any one of the $g+1$ MUBs,
we have,
\begin{equation}
F_B=\tilde{F}_B^{(0)}=...=\tilde{F}_B^{(g-1)}.
\end{equation}
These constraints can determine the optimal cloner. Eve could maximize all these $g+1$ fidelities
simultaneously and and let them equal.
This is can be realized by ``vectorization'' of the matrix elements of $(a_{mn})$.
Define,
\begin{eqnarray}
&&\vec{\alpha}_i=(a_{1,1i},...,a_{d-1,(d-1)i}),~~~(i=0,1,...,g-1),\\
&&\vec{A}_i=(A_1,...,A_{d-1}),\\
&&A_i=\sum_{j\neq0,i,...,(g-1)i}^{d-1}a_{ij}(i=1,2,...,d-1),
\end{eqnarray}
and the rest elements are restricted by the following equations:
\begin{eqnarray}
\sum_{j=1}^{d-1}|a_{0j}|^2&=&F_B-|a_{00}|^2,\\
||\vec{\alpha}_i||^2&=&F_B-|a_{00}|^2,~~~ (i=0,1,...,g-1).
\end{eqnarray}
Finally, Eve's fidelity can be expressed as
\begin{equation}
F_E=\frac{1}{d}(|\sum_{j=0}^{d-1}a_{0j}|^2+||\sum_{i=0}^{g-1}\vec{\alpha}_i+\vec{A}||^2).
\end{equation}
By maximizing Eve's fidelity,
the above result can be further simplified by some algebraic considerations,
\begin{eqnarray}
a_{mn}=
\begin{cases}
v,~m=n=0,\\
x,~m=0,n\neq0~{\rm or}~m\neq0,n=km,\\
y,{\rm ~~otherwise},
\end{cases}\label{gmatrix}
\end{eqnarray}
where $k=0,...,g-1$, and $v$ is a real number to be determined and
$x=\sqrt{\frac{F_B-v^2}{d-1}}$,
$y=\sqrt{\frac{1+gv^2-(g+1)F_B}{(d-1)(d-g)}}$.
Now we reach our conclusion that the fidelity of Eve is,
\begin{eqnarray}
F_E=\frac{1}{d}\{[v+(d-1)x]^2+(d-1)[gx+(d-g)y]^2\}.\label{gfidelity}
\end{eqnarray}
The only undetermined variable is $v$, we can change it so that
the fidelity of Eve $F_E$ reaches the maximum depending
on the fixed fidelity of Bob. The fidelities of Bob and Eve
are presented in FIG.\ref{Figure-xiong} for some special cases \cite{PhysRevA.85.012334}.

From this conclusion, we may easily find out the results for $g=1$ and $g=d$ which
lead to results in \cite{PhysRevLett.88.127902}. Also we are interested in the
condition that $F_B=F_E$ which is the symmetric cloning, and the remained variable $v$
is fixed in this case which actually takes a rather complicated form, we finally have,
\begin{eqnarray}
F=\frac{2}{d}\frac{d-g}{(g+3)-\sqrt{(g+3)^2-8\frac{(d-g)(g+1)}{d}}}
\label{gsymmfidelity}.
\end{eqnarray}

As we know, the optimal cloner for $d+1$ MUBs is actually equivalent to universal quantum cloning machine.
It is interesting to know which of the above quantum cloning machine
is equivalent to the phase-covariant quantum cloning machine. Stimulated by the fact
that $d$ MUBs presented in Eq.(\ref{mubphase}) only contain phase parameters but the
amplitude parameters are fixed, we may suppose that the $d$-dimension
phase-covariant quantum cloning should be equal to the cloning of $d$ MUBs.
Indeed, let $g=d-1$, the fidelity (\ref{gsymmfidelity}) coincides with the
phase-covariant fidelity in (\ref{dfidelity}). To further check that those
two cloning machines are the same, we need also consider the asymmetric case.

Let us consider the equatorial qudit as,
$|\psi \rangle =\frac{1}{\sqrt d}\sum_{j=0}^{d-1}e^{i\phi_j}|j\rangle $,
where $\phi_j$ are phase parameters.
One can assume that the asymmetric cloning transformation is given as,
\begin{equation}
|i\rangle\rightarrow\alpha|ii\rangle|i\rangle+\frac{\beta}{\sqrt{d-1}}\sum_{j\neq i}(\cos\theta|ij\rangle+\sin\theta|ji\rangle|j\rangle),
\end{equation}
where $\theta$ is the asymmetric parameter. Therefore one can derive two fidelities for Bob and Eve, respectively,
\begin{eqnarray}
F_1=\frac{1}{d}+\frac{2\alpha\beta}{d}\sqrt{d-1}\cos\theta+\frac{\beta^2(d-2)}{d}\cos^2\theta,\\
F_2=\frac{1}{d}+\frac{2\alpha\beta}{d}\sqrt{d-1}\sin\theta+\frac{\beta^2(d-2)}{d}\sin^2\theta,
\end{eqnarray}
where we still have $\alpha^2+\beta^2=1$. Here we would like to emphasize,
the exact value of $\alpha ,\beta $ should depend on the parameter $\theta $.
For symmetric case, $\theta =\pi /4$, their values can be found in Eqs.(\ref{ab}).
When $\theta $ changes, the values of $\alpha ,\beta $ also change.
Numerical evidences show that those fidelities are the same with the fidelities in Eq.(\ref{gfidelity}).

We know that MUBs may determine some specified quantum cloning machines which
in general are better than the universal cloning machine which admits arbitrary input states.
On the other hand, we may wonder whether those MUBs are the minimal input sets which
can be copied optimally by the corresponding cloning machines. In qubit case, we know
that the number of states in the minimal input set can be reduced \cite{Jing2013}.
It is not clear what are the minimal input sets for the optimal cloning machines of those MUBs in
general $d$ dimension.

\subsection{Quantum cloning in mean king problem as a quantum key distribution protocol}

Quantum key distribution (QKD) protocols allow two parties, called Alice (the sender) and Bob (the receiver) conventionally,
to generate shared secret keys for them to communicate securely.
In BB84 protocol\cite{Bennett1984}, we send states
by exploiting two mutually unbiased bases of qubit.
Ekert proposed a QKD protocol based on Bell theorem by using the entangled pairs in 1991 (E91) \cite{PhysRevLett.67.661}.
As we already know that the BB84 protocol can also be generalized by using a six-state protocol \cite{PhysRevLett.81.3018}.
It is also possible to propose a QKD protocol by combination of BB84
protocol and E91 protocol. This protocol is based on the so-called mean king problem \cite{meanking} since
its description is usually like a tale \cite{tale}.

The protocol of mean king problem,
can be considered as two steps: The first step is the same as E91 protocol except without
classical announcement of measurement bases and the second step is like BB84 protocol.
In this protocol, Bob needs to retrodict the outcome of a projective measurement by Alice without knowing the bases she used.
For qubit first \cite{meanking} and higher dimension latter \cite{LatinSquare,OrthogonalArray},
it is shown that Bob has a 100\% winning strategy.
So it is realized that this quantum retrodiction protocol might be applied as a QKD in quantum cryptography \cite{Bub,Werner,Yoshida2010}.

In this protocol, Alice may exploit bases in a ``meaner'' way by utilizing biased (nondegenerate) bases \cite{Biased}.
The security of the QKD protocol is analyzed by considering a full coherent attack
on both quantum channels \cite{Werner}.
To be explicit, Eve controls completely the preparation of entangled pairs, which are used by Bob before sending
 one part of them to Alice, as well as the feedback channel which is used for transmitting
 back the quantum state after a measurement by Alice.
 To specify the attack, in the former scenario, Eve initially prepares a maximally entangled
 state $|\Phi^+ \rangle_{BB'}$ which will be shared for Alice and Bob.
But she adversarial prepares another completely same entangled pair $|\Phi ^+\rangle _{EE\rq{}}$,
partially swaps her qubit with the providing entangled pair.
Consequently, Alice and Bob both are partially entangled with Eve, in contrast, they are
maximally entangled with each other if no Eve exists. So the whole system with Alice, Bob and Eve
possesses a superposition of two pairs maximally entangled states. For the second channel,
Eve is confined to only perform a cloning-based individual attack on the particle Alice sends to
Bob after her projective measurement \cite{PhysRevLett.81.3018,PhysRevLett.88.127902,Cerf1998,PhysRevA.85.012334}.
The attack on this retrodiction protocol
on both steps of the entangled pair preparation and quantum state transmission can be understood from a general
viewpoint by
the unified quantum cloning machine \cite{PhysRevA.84.034302}.
A QKD protocol is secure when mutual information
between Alice and Bob is larger than that between Alice and Eve,
under this condition can Alice and Bob use classical error correction and privacy amplification methods \cite{QKDreview,PhysRevLett.88.127902} to
guarantee a secure communication. Alternatively, we may also compare the fidelity between
Eve and Bob to see which one is closer with the ideal case.

\begin{figure}
\centering
\includegraphics[width =8cm ]{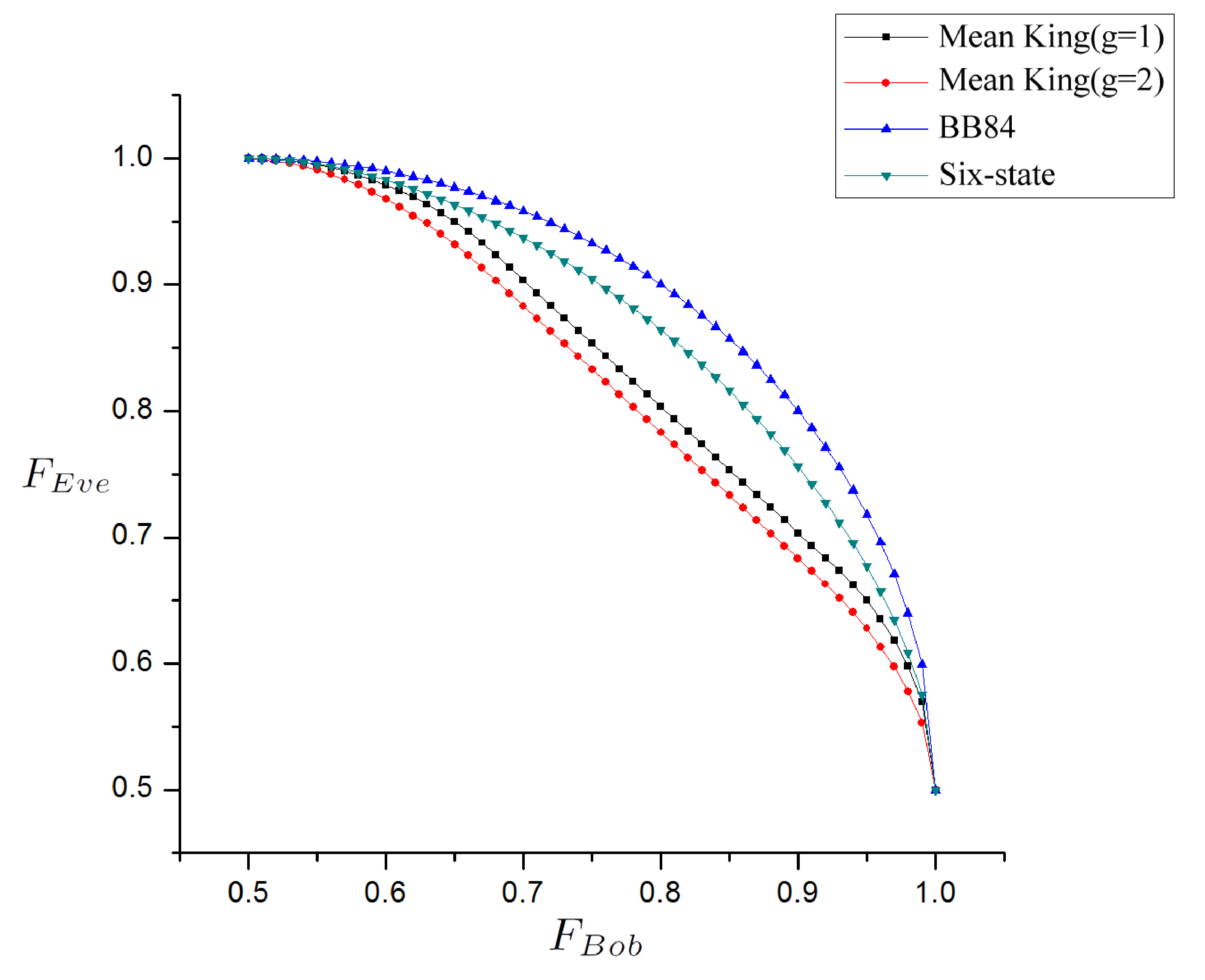}
\caption{(color online) $F_{Eve}$ vs $F_{Bob}$ curve for $d=2$. In our scenario, the mean king retrodiction QKD has higher
security than both BB84 and six-state protocols.}
\label{d2}
\end{figure}

Here let us review the comparison between different protocols in FIG.\ref{d2} which
includes four cases, the standard QKD by BB84 states and six-state, and their correspondences
by retrodiction protocol. Interestingly, it is clear that
the retrodiction QKD protocol presented here is more secure than BB84 protocol
and six-state protocol, i.e., with fixed disturbance ($F_{Bob}$ is fixed), Eve's probability
to figure out the correct result is lower.
And using 3 bases ($g=2$) is even more secure than 2 bases($g=1$).

The efficiency of the mean king retrodiction has the advantage of generating a raw key in \emph{every single run}
no matter how many mutually unbiased bases are utilized.
For comparison, in standard QKD by exploiting $g+1$ mutually unbiased bases,
 there would only have a raw key in $g+1$ runs on average for Alice and Bob.

\subsection{Other developments and related topics}

No-cloning is a fundamental principle of quantum mechanics.
On the other hand, the quantum cloning machine is concerned about to clone
quantum states,
approximately or probabilistically with both cases not violating
the principle of quantum mechanics, but with the highest quality measured by
different figures of merit. If we know nothing about the input quantum states,
the cloning machine should be in the sense of universal as we have reviewed in last
section. In case we know partial information of the input states, we can
use state-dependent cloning which performs, at least, as good as the universal
cloning machine. It is naturally expected that we can do better in most cases.
The phase-covariant cloning machine belongs to the class of state-dependent cloning, however,
we usually listed it independently.
The phase-covariant is in the sense that the density matrix of
single output copy has the same form of phase with the input state density matrix,
to be more explicit, the difference of those two density matrices is a mixture
of identity with a probability. This property ensures that the fidelity of this
cloning machine is independent of input states which differs only in phase parameters.

The phase-covariant quantum cloning was initiated by Bru\ss ~\emph{et al.}
for considering the 1 to 2 cloning of equatorial qubit,
which is a qubit located in the equator of the Bloch sphere \cite{Bruss2000}.
It is also shown that the cloning of equatorial qubits is optimal if
the input is restricted to only four states corresponding to BB84 states.
The minimal input set which can determine completely this optimal
quantum cloning machine includes only three states located symmetrically
on the equator of the Bloch sphere \cite{Jing2013}.
The more general 1 to 3 phase cloning case \cite{PhysRevA.64.042308}
and 1 to many case are also studied \cite{PhysRevA.65.012304}.
For higher dimensional case, the three-dimension phase-covariant
quantum cloning is studied in \cite{PhysRevA.64.042308,688607420020710},
the general $d$-dimension phase cloning is presented in \cite{PhysRevA.67.022317}.
Various kinds of phase-covariant and state-dependent cloning are proposed.
Next, we list some of those developments below.

\begin{itemize}

\item
The input states for cloning are limited to some conditions, which
in general can be described by some symmetries.
The cloning of states in higher-dimension, but with only real parameters is studied in \cite{PhysRevA.68.032313}.
The asymmetric qudit phase-covariant quantum cloning
is studied in \cite{ISI:000228015900003}.
The quantum cloning of set of states which is invariant under the Weyl-Heisenberg group
is studied by the extremal cloning machine \cite{extremalcloning}.
The quantum cloning
of states with fixed amplitudes but arbitrary phase is studied in \cite{Karimipour-phase},
which is suboptimal while the experimental scheme uses the optimal one \cite{PhysRevLett.94.040505}.
The cloning of states in a belt of Bloch sphere is studied in \cite{yuzw}. The case
of distribution with mirror like symmetry, i.e., with known modulus of expectation of Pauli $\sigma _z$
matrix is studied in \cite{mirror}, the case of arbitrary
axisymmetric distribution on the Bloch sphere is studied in \cite{axi-bloch},
see also \cite{Bartkiewicz2012}. The cloning of a pair of orthogonally polarized photons is studied in \cite{Fiurasek2008}.
The optimal broadcasting of mixed equatorial qubits
is studied in \cite{Yu2009}.
A hybrid quantum cloning machine combines
universal and state-dependent cases together is presented in \cite{Adhikari2007}.

\item
The estimation of states or phases for finite quantum states.
We have known that UQCM and phase-covariant cloning machine are for states with some
parameters, amplitude or phase, which can be assumed to be continuous. On the other hand,
we already know that we cannot do better for cases even the number of input states are finite,
for example for some sets of MUBs.
We remark that the quantum cloning of sets of MUBs should be related with state estimation.
The results of arbitrary state estimation and phase estimation are available which correspond to $g=d, d-1$,
however, the general $g+1$ MUBs estimations are not yet studied.
The phase estimation of qubits is studied in \cite{prl80.1571},
the case of qubits in mixed states is presented in \cite{mixedstate-phaseestimation}.
The phase estimation of multiple phases is studied in \cite{multiphase-esti}.
The criterion for estimation and the quality of state-dependent cloning is
analyzed in \cite{ISI:000179502200033}.

\item
State-dependent cloning related with QKD protocols.
We should note that the security of QKD is generally defined by various criteria \cite{QKDreview},
in this review, we consider the attack by the scheme of quantum cloning.
Quantum copying of two states is studied in \cite{hillery-buzek}.
The QKD in three dimensions
is studied in \cite{prl88.127901}. The four-dimensional case is studied in \cite{4dim2000,PhysRevA.68.042323},
The optimal eavesdropping of BB84 states is studied in \cite{fuchs97},
higher-dimensional case and some related results are presented by some other groups \cite{Karimipour,Acin2003,Nikolopoulos2005,Nikolopoulos2006,Bae2007,prl95.080501,BourennaneKarlsson}.
The extension of BB84 states for qubits is also studied as the spherical-code \cite{Renes2004}.
The comparison between photon-number-splitting attack and quantum cloning attack of BB84 states
is studied in \cite{pns-clone}. The extension of phase-covariant cloning to multipartite
quantum key distribution is studied in \cite{scarani2001}. The cloning network of generalized BB84 states constituted by two pairs of orthogonal states is presented in \cite{ISI:000221546500001}.

\item Concepts related with phase-covariant and state-dependent cloning.
The state-dependent cloning machine and the relation
with completely positive trace-preserving maps is studied in \cite{ISI:000187004700046}.
Relation of state-dependent cloning with quantum tracking is studied in \cite{ISI:000258180300064}.
The assisted phase cloning of qudit by remote
state preparation is presented in \cite{ISI:000263307800011}.
The network of state-dependent quantum cloning is studied in \cite{chefles},
see also \cite{Zhou2011}.
The relations between teleportation and dissipative channels with the
universal and phase-covariant cloning machine are analyzed in \cite{Ozdemir2007}.
The phenomena of superbroadcasting is also studied for phase-covariant case \cite{ISI:000241723100032}.
The no-cloning theorem for a single POVM is
presented in \cite{ISI:000282921000007}.
It is found that while equatorial qubit contains only one arbitrary parameter, the phase information
cannot be compressed \cite{1751-8121-45-2-025304}.

Quantum cloning is generally not concerned with
relativity. However with relativistic covariance requirement, the state-dependent cloning of photons and
the BB84 states are studied in \cite{relative-cloning}.
It is shown by phase-covariant quantum cloning that the cloned quantum states
are not macroscopic in the spirit of Schrodinger's cat \cite{macro-clone}.

\item Implementation theoretically and experimentally.
Various proposals of implementation have been put up.
The economic realization of phase-covariant devices in arbitrary dimension,
where phase cloning as a special case, is studied in \cite{Buscemi2007}.
The scheme of one to three economic phase-covariant quantum cloning
machine is proposed implementing by linear optics system \cite{Zou2005-M}.
The one to many symmetric economic phase cloning is proposed in \cite{Zhang2007}, see
also \cite{Zhang2009}.
The scheme to realize economic one to many phase cloning for qubit and qutrit
is proposed in \cite{Zou2006}.
The realization of phase-covariant and real qubit states quantum cloning
are presented in \cite{Fang2010}. The phase cloning in spin networks
is proposed in \cite{ChiaraFazio}.
The proposal of optical implementation of phase cloning of qubits is presented in \cite{fiurasek2003},
the cloning of real state is studied in \cite{Hu2010}.
The one to many phase-covariant quantum cloning
is also analyzed by the general angular momentum formalism \cite{Sciarrino2007}.
Quantum circuits for both entanglement manipulation
and asymmetric phase-covariant cloning are studied in \cite{Levente2010}.

Experimentally, the asymmetric phase cloning is realized in optical system \cite{bartphase},
and in \cite{Soubusta2008}. The ancilla-free phase-covariant cloning
through Hong-Ou-Mandel interference is realized in experiment by Khan and Howell \cite{Khan-Howell}. The one to three economic quantum cloning of equatorial qubits
encoded by polarization states of photons and the universal
cloning are realized experimentally in \cite{XuLiChen}.
Realization by NMR system can be found in\cite{PhysRevLett.94.040505,chendujfphase}.
In optical parametric amplification of a single photon
in the high gain-regime, experiment is performed to distribute the
photon polarization state to a large number of particles which corresponds
to the phase-covariant quantum cloning \cite{Nagali2007}.
The phase-covariant quantum cloning is also implemented in nitrogen-vacancy center
of diamond by using three energy levels \cite{Pan2011}, in nanodiamond with
full coherent control of phases is reported \cite{Pan2013}, this will be reviewed
in detail later.
The experimental implementation of eavesdropping of BB84 states and
trine states by optimal cloning is studied in \cite{Bartkiewicz-experiment12}.

\end{itemize}

\newpage

\section{Local cloning of entangled states, entanglement in quantum cloning}

Quantum cloning is generally to find the quantum operations
to realize the optimal cloning. The only restriction is that
the operations should satisfy quantum mechanics.
We next study the local cloning of entangled
states, in this case, the operations are additionally restricted to be local.
In principle, there is also a no-cloning theorem for entangled states \cite{prl81.4264}.

In addition, since the crucial
role of quantum entanglement in quantum information, we will also study
the entanglement properties in quantum cloning machines.

\subsection{Local cloning of Bell states}

Quantum entanglement plays a key role in quantum computation
and quantum information. It is the precious resource in quantum
information processing. Also entanglement is a unique property of quantum system
which does not have any classical correspondence. In this sense, quantum entanglement
has already become a common concept and has many
applications in various quantum systems.
The study of entanglement
is generally under the condition of
local (quantum) operations and classical communication (LOCC).
This is due to the consideration that entanglement does not increase
under LOCC.

The local cloning of entangled states
is an interesting topic \cite{OH06,buzek-lcoal-cloning,GKRlocal-cloning,njp6.164}. First let us raise
the problem: Suppose two spatially separated parties, Alice (A)
and Bob (B), share some entangled states, by LOCC, they want to copy the shared
entangled states. As an example, let us study the following
problem \cite{GKRlocal-cloning}, the four Bell states are defined as usual
as the following,
\begin{eqnarray}
|\Phi ^+\rangle &=&\frac {1}{\sqrt {2}}(|00\rangle +|11\rangle ),
\nonumber \\
|\Phi ^-\rangle &=&\frac {1}{\sqrt {2}}(|00\rangle -|11\rangle )
=(I\otimes Z)|\Phi ^+\rangle ,
\nonumber \\
|\Psi ^+\rangle &=&\frac {1}{\sqrt {2}}(|01\rangle +|10\rangle )
=(I\otimes X)|\Phi ^+\rangle ,\nonumber \\
|\Psi ^-\rangle &=&\frac {1}{\sqrt {2}}(|01\rangle -|10\rangle ).
=(I\otimes XZ)|\Phi ^+\rangle .
\label{Bell}
\end{eqnarray}
Alice and Bob share one Bell states from a known subset, say
$\{ |\Psi ^+\rangle , |\Phi ^+\rangle \}$,
they want to copy this state by LOCC.
Several problems should be considered before
to study this problem: (1).The entanglement between $A$
and $B$ does not increase under LOCC. So to copy locally this
state,
we generally assume that some known entangled
states, for example $|\Phi ^+\rangle $,
are shared between $A$ and $B$ which can be used
as ancilla. (2).The entanglement resource used by local
copying should be minimum. Otherwise,  we can use
the teleportation scheme\cite{PhysRevLett.70.1895}, let Bob (Alice) obtain
the full state $\{ |\Psi ^+\rangle , |\Phi ^+\rangle \}$,
he knows the state exactly by measurement, and copies of the
entangled state between $A$ and $B$ can
be obtained easily by local unitary operations which are
shown explicitly above in (\ref{Bell}).
Actually Alice and Bob can discriminate any two Bell states by
LOCC \cite{walgateshort,walgate,prl87.277902}.

In these conditions, the problem can be explicitly stated as:
Alice and Bob share either of two maximally entangled states
$\{ |\Psi ^+\rangle , |\Phi ^+\rangle \}$, but they do not know
which one it is. Additionally, they share known maximally entangled
states in form $|\Phi ^+\rangle $ as the resource which are used as ancillary states.
The question is: can they obtain the state
$|\Psi ^+\rangle ^{\otimes 2}$ or
$|\Phi ^+\rangle ^{\otimes 2}$ by LOCC? The answer is `yes': both Alice
and Bob do CNOT gate with the unknown qubit as
the controlled qubit and the ancilla as the target qubit,
they can achieve their aim.
We name this method as CNOT scheme.
The key point here is that Alice and Bob do not need to know which
state they share, they can finally obtain two copies of this
state, and only one known state $|\Phi ^+\rangle $ (resource)
is consumed.

Let us next analyze the advantages of this scheme by comparing with
the teleportation scheme.
By using the available resource $|\Phi ^+\rangle $ for teleportation,
the unknown state, which is either $|\Psi ^+\rangle $ or $|\Phi ^+\rangle $,
can be teleported to either Alice or Bob's side, here we suppose Alice receives
this unknown state. Alice can find the exact form of this state by using Bell measurement.
Now according to the obtained information, Alice and Bob use additional two
entangled resource, and can share two copies of the previous unknown maximally entangled state.
In this process, three maximally entangled state are consumed, where one is for teleportation,
and another two are used to share between Alice and Bob.
Since three entanglement resource is used, this scheme
is not as efficient as the CNOT scheme.
If we use the local discrimination scheme , i.e., by local measurement
in $\{ |0\rangle , |1\rangle \}$ basis on the unknown entangled state,
then with assistance of classical communication,
we know the exact form
of the shared state \cite{walgateshort,walgate,prl87.277902}.
Since resource of maximally entangled states $|\Phi ^+\rangle $
are available, by local unitary transformation, we can change
two resource entangled states to the detected known form.
We still achieve the aim that
two copies of an entangled state are shared between Alice and Bob.
In this scheme, two known Bell states (resource) are consumed which
is not as efficiency as the CNOT scheme.
On the other hand, in order to obtain two copies of
$|\Psi ^+\rangle $ or  $|\Phi ^+\rangle $, at least, one
entangled state (resource) should be used.
We already know that the CNOT scheme uses only one entangled state (resource),
it is thus optimal.

\subsection{Local cloning and local discrimination}
If a set of quantum states can be perfected discriminated,
they can be copied perfectly since we can discriminate
them first, then prepare many copies of these states by using
the available entanglement resource.
For example, two orthogonal states can be copied perfectly.
We know that two Bell states can be locally discriminated,
as shown in the last subsection, they can be local cloned
perfectly if a {\it priori} Bell state resource is available.
Is it generally true that local discrimination means local
cloning being possible? In \cite{OH06}, it is stated that,
in general, the local copying is more
difficult than local discrimination.

However, local cloning and local discrimination are closely
related \cite{OH06}. The following result was
obtained in \cite{OH06}:
For d-dimensional system, and
suppose $d$ is prime, a set of maximally entangled states $\{ |\Psi _j\rangle \} _{j=0}^{N-1} $
are defined as
\begin{eqnarray}
|\Psi _j\rangle =(U_j\otimes I)|\Phi ^+\rangle ,
\end{eqnarray}
and
\begin{eqnarray}
U_j=\sum _{j=0}^{D-1}\omega ^{jk}|k\rangle \langle k|,
\end{eqnarray}
then the set  $\{ |\Psi _j\rangle \} _{j=0}^{N-1} $ can
be locally copied.

Here let us point out that the states of this set
can be local discriminated perfectly according to the
criteria proposed in \cite{Fan04}. The scheme can be like the following. It is known that $U_j=\sigma _z^j$, where the generalized Pauli
matrix $\sigma _z|k\rangle =\omega ^{k}|k\rangle $. We define
a class generalized Hadamard transformations as, up to an unimportant factor,
\begin{eqnarray}
(H_{\alpha })_{jk}=\omega ^{-jk}\omega ^{-\alpha s_k},
\end{eqnarray}
where $s_k=k+...+(d-1)$. We remark that those transformations correspond
to the $d+1$ mutually unbiased states. By applying those Hadamard transformations,
the generalized Pauli matrices transform as,
\begin{eqnarray}
H_{\alpha }\sigma _x^m\sigma _z^nH_{\alpha }^{\dagger }=
\sigma _x^{m\alpha +n}\sigma _z^{-m}.
\end{eqnarray}
Now we know that $\sigma _z$ matrix can be transformed to $\sigma _x$ matrix,
$U_j\rightarrow \sigma _x^j$. Since $(\sigma _x^j\otimes I)\sum |kk\rangle =\sum |k+j,k\rangle $,
that means those states can be distinguished by LOCC, also the above transformations
correspond to local unitary operations, we now conclude that states in set $\{ |\Psi _j\rangle \} _{j=0}^{N-1} $
can be distinguished by LOCC.

We next see how those states can be cloned locally, define generalized CNOT gate as,
\begin{eqnarray}
CNOT: |a\rangle |b\rangle \rightarrow |a\rangle |b+ a\rangle ,
\end{eqnarray}
where $|a+b\rangle $ modula $d$ is assumed.
We suppose an ancilla state $|\Phi ^+\rangle $ is shared
between Alice and Bob. Let both Alice and Bob perform
the generalized CNOT gate, we obtain the perfect copies
$|\Psi _j\rangle ^{\otimes 2}$.
This result can be derived as follows, according
the definition of the CNOT gate, we know that
\begin{eqnarray}
CNOT^{\dagger }: |a\rangle |b\rangle \rightarrow |a\rangle |b-a\rangle ,
\end{eqnarray}
It is straightforward to check that we have the
following properties
\begin{eqnarray}
|\Phi ^+\rangle _{12}|\Phi ^+\rangle _{34}&=&
CNOT_{13}^{\dagger }\otimes CNOT^{\dagger }_{24}
|\Phi ^+\rangle _{12}|\Phi ^+\rangle _{34}
\nonumber \\
&=&CNOT_{13}\otimes CNOT_{24}
|\Phi ^+\rangle _{12}|\Phi ^+\rangle _{34}.
\end{eqnarray}
Then we can find
\begin{eqnarray}
CNOT_{13}\otimes CNOT_{24}
|\Psi _j\rangle _{12}|\Phi ^+\rangle _{34}.
&=&CNOT_{13}\otimes CNOT_{24}(U_j\otimes I)_{13}
|\Phi ^+\rangle _{12}|\Phi ^+\rangle _{34}
\nonumber \\
&=&CNOT_{13}(U_j\otimes I)_{13}CNOT_{13}^{\dagger }
|\Phi ^+\rangle _{12}|\Phi ^+\rangle _{34}.
\end{eqnarray}
And we know the following result:
\begin{eqnarray}
CNOT(U_j\otimes I)CNOT^{\dagger }=U_j\otimes U_j.
\end{eqnarray}
The operator $U_j$ is copied. Thus by this method,
a set of maximally entangled states $\{ |\Psi _j\rangle \} _{j=0}^{N-1} $
are locally copied. This interesting phenomenon means that some unitary
operators can be cloned perfectly in the above framework.

In 2-dimensional system, we have presented
relations (\ref{opnoclone}) for CNOT gate previously \cite{gottesman98},
$(\sigma _x\otimes I)\rightarrow \sigma _x\otimes \sigma _x,
(\sigma _z\otimes I)\rightarrow \sigma _z\otimes I,
(I\otimes \sigma _x)\rightarrow I\otimes \sigma _x,
(I\otimes \sigma _z)\rightarrow \sigma _z\otimes \sigma _z$.
Those results imply that the bit flip errors are copied forwards while the phase
errors are copied backwards. But we cannot copy simultaneously the
bit flip errors and phase flip errors. This is a kind of no-cloning theorem.

Some other results about the cloning of entanglement are listed in the following.
The local cloning of product states without the shared entanglement ancilla
is studied in \cite{yingms05}. Distinguishing states locally
is also studied in \cite{walgate,walgateshort,chenyx1,chenyx2}.
The entangled states studied are generally pure states, however, it is
shown that maximally entangled states can also be mixed which is constituted
by very special structures. A subset of
those mixed maximally entangled states has similar properties
as those of pure maximally entangled states, and can be local distinguished perfectly \cite{Lizongguo}.
The local cloning of other cases are also studied,
including three-qubit case \cite{PhysRevA.74.032323.2006},
the continuous-variable case \cite{pra77.042301}, orthogonal entangled states
and catalytic copying \cite{njp6.164}. The local cloning of partially entangled
pure states in higher dimension is studied in \cite{ISI:000269858900005}.
Some results of local cloning with entanglement resource are presented in
\cite{EisertJacobs,CheflesGilson,CollinsLinden}.

Various schemes of quantum cloning of entanglement are studied in \cite{c-entagle0,c-entangle}. Quantum cloning of continuous-variable entangled states is studied in \cite{Weedbrook2008}.
The cloning of entangled photons to large scales which might be see by human eye
is analyzed in \cite{Sekatski2010}. The scheme of cloning unknown entangled state
and its orthogonal-complement state with some assistances is studied in \cite{ISI:000267510900005}
and also in \cite{ISI:000227361300007} and \cite{ISI:000242357000016}, the case of arbitrary
unknown two-qubit entangled
state is studied in \cite{ISI:000269486400010}.
The partial quantum cloning of bipartite state, i.e., only
part of the of the two-particle state is cloned, and the cloning of mixed states
are studied in \cite{ISI:000278194800002}.
 Coherent states cloning and local cloning are
presented in \cite{Dong2008}. The disentanglement is to preserve the local properties of an entangled state but erase the entanglement between the
subsystems, it is closely related with quantum cloning and the
broadcasting \cite{ISI:000084148000024}. The cloning machine used as
 approximate disentanglement is presented in \cite{ISI:000221638800002}.
 The two-qubit disentanglement and
inseparability correlation are presented in \cite{ISI:000085836300022}.

\subsection{Entanglement of quantum cloning}

It is also of interest to know the entanglement structure of states in the quantum cloning machines.
Potentially, those properties can be used to distinguish quantum from classical since entanglement is
considered to be one unique property of quantum world.

There are much progress about the theory of entanglement, see \cite{RMP-Horodecki} for a nice review.
For example, Peres-Horodeckis criteria \cite{Peres96,Horodecki19961} is simple to detect the entangled state.
Since the output states of the quantum cloning machines are generally available, we can use various techniques
to study the entanglement properties of the sole copies, or the whole output state of the cloning machine,
or the copies with the ancillary states, etc.

In \cite{PhysRevA.65.012304}, it is shown that for the $1\rightarrow 2$ cloning machines, the two copies of the
UQCM are entangled, while the two copies for the phase-covariant cloning machine are separable.
Further, we can use some measures of entanglement to quantify the entanglement.
The entanglement structure or separability of the asymmetric phase-covariant
quantum cloning is studied in \cite{Rezakhani2005278}. The bipartite and
tripartite entanglement of the output state of cloning are
studied in \cite{Bruss2003}.

\newpage

\section{Telecloning}

Quantum telecloning, as its name suggests, combines teleportation and quantum cloning so that quantum
states are distributed to some more spatially separated parties.
In the well-known teleportation scheme in \cite{PhysRevLett.70.1895}, quantum information of an unknown d-level system is completely transmitted from a sender Alice to a remote  receiver Bob by using the resource of a maximally entangled state.
It is natural to consider ``one-to-many'' and ``many-to-many'' communication via quantum channels.
This is the generalized teleportation scheme discussed in \cite{PhysRevA.61.032311,PhysRevA.67.012323}.
Of course, it is impossible to transmit quantum information with perfect fidelities for many copies,
because the no-cloning theorem \cite{Wootters1982} claims that an unknown quantum state can not be cloned perfectly.
However, as we already shown, we can try to quantum clone those quantum states approximately or probabilistically
which are allowed by quantum mechanics.

As we already presented, there are various quantum cloning machines which create optimal copies.
The aim of teleclone is to create optimal copies which is the same as that of the cloning machines,
in addition, we need to create optimal copies remotely by teleportation. Those remote copies themselves
may be spatially separated with each other. One may imagine that we can use first
quantum cloning machines to create optimal copies locally, then send those copies to their destination
points. The aim of teleclone can indeed be realized by this way. In this point, the importance of teleclone is like teleportation. Instead of teleportation, we can surely use flying qubits for states transportation. However, teleportation in the one
hand provides an alternative method. On the other hand, in case the quantum channel is noisy, the flying
qubits may experience inevitable decoherence which will induce errors. The teleportation scheme can avoid
this disadvantage by using the maximally entangled state resource. Even when the entanglement resource is
not perfect, the non-maximally entangled states can be purified locally to create maximally entangled states.
Now we are ready to study teleclone which combines together the quantum cloning and the quantum teleportation.
Still the resource of entanglement is necessary, however, its exact
form depends on our specially designed scheme.

Murao \emph{et al.} studied the optimal telecloning of 1 qubit to M qubits by using maximally entangled state \cite{PhysRevA.59.156}.
Telecloning which transmites an unknown d-level state to $M$ spatially separated receivers is studied in \cite{PhysRevA.61.032311}. And the telecloning of $N$ qubits to $M$ qubits, $M>N$, that requires positive valued operator measure (POVM) was proposed in \cite{Dur2000}. These telecloning are also called reversible telecloning because there is no loss of quantum information. The $1\rightarrow 2$ telecloning which uses nonmaximum entanglement (it is named irreversible telecloning, in comparison),  is studied in \cite{Bruss1998}, and the generalized case, $1\rightarrow M$ irreversible telecloning, is given in \cite{PhysRevA.63.020303}.
Quantum information can be encoded by states of continuous variables (CV)\cite{RevModPhys.77.513}.
The teleportation of CV is presented in \cite{prl84.3482}. The optimal 1 to $M$ telecloning of CV coherent states using a $(M+1)$-partite entangled state as a multiuser quantum channel is shown in \cite{PhysRevLett.87.247901}. This optimal telecloning could be achieved by exploiting nonmaximum entanglement between the sender and receivers. So this protocol was regarded as a CV irreversible telecloning. A scheme of CV reversible $N\rightarrow M$ telecloning, $M>N$, which distributes information without loss, is presented in \cite{PhysRevA.73.042315}. In this scheme, besides $M$ clones, additional $M-N$ anti-clones
are obtained at the same time by using $2M$-partite entanglement generalizing the scheme presented in Ref.\cite{PhysRevA.59.156}.

\subsection{Teleportation}

Let us review the original teleportation protocol and its generalization, i.e.,
the many-to-many scheme for transmitting quantum information. The teleportation scheme is proposed in \cite{PhysRevLett.70.1895}.
Alice wants to send an unknown state of a d-level particle to a spatially separated observer Bob with the help of quantum channel and classical communication. Alice's initial unknown state is,
\begin{equation}
|\psi\rangle=\sum_{k=0}^{d-1} \alpha_k |k\rangle_A,
\end{equation}
where $\sum_{k=0}^{d-1}|\alpha_k|^2=1$ and $\{|k\rangle\}$ is a complete orthogonal basis. In order to achieve the teleportation,
Alice and Bob are assumed to share a prior maximally entangled state, $|\xi \rangle =|\Phi ^+\rangle $,
\begin{align}
|\xi\rangle=\frac{1}{\sqrt{d}}\sum_{j=0}^{d-1}|j\rangle_P|j\rangle_B.
\end{align}
The total system is, $|\Psi\rangle=|\psi\rangle\otimes|\xi\rangle$, which can be rewritten as,
\begin{align}
|\Psi\rangle=|\psi\rangle _A\otimes|\xi\rangle _{PB}=
\frac{1}{d}\sum_{m,n=0}^{d-1}|\Phi_{mn}\rangle_{AP}\sum_{k=0}^{d-1}\exp\Big(-i\frac{2\pi nk}{d}\Big)\alpha_k|k+m\rangle_B,
\end{align}
where $k+m$ is assumed to module d. As standard, the generalized Bell basis are,
\begin{equation}
|\Phi_{mn}\rangle=\frac{1}{\sqrt{d}}\sum_{k=0}^{d-1} \exp\Big(i\frac{2\pi nk}{d}\Big)|k\rangle|k+m\rangle .
\end{equation}
Alice performs a joint Bell-type measurement on the input and port particles,
sends the measurement result $m,n$ to receiver Bob via classical communication.
The unitary transformation which brings Bob's particle to the original state of Alice's is
\begin{equation}
U_{mn}=\sum_{j=0}^{d-1}exp\Big(i\frac{2\pi jn}{d}\Big)|j\rangle\langle j+m|.
\end{equation}
As we already know, they are the generalized Pauli matrices in $d$ dimension.

Next, we will review the generalized symmetric teleportation scheme of N senders and $M, (M>N)$, receivers proposed in \cite{PhysRevA.67.012323}.
Assuming the senders $X_1, X_2, \cdots,X_N$ share an unknown, but with fixed form of entangled state $|\psi\rangle_X=\sum_{k=0}^{d-1}\alpha_k|\psi_k\rangle_{X_1}|\psi_k\rangle_{X_2}\cdots|\psi_k\rangle_{X_N}$, where $\{|\psi_k\rangle\}$ is an orthonormal basis of d-dimensional space. This state is a generalization of GHZ state.
The quantum entangled state which is the resource, consisting of N ``port'' particles $P_k(k=1,\cdots,N)$ and M receivers $C_k(k=1,\cdots,M)$,
takes a special form which is a $(N+M)$-partite state,
\begin{equation}
|\xi\rangle=\frac{1}{\sqrt{d}}\sum_{j=0}^{d-1}|\pi_j\rangle_{P_1}|\pi_j\rangle_{P_2}\cdots|\pi_j\rangle_{P_N}
|\phi_j\rangle_{C_1C_2\cdots C_M},
\end{equation}
where $\{|\pi_j\rangle\}$ denotes a d-dimensional orthonormal basis. The complete state of the system is,
\begin{align}
|\psi\rangle|\xi\rangle &=\frac{1}{\sqrt{d}}\sum_{k,j=0}^{d-1}\alpha _k |\psi_k\rangle_{X_1}|\pi_j\rangle_{P_1}|\psi_k\rangle_{X_2}|\pi_j\rangle_{P_2}\cdots|\psi_k\rangle_{X_N}|\pi_j\rangle_{P_N}
|\phi_j\rangle_{C_1C_2\cdots C_M}\nonumber\\
&=\frac{1}{d^{(N+1)/2}}\sum^{d-1}_{m,n_1,n_2,\cdots,n_N}|\Phi_{m,n_1}\rangle|\Phi_{m,n_2}\rangle\cdots|\Phi_{m,n_N}\rangle\otimes
\sum_k^{d-1}exp\Big[-i\frac{2\pi k}{d}(n_1+n_2+\cdots+n_N)\Big] \alpha_k|\phi_{k+m}\rangle
\end{align}
The following steps are involved in this protocol:
\begin{enumerate}
  \item The senders perform a joint Bell-type measurement on particles $X_j$ and $P_j$ and get the outcomes, $|\Phi_{m,n_1}\rangle,|\Phi_{m,n_2}\rangle,\cdots,|\Phi_{m,n_1}\rangle $. Here the generalized
      Bell states take the form $|\Phi_{m,n}\rangle =\frac {1}{\sqrt {d}}\sum _{k=0}^{d-1}exp\left( \frac {2\pi ikn}{d}\right)
      |\psi _k\rangle |\pi _{k+m}\rangle $, where modulo $d$ is assumed.
  \item The outcomes are sent to the receivers by using classical communication,
  \item Then, the receivers perform a local recovery unitary operator(LRUO) that satisfies $U_{m;n_1,n_2,\cdots,n_N}
  |\phi_{k+m}\rangle=exp\Big[i\frac{2\pi k}{d}(n_1+n_2+\cdots+n_N)\Big]|\phi_{k}\rangle$ .
\end{enumerate}
Several remarks are here: (i). In case that local operations are allowed, state $|\psi\rangle_X$ can be transformed
locally to just one qudit, $\sum_{k=0}^{d-1}\alpha_k|\psi_k \rangle $. (ii). The $M$ receivers are located in spatially
separated places, otherwise if they are in the same port, local quantum operations can reversely change the qudit
$\sum_{k=0}^{d-1}\alpha_k|\psi_k \rangle $ to a generalized GHZ like state shared by $M$ parties. (iii). The
scheme presented above combines the quantum information distribution and the teleportation together.

\subsection{Symmetric $1\rightarrow M$ telecloning}

In this subsection, we study the $1\rightarrow M$ generalized telecloning of qudit which is
studied in \cite{PhysRevA.61.032311}. In that scenario, the quantum information of d-level particle is transmitted optimally from one sender $X$ to $M$ receivers $C_1,C_2,\cdots,C_M$. One ``port'' and $(M-1)$ ancillary particles were involved. The resource, including the port particle and $(2M-1)$ output states ($M$ receivers and $(M-1)$ ancillas), is the maximally entangled state,
\begin{align}
|\xi\rangle &=\frac{1}{\sqrt{d[M]}}\sum_{k=0}^{d[M]-1}|\xi_k^M\rangle_{PA}|\xi_k^M\rangle_C \nonumber\\
&=\frac{1}{\sqrt{d}}\sum_{j=0}^{d-1}|j\rangle_P\otimes\Big(\frac{\sqrt{d}}{\sqrt{d[M]}}\sum_{k=0}^{d[M]-1}
  \sideset{_P}{}{\mathop{\langle}} j|\xi_k^M\rangle_{PA}|\xi_k^M\rangle_C\Big)\nonumber\\
  &=\frac{1}{\sqrt{d}}\sum_{j=0}^{d-1}|j\rangle_P\otimes|\phi_j\rangle,
\end{align}
where, $d[M]=C^{N}_{N+d-1}$, we denote the normalized symmetric state as,
$|\xi_k^M\rangle=\frac{1}{\sqrt{\mathcal{N}(\xi_k^M)}}|\mathcal{P}(a_0,a_1,\cdots,a_{M-1})\rangle $, ($\mathcal{P}$ denotes the sum of all possible permutation of the elements $\{a_0,a_1,\cdots,a_{M-1}\}$ for $a_j\in\{0,1,\cdots, d-1\}$ and $a_{j+1}>a_j$), $\{|\phi_j\rangle=\frac{\sqrt{d}}{\sqrt{d[M]}}\sum_{k=0}^{d[M]-1}
  \sideset{_P}{}{\mathop{\langle}} j|\xi_k^M\rangle_{PA} |\xi_k^M\rangle_C \}$ is a basis of the output state. The LRUO that satisfies $U_{mn}|\phi_{j+m}\rangle=e^{i\frac{2\pi nj}{d}}|\phi_j\rangle$ for the output state $\{|\phi_j\rangle\}$ is
\begin{equation}
U_{mn}=\underbrace{U_{mn}^A\otimes \cdots \otimes U_{mn}^A}_{M-1}\otimes\underbrace{ U_{mn}^C\otimes\cdots \otimes U_{mn}^c}_{M},
\end{equation}
where
\begin{align}
U_{mn}^A=\sum_{j=0}^{d-1}e^{-i\frac{2\pi jn}{d}}|j\rangle\langle j+m|,
U_{mn}^C=\sum_{j=0}^{d-1}e^{i\frac{2\pi jn}{d}}|j\rangle\langle j+m|
\end{align}
The initial state $|\psi\rangle_X=\sum_{j=0}^{d-1}\alpha_j |j\rangle $ of the sender X is ``encoded" to the separated output state  $|\phi\rangle_X=\sum_{j=0}^{d-1}\alpha_j |\phi_j\rangle $ held by the (M-1) ancillas and M receivers.
\begin{align}
|\xi_k^M\rangle=\frac{1}{\sqrt{\mathcal{N}(\xi_k^M)}}|\mathcal{P}(a_0,a_1,\cdots,a_{M-1})\rangle
=\frac{1}{\sqrt{\mathcal{N}(\xi_k^M)}}\sum_{a_j} \sqrt{\mathcal{N}(\xi_k'^{M-1})}|a_j\rangle|\xi_{k'}^{M-1}\rangle,
\end{align}
where $k'=f_M(a_0,\cdots,a_{j-1},a_j,\cdots, a_{M-1})$. There is a relationship between index k and k': $k=g(a_j,k')$, then
the total system takes the form,
\begin{eqnarray}
|\phi_j\rangle=\frac{\sqrt{d}}{\sqrt{d[M]}}\sum_{k'=0}^{d[M-1]-1}R^{k'}_j|\xi_{k'}^{M-1}\rangle_A\otimes|\xi
_{g(j,k')}^M\rangle_C,
\end{eqnarray}
where we use the notation,
$R^{k'}_j=\frac{\sqrt{\mathcal{N}(\xi_{k'}^{M-1})}}{\sqrt{\mathcal{N}(\xi_{g(j,k')}^M)}}$. By tracing out the
ancillary states $A$, we obtain the output state of $M$ qudits,
\begin{align}
\rho_C=&tr_A(|\phi\rangle\langle\phi|)=\sum_{l=0}^{d[M-1]-1} \sideset{_A}{}{\mathop{\langle}}\xi_l^{M-1}|\phi\rangle\langle\phi|\xi_l^{M-1}\rangle_A\nonumber\\
&=\frac{d}{d[M]}\sum_{j,j'=0}^{d-1}\sum_{k'=0}^{d[M-1]-1}\alpha_j\alpha_{j'}^\ast R_j^{k'}R_{j'}^{k'}|\xi_{g(j,k')}^M\rangle_C\langle\xi_{g(j,k')}^M|\nonumber\\
&=\frac{d[1]}{d[M]}\Big(\sum_{k=0}^{d[M]-1}|\xi_k^M\rangle\langle\xi_k^M|\Big)\Big(|\psi\rangle\langle\psi|\otimes
\mathbb{I}^{\otimes(M-1)}\Big)\Big(\sum_{k'=0}^{d[M]-1}|\xi_{k'}^M\rangle\langle\xi_{'k}^M|\Big)\nonumber\\
&=\frac{d[1]}{d[M]}s_M(|\psi\rangle\langle\psi|\otimes
\mathbb{I}^{\otimes(M-1)})s_M=\hat{T}(|\psi\rangle\langle\psi|)
\end{align}
 This reduced density matrix of the receivers is consistent with the density matrix for $1\rightarrow M$ d-level optimal clones
 \cite{PhysRevA.58.1827,PhysRevA.84.034302}. Let us emphasize that the output of $M$ qudits are consistent
 with optimal cloning, moreover, they are spatially separated in different places. The $1\rightarrow M$ telecloning
 is also related with programming protocol which is studied in \cite{Ishizaka-Hiroshima}.

\subsection{Economical phase-covariant telecloning}

Quantum cloning machines have economic and non-economic cases. Similarly, we also
have the economic telecloning \cite{Wang2009,Wang2009a}.
We know that phase-covariant cloning has been studied in \cite{Bruss2000,PhysRevA.65.012304,PhysRevA.67.042306,PhysRevA.67.022317}.
The $1\rightarrow M$ optimal economical phase-covariant cloning for qubits was proposed in \cite{PhysRevA.76.034303},
and the $1\rightarrow 2$ economic map (non-optimal) for qudits also was studied \cite{PhysRevA.72.052322}.
For special value $M=k d+N$, the optimal $N\rightarrow M$ economical cloning for qudits has been introduced \cite{PhysRevA.71.042327}.
A protocol for the $1\rightarrow M$ economical phase-covariant telecloning of qubits has been demonstrated
in \cite{Wang2009}, and the $1\rightarrow 2$ economical phase cloning of qudits has been derived in \cite{Wang2009a}.

We next see the $1\rightarrow M $ economical phase cloning of qubits, the input state is, $|\psi\rangle_X=\cos\frac{\theta}{2}|0\rangle_X+e^{i\phi}\sin\frac{\theta}{2}|1\rangle_X$:
\begin{align}
\begin{cases}
U |0\rangle_1|R_{2\cdots M}\rangle &=|\phi_0\rangle=|00\cdots0\rangle_M\nonumber\\
U |1\rangle_1|R_{2\cdots M}\rangle &=|\phi_1\rangle=\frac{1}{\sqrt{M}}\sum_{j=1}^{M}|0\cdots 1_j \cdots 0\rangle_M
\end{cases}
\end{align}
We get the output state which is $|\psi\rangle^{out}_M=\cos\frac{\theta}{2}|\phi_0\rangle_X+e^{i\phi}\sin\frac{\theta}{2}|\phi_1\rangle_X$, and fidelity $ F=\sideset{_X}{}{\mathop{\langle}}\psi|tr(|\psi\rangle_{out}\langle\psi|)|\psi\rangle_X=\frac{1}{M}\sin^4\frac{\theta}{2}
+\cos^4\frac{\theta}{2}+\sin^2\frac{\theta}{2}\cos^2\frac{\theta}{2}\Big(\frac{2}{\sqrt{M}}+\frac{M-1}{M}\Big)$.
Second, the telecloning scheme is that the sender X prepares the quantum information channel $|\xi\rangle_{PC}=\frac{1}{\sqrt{2}}\big(|0\rangle_P|\phi_0\rangle_C+|1\rangle_P|\phi_1\rangle_C\big)$. The total state can be expressed as
\begin{align}
|\Psi\rangle_{XPC}=|\psi\rangle_X|\xi\rangle_{PC}=\frac{1}{2}
\Big[|\Phi^0\rangle_{XP}\otimes(\cos\frac{\theta}{2}|\phi_0\rangle+e^{i\phi}\sin\frac{\theta}{2}|\phi_1\rangle)
+|\Phi^1\rangle_{XP}\otimes(\cos\frac{\theta}{2}|\phi_0\rangle-e^{i\phi}\sin\frac{\theta}{2}|\phi_1\rangle)\nonumber\\
+|\Phi^2\rangle_{XP}\otimes(e^{i\phi}\sin\frac{\theta}{2}|\phi_0\rangle+\cos\frac{\theta}{2}|\phi_1\rangle)
+|\Phi^3\rangle_{XP}\otimes(e^{i\phi}\sin\frac{\theta}{2}|\phi_0\rangle-\cos\frac{\theta}{2}|\phi_1\rangle)\Big],
\end{align}
where $ \{|\Phi^0\rangle=|\Phi^+\rangle,|\Phi^1\rangle=|\Phi^-\rangle,|\Phi^2\rangle=|\Psi^+\rangle,|\Phi^3\rangle=|\Psi^-\rangle\}$ are the Bell basis. The next steps have been reviewed above in the generalized telecloning.

For the input state $|\psi\rangle_X=\frac{1}{\sqrt{d}}\sum_{j=0}^{d-1} e^{i\theta_j}|j\rangle_X$, the $1\rightarrow 2$
economical phase-covariant cloning machine was demonstrated in \cite{PhysRevA.72.052322}. It takes the form,
\begin{align}\begin{cases}
U |0\rangle_X|R\rangle=|\phi_0\rangle=|00\rangle \\
U |j\rangle_X|R\rangle=|\phi_j\rangle=\frac{1}{\sqrt{2}}(|j0\rangle+|0j\rangle), \quad(j\neq0)
\end{cases}
\end{align}
The fidelity is $F_{econ}=\sideset{_X}{}{\mathop{\langle}}\psi|tr(|\psi\rangle_{out}\langle\psi|)|\psi\rangle_X=\frac
{1}{2d^2}[(d-1)^2+(1+2\sqrt{2})(d-1)+2]$. However, the optimal fidelity of $1\rightarrow2$ phase-covariant (with an ancilla)
presented in \cite{PhysRevA.67.022317} is,
$F_{opt}=\frac{1}{4d}(d+2+\sqrt{d^2+4d-4})$. When $d=2$, $F_{econ}=F_{opt}$ and otherwise $d>2$, $F_{econ}<F_{opt}$.
It is possible to achieve with optimal fidelity $F_{opt}$  probabilistically \cite{Wang2009a}.
In this scheme, the entangled state used
is $|\xi\rangle_{PC}=\sum_{j=0}^{d-1}x_j|j\rangle_P|\phi_j\rangle_C,$ where the coefficients $x_j$,
that are assumed to be real numbers, satisfy the normalization condition $\sum_{j=0}^{d-1}x_j^2=1$.
The quantum state of the whole system is,
\begin{align}
|\Psi\rangle_{XPC}=|\psi\rangle_X\otimes|\xi\rangle_{PC}
=\frac{1}{d}\sum_{m,n=0}^{d-1}|\Phi_{mn}\rangle_{XP}\sum_{j=0}^{d-1}\exp\big(i\frac{2\pi nj}{d}\big)x_{j+m}
e^{i\theta_j}|\phi_{j+m}\rangle_C
\end{align}
Only when the outcome of the Bell-type joint measurement is $\{m=0,n\}$( with probability $1/d$), the receivers can obtain the clones $|\psi\rangle_{out}=\sum_{j=0}^{d-1}x_j e^{i\theta_j}|\phi_j\rangle $ by using the LRUO $U=
U_{0n}\otimes\mathbb{I}$. The fidelity of this clones is $F^t_{econ}=\frac{1}{d}\Big(1+\sqrt{2}x_0\sum_{j=0}^{d-1}
x_j+\sum_{i=1}^{d-2}\sum_{j=i+1}^{d-1}x_i x_j\Big)$. We set $\{x_j\}$ as
\begin{eqnarray}
&&x_0=X(d)=\sqrt{\frac{4(d-1)}{D(D+d-2)}}, \nonumber \\
&&x_j=Y(d)=\sqrt{\frac{d^2+(d-2)D}{D(D+d-2)(d-1)}},~~(j\neq0),
\end{eqnarray}
where $D=\sqrt{d^2+4d-4}$. It's not difficult to verify that $F^t_{econ}=F_{opt}$ for any $d$.
Actually, the output state of this telecloning scheme is equivalent to the $\rho^C_{opt}$ of the optimal phase-covariant  cloning after tracing out of the ancilla \cite{PhysRevA.67.022317}. For $d>2$, the von Neumann entropy $S(|\xi\rangle\langle\xi|)=-X^2(d)\log_2 X^2(d)-(d-1)Y^2(d)\log_2Y^2(d)<\log_2 d$, which implies $|\xi\rangle$ is only partially entangled. Thus, we can conclude that the suitable quantum entanglement in realizing the optimal $1\rightarrow 2$
cloning of qudits with a certain probability $1/d$ are special configurations of nonmaximally entangled states rather than the maximally entangled states.

\subsection{Asymmetric telecloning}

Quantum telecloning described in the previous section evenly distributes information of the unknown input state to the distant receivers. However, it may be desirable to transmit information to several different receivers with different fidelities. For example, the sender Alice trusts Bob more than Claire hope
Bob's fidelity is larger. These schemes are asymmetric telecloning.
The 1 to 2 optimal asymmetric quantum cloning of qubits was introduced in \cite{Cerf1998, ISI:000074966200008,PhysRevA.58.4377, PhysRevLett.84.4497}. The 1 to 2 asymmetric cloning machine was generalized to d-dimension case in \cite{383952220000215, Braunstein2001},
and recently in \cite{PhysRevA.84.034302}.

Here, we briefly review $1\rightarrow 2$ asymmetric telecloning for qubits \cite{PhysRevA.61.032311} as an example.
The entanglement state resource is,
$|\xi\rangle=\frac{1}{\sqrt{2}}(|0\rangle_P|\phi_0\rangle +|1\rangle_P|\phi_1\rangle,$ where
\begin{align}\begin{cases}
|\phi_0\rangle &=\frac{1}{\sqrt{1+p^2+q^2}}(|0\rangle|0\rangle_B|0\rangle_C+p|1\rangle|0\rangle_B
1\rangle_C+q|1\rangle|1\rangle_B|0\rangle_C) \\
|\phi_1\rangle &=\frac{1}{\sqrt{1+p^2+q^2}}(|1\rangle|1\rangle_B|1\rangle_C+p|0\rangle|1\rangle_B
0\rangle_C+q|0\rangle|0\rangle_B|1\rangle_C) \quad(p+q=1)
\end{cases}\end{align}
The LRUOs satisfy the conditions,
$\sigma_z\otimes\sigma_z\otimes\sigma_z|\phi_{0(1)}\rangle =(-)|\phi_{0(1)}\rangle,
\sigma_x\otimes\sigma_x\otimes\sigma_x|\phi_{0(1)}\rangle =|\phi_{1(0)}\rangle$. And the final output state is $|\psi\rangle
_{out}=\alpha_0|\phi_0\rangle+\alpha_1|\phi_1\rangle$ while the input state being $|\psi\rangle
_{X}=\alpha_0|0\rangle+\alpha_1|1\rangle$. The fidelities of Bob and Claire, which satisfy the
trade-off relation, $\sqrt{(1-F_B)(1-F_C)}= F_B+F_C-\frac{3}{2}$, respectively are
\begin{equation}
F_B=\frac{1+p^2}{1+p^2+q^2} \quad F_C=\frac{1+q^2}{1+p^2+q^2}.
\end{equation}
Next, we show the results of 1 to 2 asymmetric telecloning of qudits. The asymmetric cloning machine is
\begin{align}
U|j\rangle_{C_1}|00\rangle_{C_2 A}=|\phi_j\rangle
=\sum_{m,n=0}^{d-1}\beta_{m,n}(V_{m,n}|j\rangle_{C_1})\otimes|\Phi_{m,-n}\rangle_{C_2 A}\\
=\sum_{m,r=0}^{d-1}b_{m,r}|j+m\rangle_{C_1}|j+r\rangle_{C_2}|j+m+r\rangle_{A},
\end{align}
where $V_{m,n}=\sum_{j=0}^{d-1}e^{2\pi jn/d}|j+m\rangle \langle j|$ are generalized Pauli matrices  and
$b_{m,r}=\frac {1}{\sqrt {d}}\sum_{n=0}^{d-1}e^{-i2\pi nr/d}\beta_{m,n}$. We have a mathematical equation,
\begin{equation}
\sum_{m,n=0}^{d-1}\beta_{m,n}|\Phi_{m,n}\rangle_{R C_1}|\Phi_{m,-n}\rangle_{C_2 A}
=\sum_{m,n=0}^{d-1}\gamma_{m,n}|\Phi_{m,n}\rangle_{R C_2}|\Phi_{m,-n}\rangle_{C_1 A}
\end{equation}
where $\sum_{m,n=0}^{d-1}|\beta_{m,n}|^2=1$, $\gamma_{m,n}=\frac{1}{d}\sum_{x,y=0}^{d-1}e^{i2\pi (nx-my)/d}\beta_{x,y}$. Then we project this equation on $|j\rangle_R$ and get $|\phi_j\rangle=\sum_{m,n=0}^{d-1}\gamma_{m,n}(V_{m,n}|j\rangle_{C_2})\otimes|\Phi_{m,-n}\rangle_{C_1 A}$. The input state $|\psi\rangle_X=\sum_{j=0}^{d-1}\alpha_j |j\rangle$  is copied into the output states $|\psi\rangle_{out}=\sum_{j=0}^{d-1}\alpha_j |\phi_j\rangle_{C_1 C_2 A}$. This output states are described by the reduced density matrices, respectively,
\begin{align}
\rho_{C_1}=tr_{C_2 A}(|\psi\rangle_{out}\langle\psi|)=
\sum_{m,n=0}^{d-1}|\beta_{m,n}|^2 V_{m,n}|\psi\rangle_X\langle\psi|V^\dag_{m,n},\\
\rho_{C_2}=tr_{C_1 A}(|\psi\rangle_{out}\langle\psi|)=
\sum_{m,n=0}^{d-1}|\gamma_{m,n}|^2 V_{m,n}|\psi\rangle_X\langle\psi|V^\dag_{m,n}.
\end{align}
In order to generate the clones that are characterized by the optimal fidelities which are independent of the input state, the following condition should be satisfied \cite{383952220000215,688607420020710},
\begin{eqnarray}
b_{0,0}=\frac{1}{\sqrt{d}}[\nu+(d-1)\mu],&&b_{m,0}=\sqrt{d}\mu, \nonumber \\
b_{0,r}=\frac{1}{\sqrt{d}}(\nu-\mu),&&b_{m,r}=0, \quad (m\neq0,r\neq0)
\end{eqnarray}
And we get the fidelities of two clones
\begin{align}
F_{C_1}=\frac{1+(d-1)p^2}{1+(d-1)(p^2+q^2)},\quad F_{C_2}=\frac{1+(d-1)q^2}{1+(d-1)(p^2+q^2)},
\end{align}
where $p=\frac{\nu-\mu}{\nu+(d-1)\mu}, q=1-p$. When $p=q=1/2$, the fidelities are in agreement with $F=\frac{N(d-1+M)+M}
{M(d+N)}(N=1,M=2)$ obtained by Werner \cite{PhysRevA.58.1827}. The telecloning scheme requires the quantum entanglement, shared by the port, ancilla, and the receivers $C_1,C_2$, is given as $|\xi\rangle=\frac{1}{\sqrt{d}}\sum_{j=0}^{d-1}|j\rangle_P|\phi_j\rangle_{C_1 C_2 A}$. After the sender performs a Bell-type joint measurement on the input and port particles, and gets the result $m,n$, the ancilla and the receivers $C_1,C_2$ perform the LRUO $U_{m,n}^{local}=\sum_{j_1,j_2,j_3}e^{i2\pi n(j_1+j_2-j_3)/d}|j_1\rangle\langle j_1+m|_{C_1}
\otimes |j_2\rangle\langle j_2+m|_{C_2}\otimes |j_3\rangle\langle j_3+m|_{A}$ and gets the output state $|\psi\rangle_{out}=\sum_{j=0}^{d-1}\alpha_j |\phi_j\rangle_{C_1 C_2 A}$.

\subsection{General telecloning}

In the general case \cite{Zhangnew}, the sender hold the $N$ identical input states $|\varphi \rangle ^{\otimes N}=(\sum _jx_j|j\rangle )^{\otimes N}$ at the same location $X$, so we have the state, $|\psi\rangle_X=|\varphi \rangle ^{\otimes N}$. One may find that this state
belongs to the symmetric subspace,
\begin{eqnarray}
|\psi\rangle_X=\sum_{\overrightarrow{n}}^{N}(\sqrt{N!}\prod_j \frac{\alpha_j^{n_j}}{\sqrt{n_j !}})|\overrightarrow{n}\rangle ,
\label{teleinitial}
\end{eqnarray}
where $|\overrightarrow{n}\rangle $ are the basis of the symmetric subspace $\mathcal{H}^{\otimes N}_+$ in the
standard representation,
each element of vector $\overrightarrow{n}$ corresponding to the number of state $|i\rangle $ as shown
explicitly in (\ref{expansion}).
For convenience, we introduce the notation $y_{\overrightarrow{n}}\equiv (\sqrt{N!}\prod_j \frac{\alpha_j^{n_j}}{\sqrt{n_j !}})$
and the initial state in Eq.(\ref{teleinitial}) can be rewritten as,
\begin{eqnarray}
|\psi\rangle_X=\sum_{\overrightarrow{n}}^{N} y_{\overrightarrow{n}}|\overrightarrow{n}\rangle .
\end{eqnarray}
The sender would like to distribute these states to spatially separated $M$ receivers, $M\ge N$.
Following the teleportation procedure,
the sender performs a joint measurement on input particles $X_1,X_2,\cdots, X_N$ and port particles $P_1,P_2,\cdots,P_N$
which are acting as ancillary states,
then announce the outcome to the $M-N$ ancillas and $M$ receivers
via classical communication. Next,
the ancillas and receivers get the optimal clones after applying the specific Local Recovery Unitary Operator (LRUO).
In order to achieve this aim, instead of using joint Bell-type measurement, the sender performs a
more general positive operator-valued measure (POVM) defined by $|\chi(\overrightarrow{x})\rangle $
on the system,
\begin{eqnarray}
|\chi(\overrightarrow{x})\rangle=[\mathbb{I}^{\otimes N}_X\otimes U(\overrightarrow{x})^{\otimes N}_P]
\frac{1}{\sqrt{d[N]}}\sum_{\overrightarrow{n}}^{N}|\overrightarrow{n}\rangle_{X}|\overrightarrow{n}\rangle_{P}.
\end{eqnarray}
We remark that the state $|\chi(\overrightarrow{x})\rangle $ in the projection corresponds
to a bipartite maximally entangled state $\frac{1}{\sqrt{d[N]}}\sum_{\overrightarrow{n}}^{N}|\overrightarrow{n}\rangle_{X}|\overrightarrow{n}\rangle_{P}$
with tensor product of $N$ identical unitary operators on one party. This POVM
can be confirmed by the following property,
\begin{eqnarray}
\int d\overrightarrow{x}F_{\overrightarrow{x}}=\int d\overrightarrow{x}\lambda(\overrightarrow{x})|
 \chi(\overrightarrow{x})\rangle\langle \chi(\overrightarrow{x})| =S^N_X\otimes S^N_P,
 \label{POVM}
\end{eqnarray}
where $S^N\otimes S^N$ is the identity in the space  $\mathcal{H}^{\otimes N}_+\otimes\mathcal{H}^{\otimes N}_+$, $ U(\overrightarrow{x})$ is an element of Lie group $SU(d)$, and the vector $\overrightarrow{x}$ consisting $(d^2-1)$ parameters
which can determine the unitary operator. Next we show that the latter equation can be satisfied.
According to the theorem of Weyl Reciprocity \cite{ma2007group},
the unitary transformation $U^{\otimes N}$ and permutation permutation $P_\alpha$ can be exchanged.
If $\mathcal{Y}^{[\lambda]}_{\mu}$ is a standard Young operator corresponding to the standard Young tableau with $N$ boxes, the subspace $\mathcal{Y}^{[\lambda]}_{\mu}\mathcal{H}^{\otimes N}$ will be invariant under transformation $U^{\otimes N}$.
Considering that the symmetric projection $S^N$ is equal to the standard Young operator $\frac{1}{N!}\mathcal{Y}^{[N]}$, we have
\begin{eqnarray}
U(\overrightarrow{x})^{\otimes N} S^N&=&S^N U(\overrightarrow{x})^{\otimes N},\\
U(\overrightarrow{x})^{\otimes N}|\overrightarrow{n_1}\rangle &=&
\sum_{\overrightarrow{n_2}}D_{\overrightarrow{n_2},\overrightarrow{n_1}}(\overrightarrow{x})|\overrightarrow{n_2}\rangle,
\end{eqnarray}
where $D(\overrightarrow{x})$ is a representation of Lie group $SU(d)$. A group theorem  states that an irreducible representation of group $SU(d)$ will be induced when $U(\overrightarrow{x})^{\otimes N}$ operates on invariant subspace $\mathcal{Y}^{[\lambda]}_\mu\mathcal{H}^{\otimes N}$  when $\mathcal{Y}^{[\lambda]}_\mu$ is a standard Young operator, see \cite{ma2007group}. Thus $D(\overrightarrow{x})$ is an irreducible representation of group $SU(d)$. Then according to Schur's lemmas and the orthogonality relations \cite{ma2007group}, we obtain,
\begin{equation}\label{schur}
\frac{1}{d[N]}\int d\overrightarrow{x}\lambda(\overrightarrow{x})D_{\overrightarrow{n_1},\overrightarrow{n_2}}(\overrightarrow{x})
D^*_{\overrightarrow{n_3},\overrightarrow{n_4}}(\overrightarrow{x})=\delta_{\overrightarrow{n_1},\overrightarrow{n_3}}
\delta_{\overrightarrow{n_2},\overrightarrow{n_4}}
\end{equation}
This formula ensures that the integral of the projectors $F_{\overrightarrow{x}}$ is equal to the identity operator in the space  $\mathcal{H}^{\otimes N}_+\otimes\mathcal{H}^{\otimes N}_+$ which should be satisfied
for a POVM. In special case $d=2$, because we know the analytical expression of the unitary matrix $U(\overrightarrow{x})$ and its irreducible representation $D(\overrightarrow{x})$, an appropriate finite POVM can be constructed, then the integral reduces to summation.
The importance to construct finite POVM is that its explicit form is necessary for experimental implementation.

The total system can be expressed as
\begin{align}
|\psi\rangle_X |\xi\rangle_{PAC}
&=\frac{1}{d[N]} \sum_{\overrightarrow{x}}\lambda(\overrightarrow{x})|\chi({\overrightarrow{x})}\rangle_{XP}
[U^\dag(\overrightarrow{x})^{\otimes (M-N)}_A\otimes U^T(\overrightarrow{x})^{\otimes M}_C]^\dag
\sqrt{\frac{d[N]}{d[M]}}\Big(\sum_{\overrightarrow{n}}^{N}y_{\overrightarrow{n}}
\sideset{_P}{}{\mathop{\langle}}\overrightarrow{n}|\Big)
\Big(\sum_{\overrightarrow{m}}^{M}|\overrightarrow{m}\rangle_{PA}|\overrightarrow{m}\rangle_{C}\Big)\nonumber\\
&=\frac{1}{d[N]} \sum_{\overrightarrow{x}}\lambda(\overrightarrow{x})|\chi({\overrightarrow{x})}\rangle_{XP} [U^{local}(\overrightarrow{x})]^\dag |\psi_c\rangle_{AC}
\end{align}

\begin{figure}
\includegraphics[height=9cm]{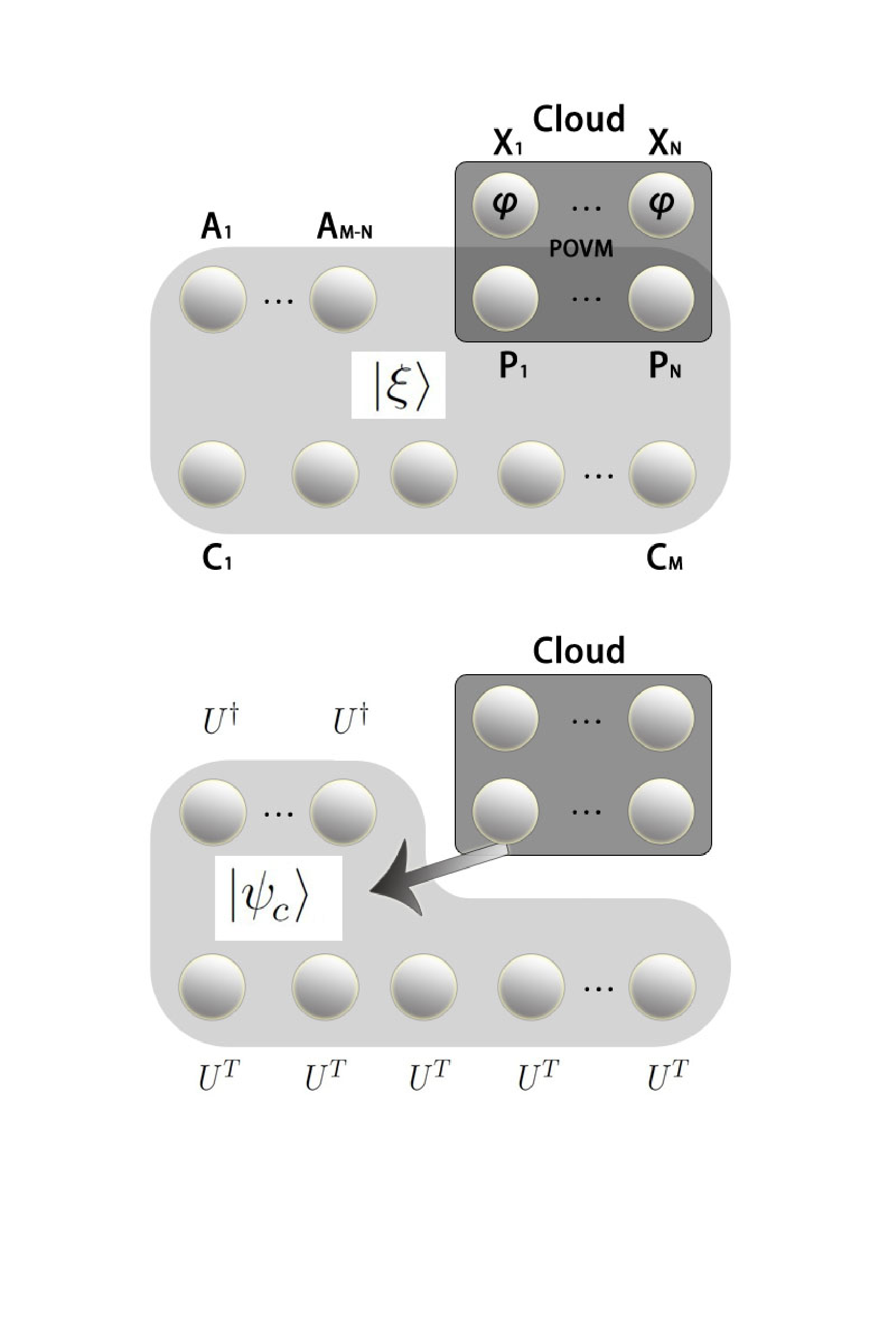}
\caption{Scheme of telecloning. The sender who may possess several ports
share a maximally entangled state with several receivers who are spatially separated, and
possibly assisted with ancillary states. The sender (Cloud) performs a POVM and announces the measurement
result, the receiver can recover their states locally, see \cite{Zhangnew}.}
\label{figure-yinan-network}
\end{figure}

The LRUO is $U^{local}(\overrightarrow{x})=U^\dag(\overrightarrow{x})^{\otimes (M-N)}_A\otimes U^T(\overrightarrow{x})^{\otimes M}_C$. As we expect, the sender distributes the universal cloning state $|\psi_c\rangle_{AC}$ to spatially separated $M$ receivers assisted by
$(M-N)$ ancillas. The scheme of the telecloning can be represented in FIG.\ref{figure-yinan-network}.

The asymmetric quantum telecloning for multiqubit states with various figures of merit
are investigated by Chen and Chen \cite{ISI:000251336100002}.
The reverse processing of telecloning is the remote state concentration. Roughly speaking,
the final state of the information concentration is the initial state of the telecloning.
It is shown that in the concentration processing, the bound entangled state can be used as
a resource \cite{muraovedral}. The standard entangled state possesses
similar capability in the quantum information concentration \cite{Zhangnew}.
This remote quantum information concentration is also studied in \cite{Wang2011}.

Telecloning is a combination of teleportation and quantum cloning, its reverse process
is remote quantum information concentration. The quantum  information distribution
and concentration are expected to be the fundamental functions of quantum networks.
Those functions can be potentially used for clocks synchronization with quantum
advantage \cite{zhangnew2}. We can expect that the network quantum computation
will be an important subject for further explorations.
Some experimental and implementation schemes of the telecloning are reported as in the
following.
The experimental realization of telecloning is performed by partial teleportation scheme \cite{Zhaozhi05}.
A proposal of distance cloning is in \cite{filip2}.
The entanglement resource of up to six qubits of Dicke states
is created experimentally \cite{dickestate-experiment}.
The experimental implementation of telecloning of optical
coherent states is demonstrated in \cite{KoikeTakahashi}. The experimental telecloning
of phase-conjugate inputs is presented in \cite{ZhangJing08PRA}.
Telecloning of
entanglement is presented in \cite{ISI:000232228300059}, the telecloning
of W state is studied in \cite{Yan2009}.
A scheme to implement an economical phase-covariant quantum telecloning is separate
cavities is proposed in \cite{Fang2012a}. Implementation of telecloning of economic phase-covariant
about bipartite entangled state is studied in \cite{Meng2009}.
The continuous variable telecloning with bright entangled beams is studied in \cite{Olivares2008}.
The controlled telecloning and teleflipping
for one pure qubit is studied in \cite{Zhan2009}.

\newpage

\section{Quantum cloning for continuous variable systems}

This section is devoted to the issue of quantum cloning machine for continuous variable (CV) systems. The available reviews of this topic can be found in \cite{RevModPhys.77.1225,CerfGrangier}. The photonic state of optical system is usually described by CV. Most schemes and protocols of quantum computation and quantum information can be realized and demonstrated by photonic state of CV, it may possesses unique advantage other than other systems.
The reviews of CV quantum information can be found in \cite{RevModPhys.77.513,Wangphysreport,RMP-gaussian}, see spin squeezing in \cite{Majian}.
An example of continuous systems is simply to consider the position and momentum of a particle, or the two quadratures of a quantized electromagnetic field. Instead of universal cloning, we only study the case of $N\rightarrow M$ Gaussian cloning for coherent states, whose precise definition will be given in the context below. We shall first get the fidelity bound for $N\rightarrow M$ Gaussian cloning\cite{PhysRevA.62.040301}, and then give an explicit implementation using a linear amplifier and beam splitters\cite{PhysRevLett.86.4938}. Note that the same procedure is also suitable if the input states are squeezed states, provided little change of parameters of the devices is made\cite{PhysRevA.62.040301,PhysRevLett.86.4938}.

\subsection{Optimal bounds for Gaussian cloners of coherent states}

We deal with a quantum system described in terms of two canonically conjugated operators $\hat{x}$ and $\hat{p}$, which respectively has a continuous spectra. Since $\hat{x}$ and $\hat{p}$ are conjugated, they cannot both be copied perfectly, so we hope to find a cloning machine which makes an approximate cloning and obtain an ``optimal'' result. Corresponding to universal cloning, we here focus on cloning transformations which take only coherent states as input.
That is, the input states of the cloning machine form a set $\mathcal{S}$, which can be parametrized as,
\begin{equation}
\mathcal{S}=\left\{ |\alpha\rangle:\alpha=\frac{1}{\sqrt{2}}(x+ip),x,p\in\bm{R}\right\},
\end{equation}
where $\langle \alpha|\hat{x}|\alpha\rangle = x$ and $\langle \alpha|\hat{p}|\alpha\rangle = p$. Moreover we shall only consider $N\rightarrow M$ symmetric Gaussian cloners(SGCs) which can be defined as a linear completely positive map:$C_{N,M}:\mathcal{H}^{\otimes N}\rightarrow\mathcal{H}^{\otimes M}$, where $\mathcal{H}$ stands for an infinite-dimensional Hilbert space. So after the transformation we shall get $\rho_M=C_{N,M}(|\alpha\rangle\langle\alpha|^{\otimes N})$. To mean Gaussian, the reduced state of a single clone needs to satisfy:
\begin{align}
\rho_1 &= Tr_{M-1}(\rho_M) \nonumber\\
&= \frac{1}{\pi\sigma_{N,M}^2}\int d^2 \beta e^{-|\beta|^2/\sigma_{N,M}^2}D(\beta)|\alpha \rangle\langle\alpha |D^{\dagger}(\beta),
\end{align}
where the integral is performed over all values of $\beta=(x+ip)/\sqrt{2}$ in the complex plane, note that we have set $\hbar=1$, and $D(\beta)=exp(\beta\hat{a}^{\dagger}-\beta^{*}\hat{a})$ is a displacement operator which shifts a state of $x$ in position and $p$ in momentum, $\hat{a}$ and $\hat{a}^{\dagger}$ denote annihilation and creation operators respectively. As a result, after cloning for each copy an extra noise $\sigma_x^2=\sigma_p^2=\sigma_{N,M}^2 $ on the conjugate variables $x$ and $p$ are added. It is readily checked that the cloning fidelity $f_{N,M}=\langle\alpha|\rho_1|\alpha\rangle$ is the same for any coherent input state $|\alpha\rangle$, provided $\sigma_{N,M}$ remains invariant, which means, our cloner is symmetric.
Through simple computation one finds,
\begin{equation}
f_{N,M}=\langle\alpha|\rho_1|\alpha\rangle=\frac{1}{1+\sigma_{N,M}^2}.
\end{equation}
Now we shall make the proposition that the lower bound of $\sigma_{N,M}$ is
\begin{equation}\label{lowerboundfornoise}
\bar{\sigma}_{N,M}^2=\frac{M-N}{MN},
\end{equation}
which implies the optimal fidelity for $N\rightarrow M$ cloning machine is
\begin{equation}
f_{N,M}=\frac{1}{1+\bar{\sigma}_{N,M}^2}=\frac{MN}{MN+M-N}.
\end{equation}
Next we will prove (\ref{lowerboundfornoise}). As first step we shall come up with a lemma.

\textit{Lemma 1}. Cascading a $N\rightarrow M$ cloner with an $M\rightarrow L$ cloner cannot be better than the optimal $N\rightarrow L$ cloner. In our case, two cascading $N\rightarrow M$ and $M\rightarrow L$ SGCs result in a single $N\rightarrow L$ SGC whose variance is simply the sum of variances of the two cascading SGCs. Hence we have
\begin{equation}\label{noisecascading}
\bar{\sigma}_{N,L}^2\le \sigma_{N,M}^2+\sigma_{M,L}^2,
\end{equation}
where $\bar{\sigma}_{N,M}^2$ stands for the low variance bound of $N\rightarrow L$ cloner.

The proof for Lemma 1 can be found in \cite{PhysRevA.62.040301}. We will use lemma 1 to reach (\ref{lowerboundfornoise}). From (\ref{noisecascading}), setting $L\rightarrow \infty$ we get
\begin{equation}
\bar{\sigma}_{N,\infty}^2\le \sigma_{N,M}^2+\bar{\sigma}_{M,\infty}^2.
\end{equation}
Then we can use quantum estimation theory to analyze $\bar{\sigma}_{N,\infty}^2$, which is the variance of an optimal joint measurement of $\hat{x}$ and $\hat{p}$ on N replicas of a system. We have \cite{Holevo1982},
\begin{equation}\label{inequalestimation}
g_x\sigma_x^2(1)+g_p\sigma_p^2(1)\ge g_x\Delta\hat{x}^2
+g_p\Delta\hat{p}^2+\sqrt{g_xg_p},
\end{equation}
for all values of the constants $g_x,g_p>0$, where $\sigma_x^2(1)$ and $\sigma_p^2(1)$ denote the variance of the measured values of $\hat{x}$ and $\hat{p}$, while $\Delta \hat{x}^2$ and $\Delta \hat{p}^2$ denote the intrinsic variance of observables $\hat{x}$ and $\hat{p}$, respectively. For each value of $g_x$ and $g_p$, we have a specific positive-operator-valued measure(POVM) which achieves the bound. Also, as in classical statistics, we have \cite{Holevo1982},
\begin{equation}
\sigma_x^2(N)=\frac{\sigma_x^2(1)}{N},
\sigma_p^2(N)=\frac{\sigma_p^2(1)}{N},
\end{equation}
where $\sigma_x^2{N}$ or $\sigma_p^2{N}$ is the measured variance of $\hat{x}$ or $\hat{p}$ if we perform the measurement on N independent and identical systems. In the context of coherent states, $\Delta \hat{x}^2=\Delta \hat{p}^2=1/2$, if we further require $\sigma_x^2(N)=\sigma_p^2(N)$, the tight bound of (\ref{inequalestimation}) is reached for $g_x=g_p$. Then it yields from (\ref{inequalestimation})
\begin{equation}\label{varianceboundinf}
\bar{\sigma}_{N,\infty}^2=1/N.
\end{equation}
Combine (\ref{varianceboundinf}) and (\ref{noisecascading}), we have completed our proof.

\subsection{Implementation of optimal Gaussian QCM with a linear amplifier and beam splitters}

In this section, we shall give the explicit transformation for the optimal Gaussian $N\rightarrow M$ cloning of coherent states, and show that the transformation can be implemented through the common devices used in quantum optics experiments: a phase-insensitive linear amplifier and a network of beam splitters\cite{PhysRevLett.86.4938}. Thus we can prove that the optimal bounds of fidelity derived in the previous section can actually be achieved. Note also other implementations may be possible as well, for example, a scheme using a circuit of CNOT gates is proposed to be an implementation for the $1\rightarrow 2$ Gaussian cloning\cite{PhysRevLett.85.1754}.

Assume the state to be cloned is $|\alpha\rangle$, we denote the initial input state of the cloning machine as $|\Psi\rangle = |\alpha\rangle^{\otimes N}\otimes |0\rangle^{\otimes M-N}\otimes|0\rangle_z$, where except the $N$ input modes to be cloned, we have $M-N$ blank modes and an ancillary mode $z$. The blank modes and the ancilla are prepared initially in the vacuum state $|0\rangle$. Let $\{ x_k, p_k\}$ denote the pair of quadrature operators associated with each mode $k$ involved in the cloning transformation, where $k=0,...,M-1$(for simplicity, we sometimes omit the hats for operators when the context is unambiguous). As usual, for cloning we mean a quantum operation $U:\mathcal{H}^{\otimes M-1} \rightarrow \mathcal{H}^{\otimes M-1} $ performed on the initial state $|\Psi\rangle$, and the output state becomes $|\Psi^{''}\rangle = U|\Psi\rangle$.

For simplicity of analysis and calculation which shall be shown below, we work in the Heisenberg picture, then U can be described by a canonical transformation acting on the operators $\{x_k,p_k\}$:
\begin{equation}
x_k^{''}=U^{\dagger}x_k U,\ \ p_k^{''}=U^{\dagger}p_k U,
\end{equation}
while the state $|\Psi\rangle$ is left invariant. We will now impose several requirements for the transformation U which establish some expected properties of the state after cloning:
\begin{enumerate}

\item The expected values of x and k for the M output modes be:
    \begin{equation}
    \langle x_k^{''}\rangle = \langle\alpha|x_0|\alpha\rangle,\ \
    \langle p_k^{''}\rangle = \langle\alpha|p_0|\alpha\rangle,
    \end{equation}
    which means the state of the clones is centered on the original coherent state.

\item Note that for a coherent state, we have $\sigma_x^2 = \sigma_p^2=\Delta x_{vac}^2=\frac{1}{2}$, and also by a rotation in the phase space, we get the operator $v=cx+dp$, (where $c$ and $d$ are complex numbers satisfying $|c|^2+|d|^2=1$), the error variance of which is the same:
    \begin{equation}
    \sigma_v^2=\sigma_x^2=\sigma_p^2=\Delta x_{vac}^2=\frac{1}{2}.
    \end{equation}
    We then require that the invariance property under rotation is preserved by the transformation $U$, which yields
    \begin{equation}
    \sigma_{v_k^{''}}^2 = \sigma_{x_k^{''}}^2 = \sigma_{p_k^{''}}^2 = (1+\frac{2}{N}-\frac{2}{M})\Delta x_{vac}^2,
    \end{equation}
    where $v_k^{''}=cx_k^{''}+dp_k^{''}$.

\item
$U$ is unitary, which in the Heisenberg picture is equivalent to demand that the commutation relations are preserved through the transformation:
\begin{equation}
[x_j^{''},x_k^{''}]=[p_j^{''},p_k^{''}]=0,\ \ [x_j^{''},p_k^{''}]=i\delta_{jk},
\end{equation}
for $j,k=0,...,M-1$ and for the ancilla.
\end{enumerate}
Based on the above requirements we shall then give the explicit implementation of the cloning machine.

\subsection{Optimal $1\rightarrow 2$ Gaussian QCM}
We first consider the simple case of duplication ($N=1, M=2$). An explicit transformation can be found:
\begin{align}\label{1to2transformation}
x_0^{''}=x_0+\frac{x_1}{\sqrt{2}}+\frac{x_z}{\sqrt{2}},\ \
p_0^{''}=p_0+\frac{p_1}{\sqrt{2}}-\frac{p_z}{\sqrt{2}},\nonumber\\
x_1^{''}=x_0-\frac{x_1}{\sqrt{2}}+\frac{x_z}{\sqrt{2}},\ \
p_1^{''}=p_0-\frac{p_1}{\sqrt{2}}-\frac{p_z}{\sqrt{2}},\nonumber\\
x_z^{'}=x_0+\sqrt{2}x_z, \ \ p_z^{'}=-p_0+\sqrt{2}p_z,
\end{align}
for which one can check that all the three requirements are satisfied.

\begin{figure}
\includegraphics[height=4cm]{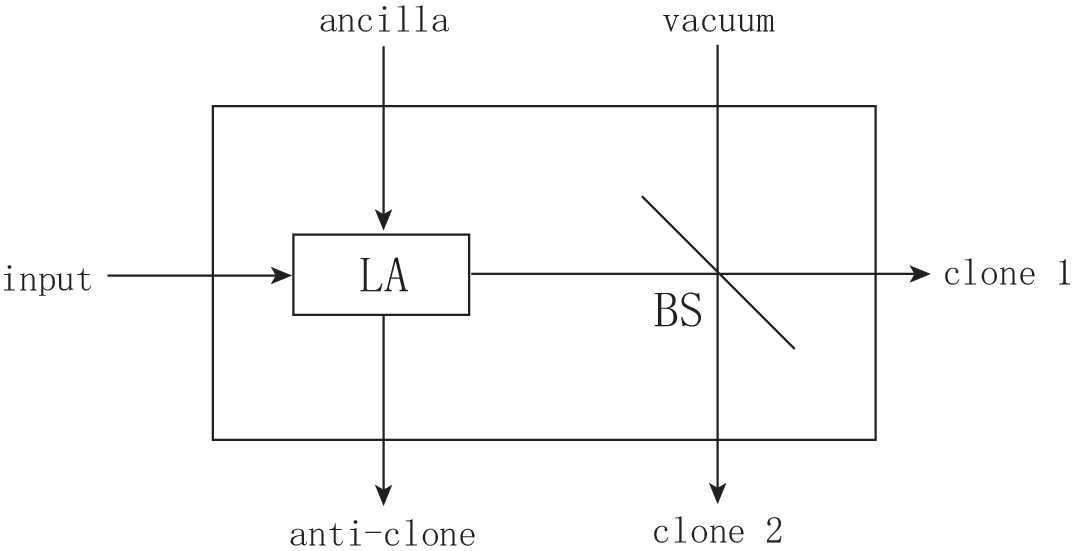}
\caption{Implementation of the optimal Gaussian $1\rightarrow 2$ QCM for light modes.
LA stands for linear amplifier and BS represents a balanced beam splitter, see \cite{PhysRevLett.86.4938}.}
\label{gaussianQCM1to2pic}
\end{figure}

Next we proceed to see how to implement the above duplicator in practice. First interpret (\ref{1to2transformation}) as a sequence of two canonical transformations:
\begin{align}
a_0^{'}=\sqrt{2}a_0+a_z^{\dagger},\ \ a_z^{'}=a_0^{\dagger}+\sqrt{2}a_z \nonumber\\
a_0^{''}=\frac{1}{\sqrt{2}}(a_0^{'}+a_1), \ \
a_1^{''}=\frac{1}{\sqrt{2}}(a_0^{'}-a_1),
\end{align}
where $a_k=(x_k+ip_k)/\sqrt{2}$ and $a_k^{\dagger}=(x_k-ip_k)/\sqrt{2}$ denote the annihilation and creation operators for mode k. We then can immediately come up with a practical scheme which has two steps to  have the desired transformation realized. Step 1 is a phase-insensitive amplifier whose gain G is equal to 2, while step 2 is a phase-free 50:50 beam splitter(see Fig. \ref{gaussianQCM1to2pic}). To see the cloner is optimal, we note from \cite{PhysRevD.26.1817}, for an amplifier of gain $G$, each quadrature's excess noise variance is bounded by
\begin{equation}\label{variancegainrelation}
\sigma_{LA}^2\ge (G-1)/2.
\end{equation}
Since we have chosen $G$ to be 2, it yields $\sigma_{LA}^2=1/2$, which proves the optimality of the cloning transformation.

\subsection{Optimal Gaussian $N\rightarrow M$ QCM}

Now we continue to study the case of $N\rightarrow M$ Gaussian cloning, this time we shall again use linear amplifier to achieve the transformation. Due to the relation of extra variance and gain from (\ref{variancegainrelation}), we need to make $G$ as low as possible in order to reach the optimal limit of $\sigma_{N,M}^2$. The cloning procedure is as follows: (i) concentrate the $N$ input modes to one single mode, which is then amplified. (ii) distribute the concentrated mode symmetrically among the
$M$ output modes. Obviously an easy method to realize the processes is through discrete Fourier transform (DFT), with which we can write out the detailed steps of the cloning procedure.

Step 1: concentration of the $N$ input modes by a DFT:
\begin{equation}
a_k^{'}=\frac{1}{\sqrt{N}}\sum_{l=0}^{N-1}exp(ikl2\pi/N)a_l,
\label{DFT}
\end{equation}
where $k=0,...,N-1$. After the concentration, the energy of the $N$ input modes is put together on one single mode, which we shall rename as $a_0$, while every other mode becomes a vacuum state. To see this more clearly, we note that the energy of one mode is $E_k = \hbar\omega(\langle {a}_k^{\dagger}{a}_k\rangle + \frac{1}{2}), k = 0, ..., N-1.$  Since the input state is $|\alpha\rangle^{\otimes N}$, $\langle {a}_k^{\dagger}{a}_l\rangle = |\alpha|^2$ for all $k$ and $l$, and the total energy is $E = \sum_{k=0}^{N-1} E_k = N\hbar\omega(|\alpha|^2+\frac{1}{2})$. On the other hand, after the DFT process, all modes except one become vacuum states. From Eq.(\ref{DFT}), for any $k\neq 0$, we have
\begin{eqnarray}
\langle {a}_k^{'\dagger} {a}_k^{'} \rangle &=& \frac{1}{N} \sum_{l = 0}^{N-1}\sum_{m=0}^{N-1} exp(i\frac{k2\pi}{N}(m-l))\langle {a}_l^{\dagger} {a}_m\rangle \nonumber \\
&=& |\alpha|^2 \frac{1}{N} \sum_{l = 0}^{N-1}\sum_{m=0}^{N-1} exp(i\frac{k2\pi}{N}(m-l)) \nonumber \\
&=& 0.
\end{eqnarray}
So the new mode $k \neq 0$ is a vacuum state. On the other hand for $k = 0$, we have
\begin{eqnarray}
\langle {a}_0^{'\dagger} {a}_0^{'} \rangle = \frac{1}{N}  \sum_{l = 0}^{N-1}\sum_{m=0}^{N-1} \langle  {a}_l^{\dagger} {a}_m\rangle = N|\alpha|^2.
\end{eqnarray}
Now we see energy is concentrated in the new mode $0$.

Step 2: take the mode $a_0$ together with the ancilla as the input of a linear amplifier of gain $G=M/N$, which results,
\begin{align}
a_0^{'}=\sqrt{\frac{M}{N}}a_0+\sqrt{\frac{M}{N}-1}a_z^{\dagger}, \nonumber \\
a_z^{'}=\sqrt{\frac{M}{N}-1}a_0^{\dagger}+\sqrt{\frac{M}{N}}a_z.
\end{align}

Step 3: distribute energy symmetrically onto the $M$ outputs by performing a DFT on $a_0^{'}$ and the $M-1$ vaccum modes produced in step 1:
\begin{equation}
a_k^{''}=\frac{1}{\sqrt{M}}\sum_{l=0}^{M-1}exp(ikl2\pi/M)a_l^{'},
\end{equation}
It's readily checked that the procedure can meet our three requirements. Moreover if we choose $\sigma_{LA}^2=[M/N-1]/2$, the optimality is then confirmed.

\begin{figure}
\includegraphics[height=4cm]{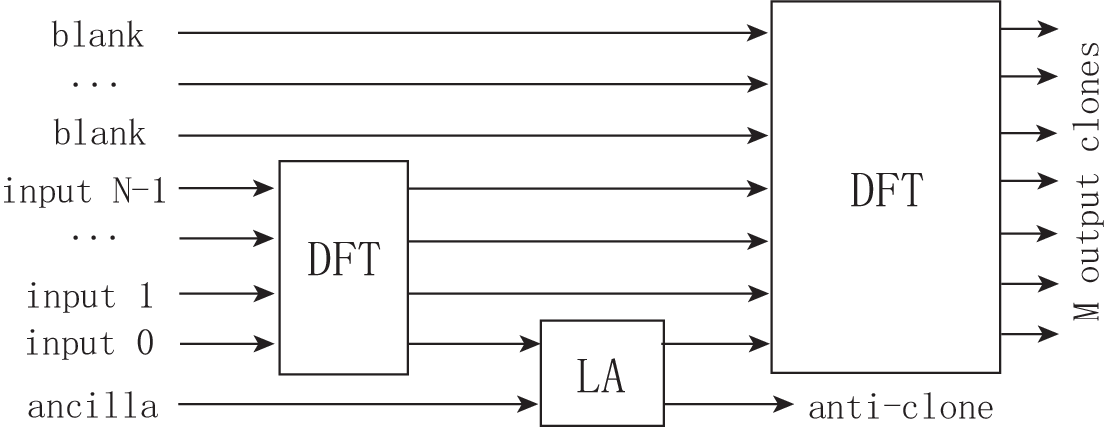}
\caption{Implementation of the optimal Gaussian $N\rightarrow M$ QCM for light modes. LA represents linear amplifier,
DFT stands for discrete Fourier transform, see \cite{PhysRevLett.86.4938}.}
\label{gaussianQCMntompic}
\end{figure}

Like the case of $1\rightarrow 2$ cloning, we shall also use a network of beam splitters to construct the required DFT. It is shown that any discrete unitary operator can be experimentally realized by a sequence of beam splitters and phase shifters \cite{Reck1994}. An explicit construction is given in \cite{PhysRevLett.86.4938} as shown in FIG. \ref{gaussianQCMntompic}.

\subsection{Other developments and related topics}

Similarities and differences exist between cloning in discrete space and CV space.
Similar as in discrete space case,
if we know partial information of the input state in CV system, the fidelity can also be improved \cite{pra73.045801}.
Without analog in discrete case, the quantum cloning with phase-conjugate input modes is studied in \cite{prl87.247903}.

One can expect that many proposals in discrete case can be extended to CV case.
Next, we list those results in areas of QKD, CV quantum cloning
and implementation schemes in the following.

\begin{itemize}
\item
In relation with QKD, the application of CV cloning machine in key distribution is studied in \cite{cerf-cvqkd-squeezed}.
By some figure of merit, the optimal cloning of coherent states
with non-Gaussian setting may be better than a Gaussian setting \cite{prl95.070501}.
This may pose a question about whether the security of a QKD can be challenged or not
in a more general condition.
However, CV cryptography \cite{prl88.057902} is shown to be still secure under non-Gaussian attack \cite{prl92.047905}.
The asymmetry CV cloning used for security analysis of
cryptography is also discussed in \cite{CerfIblisdirAssche}. The review of CV cloning
and QKD can be found in \cite{CerfGrangier}.

\item
The CV universal NOT
gate is studied in \cite{cerf-cv-not}.
The optimal cloning of mixed Gaussian states
is studied in \cite{ISI:000241067100032}. The superbroadcasting of CV mixed states
is studied in \cite{ISI:000238292400001}.
The quantum cloning limits for finite distributions of coherent states
are studied in \cite{finite-coherent}, and also in \cite{ISI:000189304700015}.
A proposal to test quantum
limits of a Gaussian-distributed set of coherent state related with
cloning is presented in \cite{Namiki2011}. The cloning of CV entangled state is
studied in \cite{Weedbrook2008}.

\item
The implementation of CV quantum cloning via various schemes is proposed in \cite{PhysRevLett.86.4938,prl86.914,prl86.4942}.
The multicopy Gaussian states is studied in \cite{Fiurasek2007}. The Gaussian cloning of coherent light states
into an atomic quantum memory is presented in \cite{FiurasekCerfPolzik}.

Experimentally, implementation of Gaussian cloning of coherent states with fidelity
of about $65\% $ by only linear optics is shown in \cite{prl94.240503}, the results
are further analyzed in \cite{Olivares2006}.
Experimental realization of CV cloning with phase-conjugate inputs is shown in \cite{experiementalcvcloning} and also in \cite{ZhangJing07PRA}.
The experimental realization of both CV teleportation
and cloning is reported in \cite{ZhangJing05PRL}.

\end{itemize}

As one of the most basic protocols of quantum information processing, the CV teleportation is studied in \cite{ISI:000085234900012}.
The criteria of CV cloning and teleportation
are studied in \cite{GrosshansGrangier}.
We remark again that the reviews of CV quantum information can be found in \cite{RevModPhys.77.513,Wangphysreport,RMP-gaussian}.

\newpage

\section{Sequential universal quantum cloning}

In past years, theoretical research on quantum cloning machines have progressed greatly.
At the same time, various cloning schemes have been realized experimentally, by using polarized photons\cite{667603820020426,PhysRevA.68.042306,PhysRevLett.92.047901,PhysRevLett.92.047902} or nuclear spins in NMR \cite{PhysRevLett.88.187901,PhysRevLett.94.040505}. However, these experiments are only restricted to
$1 \rightarrow 2$ or $1 \rightarrow 3$ cloning machines, leaving the general case of $N \rightarrow M$ cloning unsolved.
The difficulty of realizing $N \rightarrow M$ cloning mainly arises in preparing multipartite entangled states, since it is very difficult to perform a global unitary operation on large-dimensional systems to create multipartite entangled states. While on the other hand, using the technique of sequential cloning, one may be able to divide the big global unitary operation into small ones, each of which is only concerned with a small quantum system and as a result makes it possible to get the desired entangled state. Several quantum cloning procedures for multipartite cloning were proposed, but they are not in the sequential method\cite{PhysRevLett.84.2993,PhysRevA.67.022317}. In 2007, based on the work of Vidal \cite{PhysRevLett.91.147902},
Delgado \emph{et al.} proposed a scheme of a sequential $1 \rightarrow M$ cloning machine\cite{Delgado2007}.
Since the procedure is sequential, it significantly reduces the difficulty of its realization.
Later, a scheme of more general $N \rightarrow M$ sequential cloning is presented \cite{Dang2008}.
The case of $N \rightarrow M$ sequential cloning of qudits is also proposed briefly, yet the details are not presented.
The essential idea of sequential method is to express the desired state in the form of matrix product state (MPS),
and according to results in \cite{PhysRevLett.95.110503}, any MPS can be sequentially generated.
On the other hand, it is also pointed out that sequential unitary decompositions are not always successful for
genuine entangling operations \cite{lamatasequential}.
Here in this section, we will present in detail how the procedure of $1 \rightarrow M$ and $N \rightarrow M$ sequential cloning can work.

\subsection{$1 \rightarrow M$ sequential UQCM}

According to the method of Delgado \emph{et al.}\cite{Delgado2007}, we first need an ancilla system of dimension $D$. Let $\mathcal{H}_\mathcal{A}$ denotes the D-dimensional Hilbert space of the ancilla system, and $\mathcal{H}_\mathcal{B}$ the 2-dimensional Hilbert space of one qubit. In every step of the sequential cloning, we perform a quantum evolutional operator V on the product space of the ancilla and a single qubit. Here we suppose that
each qubit is initially state $|0\rangle$ which will not appear in the following equations. $V$ then can be represented by an isometric transformation: $V: \mathcal{H}_\mathcal{A} \rightarrow \mathcal{H}_\mathcal{A}\otimes\mathcal{H}_\mathcal{B}$, in which $V = \sum_{i,\alpha,\beta}V_{\alpha,\beta}^i|\alpha,i\rangle\langle\beta|$. Let $V^i=\sum_{\alpha,\beta}V_{\alpha,\beta}^i|\alpha\rangle\langle\beta|$, then $V^i$ is a $D{\times}D$ matrix and satisfies the isometry condition $\sum_iV^{i\dagger}V^i=I$. Let the initial state of the ancilla be $|\phi_I\rangle\in\mathcal{H}_\mathcal{A}$. We make the ancilla to interact with the qubits once a time and sequentially, after the unitary operation, we would not recover the ancilla state.
So when $n$ operations have been done, the final output state of the ancilla and all the qubits take the form, $|\Psi\rangle=V^{[n]}...V^{[2]}V^{[1]}|\phi_I\rangle$, where indices in squared brackets represent the steps of sequential generation. Now we need to decouple the aniclla from the multi-entangled qubits, and then the $n$-qubit state shall be left:
\begin{equation}\label{eqexpand1}
|\psi\rangle = \sum_{i_1,i_2,...,i_n}\langle\phi_F|V^{[n]i_n}...V^{[1]i_1}|\phi_I\rangle|i_1...i_n\rangle,
\end{equation}
where $\phi_F$ represents the final state of the ancilla.
For 2-dimensional system $\{|0\rangle,|1\rangle\}$, the cloning transformation of the optimal $1 \rightarrow M$ cloning machine is \cite{PhysRevLett.79.2153}:
\begin{eqnarray}
|0\rangle\otimes|R\rangle\rightarrow|\Psi_M^{(0)}\rangle=\sum_{j=0}^{M-1}\beta_j|(M-j)0,j1\rangle\otimes|(M-j-1)1,j0\rangle_R,\\
|1\rangle\otimes|R\rangle\rightarrow|\Psi_M^{(1)}\rangle=\sum_{j=0}^{M-1}\beta_{M-j-1}|(M-j-1)0,(j+1)1\rangle\otimes|(M-j-1)1,j0\rangle_R,
\end{eqnarray}
in which $\beta_j=\sqrt{2(M-j)/M(M+1)}$, $|(M-j-1)1,j0\rangle_R$ is the final state of the cloning machine, and $|(M-j)0,j1\rangle$ denotes the normalized completely symmetric $M$-qubit state with ($M-j$) qubits in $|0\rangle$ and $j$ qubits in $|1\rangle$.
In order to clone a general state $|\phi\rangle$, it is necessary to know how to sequentially generate the states $|\Psi_M^{(0)}\rangle$ and $|\Psi_M^{(1)}\rangle$, as a result of which we need to express these two states in the MPS form:
\begin{eqnarray}
|\Psi_M^{(0)}\rangle=\sum_{i_1,...i_n}\langle\phi_F^{(0)}|V_0^{[n]i_n}...V_0^{[1]i_1}|0\rangle_D|i_1...i_n\rangle,\\
|\Psi_M^{(1)}\rangle=\sum_{i_1,...i_n}\langle\phi_F^{(1)}|V_1^{[n]i_n}...V_1^{[1]i_1}|0\rangle_D|i_1...i_n\rangle.
\end{eqnarray}
Now we aim to get the explicit expression of the matrices $V_0^{[k]i_k}$ and $V_1^{[k]i_k}$. A way to do so is by using Schmidt decomposition(SD) \cite{PhysRevLett.91.147902}, see textbook \cite{Nielsen2000}. Consider an arbitrary state $|\Psi\rangle$ in Hilbert space $\mathcal{H}_2^{\otimes{n}}$, the SD of $|\Psi\rangle$ according to the bipartition $A:B$ is
\begin{equation}
|\Psi\rangle = \sum_{\alpha} \lambda_{\alpha}|\Phi_{\alpha}^{[A]}\rangle|\phi_{\alpha}^{[B]}\rangle,
\end{equation}
where $|\Phi_{\alpha}^{[A]}\rangle(|\Phi_{\alpha}^{[B]}\rangle)$ is an eigenvector of the reduced density matrix $\rho^{[A]}(\rho^{[B]})$ with eigenvalue $|\lambda_{\alpha}|^2\geq 0$, and the Schmidt coefficient $\lambda_{\alpha}$ satisfies $\langle \Phi_{\alpha}^{[A]}|\Psi\rangle = \lambda_{\alpha}|\Phi_{\alpha}^{[B]}\rangle$.

With the help of SD, we proceed the following protocol:
\begin{enumerate}
\item Compute the SD of $|\Psi\rangle$ according to the bipartite $1:n-1$ splitting of the $n$-qubit system, which is
\begin{align}\label{eq1}
|\Psi\rangle &= \sum_{\alpha_1}\lambda_{\alpha_1}^{[1]}|\Phi_{\alpha_1}^{[1]}\rangle|\Phi_{\alpha_1}^{[2...n]}\rangle\\
&= \sum_{i_1,\alpha_1}\Gamma_{\alpha_1}^{[1]i_1}\lambda_{\alpha_1}^{[1]}|i_1\rangle|\Phi_{\alpha_1}^{[2...n]}\rangle,
\end{align}
where in the second line we have expressed the Schmidt vector $|\Phi_{\alpha_1}^{[1]}\rangle$ in the computational basis $\{|0\rangle,|1\rangle\}$: $|\Phi_{\alpha_1}^{[1]}\rangle = \sum_{i_1}\Gamma_{\alpha_1}^{[1]i_1}|i_1\rangle.$

\item Expand $|\Phi_{\alpha_1}^{[2...n]}\rangle$ in local basis for qubit 2,
\begin{equation}\label{eq2}
|\Phi_{\alpha_1}^{[2...n]}\rangle = \sum_{i_2}|i_2\rangle|\tau_{\alpha_1i_2}^{[3...n]}\rangle.
\end{equation}
\item Express $|\tau_{\alpha_1i_2}^{[3...n]}\rangle$ by at most $\chi$ Schmidt vectors $|\Phi_{\alpha_2}^{[3...n]}\rangle$(the eigenvectors of $\rho^{[3...n]}$), where $\alpha_2$ ranges from 1 to $\chi$ and $\chi = \max_{A}\chi_A,$ here $\chi_A$ denotes the rank of the reduced density matrix $\rho_A$ for a particular partition $A:B$ of the $n$-qubit state:
    \begin{equation}\label{eq3}
    |\tau_{\alpha_1i_2}^{[3...n]}\rangle = \sum_{\alpha_2}\Gamma_{\alpha_1\alpha_2}^{[2]i_2}\lambda_{\alpha_2}^{[2]}|\Phi_{\alpha_2}^{[3...n]}\rangle,
    \end{equation}
    where $\lambda_{\alpha_2}^{[2]}$'s are the corresponding Schmidt coefficients.
\item Subsitute (\ref{eq2}) and (\ref{eq3}) into (\ref{eq1}), we get
\begin{equation}
|\Psi\rangle = \sum_{i_1,\alpha_1,i_2,\alpha_2}\Gamma_{\alpha_1}^{[1]i_1}\lambda_{\alpha_1}^{[1]}
\Gamma_{\alpha_1\alpha_2}^{[2]i_2}\lambda_{\alpha_2}^{[2]}|i_1i_2\rangle|\Phi_{\alpha_2}^{[3...n]}\rangle.
\end{equation}
Now it's easy to see if we repeat the steps 2-4, we get the expansion of $|\Psi\rangle$ in the computational basis:
\begin{equation}
|\Psi\rangle = \sum_{i_1}...\sum_{i_n}c_{i_1...i_n}|i_1\rangle...|i_n\rangle,
\end{equation}
where the coefficients $c_{i_1...i_n}$ are
\begin{equation}\label{eqexpand2}
c_{i_1...i_n} = \sum_{\alpha_1,...,\alpha_{n-1}}\Gamma_{\alpha_1}^{[1]i_1}\lambda_{\alpha_1}^{[1]}
\Gamma_{\alpha_1\alpha_2}^{[2]i_2}\lambda_{\alpha_2}^{[2]}...\Gamma_{\alpha_{n-1}}^{[n]i_n}.
\end{equation}
\end{enumerate}

Through comparing equations (\ref{eqexpand1}) and (\ref{eqexpand2}), we are able to construct $V_0^{[k]i_k}$ and $V_1^{[k]i_k}$ explicitly. The detailed work is omitted here since in next section about the more general $N\rightarrow M$ sequential cloning case, each step of getting the matrix is provided.

When the input state of the cloning machine is an arbitrary state $|\psi\rangle = x_0|0\rangle+x_1|1\rangle$(normalization condition is satisfied: $|x_0|^2+|x_1|^2=1$.), according to the linearity principle of quantum mechanics, the state after cloning transformation is $x_0|\Psi_M^{(0)}\rangle+x_1|\Psi_M^{(1)}\rangle$, which can also be sequentially generated. First, view the arbitrary state $|\psi\rangle$ and the ancilla's initial state $|0\rangle_D$ as a unified state: $|\phi_I\rangle = |\psi\rangle\otimes|0\rangle_D$. Then use the qubit $k$,($k=1,2,...,n$), sequentially to interact with the ancilla according to the 2D-dimensional isometric operators $V^{[k]i_k} = |0\rangle\langle0|\otimes{V_0^{[k]i_k}}+|1\rangle\langle1|\otimes{V_1^{[k]i_k}}$. After all qubits have interacted with the ancilla, perform a generalized Hadamard transformation to the ancilla
\begin{eqnarray}
|0\rangle|\phi_F^{(0)}\rangle\rightarrow\frac{1}{\sqrt2}[|0\rangle|\phi_F^{(0)}\rangle+|1\rangle|\phi_F^{(1)}\rangle]\\
|1\rangle|\phi_F^{(1)}\rangle\rightarrow\frac{1}{\sqrt2}[|0\rangle|\phi_F^{(0)}\rangle-|1\rangle|\phi_F^{(1)}\rangle]
\end{eqnarray}
Now measure the ancilla with the basis $\{|0\rangle|\phi_F^{(0)}\rangle,|1\rangle|\phi_F^{(1)}\rangle\}$, either result occurs with probability 1/2. When the result is $|0\rangle|\phi_F^{(0)}\rangle$, we get the desired state $x_0|\Psi_M^{(0)}\rangle+x_1|\Psi_M^{(1)}\rangle$; while if the result is $|1\rangle|\phi_F^{(1)}\rangle$, we need to perform a $\pi$-phase gate upon each qubit, and the desired state will be obtained.

To realized the above $1 \rightarrow M$ cloning scheme, an ancilla system of dimension $2M$ is needed, while if we take a global unitary operation to accomplish the cloning, the dimension of the unitary operation will increase exponentially with $M$. So we see sequential cloning is much easier to realize experimentally.

\subsection{$N \rightarrow M$ optimal sequential UQCM}

In this section, we will discuss the more general case $N \rightarrow M$ optimal sequential UQCM. An arbitrary qubit is written $|\Psi\rangle = x_0|0\rangle+x_1|1\rangle$($|x_0|^2+|y_0|^2=1$), then N identical $|\Psi\rangle$ can be expressed as
\begin{equation}
|\Psi\rangle^{\otimes{N}} = \sum_{m=0}^{N}x_0^{N-m}x_1^m\sqrt{C_N^m}|(N-m)0,m1\rangle,
\end{equation}
where $C_N^m=\frac{N!}{m!(N-m)!}$, and $|(N-m)0,m1\rangle$ denotes the normalized completely symmetric N-qubit state with (N-m) qubits in state $|0\rangle$ and m qubits in state $|1\rangle$.

It is well known that the optimal UQCM transformation for completely symmetric states\cite{PhysRevLett.79.2153} is
\begin{eqnarray}
|(N-m)0,m1\rangle\otimes|R\rangle \rightarrow |\Psi_M^{(m)}\rangle
 = \sum_{j=0}^{M-N}\beta_{mj}|(M-m-j)0,(m+j)1\rangle\otimes|R_j\rangle,
\end{eqnarray}
where $\beta_{mj}=\sqrt{C_{M-m-j}^{M-N-j}C_{m+j}^j/C_{M+1}^{N+1}}$, $R_j$ denotes the final states of the cloning machine, and for different j, $|R_j\rangle$'s are orthogonal to each other. Here we can choose $|R_j\rangle=|(M-N-j)1,j0\rangle_R$. Since we have found the cloning transformation of any state in the form $|((N-m)0,m1\rangle$, according to the linearity principle of quantum mechanics, the transformation for N arbitrary state $|\Psi\rangle$ is
\begin{equation}
|\Psi\rangle^{\otimes{N}}\otimes|R\rangle \rightarrow |\Psi_M\rangle=\sum_{m=0}^{N}x_0^{N-m}x_1^m\sqrt{C_N^m}|\Psi_M^{(m)}\rangle,
\end{equation}
here $|\Psi_M\rangle$ is the final state of all the qubits and the cloning machine we hope to obtain, like the $1 \rightarrow N$ sequential UQCM case, we first need to show how $|\Psi_M^{(m)}\rangle$ can be sequentially generated. Hence it's necessary to know the MPS form of $|\Psi_M^{(m)}\rangle$:
\begin{equation}\label{NMeq1}
|\Psi_M^{(m)}\rangle = \sum_{i_1...i_{2M-N}}\langle\phi_F|V^{[2M-N]i_{2M-N}}...V^{[1]i_1}|\phi_I\rangle|i_1...i_{2M-N}\rangle,
\end{equation}
where $V^{[n]i_n}(1\leq n\leq 2M-N)$ is a $D\times D$ dimensional matrix, and satisfies the isometry condition:$\sum_{i_n} (V^{[n]i_n})^{\dagger}V^{[n]i_n}=I$. Now we shall follow the idea of SD, and give detailed elaboration on how to get the explicit form of $V^{[n]i_n}$.

\begin{enumerate}
\item Case $n = 1$.

Compute the SD of $|\Psi_M^{(m)}\rangle$ according to partition 1:2...(2M-N):
\begin{eqnarray}\label{SD1}
\left\vert \Psi _{M}^{(m)}\right\rangle &=&\overset{M-N}{\underset{j=0}{\sum }}%
\beta _{mj}\left\vert \left( M-m-j\right) 0,\left( m+j\right) 1\right\rangle
\otimes |R_j\rangle \nonumber\\
&=& \sum_{\alpha_1}\lambda_{\alpha_1}^{[1]}|\phi_{\alpha_1}^{[1]}\rangle\otimes
|\phi_{\alpha_1}^{[2...(2M-N)]}\rangle\nonumber\\
&=& \sum_{\alpha_1,i_1}\Gamma_{\alpha1}^{[1]i_1}\lambda_{\alpha_1}^{[1]}|i_1\rangle\otimes
|\phi_{\alpha_1}^{[2...(2M-N)]}\rangle\nonumber\\
&=&\left\vert 0\right\rangle \otimes \lambda ^{\left[ 1\right]} _{1}\left\vert
\phi _{1}^{\left[ 2...\left( 2M-N\right) \right] }\right\rangle +\left\vert
1\right\rangle \otimes \lambda ^{\left[ 1\right]} _{2}\left\vert \phi _{2}^{%
\left[ 2...\left( 2M-N\right) \right] }\right\rangle ,
\end{eqnarray}%
where through comparing the first and last lines, we get%
\begin{equation*}
\left\vert \phi _{1}^{\left[ 2...\left( 2M-N\right) \right] }\right\rangle =%
\overset{M-m-1}{\underset{k=-m}{\sum }}\beta _{mk}\sqrt{\frac{C_{M-1}^{m+k}}{%
C_{M}^{m+k}}}\left\vert \left( M-m-k-1\right) 0,\left( m+k\right)
1\right\rangle \otimes |R_k\rangle
/\lambda ^{\left[ 1\right]} _{1},
\end{equation*}%
\begin{equation*}
\left\vert \phi _{2}^{\left[ 2...\left( 2M-N\right) \right] }\right\rangle =%
\overset{M-m-1}{\underset{k=-m}{\sum }}\beta _{m\left( k+1\right) }\sqrt{%
\frac{C_{M-1}^{m+k}}{C_{M}^{m+k+1}}}\left\vert \left( M-m-k-1\right)
0,\left( m+k\right) 1\right\rangle \otimes |R_{k+1}\rangle /\lambda ^{\left[ 1\right]} _{2}.
\end{equation*}%
Compare the last two lines, we also have%
\begin{equation*}
\Gamma  _{\alpha _{1}}^{\left[ 1\right]0}=\delta _{\alpha _{1},1},\text{ }%
\Gamma  _{\alpha _{1}}^{\left[ 1\right]1}=\delta _{\alpha _{1},2},\text{ }%
\alpha _{1}=1,2.
\end{equation*}%

Now use the condition of normalization, Schmidt coefficients could be calculated,
\begin{equation*}
\lambda ^{\left[ 1\right]} _{1}=\sqrt{\overset{M-m-1}{\underset{k=-m}{\sum }}%
\beta _{mk}^{2}\frac{C_{M-1}^{m+k}}{C_{M}^{m+k}}},\text{ }\lambda ^{\left[ 1%
\right]} _{2}=\sqrt{\overset{M-m-1}{\underset{k=-m}{\sum }}\beta _{m\left(
k+1\right) }^{2}\frac{C_{M-1}^{m+k}}{C_{M}^{m+k+1}}}.
\end{equation*}%

Then we have%
\begin{equation*}
V _{\alpha _{1}}^{\left[ 1\right]i_{1}}=\Gamma  _{\alpha
_{1}}^{\left[ 1\right]i_{1}}\lambda ^{\left[ 1\right]} _{\alpha _{1}}.
\end{equation*}%
The explicit form of $V^{[1]i_1}$ is given in the appendix, so is other $V^{[i]i_n}$.

Next, we will not present the detailed calculations for other cases since the method
is almost the same, but only list the results.

\item For $1<n\le M-1$:
We calculate the SD of $|\Psi_M^{(m)}\rangle$ according to partitions.
The results are:
\begin{equation*}
\lambda ^{\left[ n\right]} _{j+1}=\sqrt{C_{n}^{j}\overset{M-m-n}{\underset{k=-m}%
{\sum }}\beta _{m\left( j+k\right) }^{2}\frac{C_{M-n}^{m+k}}{C_{M}^{m+j+k}}},
~~~\lambda ^{\left[ n-1\right]} _{j+1}=\sqrt{C_{n-1}^{j}\overset{M-m-n+1}{\underset%
{k=-m}{\sum }}\beta _{m\left( j+k\right) }^{2}\frac{C_{M-n+1}^{m+k}}{%
C_{M}^{m+j+k}}}.
\end{equation*}%
\begin{equation*}
\Gamma _{\left( j+1\right) \alpha _{n}}^{ \left[ n\right]0}=\delta _{\left(
j+1\right) \alpha _{n}}\frac{\sqrt{C_{n-1}^{j}}}{\lambda ^{\left[ n-1\right]}
_{j+1}\sqrt{C_{n}^{j}}},
~~~\Gamma  _{\left( j+1\right) \alpha _{n}}^{\left[ n\right]1}=\delta _{\left(
j+2\right) \alpha _{n}}\frac{\sqrt{C_{n-1}^{j}}}{\lambda ^{\left[ n-1\right]}
_{j+1}\sqrt{C_{n}^{j+1}}}.
\end{equation*}

And for this case, the summarized form is $V _{\alpha _{n}\alpha _{n-1}}^{\left[ n\right]i_{n}}=\Gamma
_{\alpha _{n-1}\alpha _{n}}^{\left[ n\right]i_{n}}\lambda ^{\left[ n\right]} _{\alpha _{n}}$.

\item Case $n=M$: We have,
\begin{equation*}
\lambda ^{\left[ M\right]} _{j+1}=\beta _{m\left( j-m\right) },
~~~\lambda ^{\left[ M-1\right]} _{j+1}=\sqrt{C_{M-1}^{j}\overset{-m+1}{\underset{%
k=-m}{\sum }}\beta _{m\left( j+k\right) }^{2}\frac{C_{1}^{m+k}}{C_{M}^{m+j+k}%
}}.
\end{equation*}%
And also,%
\begin{equation*}
\Gamma  _{\left( j+1\right) \alpha _{M}}^{\left[ M\right]0}=\delta _{\alpha
_{M}\left( j+1\right) }\frac{\sqrt{C_{M-1}^{j}}}{\lambda ^{\left[ M-1\right]}
_{j+1}\sqrt{C_{M}^{j}}},
\end{equation*}%
\begin{equation*}
\Gamma  _{\left( j+1\right) \alpha _{M}}^{\left[ M\right]1}=\delta _{\alpha
_{M}\left( j+2\right) }\frac{\sqrt{C_{M-1}^{j}}}{\lambda ^{\left[ M-1\right]}
_{j+1}\sqrt{C_{M}^{j+1}}}.
\end{equation*}
Similarly, for this case $V_{\alpha_M\alpha_{M-1}}^{[M]i_M}=\Gamma_{\alpha_{M-1}\alpha_M}^{[M]i_M}
\lambda_{\alpha_M}^{[M]}$.

\item Case $M+l$ $\left( 1\le l\le M-N\right) $: We have,
\begin{equation*}
\lambda ^{\left[ M+l\right]} _{j+1}=\sqrt{C_{M-N-l}^{j-m}\overset{l}{\underset{%
k=0}{\sum }}\beta _{m\left( j+k-m\right) }^{2}\frac{C_{l}^{k}}{%
C_{M-N}^{j+k-m}}}.
\end{equation*}%
\begin{equation*}
\lambda ^{\left[ M+l-1\right]} _{j+1}=\sqrt{C_{M-N-l+1}^{j-m}\overset{l-1}{%
\underset{k=0}{\sum }}\beta _{m\left( j+k-m\right) }^{2}\frac{C_{l-1}^{k}}{%
C_{M-N}^{j+k-m}}},
\end{equation*}%
\begin{equation*}
\Gamma  _{\left( j+1\right) \alpha _{M+l}}^{\left[ M+l\right]0}=\delta
_{\alpha _{M+l}j}\sqrt{\frac{C_{M-N-l}^{j-m-1}}{C_{M-N-l+1}^{j-m}}}/\lambda %
^{\left[ M+l\right]} _{\alpha _{M+l}},
\end{equation*}%
\begin{equation*}
\Gamma  _{\left( j+1\right) \alpha _{M+l}}^{\left[ M+l\right]1}=\delta
_{\alpha _{M+l}\left( j+1\right) }\sqrt{\frac{C_{M-N-l}^{j-m}}{%
C_{M-N-l+1}^{j-m}}}/\lambda ^{\left[ M+l\right]} _{\alpha _{M+l}}.
\end{equation*}%
Then we have $V _{\alpha _{M+l}\alpha
_{M+l-1}}^{\left[ M+l\right]i_{M+l}}=\Gamma  _{\alpha _{M+l-1}\alpha
_{M}+l}^{\left[ M+l\right]i_{M+l}}\lambda ^{\left[ M+l\right]} _{\alpha _{M}+l}$.

\end{enumerate}

Up till now, we have calculated out the explicit form of every $V^{[k]i_k}$, and since $V^{[k]i_k}$ depends on m, we denote it as $V_{(m)}^{[k]i_k}$ here after.
 Through computation, we can get the smallest dimension needed for the isometric operator $V_(m)^{[k]i_k}$,
 \begin{equation}
 D = \{
 \begin{array}{ll}
 M-N/2+1 & \text{if N is even;}\\
 M-(N-1)/2+1 & \text{if N is odd.}
 \end{array}
 \end{equation}
 So we see D increases linearly with $M$, which shall significantly ease the difficulty of sequential cloning.

Based on the above computation, we have known that the state $|\Psi_M^{(m)}\rangle$ can be expressed in the MPS form, so the $N$-qubit pure state $|(N-m)0,m1\rangle$ can be sequentially transformed to $|\Psi_M^{(m)}\rangle$. Now in order to sequentially clone the $N$-qubit $|\Psi\rangle^{\otimes N}$ to $M$ qubits, the scheme is as follows\cite{Dang2008}.

1). Encode the N-qubit $|\Psi\rangle^{\otimes N}$ in the ancilla, which makes the initial state of the united ancilla
\begin{equation}
|\phi_I^{'}\rangle = \sum_{m=0}^N x_0^{N-m} x_1^m \sqrt{C_N^m} |(N-m)0,m1\rangle\otimes|0\rangle_D.
\end{equation}

2). Build the operators
\begin{equation}
V^{[k]i_k}=\sum_{m=0}^N (\sqrt{C_N^m})^{\frac{1}{2M-N}}(|0\rangle\langle0|)^{\otimes N-m}(|1\rangle\langle1|)^{\otimes m} \otimes V_{(m)}^{[n]i_n}.
\end{equation}

3). Let all the qubits interact sequentially with the united ancilla according to the operator $V^{[k]i_k}$, we get the final state of the whole system
\begin{align*}
|\Psi_{out}\rangle &= \sum_{i_1...i_{2M-N}} V^{[2M-N]i_{2M-N}}...V^{[1]i_1}|\varphi_i^{'}\rangle\otimes|i_1...i_{2M-N}\rangle\\
&= \sum_{m=0}^N x_0^{N-m} x_1^m \sqrt{C_N^m} |0\rangle^{\otimes N-m}|1\rangle^{\otimes m}\otimes|\varphi_F^{(m)}\rangle\otimes|\Psi_M^{(m)}\rangle,
\end{align*}
where $|\varphi_F^{(m)}\rangle$ is the final state of the ancilla when the input state is $|(N-m)0,m1\rangle$.

4). Perform a generalized Hadamard gate on the ancilla (quantum fourier transformation)
\begin{equation}
|0\rangle^{\otimes N-m}|1\rangle^{\otimes m}\otimes |\varphi_F^{(m)}\rangle \rightarrow \frac{1}{\sqrt{N+1}}\sum_{m'=0}^N e^{\frac{i2\pi mm'}{N+1}}
|0\rangle^{\otimes N-m'}|1\rangle^{\otimes m'}\otimes |\varphi_F^{(m')}\rangle,
\end{equation}
after which the final state becomes
\begin{equation}
|\Psi_{out}^{'}\rangle = \frac{1}{\sqrt{N+1}}\sum_{m'=0}^N |0\rangle^{\otimes N-m'} |1\rangle^{\otimes m'} \otimes |\varphi_F^{(m')}\rangle \otimes |\Psi_M^{'}\rangle,
\end{equation}
where
\begin{equation}
|\Psi_M^{'}\rangle = \sum_{m=0}^N e^{\frac{i2\pi mm'}{N+1}} x_0^{N-m} x_1^m \sqrt{C_N^m} |\Psi_M^{(m)}\rangle.
\end{equation}

5). Make measurement on the whole ancilla with the basis $\{ |0\rangle^{\otimes N-m'}|1\rangle^{\otimes m'}\otimes|\varphi_F^{(m')}\rangle\}_{m'=0}^N$. When the result is $m'=0$, the desired state $|\Psi_M\rangle = \sum_{m=0}^N x_0^{N-m} x_1^m \sqrt{C_N^m} |\Psi_M^{(m)}\rangle$ is directly obtained. If the measured $m'\neq0$, then we need to act a local phase gate $U_S$ on every qubit. Through computation, a proper phase gate is
\begin{equation}
U_S = |0\rangle\langle0|+e^{i\theta}|1\rangle\langle1|,
\end{equation}
where $\theta=-\frac{2\pi m'}{N+1}$.
With the effect of the phase gate, the output state becomes
\begin{equation}
|\Psi_M^{''}\rangle = e^{i(M-N)\theta}|\Psi_M\rangle.
\end{equation}
Since the phase $e^{i(M-N)\theta}$ won't affect, the output state is what we want.
Now it can be seen we have realized the sequential $N \rightarrow M$ UQCM. When $N=1$, all the results coincide with the $1 \rightarrow M$ case in last section.

Recently, the sequential cloning concerning about the real-life experimental condition
is investigated in \cite{Saberi2012}.

\subsection{Sequential UQCM in d dimensions}
We now further proceed to a more general case where qubit is extended to qudit. In the space of d dimensions, an arbitrary quantum pure state can be expressed as
\begin{equation}
|\Psi\rangle=\sum_{i=0}^{d-1}x_i|i\rangle, \sum_{i=0}^{d-1}|x_i|^2=1.
\end{equation}
Then N identical such qudits will be expanded in symmetric space as\cite{PhysRevA.58.1827}
\begin{equation}
|\Psi\rangle^{\otimes N} = \sum_{\bm{m}=0}^N \sqrt{\frac{N!}{m_1!...m_d!}}x_0^{m_1}...x_{d-1}^{m_d}|\bm{m}\rangle,
\end{equation}
where $|\bm{m}\rangle$ denotes the symmetric state, whose form is
\begin{equation}
|\bm{m}\rangle = |m_1 0,m_2 1, ..., m_d (d-1)\rangle,
\end{equation}
which means the symmetric state $|\bm{m}\rangle$ has $m_i$ qudits in the computational base $|i\rangle$(i=1,...,d), and the sum of qubits in each base satisfies $\sum_{i=1}^d m_i=N$.

Take the symmetric state $|\bm{m}\rangle$ as the input state of the optimal d-level UQCM according to Fan et al's scheme\cite{PhysRevA.64.064301}, the corresponding M-qudit output state will be
\begin{equation}
|\Psi_M^{(\bm{m})}\rangle = \sum_{\bm{j}=0}^{M-N} \beta_{\bm{m}\bm{j}}|\bm{m}+\bm{j}\rangle\otimes|R_{\bm{j}}\rangle,
\end{equation}
where the vector $\bm{j}=(j_1,j_2,...j_d)$ satisfies $\sum_{i=1}^d j_i = M-N$, $|R_{\bm{j}}\rangle=|\bm{j}\rangle_R$ denotes the state of the cloning machine, and $\beta_{\bm{m}\bm{j}}=\sqrt{\prod_{i=1}^d C_{m_i+j_i}^{m_i}/C_{M+d-1}^{M-N}}$.
The following steps are similar to the 2-level case presented previously. We still need to find the MPS form of the state $|\Psi_M^{(\bm{m})}\rangle$, and the method is through SD as well. Express $|\Psi_M^{(\bm{m})}\rangle$ in the computational basis
\begin{equation}
|\Psi_M^{(\bm{m})}\rangle = \sum_{i_1...i_{2M-N}}\langle \varphi_F^{(\bm{m})}|V^{[2M-N]i_{2M-N}}...V^{[1]i_1}|0\rangle_D\otimes|i_1...i_{2M-N}\rangle
\end{equation}
Through computation, $V^{[k]i_k}$ can be obtained \cite{Dang2008}. The detailed process is just a direct extension of the 2-level case and shall be omitted here, see Appendix for the details. Besides, we also know that the necessary dimension of the ancilla is $D_d = C_{M-\lfloor\frac{N+1}{2}\rfloor+d-1}^{M-\lfloor\frac{N+1}{2}\rfloor}$,
where the symbol $\lfloor X \rfloor$ denotes the floor function.
So we may observe that $d\geq2$ $D_d$ is far smaller than $d_M$, that simplification shows the advantage of sequential cloning of qudits.

\newpage

\section{Implementation of quantum cloning machines in physical systems}

In general, the cloning machines can be realized by the corresponding quantum circuits
constituted by single qubit rotation gates and CNOT gates
just like other quantum computations.
This is guaranteed by the universal quantum computation \cite{elementarygates} which
can be realized by a complete set of universal gates.

\subsection{A unified quantum cloning circuit}

\begin{figure}
\centering
\includegraphics[height=5.5cm]{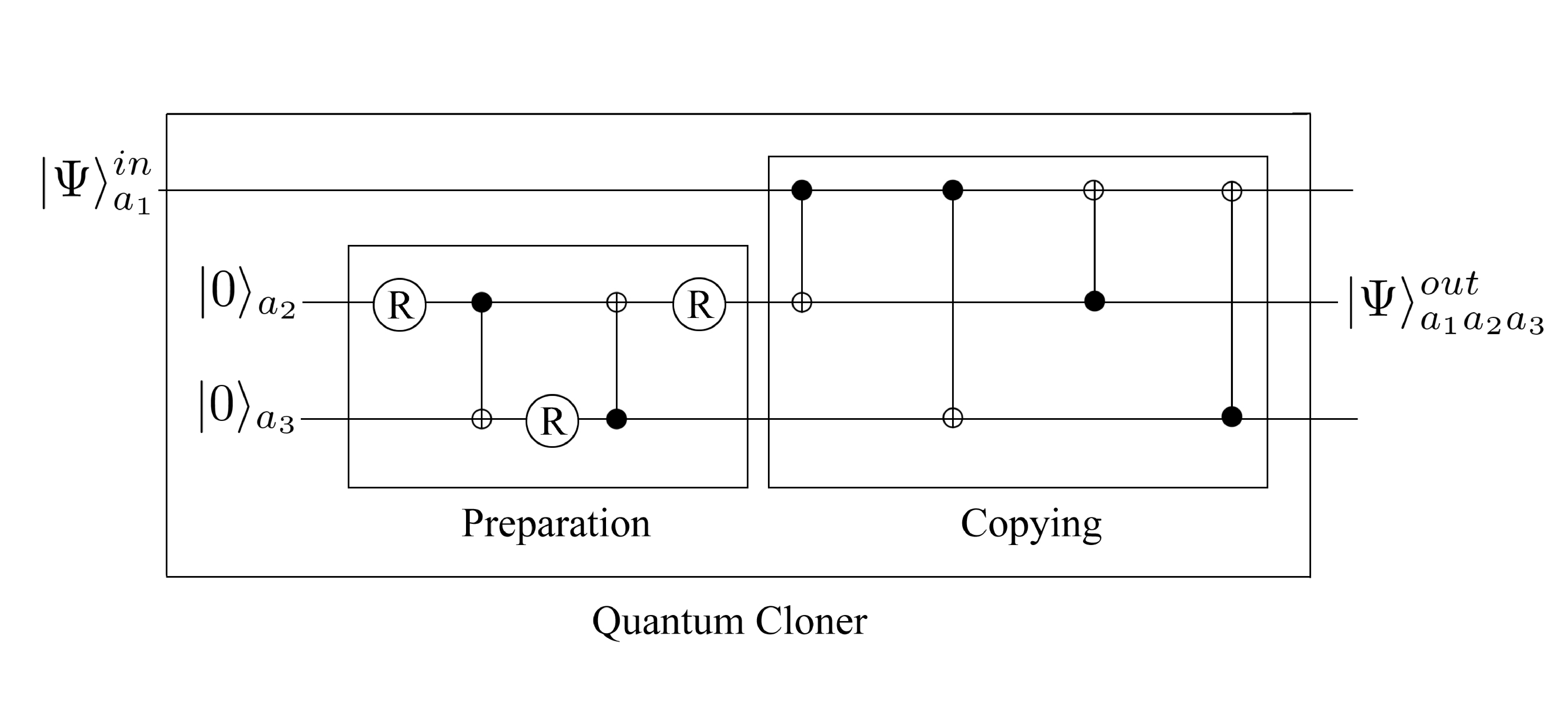}
\caption{Quantum circuit implementing quantum cloning machines. This quantum circuit can realize
both universal cloning machine and phase-covariant quantum cloning machine by adjusting parameters
in the single qubit rotation gates. This circuit is the same as the one in \cite{BBHBnetwork}.}
\label{quantumcircuit}
\end{figure}

It is interesting that the UQCM and the phase-covariant QCM
can be realized by a unified quantum cloning circuit by adjusting
angles in the single qubit rotation gates,
as shown in FIG.\ref{quantumcircuit} first presented by
Bu\v{z}ek \emph{et~al.} \cite{BBHBnetwork}.
Let us consider the definition of single qubit rotation gate (\ref{singlegate}) with a fixed
phase parameter which can be omitted,
it can be written in matrix form as,
\begin{eqnarray}
\hat {R}(\vartheta )=\left( \begin{array}{cc}
\cos \vartheta &\sin \vartheta\\
-\sin \vartheta &\cos \vartheta \end{array}\right) .
\end{eqnarray}
The form of CNOT gate is in (\ref{CNOT}). Here we use subindices in a CNOT gate,
$CNOT_{jk}$, to specify that the controlled qubit is $j$ and the target qubit
is $k$.
Following the copying scheme in \cite{BBHBnetwork}, the cloning procession
is divided into two unitary transformations,
\begin{eqnarray}
|\Psi _{a_1}^{(in)}|0\rangle _{a_2}|0\rangle _{a_3}
\rightarrow |\Psi \rangle _{a_1}^{(in)}|\Psi \rangle _{a_1a_2}^{(prep)}
\rightarrow |\Psi \rangle _{a_1a_2a_3}^{(out)}.
\end{eqnarray}
The preparation state is constructed as follows,
\begin{eqnarray}
|\Psi \rangle _{a_2a_3}^{(prep)}=\hat {R}_2(\vartheta _3)
CNOT_{32}\hat {R}_3(\vartheta _2)CNOT_{23}
\hat {R}_2(\vartheta _1)|0\rangle _{a_2}|0\rangle _{a_3}.
\end{eqnarray}
The second step is as,
\begin{eqnarray}
|\Psi \rangle _{a_1a_2a_2}^{(out)}=CNOT_{a_3a_1}CNOT_{a_2a_1}
CNOT_{a_1a_3}CNOT_{a_1a_2}|\psi \rangle _{a_1}^{(in)}
|\psi \rangle _{a_2a_3}^{(prep)}.
\label{copy}
\end{eqnarray}
We may find that two copies are in $a_2, a_3$ qubits.
For UQCM, the angles in the single qubit rotations are chosen as,
\begin{eqnarray}
\vartheta _1=\vartheta _3=\frac {\pi }{8},
~~\vartheta _2=-\arcsin \left( \frac {1}{2}-\frac {\sqrt{2}}{3}\right) ^{1/2}.
\end{eqnarray}

This scheme is flexible and can be adjusted for phase-covariant quantum cloning.
We only need to choose different angles for the single qubit rotations, and those
angles are shown to be as follows \cite{PhysRevA.65.012304},
\begin{eqnarray}
\vartheta _1=\vartheta _3=\arcsin \left( \frac {1}{2}
-\frac {1}{2\sqrt{3}}\right)^{\frac {1}{2}},
~~\vartheta _2=-\arcsin \left(
\frac {1}{2}-\frac {\sqrt{3}}{4}\right) ^{\frac {1}{2}}.
\end{eqnarray}
To be explicit, we may find that the preparation state takes the form,
\begin{eqnarray}
|\Psi \rangle _{a_2a_3}^{(perp)}
=\frac {1}{\sqrt{2}}|00\rangle _{a_2a_3}
+\frac {1}{2}(|01\rangle _{a_2a_3}+|10\rangle _{a_2a_3}).
\end{eqnarray}
The second step for phase-covariant quantum cloning is the same
as the that of the UQCM. So this cloning circuit is general and
can be applied for both universal cloning and phase-covariant cloning.

\subsection{A simple scheme of realization of UQCM and Valence-Bond Solid state}

We already know that the universal cloning machine can be
realized by a symmetric projection. This fact can let us find a simple scheme
for the implementation of the universal cloning machine.
The simplest universal cloning machine can be obtained by a symmetric
projection on the input qubit and one part of a maximally entangled state. This symmetric projection can be naturally realized by bosonic operators in the Fock space representation. Suppose the input
state is $a_H^{\dagger }$, the available maximally entangled state is $a_H^{\dagger }a_H^{\dagger }+
a_V^{\dagger }a_V^{\dagger }$, here $H,V$ can be horizontal and vertical polarizations of a photon, or
any other degrees of freedom of the bosonic operator.
We also suppose that those operators are acting on the vacuum state,
then we have,
\begin{eqnarray}
a_H^{\dagger }\left( a_H^{\dagger }a_H^{\dagger }+
a_V^{\dagger }a_V^{\dagger }\right) =\left[ \sqrt {\frac {2}{3}}\frac {(a_H^{\dagger })^2}{\sqrt {2!}}a_H^{\dagger }
+\sqrt {\frac {1}{3}}(a_H^{\dagger }a_V^{\dagger })a_V^{\dagger }\right] \sqrt {3}.
\label{fockrep}
\end{eqnarray}
Now we consider that the last bosonic operators are acting as ancillary qubits, in Fock space representation, $\frac {(a_H^{\dagger })^2}{\sqrt {2}}$ corresponds to two photons in horizontal polarization, while
$a_H^{\dagger }a_V^{\dagger }$ is a symmetric state with one horizontal photon and one vertical photon. By a whole normalization  factor $\sqrt {3}$, the above formula then takes the following form, with initial state $|H\rangle $,
\begin{eqnarray}
|H\rangle \rightarrow \sqrt {\frac {2}{3}} |2H\rangle |H\rangle _a+\sqrt {\frac {1}{3}} |H,V\rangle \rangle |V\rangle _a.
\label{photonclone1}
\end{eqnarray}
Similarly for a vertical photon, we have
\begin{eqnarray}
|V\rangle \rightarrow \sqrt {\frac {2}{3}} |2V\rangle |V\rangle _a+\sqrt {\frac {1}{3}} |H,V\rangle \rangle |H\rangle _a.
\label{photonclone2}
\end{eqnarray}
It is now clear that those two transformations constitute exactly a UQCM.  This fact is noticed by Simon \emph{et al.}. Actually it also provides a natural realization of the UQCM by photon stimulated emission.
Based on the experiment of the preparation of maximally entangled state, the UQCM can be
realized by the above scheme which we will present later.

For no-cloning theorem,  a frequently misunderstanding point may be that, it seems that ``laser'' itself can provide a perfect cloning machine, one photon can be cloned perfectly to have many completely  same photons. This seems contradict with no-cloning theorem. The point is that we can for sure clone
a photon to have may copies. However, the no-cloning theorem states that if we clone horizontal and vertical photons perfectly, we can not clone perfectly the photon superposition state of horizontal and vertical. Thus ``laser'' does not conflict with no-cloning theorem.

When a maximally entangled state is available, it seems that a UQCM can be realized. In condensed matter physics, the Valence-Bond Solid state is constructed by a series of singlet states, see for example \cite{VBSPRL},
\begin{eqnarray}
|VBS\rangle =\prod _{i=0}^N\left( a_i^{\dagger }b_{i+1}^{\dagger }-a_{i+1}^{\dagger }b_{i}^{\dagger }\right)|{\rm vacuum}\rangle ,
\end{eqnarray}
where sites 0 and $N+1$ are two ends. We remark that the sites in the bulk will be restricted to the
symmetric subspace also by the reason of Fock space representation as shown schematically in the following:

\begin{center}
\unitlength=1mm
\begin{picture}(80,5)(0,3)
\put(0,0){\makebox(3,3){0}} \put(9,0){\makebox(3,3){1}}
\put(19,0){\makebox(3,3){2}}
\put(27,0){\makebox(3,3){...}} \put(30,0){\makebox(3,3){...}}
\put(37,0){\makebox(3,3){...}} \put(40,0){\makebox(3,3){...}}
\put(47,0){\makebox(3,3){...}} \put(50,0){\makebox(3,3){...}}
\put(59,0){\makebox(3,3){N}}
\put(67,0){\makebox(3,3){N+1}}
\end{picture}

\begin{picture}(80,5)(0,0)
\put(0,0){\line(1,0){7}} \put(10,0){\line(1,0){7}}
\put(20,0){\line(1,0){7}} \put(30,0){\line(1,0){7}}
\put(40,0){\line(1,0){7}} \put(50,0){\line(1,0){7}}
\put(60,0){\line(1,0){7}} \put(0,0){\circle*{1}}
\put(7,0){\circle*{1}} \put(10,0){\circle*{1}}
\put(17,0){\circle*{1}} \put(20,0){\circle*{1}}
\put(27,0){\circle*{1}} \put(30,0){\circle*{1}}
\put(37,0){\circle*{1}} \put(40,0){\circle*{1}}
\put(47,0){\circle*{1}} \put(50,0){\circle*{1}}
\put(57,0){\circle*{1}} \put(60,0){\circle*{1}}
\put(67,0){\circle*{1}} \put(8.5,0){\circle{5}}
\put(18.5,0){\circle{5}} \put(28.5,0){\circle{5}}
\put(38.5,0){\circle{5}} \put(48.5,0){\circle{5}}
\put(58.5,0){\circle{5}}
\end{picture}
\end{center}

By the same consideration as presented above, since one singlet state is a maximally entangled state, the UQCM can be realized if the input state is put in site $0$, $(\alpha a_0^{\dagger }+\beta b_0^{\dagger })\left( a_0^{\dagger }b_{1}^{\dagger }-a_{1}^{\dagger }b_{0}^{\dagger }\right)$.
Further one may notice we do not need to restrict just a singlet state where only two sites are involved, a whole one-dimensional Valence-Bond Solid state can be dealed as a maximally entangled state so that
a UQCM can be realized like the following:
\begin{eqnarray}
(\alpha a_0^{\dagger }+\beta b_0^{\dagger })|VBS\rangle .
\end{eqnarray}
The state of input $\alpha a_0^{\dagger }+\beta b_0^{\dagger }$ is like the open boundary operator.
One feature of this universal cloning machine may be that the ancillary states are at one end
of this 1D state, the two copies are located on another end. This system is like a majorana fermion quantum wire proposed by Kitaev \cite{Kitaev} where the encoded qubit is topologically protected.
It is also pointed out that the cloning machine can be realized by networks of spin chains \cite{cloning-spinnetwork}.

\subsection{UQCM realized by photon stimulated emission and the experiment}

With the results in last section, one may realize that
a maximally entanglement source may provide a mechanism for quantum cloning.
The corresponding fidelity is optimal.
It seems that photon stimulated emission possesses such a property and can
give a realization of the UQCM.
This is first proposed in \cite{PhysRevLett.84.2993,Kempe00} and realized experimentally \cite{667603820020426,prl89.107901}.
In this scheme,
certain types of three-level atoms can be used to optimality clone
qubit that is encoded as an arbitrary superposition of
excitations in the photonic modes corresponding to the atomic
transitions.  Next, we shall first review briefly the qubit case
followed by a general d-dimensional result.

For qubit case \cite{PhysRevLett.84.2993,Kempe00}, we consider the inverted medium that consists of an ensemble of
$\Lambda $ atoms with three energy levels.
These three levels correspond to two degenerate ground
states $|g_1\rangle $ and $|g_2\rangle $ and an excited level
$|e\rangle $. The ground states are coupled to
the excited state by two modes of the electromagnetic field $a_1$
and $a_2$, respectively.
The Hamiltonian of this system takes the form,
\begin{eqnarray}
H=\gamma \left( a_1\sum _{k=1}^N|e^k\rangle \langle g_1^k|
+a_2\sum _{k=1}^N|e^k\rangle \langle g_2^k| \right) + H.c.
\label{Hami}
\end{eqnarray}
We then introduce the operator
as $b_rc^{\dagger }\equiv \sum _{k=1}^N|e^k\rangle \langle g_r^k|,
~~~r=1,2$, where $c^{\dagger} $ is a creation operator of ``e-type''
excitation, $b_r$ is a annihilation operator of $g_r$ ground states,
$r=1,2$. Now the Hamiltonian (\ref{Hami}) becomes as,
\begin{eqnarray}
{\cal {H}}=\gamma (a_1b_1+a_2b_2)c^{\dagger }+H.c.
\label{Hami-osc}
\end{eqnarray}
Now we find that the source of maximally entangled states is available.
The input state can be considered as the form
$(\alpha a_1^{\dagger }+\beta a_2^{\dagger })|0,0\rangle =\alpha
|1,0\rangle +\beta |0,1\rangle $. The number of copies in this cloning
system is restricted by the number of atoms in excited states
$\otimes _{k=1}^N|e^k\rangle $ which are represented as
$(c^{\dagger })^N/\sqrt {N!}$. We may consider that initially
there are $i+j$ qubits in $a_1^{\dagger }$ and $a_2^{\dagger }$ which
corresponds to a completely symmetric state with $i,j$ states in
two different levels of qubits,
\begin{eqnarray}
|\Psi _{in},(i,j)\rangle &=&
\frac {(a_1^{\dagger })^{i}(a_2^{\dagger })^j(c^{\dagger })^N}
{\sqrt {i!j!N!}}|0\rangle
\nonumber \\
&=&|i_{a_1},j_{a_2}\rangle |0_{b_1},0_{b_2}\rangle |N_c\rangle
\nonumber \\
&\equiv &|i,j\rangle _a|0,0\rangle _b|N\rangle _c.
\label{g-input}
\end{eqnarray}

With the Hamiltonian (\ref{Hami-osc}), the time evolution of the
state starting from the initial state (\ref{g-input}) becomes as follows,
\begin{eqnarray}
&&|\Psi (t),(i,j)\rangle =e^{-iHt}|\Psi _{in},(i,j)\rangle
\nonumber \\
&=&\sum _p(-iHt)^p/p!|\Psi _{in},(i,j)\rangle
=\sum _{l=0}^Nf_l(t)|F_l,(i,j)\rangle ,
\label{time}
\end{eqnarray}
where $|\Psi _{in},(i,j)\rangle =|F_0,(i,j)\rangle$, $l$
is the additional photons emitted corresponding to additional
copies, thus there are altogether $i+j+l$ copies in the output
which is expressed as $|F_l,(i,j)\rangle $.
In this process, the states with different copies are actually
superposed together. The amplitude parameter in the superposed
state corresponds to the probability of finding
$l$ additional copies which is $|f_l(t)|^2$.

To show that this process is exactly the realization of the optimal UQCM,
we can show that the corresponding cloning transformation with $l$
additional copes can be calculated as,
\begin{eqnarray}
|i,j\rangle _a|0,0\rangle _b|N\rangle _c\rightarrow |F_l,(i,j)\rangle =
\sum _{k=0}^l\sqrt {\frac {l!(i+j+1)!}{(i+j+l+1)!}}
\sqrt{ \frac {(i+l-k)!(j+k)!}{i!j!k!(l-k)!}}
\nonumber \\
|i+l-k,j+k\rangle _a|l-k,k\rangle _b|N-l\rangle _c.
\label{g-output}
\end{eqnarray}
This is indeed the UQCM. In addition, this provides an alternative
method to find the optimal cloning transformations.

\begin{figure}
\includegraphics[height=6cm]{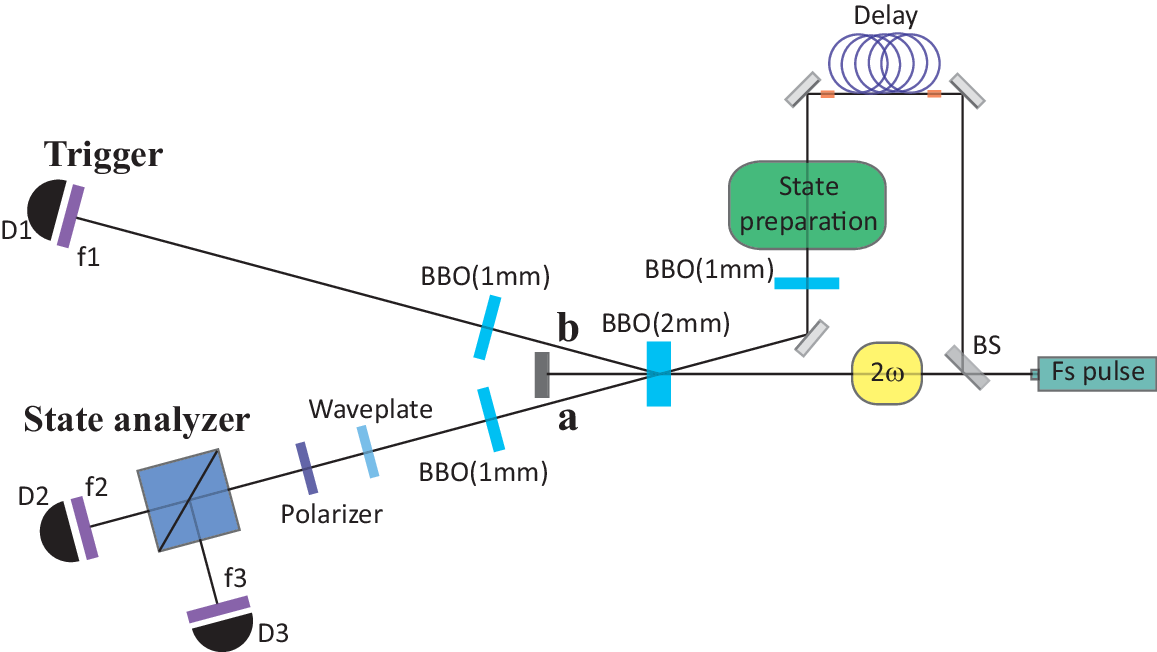}
\caption{Schematic of experimental universal quantum cloning, see Ref.\cite{667603820020426}.}
\label{clone-experiment}
\end{figure}

Experimentally, Lamas-Linares \emph{et al.} successfully performed the universal
quantum cloning by using the Hamiltonian (\ref{Hami-osc}) shown above \cite{667603820020426}.
The operators $a^{\dagger },b^{\dagger }$ are creation operators of photons in
the spatial modes $a,b$ marked in FIG.(\ref{clone-experiment})
corresponding to two different directions of emission after passing the non-linear
crystal (BBO 2mm). Photons of mode $a^{\dagger }$ with subscript $1,2$ refer to vertical
and horizontal polarization, respectively. In the state analyzer part of the experimental set up,
vertical and horizontal photons can be analyzed by polarizing beam splitter in front of
photon detectors $D2$ and $D3$.

In experiment, a laser produces light pulses of 120 fs duration (Fs pulse shown in FIG.\ref{clone-experiment}).
By beam splitter, a tiny part of each pulse is split off and attenuated below the single-photon
level resulting in probabilistically the input photon. The polarization of the input photon
can be adjusted in the state preparation part corresponding to arbitrary input pure qubit.
The major part of the pulse is frequency doubled (shown as $2\omega $ in FIG.\ref{clone-experiment})
and used to pump the non-linear crystal (BBO 2mm) where photon pairs entangled
in polarization are created, as shown in Hamiltonian (\ref{Hami-osc}).
The input photon and the $a$ photons with vertical and horizontal polarizations created
are adjusted so that they can overlap perfectly and are indistinguishable for quantum cloning.
Photon of mode $b$ severs as a trigger indicating whether the entangled state
in polarization by parametric down-conversion has created or not.
For time evolution $e^{-i{\cal {H}}t}$ with small values of $\gamma t$ for initial
state $a^{\dagger }_1|0\rangle $, the first order term corresponds to three photon state
$(a_1^{\dagger }b_1^{\dagger }+a_2^{\dagger }b_2^{\dagger })a^{\dagger }_1|0\rangle $.
Here we remark that the entangled state in vertical and horizontal polarization usually takes the form
$a_V^{\dagger }b_H^{\dagger }-a_H^{\dagger }b_V^{\dagger }$£¬which is equivalent to
what we write here. The output state has the form, $(a^{\dagger }_1)^2$ for $b_1^{\dagger }$
and $a^{\dagger }_1a^{\dagger }_2$ for $b_2^{\dagger }$ corresponding to two terms
in the universal quantum cloning transformation. By this method the information of the
input photon polarization, a qubit, is quantum cloned by universal cloning on the
down-converted photon. This process of universal cloning is exactly what we have reviewed
as symmetric projection of identical input states and one half the maximally entangled states.
We may expect that, if maximally entangled states are available, the universal quantum
cloning is possible if symmetric projection can be realized.

\subsection{Higher dimension UQCM realized by photon stimulated emission}

The higher dimensional case can be similarly studied \cite{ISI:000177872600134}.
Now the atoms have one excited state $|e\rangle $ and $d$ ($d\ge
2$) ground states $|g_n\rangle , n=1, 2,\cdots, d$, and each coupled
to a different photons $a_n$ corresponding
to modes of qudit. The Hamiltonian can also be written
as a generalized form,
\begin{eqnarray}
{\cal {H}}_d=\gamma (a_1b_1+\cdots +a_db_d)+H.c.
\label{d-Hami}
\end{eqnarray}
The initial states which are symmetric states are,
\begin{eqnarray}
|\Psi _{in},\vec {j}\rangle =
\prod _{i=1}^d\frac {(a_i^{\dagger })^{j_i}}{\sqrt{j_i!}}
\frac {(c^{\dagger })^N}{\sqrt{N!}}|0\rangle
\equiv |\vec{j}\rangle _a|\vec{0}\rangle _b|N\rangle _c,
\label{d-input}
\end{eqnarray}
where $\vec {j}=(j_1,j_2,\cdots,j_d)$.
One can find that the time evolution of states for qudits
is the same as that of qubits (\ref{time}).
So the probability to obtain additional $l$ copies is $|f_l(t)|^2$.
We use the notation $|F_0,\vec{j}\rangle \equiv |\Psi _{in},\vec{j}\rangle $,
$\sum _{i}j_i=M$, and the output of cloning with $l$ additional copies can be
obtained as,
\begin{eqnarray}
|\vec{j}\rangle _a|\vec{0}\rangle _b|N\rangle _c\rightarrow |F_l,\vec{j}\rangle &=&
\sum _{k_i}^l\sqrt{ \frac {(M+d-1)!l!}{(M+l+d-1)!}}
\prod _{i=1}^d\sqrt{ \frac {(k_i+j_i)!}{k_i!j_i!}}
\nonumber \\
&&|\vec{j}+\vec{k}\rangle _a|\vec{k}\rangle _b|N-l\rangle _c,
\label{d-output}
\end{eqnarray}
where summation $\sum _{k_i}^l$ runs for all variables
with constraint, $\sum _i^dk_i=l$. We thus realize the
optimal UQCM for qudits.

Explicitly, the action of Hamiltonian on the symmetric states takes
the following form,
\begin{eqnarray}
&&{\cal {H}}_d|F_l,\vec{j}\rangle
=\gamma (\sqrt{(l+1)(N-l)(M+l+d)}|F_{l+1},\vec{j}\rangle
\nonumber \\
&&+\sqrt{l(N-l+1)(M+l+d-1)}|F_{l-1},\vec{j}\rangle ,
\nonumber \\
&&~~~~l\le l<N,
\nonumber \\
&&{\cal {H}}_d|F_0,\vec{j}\rangle =\gamma \sqrt{N(M+d)}|F_1,\vec{j}\rangle ,
\nonumber \\
&&{\cal {H}}_d|F_N,\vec{j}\rangle =\gamma \sqrt{N(M+N+d-1)}|F_{N-1},\vec{j}
\rangle .
\label{struct1}
\end{eqnarray}

To end this subsection, we present our familiar results of UQCM but in
this photonic system.
An arbitrary qudit takes the form,
$|\Psi \rangle =\sum _{i=1}^dx_ia_i^{\dagger }|\vec{0}\rangle $,
with $\sum _{i=1}^d|x_i|^2=1$. By expansion, it corresponds to state,
\begin{eqnarray}
|\Psi \rangle ^{\otimes M}&=&
(\sum _{i=1}^dx_ia_i^{\dagger })^{\otimes M}|\vec{0}\rangle
\nonumber \\
&=&M!\sum _{j_i}^M
\prod _{i=1}^d
\frac {x_i^{j_i}}
{\sqrt{j_i!}}
\frac {(a_i^{\dagger })^{j_i}}
{\sqrt {j_i!}}|\vec{0}\rangle .
\end{eqnarray}
With the help of cloning
transformation (\ref{d-output}), the output of cloning
is,
\begin{eqnarray}
|\Psi \rangle ^{\otimes M}\rightarrow |\Psi \rangle ^{out}
&=&M!\sum _{j_i}^M
\sum _{k_i}^l\sqrt{ \frac {(M+d-1)!l!}{(L+d-1)!}}
\nonumber \\
&&\times \prod _{i=1}^d
\frac {x_i^{j_i}}
{j_i!}
\sqrt{ \frac {(k_i+j_i)!}{k_i!}}
|\vec{j}+\vec{k}\rangle _a|\vec{k}\rangle _b,
\label{out}
\end{eqnarray}
As we already know, this is the UQCM of qudits.

Quantum cloning itself is reversible since it is realized by unitary transformation.
This does not necessarily mean that the cloning realized by photon stimulated emission can be
inverted. However, it is proposed that this inverting process can succeed \cite{revsecloning}.

\subsection{Experimental implementation of phase-covariant quantum cloning by nitrogen-vacancy defect center in diamond}

The economic phase-covariant quantum cloning involves only three states of two-qubit system. Experimentally, we can
encode those three states by three energy levels in a specified physical system. The experimental implementation
of phase-covariant quantum cloning by this scheme is realized in solid state system \cite{Pan2011}.
This solid system is the nitrogen-vacancy (NV) defect center in diamond. The structure of NV center in diamond is that
a carbon atom is replaced by a nitrogen atom and additionally a vacancy is located in a nearby lattice site.
The NV center is negative charged and
can provide three states of the electronic spin one.
Those three states correspond to zero magnetic moment ($m_s=0$) with a 2.87-GHz zero field splitting, two
magnetic sub-levels induced by external magnetic field corresponding to $m_s=\pm 1$.
The experimental samples of diamond can be bulk or nanodiamond. The electronic spin in NV center
of diamond can be individually addressed by using confocal microscopy
so that we can control it exactly, however, ensemble
of NV centers can also be well controlled. The NV center can be initialized to state $m_s=0$ polarization
by a continuous 532nm laser excitation.

The superposed states of $m_s=0$ with $m_s=\pm 1$
are prepared by resonating microwaves depending on the duration time determined by
their corresponding Rabi oscillations.  The microwave radiation is sent out by a copper wire of 20 $\mu$m
diameter placed with a distance of 20 $\mu$m from the NV center.
The Rabi oscillations corresponding to different microwave frequencies show that the
prepared states are superposed states in quantum mechanics. The resonating frequencies of
the controlling microwaves are determined by the electronic-spin-resonating (ESR) spectrum of the
NV center obtained by frequency continuously changing. The readout of the electronic state
is by Rabi oscillation, the measured value depends on the intensity of the florescence
which corresponding to the amplitude of state $m_s=0$ in the superposed state.
The intensity of florescence is measured by single photon counting module connected
with a multifunction data acquisition device.
The main advantage of the NV center in diamond is its long coherence time which
is long enough for spin electronic spin manipulation for various tasks in quantum information
processing.

One key point in precisely control the electron spin state is that
it does not interact with environmental spin bath mainly constituted
by nearby nuclear spins. The fact is that when the electron spin is
in state with zero magnetic moment, $m_s=0$, it does not interact
with the nuclear spin. If the electron spin is in either of
the $m_s=\pm 1$ states, it is under the influence of the nearby
nuclear spin. We may, on the one hand, use this coherent coupling
for quantum information purposes, such as to generate entangled
state or for quantum memory. On the other hand, it causes
decoherence of the quantum state of the electron spin in the NV
center. The interaction between the electron spin and a nearby nuclear spin
in the NV center can be clearly shown by hyperfine structure in the
ESR spectrum.

The general spin
Hamiltonian of the NV center consisting of an electron spin,
$\textbf{S}$, coupled with nearby nuclear spins, $\textbf{I}_k$, is
given as,
\begin{eqnarray}
H_{spin}=H_{zf}+H_{eZeeman}+H_{hf}+H_q+H_{nZeeman},
\end{eqnarray}
where the terms in spin Hamiltonian describe: the electron spin zero
field splitting, $H_{ZF}=\textbf{S}\bar {D}\textbf{S}$, the electron
Zeeman interaction, $H_{eZeeman}=\beta _e\vec{B}_0\bar
{g}_e\textbf{S}$, hyperfine interactions between the electron spin
and nuclear spins $H_{hf}=\sum _k\textbf{S}\bar {A}_k\textbf{I}_k$,
the quadrupole interactions for nuclei with $I>1/2$, $H_q=\sum
_{I_k>1}\textbf{I}_k\bar {P}_k\textbf{I}_k $, and the nuclear Zeeman
interactions, $H_{nZeeman}=-\beta _n\sum _kg_{n,k}\vec
{B}_0\textbf{I}_k$, also $g_e$ and $g_n$ are the $g$ factors for the
electron and nuclei respectively, $\beta _e,\beta _g$ are Bohr
magnetons for electron and nucleus, $\bar {A}$ and $\bar {P}$ are
coupling tensors of hyperfine and quadrupole, and $\vec {B}_0$ is
the applied magnetic field.

In implementation of economic phase-covariant quantum cloning, we use four equatorial
qubits equivalent to BB84 states. However, according to result of minimal input set,
it is also possible to check just three equatorial qubits \cite{Jing2013}.
Since only three orthogonal states are involved in the economic phase-covariant cloning,
the scheme is to use three physical states $m_s=0, m_s=\pm 1$ to represent logic states of qubits.
In the experimental scheme, the encode scheme is that: $|10\rangle \rightarrow m_s=0$,
$|00\rangle $ and $|01\rangle $ correspond to $m_s=\pm 1$ respectively.

The implementation of phase cloning is in two steps. The first step is the
initial state preparation which includes the input state preparation and cloning machine
initialization. The experimental realization of this step is to prepare the logic
qubits $\frac {1}{\sqrt {2}}(|0\rangle +e^{i\phi }|1\rangle )|0\rangle $, which is
to prepare physically a superposed state of two involved levels. It is realized by
initializing the NV center, applying a $\pi /2$ pulse microwave. The second step
of the phase cloning is to realize the quantum cloning transformation. According to the
optimal transformation $|00\rangle \rightarrow |00\rangle ,|10\rangle \rightarrow
\frac {1}{\sqrt {2}}(|10\rangle +|01\rangle )$, we can realize it by applying another
$\pi /2$ pulse microwave. So now the phase quantum cloning is realized experimentally.
To readout the result, we can use the combination of two Rabi oscillations to
find the exact value of the output state. Experimentally, it is shown that the
experimental results are very close with theoretical expectations.
In average, the experimental fidelity is about $85.2\% $ which is very close to
theoretical optimal bound $1/2+\sqrt {1/8}\approx 85.4\% $ and is clearly better than
the universal quantum cloning \cite{Pan2011}.

\begin{figure}
\includegraphics[height=10cm]{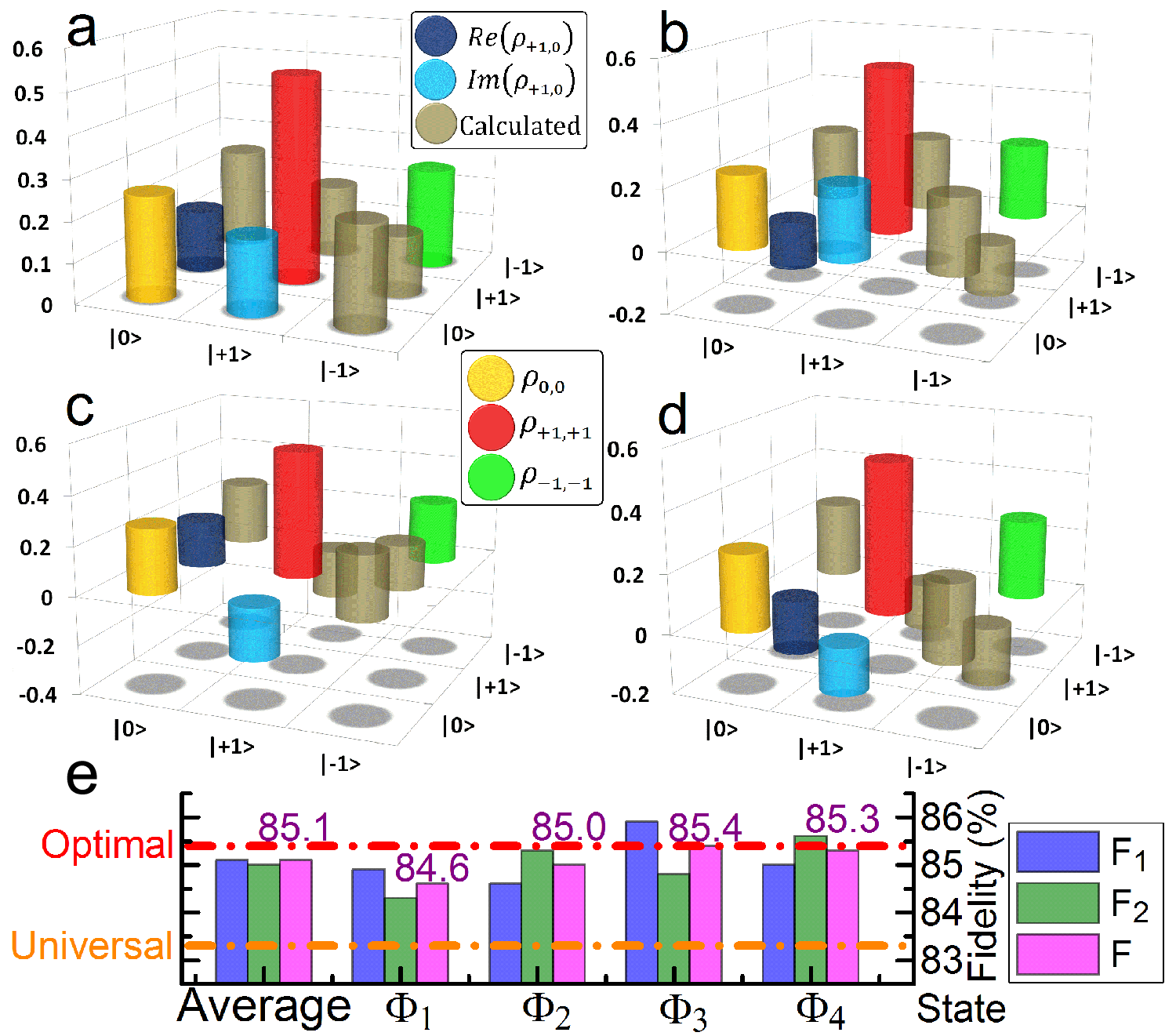}
\caption{Experimental results of phase-covariant quantum cloning in NV center of diamond at room temperature.
The output states in experiment are readout by state tomography.
The fidelities for different input states are presented and are shown to be better
than the bound $5/6$ of the universal cloning machine
The average fidelity is very close to theoretical expectation. This
figure is presented in \cite{Pan2013}.}
\label{figure-panphase}
\end{figure}

To run all values of the phase parameter, the active controlling of the phase of the input state
should be performed in experiment. This can also be realized experimentally by using two
independently microwave sources. This experiment is performed recently by using the nanodiamond \cite{Pan2013}.
The advantage by using nanodiamond instead of the bulk sample is its further integration property.
Additionally, due to the sub-wavelength size of the nanodiamond, the fluorescence collection
efficiency can increase dramatically which provides a high quality signal.
The experimental results are presented in FIG. \ref{figure-panphase}. We can find that the advantage
of the phase cloning machine than the universal cloning machine can also be demonstrated
in this experiment by using the state tomography for readout.

\subsection{Experimental developments}

Quantum cloning process have been realized by various schemes experimentally as we
already reviewed in the previous sections. Here let us present some experiments in the following.
By quantum circuit method as shown in FIG. \ref{quantumcircuit}, the universal
cloning is realized by nuclear magnetic resonance (NMR) \cite{PhysRevLett.88.187901}.
The phase-covariant cloning is also realized in NMR system with input states ranging from
the equator to the polar possessing an arbitrary phase parameter \cite{PhysRevLett.94.040505}.
The UQCM is realized experimentally by single photon with different degrees of freedoms \cite{pra64.012315}, stimulated emission with optical fiber amplifier \cite{prl89.107901}.
Also in optical system, the UQCM and the NOT gate are realized \cite{prl92.067901}.
Closely related with optical cloning, the experimental noiseless amplifier for quantum light
states is performed \cite{naturephoton2011}.
The UQCM realization in cavity QED is proposed in \cite{qed-clone}.
Experimental of various cloning machines in one set is performed recently in
optics system \cite{Lemr2012}.

Quantum cloning machine can be used for metrology. It is proposed theoretically and shown experimentally
with an all-fiber experiment at telecommunications wavelengths that the optimal
cloning machine can be used as a radiometer to measure the amount of radiated power \cite{clone-radiometry}.
The electro-optic quantum memory for light by atoms
is demonstrated experimentally and compared with the limit of no-cloning limit \cite{HetetLongdell}.

\section{Concluding Remarks}

The research of quantum cloning and the QIP are continuously developing.
In preparing this review, some new results may emerge which may not be
summarized in this review. However, we still try to include some new
results in the revision process of this review. When this review was first
posted in arXiv, a lot of colleagues informed us some related literatures
which might be missed in the previous versions. We would like to thank
those responses. At the same time, their
feedbacks let us realize that such a review is indeed necessary. We are
then encouraged to finish this review and the revision.
The anonymous referee also provided a lot of valuable suggestions for us to improve our
presentation.

\emph{Acknowlegements:} This work was supported by ``973'' program (2010CB922904), NSFC (11175248), NFFTBS (J1030310),
and Level B Primary Leading Project (through USTC) of Chinese Academy of Sciences.
H.F. would like to thank useful discussions and communications concerning about
this topic with Luigi Amico, V. Buz\v{e}k, Jun-Peng Cao, Kai Chen, Zeng-Bing Chen, I. Cirac, G. M. D'Ariano Jiang-Feng Du, Lu-Ming Duan,
Shao-Ming Fei, J. Fiurasek, N. Gisin, Guang-Can Guo, A. Hayashi,
Yun-Feng Huang, W. Y. Hwang, H. Imai, Vladimir Korepin, Le-Man Kuang, L. C. Kwek, Shu-Shen Li,
Hai-Qing Lin, Nai-Le Liu, Seth Lloyd, Gui-Lu Long, Li Lu, Shun-Long Luo, C. Macchieveallo, K. Matsumoto, Xin-Yu Pan, Jian-Wei Pan, Martin Plenio,
Xi-Jun Ren, Vwani Roychowdhury,
Chang-Pu Sun, Vlatko Vedral, Miki Wadati, Xiang-Bin Wang, Zi-Dan Wang, R. Werner, Ru-Quan Wang, Sheng-Jun Wu, Zheng-Jun Xi,
Tao Xiang, Zhao-Xi Xiong, Eli Yablonovitch, Si-Xia Yu, P. Zanardi, Jing Zhang, Zheng-Wei Zhou, Xu-Bo Zou. He also would like to thank
continuous support from En-Ge Wang, Yu-Peng Wang, Xin-Cheng Xie, Qi-Kun Xue and Lu Yu.
We thank Jun Feng and Xi Chen for their careful reading this review and for numerous suggestions.
We thank Ling-An Wu for helping us to fix the language problems.
We thank Shuai Cui and Yu-Ran Zhang for drawing some pictures.

\section{Appendix}
For sequential quantum cloning machine of qudits, some detailed results are presented here.
We shall provide the explicit form of matrix $V^{[n]i_n}$ for different kinds of sequential UQCM.

\subsection{$N\rightarrow M$ sequential UQCM of qubits.}
\begin{enumerate}
\item When $n=1$: The upper left corner of $V^{[1]i_1}$ is%
\begin{equation*}
V ^{\left[ 1\right]0}=\left(
\begin{array}{cc}
\lambda  _{\left[ 1\right]1} & 0 \\
0 & \lambda  _{\left[ 1\right]2}%
\end{array}%
\right) ,\text{ }V ^{\left[ 1\right]1}=\left(
\begin{array}{cc}
0 & \lambda  _{\left[ 1\right]1} \\
\lambda  _{\left[ 1\right]2} & 0%
\end{array}%
\right) ,
\end{equation*}%
while for $\alpha _{1}\ge 3$, set $V _{xy}^{\left[ 1\right]i_{1}}=\frac{1}{\sqrt{2}}\delta _{xy}%
$.

\item  Case $1<n\le M-N+m$.

For $1\le \alpha _{n},\alpha _{n-1}\le n$,
\begin{eqnarray}
V_{\alpha _{n}\alpha _{n-1}}^{\left[ n\right]0}=\delta _{\alpha _{n}\alpha _{n-1}}\sqrt{%
\frac{\overset{M-m-n}{\underset{k=-m}{\sum }}\beta _{m\left( \alpha
_{n-1}-1+k\right) }^{2}\frac{C_{M-n}^{m+k}}{C_{M}^{m+\alpha _{n-1}-1+k}}}{%
\overset{M-m-n+1}{\underset{k=-m}{\sum }}\beta _{m\left( \alpha
_{n-1}-1+k\right) }^{2}\frac{C_{M-n+1}^{m+k}}{C_{M}^{m+\alpha _{n-1}-1+k}}}},
\end{eqnarray}
otherwise $V _{\alpha _{n}\alpha _{n-1}}^{\left[ n\right]0}=\delta
_{\alpha _{n}\alpha _{n-1}}\frac{1}{\sqrt{2}}$.

For $2\le \alpha _{n}\le \left( n+1\right) $, $1\le \alpha
_{n-1}\le n$,
\begin{eqnarray}
V _{\alpha _{n}\alpha _{n-1}}^{\left[ n\right]1}=\delta
_{\alpha _{n}\left( \alpha _{n-1}+1\right) }\sqrt{\frac{\overset{M-m-n}{%
\underset{k=-m}{\sum }}\beta _{m\left( \alpha _{n-1}+k\right) }^{2}\frac{%
C_{M-n}^{m+k}}{C_{M}^{m+\alpha _{n-1}+k}}}{\overset{M-m-n+1}{\underset{k=-m}{%
\sum }}\beta _{m\left( \alpha _{n-1}-1+k\right) }^{2}\frac{C_{M-n+1}^{m+k}}{%
C_{M}^{m+\alpha _{n-1}-1+k}}}},
\end{eqnarray}

for $\alpha _{n}=1, \alpha _{n-1}=n+1$, $V%
 _{\alpha _{n}\alpha _{n-1}}^{\left[ n\right]1}=\frac{1}{\sqrt{2}}$,
otherwise $V _{\alpha _{n}\alpha _{n-1}}^{\left[ n\right]1}=\delta _{\alpha
_{n}\alpha _{n-1}}\frac{1}{\sqrt{2}}$.

\item Case $M-N+m<n\le M-m$.

For $1\le \alpha _{n},\alpha _{n-1}\le \left(
M-N+m+1\right) $,
\begin{eqnarray}
V _{\alpha _{n}\alpha _{n-1}}^{\left[ n\right]0}=\delta
_{\alpha _{n}\alpha _{n-1}}\sqrt{\frac{\overset{M-m-n}{\underset{k=-m}{\sum }%
}\beta _{m\left( \alpha _{n-1}-1+k\right) }^{2}\frac{C_{M-n}^{m+k}}{%
C_{M}^{m+\alpha _{n-1}-1+k}}}{\overset{M-m-n+1}{\underset{k=-m}{\sum }}\beta
_{m\left( \alpha _{n-1}-1+k\right) }^{2}\frac{C_{M-n+1}^{m+k}}{%
C_{M}^{m+\alpha _{n-1}-1+k}}}}.
\end{eqnarray}

For $2\le \alpha _{n}\le \left( M-N+m+1\right) $, $1\le
\alpha _{n-1}\le \left( M-N+m\right) $,
\begin{eqnarray}
V _{\alpha
_{n}\alpha _{n-1}}^{\left[ n\right] 1}=\delta _{\alpha _{n}\left( \alpha _{n-1}+1\right) }%
\sqrt{\frac{\overset{M-m-n}{\underset{k=-m}{\sum }}\beta _{m\left( \alpha
_{n-1}+k\right) }^{2}\frac{C_{M-n}^{m+k}}{C_{M}^{m+\alpha _{n-1}+k}}}{%
\overset{M-m-n+1}{\underset{k=-m}{\sum }}\beta _{m\left( \alpha
_{n-1}-1+k\right) }^{2}\frac{C_{M-n+1}^{m+k}}{C_{M}^{m+\alpha _{n-1}-1+k}}}},
\end{eqnarray}
otherwise, $V _{\alpha _{n}\alpha _{n-1}}^{\left[ n\right]1}=0$.

\item Case $M-m<n\le M-1$.

(1). For $\left( m+n+1-M\right) \le \alpha _{n},\alpha
_{n-1}\le \left( M-N+m+1\right) $,
\begin{eqnarray}
V _{\alpha
_{n}\alpha _{n-1}}^{\left[ n\right]0}=\delta _{\alpha _{n}\alpha _{n-1}}\sqrt{\frac{\overset%
{M-m-n}{\underset{k=-m}{\sum }}\beta _{m\left( \alpha _{n-1}-1+k\right) }^{2}%
\frac{C_{M-n}^{m+k}}{C_{M}^{m+\alpha _{n-1}-1+k}}}{\overset{M-m-n+1}{%
\underset{k=-m}{\sum }}\beta _{m\left( \alpha _{n-1}-1+k\right) }^{2}\frac{%
C_{M-n+1}^{m+k}}{C_{M}^{m+\alpha _{n-1}-1+k}}}}.
\end{eqnarray}

\bigskip For $\alpha _{n}=m+n-M$, $1\le \alpha _{n-1}\le \left(
M-N+m+1\right) $, $V _{\alpha _{M}\alpha _{M-1}}^{\left[ n\right]0}=0$.
Otherwise, $V _{\alpha _{n}\alpha _{n-1}}^{\left[ n\right]0}=\delta _{\alpha
_{n}\alpha _{n-1}}\frac{1}{\sqrt{2}}$.

(2). For $\left( m+n+1-M\right) \le \alpha
_{n}\le \left( M-N+m+1\right) $, $\left( m+n-M\right) \le \alpha
_{n-1}\le \left( M-N+m\right) $,
\begin{eqnarray}
V_{\alpha _{n}\alpha
_{n-1}}^{\left[ n\right] 1}=\delta _{\alpha _{n}\left( \alpha _{n-1}+1\right) }\sqrt{\frac{%
\overset{M-m-n}{\underset{k=-m}{\sum }}\beta _{m\left( \alpha
_{n-1}+k\right) }^{2}\frac{C_{M-n}^{m+k}}{C_{M}^{m+\alpha _{n-1}+k}}}{%
\overset{M-m-n+1}{\underset{k=-m}{\sum }}\beta _{m\left( \alpha
_{n-1}-1+k\right) }^{2}\frac{C_{M-n+1}^{m+k}}{C_{M}^{m+\alpha _{n-1}-1+k}}}}.
\end{eqnarray}

For $\alpha _{n}=m+n-M$, $1\le \alpha _{n-1}\le \left(
M-N+m+1\right) $, $V _{\alpha _{M}\alpha _{M-1}}^{\left[ n\right]1}=0$.
Otherwise, $V _{\alpha _{n}\alpha _{n-1}}^{\left[ n\right]1}=\delta _{\alpha
_{n}\alpha _{n-1}}\frac{1}{\sqrt{2}}$.

\item Case $n=M$.

(1). For $\left( m+1\right) \le \alpha _{M},\alpha
_{M-1}\le \left( M-N+m+1\right) $,
\begin{eqnarray}
V _{\alpha
_{M}\alpha _{M-1}}^{\left[ M\right]0}=\delta _{\alpha _{M}\alpha _{M-1}}\sqrt{\frac{\beta
_{m\left( \alpha _{M-1}-1-m\right) }^{2}/C_{M}^{\alpha _{M-1}-1}}{\beta
_{m\left( \alpha _{M-1}-1-m\right) }^{2}/C_{M}^{\alpha _{M-1}-1}+\beta
_{m\left( \alpha _{M-1}-m\right) }^{2}/C_{M}^{\alpha _{M-1}}}}.
\end{eqnarray}
For $\alpha _{M}=m$, $1\le \alpha _{M-1}\le \left(
M-N+m+1\right) $, $V _{\alpha _{M}\alpha _{M-1}}^{\left[ M\right]0}=0$.
Otherwise, $V _{\alpha _{M}\alpha _{M-1}}^{\left[ M\right]0}=\delta _{\alpha
_{M}\alpha _{M-1}}\frac{1}{\sqrt{2}}$.

(2). For $\left( m+1\right) \le \alpha _{M}\le
\left( M-N+m+1\right) $, $m\le \alpha _{M-1}\le \left(
M-N+m\right) $,
\begin{eqnarray}
V _{\alpha _{M}\alpha _{M-1}}^{\left[ M\right]1}=\delta
_{\alpha _{M}\left( \alpha _{M-1}+1\right) }\sqrt{\frac{\beta _{m\left(
\alpha _{M-1}-m\right) }^{2}/C_{M}^{\alpha _{M-1}}}{\beta _{m\left( \alpha
_{M-1}-1-m\right) }^{2}/C_{M}^{\alpha _{M-1}-1}+\beta _{m\left( \alpha
_{M-1}-m\right) }^{2}/C_{M}^{\alpha _{M-1}}}}.
\end{eqnarray}

For $\alpha _{M}=m$, $1\le \alpha _{M-1}\le \left(
M-N+m+1\right) $, $V _{\alpha _{M}\alpha _{M-1}}^{\left[ M\right]0}=0$.
Otherwise, $V _{\alpha _{M}\alpha _{M-1}}^{\left[ M\right]0}=\delta _{\alpha
_{M}\alpha _{M-1}}\frac{1}{\sqrt{2}}$.

\item Case $n=M+l$.

(1). For $\left( m+1\right) \le \alpha
_{M+l}\le \left( M-N+m-l+1\right) $, $\left( m+2\right) \le
\alpha _{M+l-1}\le \left( M-N+m-l+2\right) $,
\begin{eqnarray}
V_{\alpha _{M+l}\alpha _{M+l-1}}^{\left[ M+l\right]0}=\delta _{\alpha _{M+l}\left( \alpha
_{M+l-1}-1\right) }\sqrt{\frac{\alpha _{M+l-1}-m-1}{M-N-l+1}}.
\end{eqnarray}

For $\alpha _{M+l}=\left( M-N+m-l+2\right) $, $1\le \alpha
_{M+l-1}\le \left( M-N+m+1\right) $, $V _{\alpha
_{M+l}\alpha _{M+l-1}}^{\left[ M+l\right] 0}=0$.
Otherwise $V _{\alpha _{M+l}\alpha _{M+l-1}}^{\left[ M+l\right]0}=\delta
_{\alpha _{M+l}\alpha _{M+l-1}}\frac{1}{\sqrt{2}}$.

(2). For $\left( m+1\right) \le \alpha _{M+l},\alpha
_{M+l-1}\le \left( M-N+m-l+1\right) $,
\begin{eqnarray}
V _{\alpha
_{M+l}\alpha _{M+l-1}}^{\left[ M+l\right]1}=\delta _{\alpha _{M+l}\alpha _{M+l-1}}\sqrt{\frac{%
M-N-l-\alpha _{M+l-1}+m+2}{M-N-l+1}}.
\end{eqnarray}
For $\alpha _{M+l}=\left( M-N+m-l+2\right) $, $1\le \alpha
_{M+l-1}\le \left( M-N+m+1\right) $, $V _{\alpha
_{M+l}\alpha _{M+l-1}}^{\left[ M+l\right]0}=0$.
Otherwise $V _{\alpha _{M+l}\alpha _{M+l-1}}^{\left[ M+l\right]0}=\delta
_{\alpha _{M+l}\alpha _{M+l-1}}\frac{1}{\sqrt{2}}$.

Here we have got all the operators $V _{(m)}^{\left[ n%
\right]i_{n}}$ for $0\le m\le N-m$. When $N-m<m\le N
$, one can find that $V _{(m)}^{\left[ n\right]i_{n}}=V _{(N-m)}^{\left[ n\right]%
\overline{i_{n}}}$, $\overline{i_{n}}=i_{n}+1\left( \text{mod}2\right) $.

\end{enumerate}

\subsection{$N\rightarrow M$ sequential UQCM of qudits}

\begin{enumerate}
\item Case $n=1$.
\begin{equation*}
\lambda_{\alpha_1}^{[1]}=\sqrt{\sum_{\bm{k}=-\bm{j}'}^{M-N-\bm{j}'}
\beta_{\bm{m}(\bm{j}'+\bm{k})}^2\frac{m_{i_1+1}+j_{i_1+1}^{'}+k_{i_1+1}}{
M}},
\end{equation*}
where $\beta_{\bm{m}\bm{j}}=\sqrt{\prod_{i=1}^d C_{m_i+j_i}^{m_i}/C_{M+d-1}^{M-N}}$.
\begin{equation*}
\Gamma_{\alpha_1}^{[1]i_1}=\delta_{\alpha_1\bm{j}'}.
\end{equation*}
\begin{equation*}
V_{\alpha_1}^{[1]i_1}=\Gamma_{\alpha_1}^{[1]i_1}\lambda_{\alpha_1}^{[1]}.
\end{equation*}

\item Case $1< n\le M-1$.

$$
\lambda_{\bm{j}'}^{[n]}=\sqrt{\sum_{\bm{k}=-\bm{j}'}^{M-N-\bm{j}'}
\beta_{\bm{m}(\bm{j}'+\bm{k})}^2\prod_{i=1}^d C_{j_i^{'}+k_i+m_i}^{j_i^{'}}/C_M^n}.$$

\begin{equation*}
\lambda_{\bm{j}"}^{[n-1]}=
\sqrt{\sum_{\bm_{k}'=-\bm{j}"}^{M-N-\bm{j}"}\beta_{\bm{m}(\bm{j}"
+\bm{k}')}^2\prod_{i=1}^d C_{j_i^{"}+k_i^{'}+m_i}^{j_i^{"}}/C_{M}^{n-1}}.
\end{equation*}

$$
\Gamma_{\bm{j}"\alpha_n}^{[n]i_n}=\delta_{\alpha_n(\bm{j}"+\hat{e}_{i_n+1})}
\frac{1}{\lambda_{\bm{j}"}^{[n-1]}}\sqrt{\frac{\bm{j}_{i_n+1}^{"}+1}{n}}.
$$
$$
V_{\alpha_n\bm{j}"}^{[n]i_n}=\delta_{\alpha_n(\bm{j}"+\hat{e}_{i_n+1})}
\sqrt{\frac{\bm{j}_{i_n+1}^{"}+1}{n}}\frac{\lambda_{\bm{j}"+\hat{e}_{i_n+1}}^{[n]}}{
\lambda_{\bm{j}"}^{[n-1]}}.$$

\item Case $n=M$.

$$
\lambda_{\bm{j}"}^{[M]}=\beta_{\bm{m}(\bm{j}'-\bm{m})}.$$

$$
\lambda_{\bm{j}"}^{[M-1]}=\sqrt{\sum_{i_M=0}^{d-1}\beta_{\bm{m}(\bm{j}"-\bm{m}
+\hat{e}_{i_M+1})}^2\frac{\bm{j}_{i_M+1}^{"}+1}{M}}.$$

$$
\Gamma_{\bm{j}"\alpha_M}^{[M]i_M}=\delta_{\alpha_M(\bm{j}"+\hat{e}_{i_M+1})}
\frac{1}{\lambda_{\bm{j}"}^{[M-1]}}\sqrt{\frac{\bm{j}_{i_M+1}^{"}}{M}}.$$

$$
V_{\alpha_M\bm{j}"}^{[M]i_M}=\delta_{\alpha_M(\bm{j}"+\hat{e}_{i_M+1})}
\frac{\lambda_{\alpha_M}^{[M]}}{\lambda_{\bm{j}"}^{[M-1]}}\sqrt{\frac{\bm{j}_{i_M+1}^{"}}{M}}.$$

\item Case $n=M+l$.

$$
\lambda_{\bm{j}'}^{[M+l]}=\sqrt{\sum_{\bm{k}=\bm{m}-\bm{j}'}^{M-N+\bm{m}-\bm{j}'}
\beta_{\bm{m}(\bm{j}'-\bm{m}+\bm{k})}^2\prod_{i=1}^d C_{j_i^{'}-m_i+k_i}^{k_i}/C_{M-N}^l}
.$$

$$
\lambda_{\bm{j}"}^{[M+l-1]}=\sqrt{\sum_{\bm{k}=\bm{m}-\bm{j}"}^{M-N+\bm{m}-\bm{j}"}
\beta_{\bm{m}(\bm{j}"-\bm{m}+\bm{k})}^2\prod_{i=1}^d C_{j_i^{"}-m_i+k_i^{'}}^{k_i^{'}}/C_{M-N}^{l-1}}
.$$

$$
\Gamma_{\bm{j}"\alpha_n}^{[M+1]i_n}=\delta_{\alpha_n(\bm{j}"-\hat{e}_{i_n+1})}
\frac{1}{\lambda_{\alpha_n}^{[M+l]}}\sqrt{\frac{j_{i_n+1}^{"}-m_{i_n+1}}{M-N-l+1}}
.$$

$$
V_{\alpha_n\bm{j}"}^{[n]i_n}=\delta_{\alpha_n(\bm{j}"-\hat{e}_{i_n+1})}
\sqrt{\frac{j_{i_n+1}^{"}-m_{i_n+1}}{M-N-l+1}}.
$$
\end{enumerate}

With the above information, one can build the explicit form of $V^{[n]i_n}$ according to the 2-dimensional case, and the extension is direct.

\newpage



\begin{thebibliography}{438}
\expandafter\ifx\csname natexlab\endcsname\relax\def\natexlab#1{#1}\fi
\expandafter\ifx\csname bibnamefont\endcsname\relax
  \def\bibnamefont#1{#1}\fi
\expandafter\ifx\csname bibfnamefont\endcsname\relax
  \def\bibfnamefont#1{#1}\fi
\expandafter\ifx\csname citenamefont\endcsname\relax
  \def\citenamefont#1{#1}\fi
\expandafter\ifx\csname url\endcsname\relax
  \def\url#1{\texttt{#1}}\fi
\expandafter\ifx\csname urlprefix\endcsname\relax\def\urlprefix{URL }\fi
\providecommand{\bibinfo}[2]{#2}
\providecommand{\eprint}[2][]{\url{#2}}

\bibitem[{\citenamefont{Ac\'{i}n}
  \emph{et~al.}(2004{\natexlab{a}})\citenamefont{Ac\'{i}n, Gisin, Lluis, and
  Scarani}}]{Acin2004}
\bibinfo{author}{\bibnamefont{Ac\'{i}n}, \bibfnamefont{A.}},
  \bibinfo{author}{\bibfnamefont{N.}~\bibnamefont{Gisin}},
  \bibinfo{author}{\bibfnamefont{M.}~\bibnamefont{Lluis}}, and
  \bibinfo{author}{\bibfnamefont{V.}~\bibnamefont{Scarani}},
  \bibinfo{year}{2004}{\natexlab{a}}, \bibinfo{journal}{Int. J. Quant. Inf.}
  \textbf{\bibinfo{volume}{2}}, \bibinfo{pages}{23}.

\bibitem[{\citenamefont{Ac\'{i}n} \emph{et~al.}(2003)\citenamefont{Ac\'{i}n,
  Gisin, and Scarani}}]{Acin2003}
\bibinfo{author}{\bibnamefont{Ac\'{i}n}, \bibfnamefont{A.}},
  \bibinfo{author}{\bibfnamefont{N.}~\bibnamefont{Gisin}}, and
  \bibinfo{author}{\bibfnamefont{V.}~\bibnamefont{Scarani}},
  \bibinfo{year}{2003}, \bibinfo{journal}{Quant. Inf. Comput.}
  \textbf{\bibinfo{volume}{3}}, \bibinfo{pages}{563}.

\bibitem[{\citenamefont{Ac\'{i}n}
  \emph{et~al.}(2004{\natexlab{b}})\citenamefont{Ac\'{i}n, Gisin, and
  Scarani}}]{Acin2004a}
\bibinfo{author}{\bibnamefont{Ac\'{i}n}, \bibfnamefont{A.}},
  \bibinfo{author}{\bibfnamefont{N.}~\bibnamefont{Gisin}}, and
  \bibinfo{author}{\bibfnamefont{V.}~\bibnamefont{Scarani}},
  \bibinfo{year}{2004}{\natexlab{b}}, \bibinfo{journal}{Phys. Rev. A}
  \textbf{\bibinfo{volume}{69}}, \bibinfo{pages}{012309}.

\bibitem[{\citenamefont{Adhikari and
  Choudhury}(2006)}]{PhysRevA.74.032323.2006}
\bibinfo{author}{\bibnamefont{Adhikari}, \bibfnamefont{S.}}, and
  \bibinfo{author}{\bibfnamefont{B.~S.} \bibnamefont{Choudhury}},
  \bibinfo{year}{2006}, \bibinfo{journal}{Phys. Rev. A}
  \textbf{\bibinfo{volume}{74}}, \bibinfo{pages}{032323}.

\bibitem[{\citenamefont{Adhikari} \emph{et~al.}(2008)\citenamefont{Adhikari,
  Majumdar, and Nayak}}]{pra77.042301}
\bibinfo{author}{\bibnamefont{Adhikari}, \bibfnamefont{S.}},
  \bibinfo{author}{\bibfnamefont{A.~S.} \bibnamefont{Majumdar}}, and
  \bibinfo{author}{\bibfnamefont{N.}~\bibnamefont{Nayak}},
  \bibinfo{year}{2008}, \bibinfo{journal}{Phys. Rev. A}
  \textbf{\bibinfo{volume}{77}}, \bibinfo{pages}{042301}.

\bibitem[{\citenamefont{Adhikari} \emph{et~al.}(2007)\citenamefont{Adhikari,
  Pati, Chakrabarty, and Choudhury}}]{Adhikari2007}
\bibinfo{author}{\bibnamefont{Adhikari}, \bibfnamefont{S.}},
  \bibinfo{author}{\bibfnamefont{A.~K.} \bibnamefont{Pati}},
  \bibinfo{author}{\bibfnamefont{I.}~\bibnamefont{Chakrabarty}}, and
  \bibinfo{author}{\bibfnamefont{B.~S.} \bibnamefont{Choudhury}},
  \bibinfo{year}{2007}, \bibinfo{journal}{Quant. Inf. Process.}
  \textbf{\bibinfo{volume}{6}}, \bibinfo{pages}{197}.

\bibitem[{\citenamefont{Albeverio and Fei}(2000)}]{ShaomingFei2000}
\bibinfo{author}{\bibnamefont{Albeverio}, \bibfnamefont{S.}}, and
  \bibinfo{author}{\bibfnamefont{S.~M.} \bibnamefont{Fei}},
  \bibinfo{year}{2000}, \bibinfo{journal}{Eur. Phys. J. B}
  \textbf{\bibinfo{volume}{14}}, \bibinfo{pages}{669}.

\bibitem[{\citenamefont{Alexanian}(2006)}]{pra73.045801}
\bibinfo{author}{\bibnamefont{Alexanian}, \bibfnamefont{M.}},
  \bibinfo{year}{2006}, \bibinfo{journal}{Phys. Rev. A}
  \textbf{\bibinfo{volume}{73}}, \bibinfo{pages}{045801}.

\bibitem[{\citenamefont{Andersen} \emph{et~al.}(2005)\citenamefont{Andersen,
  Josse, and Leuchs}}]{prl94.240503}
\bibinfo{author}{\bibnamefont{Andersen}, \bibfnamefont{U.~L.}},
  \bibinfo{author}{\bibfnamefont{V.}~\bibnamefont{Josse}}, and
  \bibinfo{author}{\bibfnamefont{G.}~\bibnamefont{Leuchs}},
  \bibinfo{year}{2005}, \bibinfo{journal}{Phys. Rev. Lett.}
  \textbf{\bibinfo{volume}{94}}, \bibinfo{pages}{240503}.

\bibitem[{\citenamefont{Anselmi} \emph{et~al.}(2004)\citenamefont{Anselmi,
  Chefles, and Plenio}}]{njp6.164}
\bibinfo{author}{\bibnamefont{Anselmi}, \bibfnamefont{F.}},
  \bibinfo{author}{\bibfnamefont{A.}~\bibnamefont{Chefles}}, and
  \bibinfo{author}{\bibfnamefont{M.~B.} \bibnamefont{Plenio}},
  \bibinfo{year}{2004}, \bibinfo{journal}{New J. Phys.}
  \textbf{\bibinfo{volume}{6}}, \bibinfo{pages}{164}.

\bibitem[{\citenamefont{Araneda} \emph{et~al.}(2012)\citenamefont{Araneda,
  Cisternas, Jimenez, and Delgado}}]{twin-photons}
\bibinfo{author}{\bibnamefont{Araneda}, \bibfnamefont{G.}},
  \bibinfo{author}{\bibfnamefont{N.}~\bibnamefont{Cisternas}},
  \bibinfo{author}{\bibfnamefont{O.}~\bibnamefont{Jimenez}}, and
  \bibinfo{author}{\bibfnamefont{A.}~\bibnamefont{Delgado}},
  \bibinfo{year}{2012}, \bibinfo{journal}{Phys. Rev. A}
  \textbf{\bibinfo{volume}{86}}, \bibinfo{pages}{052332}.

\bibitem[{\citenamefont{Audenaert and De~Moor}(2002)}]{pra65.030302r}
\bibinfo{author}{\bibnamefont{Audenaert}, \bibfnamefont{K.}}, and
  \bibinfo{author}{\bibfnamefont{B.}~\bibnamefont{De~Moor}},
  \bibinfo{year}{2002}, \bibinfo{journal}{Phys. Rev. A}
  \textbf{\bibinfo{volume}{65}}, \bibinfo{pages}{030302}.

\bibitem[{\citenamefont{Azuma} \emph{et~al.}(2005)\citenamefont{Azuma,
  Shimamura, Koashi, and Imoto}}]{PhysRevA.72.032335}
\bibinfo{author}{\bibnamefont{Azuma}, \bibfnamefont{K.}},
  \bibinfo{author}{\bibfnamefont{J.}~\bibnamefont{Shimamura}},
  \bibinfo{author}{\bibfnamefont{M.}~\bibnamefont{Koashi}}, and
  \bibinfo{author}{\bibfnamefont{N.}~\bibnamefont{Imoto}},
  \bibinfo{year}{2005}, \bibinfo{journal}{Phys. Rev. A}
  \textbf{\bibinfo{volume}{72}}, \bibinfo{pages}{032335}.

\bibitem[{\citenamefont{Bae}(2012{\natexlab{a}})}]{Baepreprint2}
\bibinfo{author}{\bibnamefont{Bae}, \bibfnamefont{J.}},
  \bibinfo{year}{2012}{\natexlab{a}}, \bibinfo{journal}{arXiv:1210.3125} .

\bibitem[{\citenamefont{Bae}(2012{\natexlab{b}})}]{Baepreprint1}
\bibinfo{author}{\bibnamefont{Bae}, \bibfnamefont{J.}},
  \bibinfo{year}{2012}{\natexlab{b}}, \bibinfo{journal}{arXiv:1210.2845} .

\bibitem[{\citenamefont{Bae and Ac\'{i}n}(2006)}]{prl97.030402}
\bibinfo{author}{\bibnamefont{Bae}, \bibfnamefont{J.}}, and
  \bibinfo{author}{\bibfnamefont{A.}~\bibnamefont{Ac\'{i}n}},
  \bibinfo{year}{2006}, \bibinfo{journal}{Phys. Rev. Lett.}
  \textbf{\bibinfo{volume}{97}}, \bibinfo{pages}{030402}.

\bibitem[{\citenamefont{Bae and Acin}(2007)}]{Bae2007}
\bibinfo{author}{\bibnamefont{Bae}, \bibfnamefont{J.}}, and
  \bibinfo{author}{\bibfnamefont{A.}~\bibnamefont{Acin}}, \bibinfo{year}{2007},
  \bibinfo{journal}{Phys. Rev. A} \textbf{\bibinfo{volume}{75}},
  \bibinfo{pages}{012334}.

\bibitem[{\citenamefont{Bae and Hwang}(2012)}]{Baepreprint0}
\bibinfo{author}{\bibnamefont{Bae}, \bibfnamefont{J.}}, and
  \bibinfo{author}{\bibfnamefont{W.~Y.} \bibnamefont{Hwang}},
  \bibinfo{year}{2012}, \bibinfo{journal}{arXiv:1204.2313} .

\bibitem[{\citenamefont{Bae} \emph{et~al.}(2011)\citenamefont{Bae, Hwang, and
  Han}}]{BaeHwang}
\bibinfo{author}{\bibnamefont{Bae}, \bibfnamefont{J.}},
  \bibinfo{author}{\bibfnamefont{W.~Y.} \bibnamefont{Hwang}}, and
  \bibinfo{author}{\bibfnamefont{Y.~D.} \bibnamefont{Han}},
  \bibinfo{year}{2011}, \bibinfo{journal}{Phys. Rev. Lett.}
  \textbf{\bibinfo{volume}{107}}, \bibinfo{pages}{170403}.

\bibitem[{\citenamefont{Baghbanzadeh and Rezakhani}(2009)}]{pla373.821}
\bibinfo{author}{\bibnamefont{Baghbanzadeh}, \bibfnamefont{S.}}, and
  \bibinfo{author}{\bibfnamefont{A.~T.} \bibnamefont{Rezakhani}},
  \bibinfo{year}{2009}, \bibinfo{journal}{Phys. Lett. A}
  \textbf{\bibinfo{volume}{373}}, \bibinfo{pages}{821}.

\bibitem[{\citenamefont{Banaszek}(2001)}]{prl86.1366}
\bibinfo{author}{\bibnamefont{Banaszek}, \bibfnamefont{K.}},
  \bibinfo{year}{2001}, \bibinfo{journal}{Phys. Rev. Lett.}
  \textbf{\bibinfo{volume}{86}}, \bibinfo{pages}{1366}.

\bibitem[{\citenamefont{Bandyopadhyay}
  \emph{et~al.}(2002)\citenamefont{Bandyopadhyay, Boykin, Roychowdhury, and
  Vatan}}]{roychowdhuryMUB}
\bibinfo{author}{\bibnamefont{Bandyopadhyay}, \bibfnamefont{S.}},
  \bibinfo{author}{\bibfnamefont{P.}~\bibnamefont{Boykin}},
  \bibinfo{author}{\bibfnamefont{V.}~\bibnamefont{Roychowdhury}}, and
  \bibinfo{author}{\bibfnamefont{F.}~\bibnamefont{Vatan}},
  \bibinfo{year}{2002}, \bibinfo{journal}{Algorithmica}
  \textbf{\bibinfo{volume}{34}}, \bibinfo{pages}{512}.

\bibitem[{\citenamefont{Barenco} \emph{et~al.}(1995)\citenamefont{Barenco,
  Bennett, Cleve, DiVincenzo, Margolus, Shor, Sleator, Smolin, and
  Weinfurter}}]{elementarygates}
\bibinfo{author}{\bibnamefont{Barenco}, \bibfnamefont{A.}},
  \bibinfo{author}{\bibfnamefont{C.~H.} \bibnamefont{Bennett}},
  \bibinfo{author}{\bibfnamefont{R.}~\bibnamefont{Cleve}},
  \bibinfo{author}{\bibfnamefont{D.~P.} \bibnamefont{DiVincenzo}},
  \bibinfo{author}{\bibfnamefont{N.}~\bibnamefont{Margolus}},
  \bibinfo{author}{\bibfnamefont{P.}~\bibnamefont{Shor}},
  \bibinfo{author}{\bibfnamefont{T.}~\bibnamefont{Sleator}},
  \bibinfo{author}{\bibfnamefont{J.~A.} \bibnamefont{Smolin}}, and
  \bibinfo{author}{\bibfnamefont{H.}~\bibnamefont{Weinfurter}},
  \bibinfo{year}{1995}, \bibinfo{journal}{Phys. Rev. A}
  \textbf{\bibinfo{volume}{52}}, \bibinfo{pages}{3457}.

\bibitem[{\citenamefont{Barnum} \emph{et~al.}(2007)\citenamefont{Barnum,
  Barrett, Leifer, and Wilce}}]{ISI:000251674300007}
\bibinfo{author}{\bibnamefont{Barnum}, \bibfnamefont{H.}},
  \bibinfo{author}{\bibfnamefont{J.}~\bibnamefont{Barrett}},
  \bibinfo{author}{\bibfnamefont{M.}~\bibnamefont{Leifer}}, and
  \bibinfo{author}{\bibfnamefont{A.}~\bibnamefont{Wilce}},
  \bibinfo{year}{2007}, \bibinfo{journal}{Phys. Rev. Lett.}
  \textbf{\bibinfo{volume}{99}}, \bibinfo{pages}{240501}.

\bibitem[{\citenamefont{Barnum} \emph{et~al.}(1996)\citenamefont{Barnum, Caves,
  Fuchs, Jozsa, and Schumacher}}]{PhysRevLett.76.2818}
\bibinfo{author}{\bibnamefont{Barnum}, \bibfnamefont{H.}},
  \bibinfo{author}{\bibfnamefont{C.~M.} \bibnamefont{Caves}},
  \bibinfo{author}{\bibfnamefont{C.~A.} \bibnamefont{Fuchs}},
  \bibinfo{author}{\bibfnamefont{R.}~\bibnamefont{Jozsa}}, and
  \bibinfo{author}{\bibfnamefont{B.}~\bibnamefont{Schumacher}},
  \bibinfo{year}{1996}, \bibinfo{journal}{Phys. Rev. Lett.}
  \textbf{\bibinfo{volume}{76}}, \bibinfo{pages}{2818}.

\bibitem[{\citenamefont{Bartkiewicz}
  \emph{et~al.}(2013)\citenamefont{Bartkiewicz, Lemr, \v{C}ernoch, Soubusta,
  and Miranowicz}}]{Bartkiewicz-experiment12}
\bibinfo{author}{\bibnamefont{Bartkiewicz}, \bibfnamefont{K.}},
  \bibinfo{author}{\bibfnamefont{K.}~\bibnamefont{Lemr}},
  \bibinfo{author}{\bibfnamefont{A.}~\bibnamefont{\v{C}ernoch}},
  \bibinfo{author}{\bibfnamefont{J.}~\bibnamefont{Soubusta}}, and
  \bibinfo{author}{\bibfnamefont{A.}~\bibnamefont{Miranowicz}},
  \bibinfo{year}{2013}, \bibinfo{journal}{Phys. Rev. Lett.}
  \textbf{\bibinfo{volume}{110}}, \bibinfo{pages}{173601}.

\bibitem[{\citenamefont{Bartkiewicz and Miranowicz}(2010)}]{axi-bloch}
\bibinfo{author}{\bibnamefont{Bartkiewicz}, \bibfnamefont{K.}}, and
  \bibinfo{author}{\bibfnamefont{A.}~\bibnamefont{Miranowicz}},
  \bibinfo{year}{2010}, \bibinfo{journal}{Phys. Rev. A}
  \textbf{\bibinfo{volume}{82}}, \bibinfo{pages}{042330}.

\bibitem[{\citenamefont{Bartkiewicz and Miranowicz}(2012)}]{Bartkiewicz2012}
\bibinfo{author}{\bibnamefont{Bartkiewicz}, \bibfnamefont{K.}}, and
  \bibinfo{author}{\bibfnamefont{A.}~\bibnamefont{Miranowicz}},
  \bibinfo{year}{2012}, \bibinfo{journal}{Phys. Scrip.}
  \textbf{\bibinfo{volume}{T147}}, \bibinfo{pages}{014003}.

\bibitem[{\citenamefont{Bartkiewicz}
  \emph{et~al.}(2009)\citenamefont{Bartkiewicz, Miranowicz, and
  Ozdemir}}]{mirror}
\bibinfo{author}{\bibnamefont{Bartkiewicz}, \bibfnamefont{K.}},
  \bibinfo{author}{\bibfnamefont{A.}~\bibnamefont{Miranowicz}}, and
  \bibinfo{author}{\bibfnamefont{S.~K.} \bibnamefont{Ozdemir}},
  \bibinfo{year}{2009}, \bibinfo{journal}{Phys. Rev. A}
  \textbf{\bibinfo{volume}{80}}, \bibinfo{pages}{032306}.

\bibitem[{\citenamefont{Bartuskova}
  \emph{et~al.}(2007)\citenamefont{Bartuskova, Dusek, Cernoch, Soubusta, and
  Fiurasek}}]{bartphase}
\bibinfo{author}{\bibnamefont{Bartuskova}, \bibfnamefont{L.}},
  \bibinfo{author}{\bibfnamefont{M.}~\bibnamefont{Dusek}},
  \bibinfo{author}{\bibfnamefont{A.}~\bibnamefont{Cernoch}},
  \bibinfo{author}{\bibfnamefont{J.}~\bibnamefont{Soubusta}}, and
  \bibinfo{author}{\bibfnamefont{J.}~\bibnamefont{Fiurasek}},
  \bibinfo{year}{2007}, \bibinfo{journal}{Phys. Rev. Lett.}
  \textbf{\bibinfo{volume}{99}}, \bibinfo{pages}{120505}.

\bibitem[{\citenamefont{Bechmann-Pasquinucci and
  Gisin}(1999)}]{Bechmann-Pasquinucci1999}
\bibinfo{author}{\bibnamefont{Bechmann-Pasquinucci}, \bibfnamefont{H.}}, and
  \bibinfo{author}{\bibfnamefont{N.}~\bibnamefont{Gisin}},
  \bibinfo{year}{1999}, \bibinfo{journal}{Phys. Rev. A}
  \textbf{\bibinfo{volume}{59}}, \bibinfo{pages}{4238}.

\bibitem[{\citenamefont{Bechmann-Pasquinucci and Tittel}(2000)}]{4dim2000}
\bibinfo{author}{\bibnamefont{Bechmann-Pasquinucci}, \bibfnamefont{H.}}, and
  \bibinfo{author}{\bibfnamefont{W.}~\bibnamefont{Tittel}},
  \bibinfo{year}{2000}, \bibinfo{journal}{Phys. Rev. A}
  \textbf{\bibinfo{volume}{61}}, \bibinfo{pages}{062308}.

\bibitem[{\citenamefont{Bell}(1964)}]{Bell1964}
\bibinfo{author}{\bibnamefont{Bell}, \bibfnamefont{J.~S.}},
  \bibinfo{year}{1964}, \bibinfo{journal}{Phys.} \textbf{\bibinfo{volume}{1}},
  \bibinfo{pages}{195}.

\bibitem[{\citenamefont{Bennett}(1992)}]{PhysRevLett.68.3121}
\bibinfo{author}{\bibnamefont{Bennett}, \bibfnamefont{C.~H.}},
  \bibinfo{year}{1992}, \bibinfo{journal}{Phys. Rev. Lett.}
  \textbf{\bibinfo{volume}{68}}, \bibinfo{pages}{3121}.

\bibitem[{\citenamefont{Bennett and Brassard}(1984)}]{Bennett1984}
\bibinfo{author}{\bibnamefont{Bennett}, \bibfnamefont{C.~H.}}, and
  \bibinfo{author}{\bibfnamefont{G.}~\bibnamefont{Brassard}},
  \bibinfo{year}{1984}, in \emph{\bibinfo{booktitle}{In Proceeding if IEEE Int.
  Conf. on Computers, Systems, and Signal Processings (Bangalore, India,
  1984)}} (\bibinfo{publisher}{IEEE}, \bibinfo{address}{New York}), pp.
  \bibinfo{pages}{175--179}.

\bibitem[{\citenamefont{Bennett} \emph{et~al.}(1993)\citenamefont{Bennett,
  Brassard, Cr\'{e}peau, Jozsa, Peres, and Wootters}}]{PhysRevLett.70.1895}
\bibinfo{author}{\bibnamefont{Bennett}, \bibfnamefont{C.~H.}},
  \bibinfo{author}{\bibfnamefont{G.}~\bibnamefont{Brassard}},
  \bibinfo{author}{\bibfnamefont{C.}~\bibnamefont{Cr\'{e}peau}},
  \bibinfo{author}{\bibfnamefont{R.}~\bibnamefont{Jozsa}},
  \bibinfo{author}{\bibfnamefont{A.}~\bibnamefont{Peres}}, and
  \bibinfo{author}{\bibfnamefont{W.~K.} \bibnamefont{Wootters}},
  \bibinfo{year}{1993}, \bibinfo{journal}{Phys. Rev. Lett.}
  \textbf{\bibinfo{volume}{70}}, \bibinfo{pages}{1895}.

\bibitem[{\citenamefont{Bennett} \emph{et~al.}(1992)\citenamefont{Bennett,
  Brassard, and Mermin}}]{no-imprinting}
\bibinfo{author}{\bibnamefont{Bennett}, \bibfnamefont{C.~H.}},
  \bibinfo{author}{\bibfnamefont{G.}~\bibnamefont{Brassard}}, and
  \bibinfo{author}{\bibfnamefont{N.~D.} \bibnamefont{Mermin}},
  \bibinfo{year}{1992}, \bibinfo{journal}{Phys. Rev. Lett.}
  \textbf{\bibinfo{volume}{68}}, \bibinfo{pages}{557}.

\bibitem[{\citenamefont{Bisio} \emph{et~al.}(2010)\citenamefont{Bisio,
  Chiribella, D'Ariano, Facchini, and Perinotti}}]{learning}
\bibinfo{author}{\bibnamefont{Bisio}, \bibfnamefont{A.}},
  \bibinfo{author}{\bibfnamefont{G.}~\bibnamefont{Chiribella}},
  \bibinfo{author}{\bibfnamefont{G.~M.} \bibnamefont{D'Ariano}},
  \bibinfo{author}{\bibfnamefont{S.}~\bibnamefont{Facchini}}, and
  \bibinfo{author}{\bibfnamefont{P.}~\bibnamefont{Perinotti}},
  \bibinfo{year}{2010}, \bibinfo{journal}{Phys. Rev. A}
  \textbf{\bibinfo{volume}{81}}, \bibinfo{pages}{032324}.

\bibitem[{\citenamefont{Bisio} \emph{et~al.}(2011)\citenamefont{Bisio,
  D'Ariano, Perinotti, and Sedlak}}]{clonemeasure}
\bibinfo{author}{\bibnamefont{Bisio}, \bibfnamefont{A.}},
  \bibinfo{author}{\bibfnamefont{G.~M.} \bibnamefont{D'Ariano}},
  \bibinfo{author}{\bibfnamefont{P.}~\bibnamefont{Perinotti}}, and
  \bibinfo{author}{\bibfnamefont{M.}~\bibnamefont{Sedlak}},
  \bibinfo{year}{2011}, \bibinfo{journal}{Phys. Rev. A}
  \textbf{\bibinfo{volume}{84}}, \bibinfo{pages}{042330}.

\bibitem[{\citenamefont{Bourennane}
  \emph{et~al.}(2001)\citenamefont{Bourennane, Karlsson, and
  Bj\"{o}rk}}]{BourennaneKarlsson}
\bibinfo{author}{\bibnamefont{Bourennane}, \bibfnamefont{M.}},
  \bibinfo{author}{\bibfnamefont{A.}~\bibnamefont{Karlsson}}, and
  \bibinfo{author}{\bibfnamefont{G.}~\bibnamefont{Bj\"{o}rk}},
  \bibinfo{year}{2001}, \bibinfo{journal}{Phys. Rev. A}
  \textbf{\bibinfo{volume}{64}}, \bibinfo{pages}{012306}.

\bibitem[{\citenamefont{Bradler}(2011)}]{Bradler2011}
\bibinfo{author}{\bibnamefont{Bradler}, \bibfnamefont{K.}},
  \bibinfo{year}{2011}, \bibinfo{journal}{IEEE Trans. Inf. Theory}
  \textbf{\bibinfo{volume}{57}}, \bibinfo{pages}{5497}.

\bibitem[{\citenamefont{Bradler and Jauregui}(2008)}]{relative-cloning}
\bibinfo{author}{\bibnamefont{Bradler}, \bibfnamefont{K.}}, and
  \bibinfo{author}{\bibfnamefont{R.}~\bibnamefont{Jauregui}},
  \bibinfo{year}{2008}, \bibinfo{journal}{Phys. Rev. A}
  \textbf{\bibinfo{volume}{77}}, \bibinfo{pages}{042302}.

\bibitem[{\citenamefont{Braunstein}
  \emph{et~al.}(2001{\natexlab{a}})\citenamefont{Braunstein, Cerf, Iblisdir,
  van Loock, and Massar}}]{PhysRevLett.86.4938}
\bibinfo{author}{\bibnamefont{Braunstein}, \bibfnamefont{S.~L.}},
  \bibinfo{author}{\bibfnamefont{N.~J.} \bibnamefont{Cerf}},
  \bibinfo{author}{\bibfnamefont{S.}~\bibnamefont{Iblisdir}},
  \bibinfo{author}{\bibfnamefont{P.}~\bibnamefont{van Loock}}, and
  \bibinfo{author}{\bibfnamefont{S.}~\bibnamefont{Massar}},
  \bibinfo{year}{2001}{\natexlab{a}}, \bibinfo{journal}{Phys. Rev. Lett.}
  \textbf{\bibinfo{volume}{86}}, \bibinfo{pages}{4938}.

\bibitem[{\citenamefont{Braunstein}
  \emph{et~al.}(2000)\citenamefont{Braunstein, Fuchs, and
  Kimble}}]{ISI:000085234900012}
\bibinfo{author}{\bibnamefont{Braunstein}, \bibfnamefont{S.~L.}},
  \bibinfo{author}{\bibfnamefont{C.~A.} \bibnamefont{Fuchs}}, and
  \bibinfo{author}{\bibfnamefont{H.~J.} \bibnamefont{Kimble}},
  \bibinfo{year}{2000}, \bibinfo{journal}{J. Mod. Optic.}
  \textbf{\bibinfo{volume}{47}}, \bibinfo{pages}{267}.

\bibitem[{\citenamefont{Braunstein and van Loock}(2005)}]{RevModPhys.77.513}
\bibinfo{author}{\bibnamefont{Braunstein}, \bibfnamefont{S.~L.}}, and
  \bibinfo{author}{\bibfnamefont{P.}~\bibnamefont{van Loock}},
  \bibinfo{year}{2005}, \bibinfo{journal}{Rev. Mod. Phys.}
  \textbf{\bibinfo{volume}{77}}, \bibinfo{pages}{513}.

\bibitem[{\citenamefont{Braunstein}
  \emph{et~al.}(2001{\natexlab{b}})\citenamefont{Braunstein,
  Bu\ifmmode~\check{z}\else \v{z}\fi{}ek, and Hillery}}]{Braunstein2001}
\bibinfo{author}{\bibnamefont{Braunstein}, \bibfnamefont{S.~L.}},
  \bibinfo{author}{\bibfnamefont{V.}~\bibnamefont{Bu\ifmmode~\check{z}\else
  \v{z}\fi{}ek}}, and
  \bibinfo{author}{\bibfnamefont{M.}~\bibnamefont{Hillery}},
  \bibinfo{year}{2001}{\natexlab{b}}, \bibinfo{journal}{Phys. Rev. A}
  \textbf{\bibinfo{volume}{63}}, \bibinfo{pages}{052313}.

\bibitem[{\citenamefont{Brougham} \emph{et~al.}(2006)\citenamefont{Brougham,
  Andersson, and Barnett}}]{measure-incompatible}
\bibinfo{author}{\bibnamefont{Brougham}, \bibfnamefont{T.}},
  \bibinfo{author}{\bibfnamefont{E.}~\bibnamefont{Andersson}}, and
  \bibinfo{author}{\bibfnamefont{S.~M.} \bibnamefont{Barnett}},
  \bibinfo{year}{2006}, \bibinfo{journal}{Phys. Rev. A}
  \textbf{\bibinfo{volume}{73}}, \bibinfo{pages}{062319}.

\bibitem[{\citenamefont{Bru{\ss}}(1998)}]{PhysRevLett.81.3018}
\bibinfo{author}{\bibnamefont{Bru{\ss}}, \bibfnamefont{D.}},
  \bibinfo{year}{1998}, \bibinfo{journal}{Phys. Rev. Lett.}
  \textbf{\bibinfo{volume}{81}}, \bibinfo{pages}{3018}.

\bibitem[{\citenamefont{Bru{\ss}} \emph{et~al.}(2001)\citenamefont{Bru{\ss},
  Calsamiglia, and L\"utkenhaus}}]{bruss-clone-measurement}
\bibinfo{author}{\bibnamefont{Bru{\ss}}, \bibfnamefont{D.}},
  \bibinfo{author}{\bibfnamefont{J.}~\bibnamefont{Calsamiglia}}, and
  \bibinfo{author}{\bibfnamefont{N.}~\bibnamefont{L\"utkenhaus}},
  \bibinfo{year}{2001}, \bibinfo{journal}{Phys. Rev. A}
  \textbf{\bibinfo{volume}{63}}, \bibinfo{pages}{042308}.

\bibitem[{\citenamefont{Bru{\ss}}
  \emph{et~al.}(2000{\natexlab{a}})\citenamefont{Bru{\ss}, Cinchetti,
  Mauro~D'Ariano, and Macchiavello}}]{Bruss2000}
\bibinfo{author}{\bibnamefont{Bru{\ss}}, \bibfnamefont{D.}},
  \bibinfo{author}{\bibfnamefont{M.}~\bibnamefont{Cinchetti}},
  \bibinfo{author}{\bibfnamefont{G.}~\bibnamefont{Mauro~D'Ariano}}, and
  \bibinfo{author}{\bibfnamefont{C.}~\bibnamefont{Macchiavello}},
  \bibinfo{year}{2000}{\natexlab{a}}, \bibinfo{journal}{Phys. Rev. A}
  \textbf{\bibinfo{volume}{62}}, \bibinfo{pages}{012302}.

\bibitem[{\citenamefont{Bru{\ss}}
  \emph{et~al.}(2000{\natexlab{b}})\citenamefont{Bru{\ss}, D'Ariano,
  Macchiavello, and Sacchi}}]{PhysRevA.62.062302}
\bibinfo{author}{\bibnamefont{Bru{\ss}}, \bibfnamefont{D.}},
  \bibinfo{author}{\bibfnamefont{G.~M.} \bibnamefont{D'Ariano}},
  \bibinfo{author}{\bibfnamefont{C.}~\bibnamefont{Macchiavello}}, and
  \bibinfo{author}{\bibfnamefont{M.~F.} \bibnamefont{Sacchi}},
  \bibinfo{year}{2000}{\natexlab{b}}, \bibinfo{journal}{Phys. Rev. A}
  \textbf{\bibinfo{volume}{62}}, \bibinfo{pages}{062302}.

\bibitem[{\citenamefont{Bru{\ss}}
  \emph{et~al.}(1998{\natexlab{a}})\citenamefont{Bru{\ss}, DiVincenzo, Ekert,
  Fuchs, Macchiavello, and Smolin}}]{Bruss1998}
\bibinfo{author}{\bibnamefont{Bru{\ss}}, \bibfnamefont{D.}},
  \bibinfo{author}{\bibfnamefont{D.~P.} \bibnamefont{DiVincenzo}},
  \bibinfo{author}{\bibfnamefont{A.}~\bibnamefont{Ekert}},
  \bibinfo{author}{\bibfnamefont{C.~A.} \bibnamefont{Fuchs}},
  \bibinfo{author}{\bibfnamefont{C.}~\bibnamefont{Macchiavello}}, and
  \bibinfo{author}{\bibfnamefont{J.~A.} \bibnamefont{Smolin}},
  \bibinfo{year}{1998}{\natexlab{a}}, \bibinfo{journal}{Phys. Rev. A}
  \textbf{\bibinfo{volume}{57}}, \bibinfo{pages}{2368}.

\bibitem[{\citenamefont{Bru{\ss}}
  \emph{et~al.}(1998{\natexlab{b}})\citenamefont{Bru{\ss}, Ekert, and
  Macchiavello}}]{PhysRevLett.81.2598}
\bibinfo{author}{\bibnamefont{Bru{\ss}}, \bibfnamefont{D.}},
  \bibinfo{author}{\bibfnamefont{A.}~\bibnamefont{Ekert}}, and
  \bibinfo{author}{\bibfnamefont{C.}~\bibnamefont{Macchiavello}},
  \bibinfo{year}{1998}{\natexlab{b}}, \bibinfo{journal}{Phys. Rev. Lett.}
  \textbf{\bibinfo{volume}{81}}, \bibinfo{pages}{2598}.

\bibitem[{\citenamefont{Bru{\ss} and Macchiavello}(1999)}]{Bruss1999}
\bibinfo{author}{\bibnamefont{Bru{\ss}}, \bibfnamefont{D.}}, and
  \bibinfo{author}{\bibfnamefont{C.}~\bibnamefont{Macchiavello}},
  \bibinfo{year}{1999}, \bibinfo{journal}{Phys. Lett. A}
  \textbf{\bibinfo{volume}{253}}, \bibinfo{pages}{249 }.

\bibitem[{\citenamefont{Bru{\ss} and Macchiavello}(2002)}]{prl88.127901}
\bibinfo{author}{\bibnamefont{Bru{\ss}}, \bibfnamefont{D.}}, and
  \bibinfo{author}{\bibfnamefont{C.}~\bibnamefont{Macchiavello}},
  \bibinfo{year}{2002}, \bibinfo{journal}{Phys. Rev. Lett.}
  \textbf{\bibinfo{volume}{88}}, \bibinfo{pages}{127901}.

\bibitem[{\citenamefont{Bru{\ss} and Macchiavello}(2003)}]{Bruss2003}
\bibinfo{author}{\bibnamefont{Bru{\ss}}, \bibfnamefont{D.}}, and
  \bibinfo{author}{\bibfnamefont{C.}~\bibnamefont{Macchiavello}},
  \bibinfo{year}{2003}, \bibinfo{journal}{Found. Phys.}
  \textbf{\bibinfo{volume}{33}}, \bibinfo{pages}{1617}.

\bibitem[{\citenamefont{Bub}(2001)}]{Bub}
\bibinfo{author}{\bibnamefont{Bub}, \bibfnamefont{J.}}, \bibinfo{year}{2001},
  \bibinfo{journal}{Phys. Rev. A} \textbf{\bibinfo{volume}{63}},
  \bibinfo{pages}{032309}.

\bibitem[{\citenamefont{Buscemi} \emph{et~al.}(2005)\citenamefont{Buscemi,
  D'Ariano, and Macchiavello}}]{PhysRevA.71.042327}
\bibinfo{author}{\bibnamefont{Buscemi}, \bibfnamefont{F.}},
  \bibinfo{author}{\bibfnamefont{G.~M.} \bibnamefont{D'Ariano}}, and
  \bibinfo{author}{\bibfnamefont{C.}~\bibnamefont{Macchiavello}},
  \bibinfo{year}{2005}, \bibinfo{journal}{Phys. Rev. A}
  \textbf{\bibinfo{volume}{71}}, \bibinfo{pages}{042327}.

\bibitem[{\citenamefont{Buscemi} \emph{et~al.}(2007)\citenamefont{Buscemi,
  D'Ariano, and Macchiavello}}]{Buscemi2007}
\bibinfo{author}{\bibnamefont{Buscemi}, \bibfnamefont{F.}},
  \bibinfo{author}{\bibfnamefont{G.~M.} \bibnamefont{D'Ariano}}, and
  \bibinfo{author}{\bibfnamefont{C.}~\bibnamefont{Macchiavello}},
  \bibinfo{year}{2007}, \bibinfo{journal}{J. Opt. Soc. Am. B}
  \textbf{\bibinfo{volume}{24}}, \bibinfo{pages}{363}.

\bibitem[{\citenamefont{Buscemi} \emph{et~al.}(2006)\citenamefont{Buscemi,
  D'Ariano, Macchiavello, and Perinotti}}]{ISI:000241723100032}
\bibinfo{author}{\bibnamefont{Buscemi}, \bibfnamefont{F.}},
  \bibinfo{author}{\bibfnamefont{G.~M.} \bibnamefont{D'Ariano}},
  \bibinfo{author}{\bibfnamefont{C.}~\bibnamefont{Macchiavello}}, and
  \bibinfo{author}{\bibfnamefont{P.}~\bibnamefont{Perinotti}},
  \bibinfo{year}{2006}, \bibinfo{journal}{Phys. Rev. A}
  \textbf{\bibinfo{volume}{74}}, \bibinfo{pages}{042309}.

\bibitem[{\citenamefont{Buscemi} \emph{et~al.}(2003)\citenamefont{Buscemi,
  D'Ariano, Perinotti, and Sacchi}}]{Buscemi2003374}
\bibinfo{author}{\bibnamefont{Buscemi}, \bibfnamefont{F.}},
  \bibinfo{author}{\bibfnamefont{G.~M.} \bibnamefont{D'Ariano}},
  \bibinfo{author}{\bibfnamefont{P.}~\bibnamefont{Perinotti}}, and
  \bibinfo{author}{\bibfnamefont{M.~F.} \bibnamefont{Sacchi}},
  \bibinfo{year}{2003}, \bibinfo{journal}{Phys. Lett. A}
  \textbf{\bibinfo{volume}{314}}, \bibinfo{pages}{374 }.

\bibitem[{\citenamefont{Bu\v{z}ek}
  \emph{et~al.}(1997{\natexlab{a}})\citenamefont{Bu\v{z}ek, Braunstein,
  Hillery, and Bru\ss{}}}]{BBHBnetwork}
\bibinfo{author}{\bibnamefont{Bu\v{z}ek}, \bibfnamefont{V.}},
  \bibinfo{author}{\bibfnamefont{S.~L.} \bibnamefont{Braunstein}},
  \bibinfo{author}{\bibfnamefont{M.}~\bibnamefont{Hillery}}, and
  \bibinfo{author}{\bibfnamefont{D.}~\bibnamefont{Bru\ss{}}},
  \bibinfo{year}{1997}{\natexlab{a}}, \bibinfo{journal}{Phys. Rev. A}
  \textbf{\bibinfo{volume}{56}}, \bibinfo{pages}{3446}.

\bibitem[{\citenamefont{Bu\v{z}ek and Hillery}(1996)}]{PhysRevA.54.1844}
\bibinfo{author}{\bibnamefont{Bu\v{z}ek}, \bibfnamefont{V.}}, and
  \bibinfo{author}{\bibfnamefont{M.}~\bibnamefont{Hillery}},
  \bibinfo{year}{1996}, \bibinfo{journal}{Phys. Rev. A}
  \textbf{\bibinfo{volume}{54}}, \bibinfo{pages}{1844}.

\bibitem[{\citenamefont{Bu\v{z}ek and Hillery}(1998)}]{PhysRevLett.81.5003}
\bibinfo{author}{\bibnamefont{Bu\v{z}ek}, \bibfnamefont{V.}}, and
  \bibinfo{author}{\bibfnamefont{M.}~\bibnamefont{Hillery}},
  \bibinfo{year}{1998}, \bibinfo{journal}{Phys. Rev. Lett.}
  \textbf{\bibinfo{volume}{81}}, \bibinfo{pages}{5003}.

\bibitem[{\citenamefont{Bu\v{z}ek and Hillery}(1999)}]{ISI:000084208300020}
\bibinfo{author}{\bibnamefont{Bu\v{z}ek}, \bibfnamefont{V.}}, and
  \bibinfo{author}{\bibfnamefont{M.}~\bibnamefont{Hillery}},
  \bibinfo{year}{1999}, in \emph{\bibinfo{booktitle}{QUANTUM COMPUTING AND
  QUANTUM COMMUNICATIONS}}, edited by
  \bibinfo{editor}{\bibfnamefont{C.}~\bibnamefont{Williams}}, volume
  \bibinfo{volume}{1509} of \emph{\bibinfo{series}{LECTURE NOTES IN COMPUTER
  SCIENCE}}, pp. \bibinfo{pages}{235--246}.

\bibitem[{\citenamefont{Bu\v{z}ek and Hillery}(2000)}]{Buzek2000}
\bibinfo{author}{\bibnamefont{Bu\v{z}ek}, \bibfnamefont{V.}}, and
  \bibinfo{author}{\bibfnamefont{M.}~\bibnamefont{Hillery}},
  \bibinfo{year}{2000}, \bibinfo{journal}{J. Mod. Optic.}
  \textbf{\bibinfo{volume}{47}}, \bibinfo{pages}{211}.

\bibitem[{\citenamefont{Bu\v{z}ek} \emph{et~al.}(1998)\citenamefont{Bu\v{z}ek,
  Hillery, and Bednik}}]{ISI:000074966200008}
\bibinfo{author}{\bibnamefont{Bu\v{z}ek}, \bibfnamefont{V.}},
  \bibinfo{author}{\bibfnamefont{M.}~\bibnamefont{Hillery}}, and
  \bibinfo{author}{\bibfnamefont{R.}~\bibnamefont{Bednik}},
  \bibinfo{year}{1998}, \bibinfo{journal}{Acta Phys. Slovaca}
  \textbf{\bibinfo{volume}{48}}, \bibinfo{pages}{177}.

\bibitem[{\citenamefont{Bu\v{z}ek} \emph{et~al.}(1999)\citenamefont{Bu\v{z}ek,
  Hillery, and Werner}}]{PhysRevA.60.R2626}
\bibinfo{author}{\bibnamefont{Bu\v{z}ek}, \bibfnamefont{V.}},
  \bibinfo{author}{\bibfnamefont{M.}~\bibnamefont{Hillery}}, and
  \bibinfo{author}{\bibfnamefont{R.~F.} \bibnamefont{Werner}},
  \bibinfo{year}{1999}, \bibinfo{journal}{Phys. Rev. A}
  \textbf{\bibinfo{volume}{60}}, \bibinfo{pages}{2626(R)}.

\bibitem[{\citenamefont{Bu\v{z}ek}
  \emph{et~al.}(1997{\natexlab{b}})\citenamefont{Bu\v{z}ek, Vedral, Plenio,
  Knight, and Hillery}}]{buzek-lcoal-cloning}
\bibinfo{author}{\bibnamefont{Bu\v{z}ek}, \bibfnamefont{V.}},
  \bibinfo{author}{\bibfnamefont{V.}~\bibnamefont{Vedral}},
  \bibinfo{author}{\bibfnamefont{M.~B.} \bibnamefont{Plenio}},
  \bibinfo{author}{\bibfnamefont{P.~L.} \bibnamefont{Knight}}, and
  \bibinfo{author}{\bibfnamefont{M.}~\bibnamefont{Hillery}},
  \bibinfo{year}{1997}{\natexlab{b}}, \bibinfo{journal}{Phys. Rev. A}
  \textbf{\bibinfo{volume}{55}}, \bibinfo{pages}{3327}.

\bibitem[{\citenamefont{Caminati} \emph{et~al.}(2006)\citenamefont{Caminati,
  De~Martini, Perris, Sciarrino, and Secondi}}]{Caminati2006}
\bibinfo{author}{\bibnamefont{Caminati}, \bibfnamefont{M.}},
  \bibinfo{author}{\bibfnamefont{F.}~\bibnamefont{De~Martini}},
  \bibinfo{author}{\bibfnamefont{R.}~\bibnamefont{Perris}},
  \bibinfo{author}{\bibfnamefont{F.}~\bibnamefont{Sciarrino}}, and
  \bibinfo{author}{\bibfnamefont{V.}~\bibnamefont{Secondi}},
  \bibinfo{year}{2006}, \bibinfo{journal}{Phys. Rev. A}
  \textbf{\bibinfo{volume}{74}}, \bibinfo{pages}{062304}.

\bibitem[{\citenamefont{Cao and Song}(2004)}]{ISI:000221546500001}
\bibinfo{author}{\bibnamefont{Cao}, \bibfnamefont{Z.~L.}}, and
  \bibinfo{author}{\bibfnamefont{W.}~\bibnamefont{Song}}, \bibinfo{year}{2004},
  \bibinfo{journal}{Phys. Lett. A} \textbf{\bibinfo{volume}{325}},
  \bibinfo{pages}{309}.

\bibitem[{\citenamefont{Carlini and Sasaki}(2003)}]{ISI:000187004700046}
\bibinfo{author}{\bibnamefont{Carlini}, \bibfnamefont{A.}}, and
  \bibinfo{author}{\bibfnamefont{M.}~\bibnamefont{Sasaki}},
  \bibinfo{year}{2003}, \bibinfo{journal}{Phys. Rev. A}
  \textbf{\bibinfo{volume}{68}}, \bibinfo{pages}{042327}.

\bibitem[{\citenamefont{Caves}(1982)}]{PhysRevD.26.1817}
\bibinfo{author}{\bibnamefont{Caves}, \bibfnamefont{C.~M.}},
  \bibinfo{year}{1982}, \bibinfo{journal}{Phys. Rev. D}
  \textbf{\bibinfo{volume}{26}}, \bibinfo{pages}{1817}.

\bibitem[{\citenamefont{Cerf}(1998)}]{Cerf1998}
\bibinfo{author}{\bibnamefont{Cerf}, \bibfnamefont{N.~J.}},
  \bibinfo{year}{1998}, \bibinfo{journal}{Acta Phys. Slovaca}
  \textbf{\bibinfo{volume}{48}}, \bibinfo{pages}{115}.

\bibitem[{\citenamefont{Cerf}(2000{\natexlab{a}})}]{383952220000215}
\bibinfo{author}{\bibnamefont{Cerf}, \bibfnamefont{N.~J.}},
  \bibinfo{year}{2000}{\natexlab{a}}, \bibinfo{journal}{J. Mod. Optic.}
  \textbf{\bibinfo{volume}{47}}, \bibinfo{pages}{187 }.

\bibitem[{\citenamefont{Cerf}(2000{\natexlab{b}})}]{PhysRevLett.84.4497}
\bibinfo{author}{\bibnamefont{Cerf}, \bibfnamefont{N.~J.}},
  \bibinfo{year}{2000}{\natexlab{b}}, \bibinfo{journal}{Phys. Rev. Lett.}
  \textbf{\bibinfo{volume}{84}}, \bibinfo{pages}{4497}.

\bibitem[{\citenamefont{Cerf}
  \emph{et~al.}(2002{\natexlab{a}})\citenamefont{Cerf, Bourennane, Karlsson,
  and Gisin}}]{PhysRevLett.88.127902}
\bibinfo{author}{\bibnamefont{Cerf}, \bibfnamefont{N.~J.}},
  \bibinfo{author}{\bibfnamefont{M.}~\bibnamefont{Bourennane}},
  \bibinfo{author}{\bibfnamefont{A.}~\bibnamefont{Karlsson}}, and
  \bibinfo{author}{\bibfnamefont{N.}~\bibnamefont{Gisin}},
  \bibinfo{year}{2002}{\natexlab{a}}, \bibinfo{journal}{Phys. Rev. Lett.}
  \textbf{\bibinfo{volume}{88}}, \bibinfo{pages}{127902}.

\bibitem[{\citenamefont{Cerf}
  \emph{et~al.}(2002{\natexlab{b}})\citenamefont{Cerf, Durt, and
  Gisin}}]{688607420020710}
\bibinfo{author}{\bibnamefont{Cerf}, \bibfnamefont{N.~J.}},
  \bibinfo{author}{\bibfnamefont{T.}~\bibnamefont{Durt}}, and
  \bibinfo{author}{\bibfnamefont{N.}~\bibnamefont{Gisin}},
  \bibinfo{year}{2002}{\natexlab{b}}, \bibinfo{journal}{J. Mod. Optic.}
  \textbf{\bibinfo{volume}{49}}, \bibinfo{pages}{1355 }.

\bibitem[{\citenamefont{Cerf and Fiur\'{a}\v{s}ek}(2006)}]{Cerf2006455}
\bibinfo{author}{\bibnamefont{Cerf}, \bibfnamefont{N.~J.}}, and
  \bibinfo{author}{\bibfnamefont{J.}~\bibnamefont{Fiur\'{a}\v{s}ek}},
  \bibinfo{year}{2006} (\bibinfo{publisher}{Elsevier}),
  volume~\bibinfo{volume}{49} of \emph{\bibinfo{series}{Progress in Optics}},
  pp. \bibinfo{pages}{455 -- 545}.

\bibitem[{\citenamefont{Cerf and Grangier}(2007)}]{CerfGrangier}
\bibinfo{author}{\bibnamefont{Cerf}, \bibfnamefont{N.~J.}}, and
  \bibinfo{author}{\bibfnamefont{P.}~\bibnamefont{Grangier}},
  \bibinfo{year}{2007}, \bibinfo{journal}{J. Opt. Soc. Am. B}
  \textbf{\bibinfo{volume}{24}}, \bibinfo{pages}{324}.

\bibitem[{\citenamefont{Cerf and Iblisdir}(2000)}]{PhysRevA.62.040301}
\bibinfo{author}{\bibnamefont{Cerf}, \bibfnamefont{N.~J.}}, and
  \bibinfo{author}{\bibfnamefont{S.}~\bibnamefont{Iblisdir}},
  \bibinfo{year}{2000}, \bibinfo{journal}{Phys. Rev. A}
  \textbf{\bibinfo{volume}{62}}, \bibinfo{pages}{040301(R)}.

\bibitem[{\citenamefont{Cerf and Iblisdir}(2001{\natexlab{a}})}]{cerf-cv-not}
\bibinfo{author}{\bibnamefont{Cerf}, \bibfnamefont{N.~J.}}, and
  \bibinfo{author}{\bibfnamefont{S.}~\bibnamefont{Iblisdir}},
  \bibinfo{year}{2001}{\natexlab{a}}, \bibinfo{journal}{Phys. Rev. A}
  \textbf{\bibinfo{volume}{64}}, \bibinfo{pages}{032307}.

\bibitem[{\citenamefont{Cerf and Iblisdir}(2001{\natexlab{b}})}]{prl87.247903}
\bibinfo{author}{\bibnamefont{Cerf}, \bibfnamefont{N.~J.}}, and
  \bibinfo{author}{\bibfnamefont{S.}~\bibnamefont{Iblisdir}},
  \bibinfo{year}{2001}{\natexlab{b}}, \bibinfo{journal}{Phys. Rev. Lett.}
  \textbf{\bibinfo{volume}{87}}, \bibinfo{pages}{247903}.

\bibitem[{\citenamefont{Cerf}
  \emph{et~al.}(2002{\natexlab{c}})\citenamefont{Cerf, Iblisdir, and
  Van~Assche}}]{CerfIblisdirAssche}
\bibinfo{author}{\bibnamefont{Cerf}, \bibfnamefont{N.~J.}},
  \bibinfo{author}{\bibfnamefont{S.}~\bibnamefont{Iblisdir}}, and
  \bibinfo{author}{\bibfnamefont{G.}~\bibnamefont{Van~Assche}},
  \bibinfo{year}{2002}{\natexlab{c}}, \bibinfo{journal}{Eur. Phys. J. D}
  \textbf{\bibinfo{volume}{18}}, \bibinfo{pages}{211}.

\bibitem[{\citenamefont{Cerf} \emph{et~al.}(2000)\citenamefont{Cerf, Ipe, and
  Rottenberg}}]{PhysRevLett.85.1754}
\bibinfo{author}{\bibnamefont{Cerf}, \bibfnamefont{N.~J.}},
  \bibinfo{author}{\bibfnamefont{A.}~\bibnamefont{Ipe}}, and
  \bibinfo{author}{\bibfnamefont{X.}~\bibnamefont{Rottenberg}},
  \bibinfo{year}{2000}, \bibinfo{journal}{Phys. Rev. Lett.}
  \textbf{\bibinfo{volume}{85}}, \bibinfo{pages}{1754}.

\bibitem[{\citenamefont{Cerf} \emph{et~al.}(2005)\citenamefont{Cerf, Kr\"uger,
  Navez, Werner, and Wolf}}]{prl95.070501}
\bibinfo{author}{\bibnamefont{Cerf}, \bibfnamefont{N.~J.}},
  \bibinfo{author}{\bibfnamefont{O.}~\bibnamefont{Kr\"uger}},
  \bibinfo{author}{\bibfnamefont{P.}~\bibnamefont{Navez}},
  \bibinfo{author}{\bibfnamefont{R.~F.} \bibnamefont{Werner}}, and
  \bibinfo{author}{\bibfnamefont{M.~M.} \bibnamefont{Wolf}},
  \bibinfo{year}{2005}, \bibinfo{journal}{Phys. Rev. Lett.}
  \textbf{\bibinfo{volume}{95}}, \bibinfo{pages}{070501}.

\bibitem[{\citenamefont{Cerf} \emph{et~al.}(2001)\citenamefont{Cerf, L\'{e}vy,
  and Assche}}]{cerf-cvqkd-squeezed}
\bibinfo{author}{\bibnamefont{Cerf}, \bibfnamefont{N.~J.}},
  \bibinfo{author}{\bibfnamefont{M.}~\bibnamefont{L\'{e}vy}}, and
  \bibinfo{author}{\bibfnamefont{G.~V.} \bibnamefont{Assche}},
  \bibinfo{year}{2001}, \bibinfo{journal}{Phys. Rev. A}
  \textbf{\bibinfo{volume}{63}}, \bibinfo{pages}{052311}.

\bibitem[{\citenamefont{Cernoch} \emph{et~al.}(2006)\citenamefont{Cernoch,
  Bartuskova, Soubusta, Jezek, Fiurasek, and Dusek}}]{Cernoch2006}
\bibinfo{author}{\bibnamefont{Cernoch}, \bibfnamefont{A.}},
  \bibinfo{author}{\bibfnamefont{L.}~\bibnamefont{Bartuskova}},
  \bibinfo{author}{\bibfnamefont{J.}~\bibnamefont{Soubusta}},
  \bibinfo{author}{\bibfnamefont{M.}~\bibnamefont{Jezek}},
  \bibinfo{author}{\bibfnamefont{J.}~\bibnamefont{Fiurasek}}, and
  \bibinfo{author}{\bibfnamefont{M.}~\bibnamefont{Dusek}},
  \bibinfo{year}{2006}, \bibinfo{journal}{Phys. Rev. A}
  \textbf{\bibinfo{volume}{74}}, \bibinfo{pages}{042327}.

\bibitem[{\citenamefont{Cernoch} \emph{et~al.}(2009)\citenamefont{Cernoch,
  Soubusta, Celechovska, Dusek, and Fiurasek}}]{Cernoch2009a}
\bibinfo{author}{\bibnamefont{Cernoch}, \bibfnamefont{A.}},
  \bibinfo{author}{\bibfnamefont{J.}~\bibnamefont{Soubusta}},
  \bibinfo{author}{\bibfnamefont{L.}~\bibnamefont{Celechovska}},
  \bibinfo{author}{\bibfnamefont{M.}~\bibnamefont{Dusek}}, and
  \bibinfo{author}{\bibfnamefont{J.}~\bibnamefont{Fiurasek}},
  \bibinfo{year}{2009}, \bibinfo{journal}{Phys. Rev. A}
  \textbf{\bibinfo{volume}{80}}, \bibinfo{pages}{062306}.

\bibitem[{\citenamefont{Chang} \emph{et~al.}(2013)\citenamefont{Chang, Liu,
  Liu, Fan, and Pan}}]{Pan2013}
\bibinfo{author}{\bibnamefont{Chang}, \bibfnamefont{Y.~C.}},
  \bibinfo{author}{\bibfnamefont{G.~Q.} \bibnamefont{Liu}},
  \bibinfo{author}{\bibfnamefont{D.~Q.} \bibnamefont{Liu}},
  \bibinfo{author}{\bibfnamefont{H.}~\bibnamefont{Fan}}, and
  \bibinfo{author}{\bibfnamefont{X.~Y.} \bibnamefont{Pan}},
  \bibinfo{year}{2013}, \bibinfo{journal}{Sci. Rep.}
  \textbf{\bibinfo{volume}{3}}, \bibinfo{pages}{1498}.

\bibitem[{\citenamefont{Chefles and Barnett}(1998)}]{0305-4470-31-50-007}
\bibinfo{author}{\bibnamefont{Chefles}, \bibfnamefont{A.}}, and
  \bibinfo{author}{\bibfnamefont{S.~M.} \bibnamefont{Barnett}},
  \bibinfo{year}{1998}, \bibinfo{journal}{J. Phys. A-Math. Gen.}
  \textbf{\bibinfo{volume}{31}}, \bibinfo{pages}{10097}.

\bibitem[{\citenamefont{Chefles and Barnett}(1999)}]{chefles}
\bibinfo{author}{\bibnamefont{Chefles}, \bibfnamefont{A.}}, and
  \bibinfo{author}{\bibfnamefont{S.~M.} \bibnamefont{Barnett}},
  \bibinfo{year}{1999}, \bibinfo{journal}{Phys. Rev. A}
  \textbf{\bibinfo{volume}{60}}, \bibinfo{pages}{136}.

\bibitem[{\citenamefont{Chefles} \emph{et~al.}(2001)\citenamefont{Chefles,
  Gilson, and Barnett}}]{CheflesGilson}
\bibinfo{author}{\bibnamefont{Chefles}, \bibfnamefont{A.}},
  \bibinfo{author}{\bibfnamefont{C.~R.} \bibnamefont{Gilson}}, and
  \bibinfo{author}{\bibfnamefont{S.~M.} \bibnamefont{Barnett}},
  \bibinfo{year}{2001}, \bibinfo{journal}{Phys. Rev. A}
  \textbf{\bibinfo{volume}{63}}, \bibinfo{pages}{032314}.

\bibitem[{\citenamefont{Chen} \emph{et~al.}(2011)\citenamefont{Chen, Lu, Chong,
  Qin, Zhou, Peng, and Du}}]{duprobabilistic}
\bibinfo{author}{\bibnamefont{Chen}, \bibfnamefont{H.~W.}},
  \bibinfo{author}{\bibfnamefont{D.~W.} \bibnamefont{Lu}},
  \bibinfo{author}{\bibfnamefont{B.}~\bibnamefont{Chong}},
  \bibinfo{author}{\bibfnamefont{G.}~\bibnamefont{Qin}},
  \bibinfo{author}{\bibfnamefont{X.~Y.} \bibnamefont{Zhou}},
  \bibinfo{author}{\bibfnamefont{X.~H.} \bibnamefont{Peng}}, and
  \bibinfo{author}{\bibfnamefont{J.~F.} \bibnamefont{Du}},
  \bibinfo{year}{2011}, \bibinfo{journal}{Phys. Rev. Lett.}
  \textbf{\bibinfo{volume}{106}}, \bibinfo{pages}{180404}.

\bibitem[{\citenamefont{Chen} \emph{et~al.}(2007)\citenamefont{Chen, Zhou,
  Suter, and Du}}]{chendujfphase}
\bibinfo{author}{\bibnamefont{Chen}, \bibfnamefont{H.~W.}},
  \bibinfo{author}{\bibfnamefont{X.~Y.} \bibnamefont{Zhou}},
  \bibinfo{author}{\bibfnamefont{D.}~\bibnamefont{Suter}}, and
  \bibinfo{author}{\bibfnamefont{J.~F.} \bibnamefont{Du}},
  \bibinfo{year}{2007}, \bibinfo{journal}{Phys. Rev. A}
  \textbf{\bibinfo{volume}{75}}, \bibinfo{pages}{012317}.

\bibitem[{\citenamefont{Chen and Zhang}(2007)}]{ZhangJing07PRA}
\bibinfo{author}{\bibnamefont{Chen}, \bibfnamefont{H.~X.}}, and
  \bibinfo{author}{\bibfnamefont{J.}~\bibnamefont{Zhang}},
  \bibinfo{year}{2007}, \bibinfo{journal}{Phys. Rev. A}
  \textbf{\bibinfo{volume}{75}}, \bibinfo{pages}{022306}.

\bibitem[{\citenamefont{Chen and
  Chen}(2007{\natexlab{a}})}]{ISI:000251336100002}
\bibinfo{author}{\bibnamefont{Chen}, \bibfnamefont{L.}}, and
  \bibinfo{author}{\bibfnamefont{Y.~X.} \bibnamefont{Chen}},
  \bibinfo{year}{2007}{\natexlab{a}}, \bibinfo{journal}{Quant. Inf. Comput.}
  \textbf{\bibinfo{volume}{7}}, \bibinfo{pages}{716}.

\bibitem[{\citenamefont{Chen and
  Chen}(2007{\natexlab{b}})}]{ISI:000247624300060}
\bibinfo{author}{\bibnamefont{Chen}, \bibfnamefont{L.}}, and
  \bibinfo{author}{\bibfnamefont{Y.~X.} \bibnamefont{Chen}},
  \bibinfo{year}{2007}{\natexlab{b}}, \bibinfo{journal}{Phys. Rev. A}
  \textbf{\bibinfo{volume}{75}}, \bibinfo{pages}{062322}.

\bibitem[{\citenamefont{Chen and Yang}(2001{\natexlab{a}})}]{chenyx1}
\bibinfo{author}{\bibnamefont{Chen}, \bibfnamefont{Y.~X.}}, and
  \bibinfo{author}{\bibfnamefont{D.}~\bibnamefont{Yang}},
  \bibinfo{year}{2001}{\natexlab{a}}, \bibinfo{journal}{Phys. Rev. A}
  \textbf{\bibinfo{volume}{64}}, \bibinfo{pages}{064303}.

\bibitem[{\citenamefont{Chen and Yang}(2001{\natexlab{b}})}]{chenyx2}
\bibinfo{author}{\bibnamefont{Chen}, \bibfnamefont{Y.~X.}}, and
  \bibinfo{author}{\bibfnamefont{D.}~\bibnamefont{Yang}},
  \bibinfo{year}{2001}{\natexlab{b}}, \bibinfo{journal}{Phys. Rev. A}
  \textbf{\bibinfo{volume}{65}}, \bibinfo{pages}{022320}.

\bibitem[{\citenamefont{Chiara} \emph{et~al.}(2004)\citenamefont{Chiara, Fazio,
  Macchiavello, Montangero, and Palma}}]{ChiaraFazio}
\bibinfo{author}{\bibnamefont{Chiara}, \bibfnamefont{G.~D.}},
  \bibinfo{author}{\bibfnamefont{R.}~\bibnamefont{Fazio}},
  \bibinfo{author}{\bibfnamefont{C.}~\bibnamefont{Macchiavello}},
  \bibinfo{author}{\bibfnamefont{S.}~\bibnamefont{Montangero}}, and
  \bibinfo{author}{\bibfnamefont{G.~M.} \bibnamefont{Palma}},
  \bibinfo{year}{2004}, \bibinfo{journal}{Phys. Rev. A}
  \textbf{\bibinfo{volume}{70}}, \bibinfo{pages}{062308}.

\bibitem[{\citenamefont{Chiribella}
  \emph{et~al.}(2007)\citenamefont{Chiribella, D'Ariano, Macchiavello,
  Perinotti, and Buscemi}}]{ISI:000243894100042}
\bibinfo{author}{\bibnamefont{Chiribella}, \bibfnamefont{G.}},
  \bibinfo{author}{\bibfnamefont{G.~M.} \bibnamefont{D'Ariano}},
  \bibinfo{author}{\bibfnamefont{C.}~\bibnamefont{Macchiavello}},
  \bibinfo{author}{\bibfnamefont{P.}~\bibnamefont{Perinotti}}, and
  \bibinfo{author}{\bibfnamefont{F.}~\bibnamefont{Buscemi}},
  \bibinfo{year}{2007}, \bibinfo{journal}{Phys. Rev. A}
  \textbf{\bibinfo{volume}{75}}, \bibinfo{pages}{012315}.

\bibitem[{\citenamefont{Chiribella}
  \emph{et~al.}(2008)\citenamefont{Chiribella, D'Ariano, and
  Perinotti}}]{cloningunitary}
\bibinfo{author}{\bibnamefont{Chiribella}, \bibfnamefont{G.}},
  \bibinfo{author}{\bibfnamefont{G.~M.} \bibnamefont{D'Ariano}}, and
  \bibinfo{author}{\bibfnamefont{P.}~\bibnamefont{Perinotti}},
  \bibinfo{year}{2008}, \bibinfo{journal}{Phys. Rev. Lett.}
  \textbf{\bibinfo{volume}{101}}, \bibinfo{pages}{180504}.

\bibitem[{\citenamefont{Chiribella}
  \emph{et~al.}(2005)\citenamefont{Chiribella, D'Ariano, Perinotti, and
  Cerf}}]{extremalcloning}
\bibinfo{author}{\bibnamefont{Chiribella}, \bibfnamefont{G.}},
  \bibinfo{author}{\bibfnamefont{G.~M.} \bibnamefont{D'Ariano}},
  \bibinfo{author}{\bibfnamefont{P.}~\bibnamefont{Perinotti}}, and
  \bibinfo{author}{\bibfnamefont{N.~J.} \bibnamefont{Cerf}},
  \bibinfo{year}{2005}, \bibinfo{journal}{Phys. Rev. A}
  \textbf{\bibinfo{volume}{72}}, \bibinfo{pages}{042336}.

\bibitem[{\citenamefont{Chiribella}
  \emph{et~al.}(2013)\citenamefont{Chiribella, Yang, and
  Yao}}]{qreplication-yao}
\bibinfo{author}{\bibnamefont{Chiribella}, \bibfnamefont{G.}},
  \bibinfo{author}{\bibfnamefont{Y.~X.} \bibnamefont{Yang}}, and
  \bibinfo{author}{\bibfnamefont{A.~C.~C.} \bibnamefont{Yao}},
  \bibinfo{year}{2013}, \bibinfo{journal}{Nature Commun.}
  \textbf{\bibinfo{volume}{4}}, \bibinfo{pages}{2915}.

\bibitem[{\citenamefont{Cochrane} \emph{et~al.}(2004)\citenamefont{Cochrane,
  Ralph, and Doli\'{n}ska}}]{finite-coherent}
\bibinfo{author}{\bibnamefont{Cochrane}, \bibfnamefont{P.~T.}},
  \bibinfo{author}{\bibfnamefont{T.~C.} \bibnamefont{Ralph}}, and
  \bibinfo{author}{\bibfnamefont{A.}~\bibnamefont{Doli\'{n}ska}},
  \bibinfo{year}{2004}, \bibinfo{journal}{Phys. Rev. A}
  \textbf{\bibinfo{volume}{69}}, \bibinfo{pages}{042313}.

\bibitem[{\citenamefont{Coffman} \emph{et~al.}(2000)\citenamefont{Coffman,
  Kundu, and Wootters}}]{ckw2000}
\bibinfo{author}{\bibnamefont{Coffman}, \bibfnamefont{V.}},
  \bibinfo{author}{\bibfnamefont{J.}~\bibnamefont{Kundu}}, and
  \bibinfo{author}{\bibfnamefont{W.~K.} \bibnamefont{Wootters}},
  \bibinfo{year}{2000}, \bibinfo{journal}{Phys. Rev. A}
  \textbf{\bibinfo{volume}{61}}, \bibinfo{pages}{052306}.

\bibitem[{\citenamefont{Collins} \emph{et~al.}(2001)\citenamefont{Collins,
  Linden, and Popescu}}]{CollinsLinden}
\bibinfo{author}{\bibnamefont{Collins}, \bibfnamefont{D.}},
  \bibinfo{author}{\bibfnamefont{N.}~\bibnamefont{Linden}}, and
  \bibinfo{author}{\bibfnamefont{S.}~\bibnamefont{Popescu}},
  \bibinfo{year}{2001}, \bibinfo{journal}{Phys. Rev. A}
  \textbf{\bibinfo{volume}{64}}, \bibinfo{pages}{032302}.

\bibitem[{\citenamefont{Cummins} \emph{et~al.}(2002)\citenamefont{Cummins,
  Jones, Furze, Soffe, Mosca, Peach, and Jones}}]{PhysRevLett.88.187901}
\bibinfo{author}{\bibnamefont{Cummins}, \bibfnamefont{H.~K.}},
  \bibinfo{author}{\bibfnamefont{C.}~\bibnamefont{Jones}},
  \bibinfo{author}{\bibfnamefont{A.}~\bibnamefont{Furze}},
  \bibinfo{author}{\bibfnamefont{N.~F.} \bibnamefont{Soffe}},
  \bibinfo{author}{\bibfnamefont{M.}~\bibnamefont{Mosca}},
  \bibinfo{author}{\bibfnamefont{J.~M.} \bibnamefont{Peach}}, and
  \bibinfo{author}{\bibfnamefont{J.~A.} \bibnamefont{Jones}},
  \bibinfo{year}{2002}, \bibinfo{journal}{Phys. Rev. Lett.}
  \textbf{\bibinfo{volume}{88}}, \bibinfo{pages}{187901}.

\bibitem[{\citenamefont{Cwiklinski}
  \emph{et~al.}(2012)\citenamefont{Cwiklinski, Horodecki, and
  Studzinski}}]{Cwiklinski2012}
\bibinfo{author}{\bibnamefont{Cwiklinski}, \bibfnamefont{P.}},
  \bibinfo{author}{\bibfnamefont{M.}~\bibnamefont{Horodecki}}, and
  \bibinfo{author}{\bibfnamefont{M.}~\bibnamefont{Studzinski}},
  \bibinfo{year}{2012}, \bibinfo{journal}{Phys. Lett. A}
  \textbf{\bibinfo{volume}{376}}, \bibinfo{pages}{2178}.

\bibitem[{\citenamefont{Daffertshofer}
  \emph{et~al.}(2002)\citenamefont{Daffertshofer, Plastino, and
  Plastino}}]{classical-no-cloning}
\bibinfo{author}{\bibnamefont{Daffertshofer}, \bibfnamefont{A.}},
  \bibinfo{author}{\bibfnamefont{A.~R.} \bibnamefont{Plastino}}, and
  \bibinfo{author}{\bibfnamefont{A.}~\bibnamefont{Plastino}},
  \bibinfo{year}{2002}, \bibinfo{journal}{Phys. Rev. Lett.}
  \textbf{\bibinfo{volume}{88}}, \bibinfo{pages}{210601}.

\bibitem[{\citenamefont{Dang and Fan}(2007)}]{ISI:000249154900043}
\bibinfo{author}{\bibnamefont{Dang}, \bibfnamefont{G.~F.}}, and
  \bibinfo{author}{\bibfnamefont{H.}~\bibnamefont{Fan}}, \bibinfo{year}{2007},
  \bibinfo{journal}{Phys. Rev. A} \textbf{\bibinfo{volume}{76}},
  \bibinfo{pages}{022323}.

\bibitem[{\citenamefont{Dang and Fan}(2008)}]{Dang2008}
\bibinfo{author}{\bibnamefont{Dang}, \bibfnamefont{G.~F.}}, and
  \bibinfo{author}{\bibfnamefont{H.}~\bibnamefont{Fan}}, \bibinfo{year}{2008},
  \bibinfo{journal}{J. Phys. A-Math. Theor.} \textbf{\bibinfo{volume}{41}},
  \bibinfo{pages}{155303}.

\bibitem[{\citenamefont{D'Ariano} \emph{et~al.}(2001)\citenamefont{D'Ariano,
  De~Martini, and Sacchi}}]{prl86.914}
\bibinfo{author}{\bibnamefont{D'Ariano}, \bibfnamefont{G.~M.}},
  \bibinfo{author}{\bibfnamefont{F.}~\bibnamefont{De~Martini}}, and
  \bibinfo{author}{\bibfnamefont{M.~F.} \bibnamefont{Sacchi}},
  \bibinfo{year}{2001}, \bibinfo{journal}{Phys. Rev. Lett.}
  \textbf{\bibinfo{volume}{86}}, \bibinfo{pages}{914}.

\bibitem[{\citenamefont{D'Ariano and Lo~Presti}(2001)}]{PhysRevA.64.042308}
\bibinfo{author}{\bibnamefont{D'Ariano}, \bibfnamefont{G.~M.}}, and
  \bibinfo{author}{\bibfnamefont{P.}~\bibnamefont{Lo~Presti}},
  \bibinfo{year}{2001}, \bibinfo{journal}{Phys. Rev. A}
  \textbf{\bibinfo{volume}{64}}, \bibinfo{pages}{042308}.

\bibitem[{\citenamefont{D'Ariano and Macchiavello}(2003)}]{PhysRevA.67.042306}
\bibinfo{author}{\bibnamefont{D'Ariano}, \bibfnamefont{G.~M.}}, and
  \bibinfo{author}{\bibfnamefont{C.}~\bibnamefont{Macchiavello}},
  \bibinfo{year}{2003}, \bibinfo{journal}{Phys. Rev. A}
  \textbf{\bibinfo{volume}{67}}, \bibinfo{pages}{042306}.

\bibitem[{\citenamefont{D'Ariano}
  \emph{et~al.}(2005{\natexlab{a}})\citenamefont{D'Ariano, Macchiavello, and
  Perinotti}}]{mixedstate-phaseestimation}
\bibinfo{author}{\bibnamefont{D'Ariano}, \bibfnamefont{G.~M.}},
  \bibinfo{author}{\bibfnamefont{C.}~\bibnamefont{Macchiavello}}, and
  \bibinfo{author}{\bibfnamefont{P.}~\bibnamefont{Perinotti}},
  \bibinfo{year}{2005}{\natexlab{a}}, \bibinfo{journal}{Phys. Rev. A}
  \textbf{\bibinfo{volume}{72}}, \bibinfo{pages}{042327}.

\bibitem[{\citenamefont{D'Ariano}
  \emph{et~al.}(2005{\natexlab{b}})\citenamefont{D'Ariano, Macchiavello, and
  Perinotti}}]{ISI:000231017700008}
\bibinfo{author}{\bibnamefont{D'Ariano}, \bibfnamefont{G.~M.}},
  \bibinfo{author}{\bibfnamefont{C.}~\bibnamefont{Macchiavello}}, and
  \bibinfo{author}{\bibfnamefont{P.}~\bibnamefont{Perinotti}},
  \bibinfo{year}{2005}{\natexlab{b}}, \bibinfo{journal}{Phys. Rev. Lett.}
  \textbf{\bibinfo{volume}{95}}, \bibinfo{pages}{060503}.

\bibitem[{\citenamefont{D'Ariano and Perinotti}(2009)}]{DAriano2009}
\bibinfo{author}{\bibnamefont{D'Ariano}, \bibfnamefont{G.~M.}}, and
  \bibinfo{author}{\bibfnamefont{P.}~\bibnamefont{Perinotti}},
  \bibinfo{year}{2009}, \bibinfo{journal}{Phys. Lett. A}
  \textbf{\bibinfo{volume}{373}}, \bibinfo{pages}{2416}.

\bibitem[{\citenamefont{D'Ariano} \emph{et~al.}(2006)\citenamefont{D'Ariano,
  Perinotti, and Sacchi}}]{ISI:000238292400001}
\bibinfo{author}{\bibnamefont{D'Ariano}, \bibfnamefont{G.~M.}},
  \bibinfo{author}{\bibfnamefont{P.}~\bibnamefont{Perinotti}}, and
  \bibinfo{author}{\bibfnamefont{M.~F.} \bibnamefont{Sacchi}},
  \bibinfo{year}{2006}, \bibinfo{journal}{New J. Phys.}
  \textbf{\bibinfo{volume}{8}}, \bibinfo{pages}{99}.

\bibitem[{\citenamefont{Dasgupta and Agarwal}(2001)}]{ISI:000170297300031}
\bibinfo{author}{\bibnamefont{Dasgupta}, \bibfnamefont{S.}}, and
  \bibinfo{author}{\bibfnamefont{G.~S.} \bibnamefont{Agarwal}},
  \bibinfo{year}{2001}, \bibinfo{journal}{Phys. Rev. A}
  \textbf{\bibinfo{volume}{64}}, \bibinfo{pages}{022315}.

\bibitem[{\citenamefont{De~Angelis}
  \emph{et~al.}(2007)\citenamefont{De~Angelis, Nagali, Sciarrino, and
  De~Martini}}]{experimentalnosignaling}
\bibinfo{author}{\bibnamefont{De~Angelis}, \bibfnamefont{T.}},
  \bibinfo{author}{\bibfnamefont{E.}~\bibnamefont{Nagali}},
  \bibinfo{author}{\bibfnamefont{F.}~\bibnamefont{Sciarrino}}, and
  \bibinfo{author}{\bibfnamefont{F.}~\bibnamefont{De~Martini}},
  \bibinfo{year}{2007}, \bibinfo{journal}{Phys. Rev. Lett.}
  \textbf{\bibinfo{volume}{99}}, \bibinfo{pages}{193601}.

\bibitem[{\citenamefont{De~Chiara} \emph{et~al.}(2004)\citenamefont{De~Chiara,
  Fazio, Macchiavello, Montangero, and Palma}}]{cloning-spinnetwork}
\bibinfo{author}{\bibnamefont{De~Chiara}, \bibfnamefont{G.}},
  \bibinfo{author}{\bibfnamefont{R.}~\bibnamefont{Fazio}},
  \bibinfo{author}{\bibfnamefont{C.}~\bibnamefont{Macchiavello}},
  \bibinfo{author}{\bibfnamefont{S.}~\bibnamefont{Montangero}}, and
  \bibinfo{author}{\bibfnamefont{G.~M.} \bibnamefont{Palma}},
  \bibinfo{year}{2004}, \bibinfo{journal}{Phys. Rev. A}
  \textbf{\bibinfo{volume}{70}}, \bibinfo{pages}{062308}.

\bibitem[{\citenamefont{De~Martini}
  \emph{et~al.}(2000)\citenamefont{De~Martini, Mussi, and
  Bovino}}]{DeMartini2000}
\bibinfo{author}{\bibnamefont{De~Martini}, \bibfnamefont{F.}},
  \bibinfo{author}{\bibfnamefont{V.}~\bibnamefont{Mussi}}, and
  \bibinfo{author}{\bibfnamefont{F.}~\bibnamefont{Bovino}},
  \bibinfo{year}{2000}, \bibinfo{journal}{Opt. Commun.}
  \textbf{\bibinfo{volume}{179}}, \bibinfo{pages}{581}.

\bibitem[{\citenamefont{De~Martini}
  \emph{et~al.}(2002)\citenamefont{De~Martini, Bu\ifmmode~\check{z}\else
  \v{z}\fi{}ek, Sciarrino, and Sias}}]{martininature}
\bibinfo{author}{\bibnamefont{De~Martini}, \bibfnamefont{F.}},
  \bibinfo{author}{\bibfnamefont{V.}~\bibnamefont{Bu\ifmmode~\check{z}\else
  \v{z}\fi{}ek}}, \bibinfo{author}{\bibfnamefont{F.}~\bibnamefont{Sciarrino}},
  and \bibinfo{author}{\bibfnamefont{C.}~\bibnamefont{Sias}},
  \bibinfo{year}{2002}, \bibinfo{journal}{Nature}
  \textbf{\bibinfo{volume}{419}}, \bibinfo{pages}{815}.

\bibitem[{\citenamefont{Delgado} \emph{et~al.}(2007)\citenamefont{Delgado,
  Lamata, Leon, Salgado, and Solano}}]{Delgado2007}
\bibinfo{author}{\bibnamefont{Delgado}, \bibfnamefont{Y.}},
  \bibinfo{author}{\bibfnamefont{L.}~\bibnamefont{Lamata}},
  \bibinfo{author}{\bibfnamefont{J.}~\bibnamefont{Leon}},
  \bibinfo{author}{\bibfnamefont{D.}~\bibnamefont{Salgado}}, and
  \bibinfo{author}{\bibfnamefont{E.}~\bibnamefont{Solano}},
  \bibinfo{year}{2007}, \bibinfo{journal}{Phys. Rev. Lett.}
  \textbf{\bibinfo{volume}{98}}, \bibinfo{pages}{150502}.

\bibitem[{\citenamefont{Demkowicz-Dobrzanski}
  \emph{et~al.}(2004)\citenamefont{Demkowicz-Dobrzanski, Kus, and
  Wodkiewicz}}]{ISI:000189304700015}
\bibinfo{author}{\bibnamefont{Demkowicz-Dobrzanski}, \bibfnamefont{R.}},
  \bibinfo{author}{\bibfnamefont{M.}~\bibnamefont{Kus}}, and
  \bibinfo{author}{\bibfnamefont{K.}~\bibnamefont{Wodkiewicz}},
  \bibinfo{year}{2004}, \bibinfo{journal}{Phys. Rev. A}
  \textbf{\bibinfo{volume}{69}}, \bibinfo{pages}{012301}.

\bibitem[{\citenamefont{Derka} \emph{et~al.}(1998)\citenamefont{Derka,
  Buz\v{e}k, and Ekert}}]{prl80.1571}
\bibinfo{author}{\bibnamefont{Derka}, \bibfnamefont{R.}},
  \bibinfo{author}{\bibfnamefont{V.}~\bibnamefont{Buz\v{e}k}}, and
  \bibinfo{author}{\bibfnamefont{A.~K.} \bibnamefont{Ekert}},
  \bibinfo{year}{1998}, \bibinfo{journal}{Phys. Rev. Lett.}
  \textbf{\bibinfo{volume}{80}}, \bibinfo{pages}{1571}.

\bibitem[{\citenamefont{Deuar and
  Munro}(2000{\natexlab{a}})}]{ISI:000085168800007}
\bibinfo{author}{\bibnamefont{Deuar}, \bibfnamefont{P.}}, and
  \bibinfo{author}{\bibfnamefont{W.~J.} \bibnamefont{Munro}},
  \bibinfo{year}{2000}{\natexlab{a}}, \bibinfo{journal}{Phys. Rev. A}
  \textbf{\bibinfo{volume}{61}}, \bibinfo{pages}{010306(R)}.

\bibitem[{\citenamefont{Deuar and
  Munro}(2000{\natexlab{b}})}]{ISI:000087567900023}
\bibinfo{author}{\bibnamefont{Deuar}, \bibfnamefont{P.}}, and
  \bibinfo{author}{\bibfnamefont{W.~J.} \bibnamefont{Munro}},
  \bibinfo{year}{2000}{\natexlab{b}}, \bibinfo{journal}{Phys. Rev. A}
  \textbf{\bibinfo{volume}{61}}, \bibinfo{pages}{062304}.

\bibitem[{\citenamefont{Deuar and
  Munro}(2000{\natexlab{c}})}]{ISI:000089688700023}
\bibinfo{author}{\bibnamefont{Deuar}, \bibfnamefont{P.}}, and
  \bibinfo{author}{\bibfnamefont{W.~J.} \bibnamefont{Munro}},
  \bibinfo{year}{2000}{\natexlab{c}}, \bibinfo{journal}{Phys. Rev. A}
  \textbf{\bibinfo{volume}{62}}, \bibinfo{pages}{042304}.

\bibitem[{\citenamefont{Di~Franco} \emph{et~al.}(2007)\citenamefont{Di~Franco,
  Paternostro, Palma, and Kim}}]{DiFranco2007}
\bibinfo{author}{\bibnamefont{Di~Franco}, \bibfnamefont{C.}},
  \bibinfo{author}{\bibfnamefont{M.}~\bibnamefont{Paternostro}},
  \bibinfo{author}{\bibfnamefont{G.~M.} \bibnamefont{Palma}}, and
  \bibinfo{author}{\bibfnamefont{M.~S.} \bibnamefont{Kim}},
  \bibinfo{year}{2007}, \bibinfo{journal}{Phys. Rev. A}
  \textbf{\bibinfo{volume}{76}}, \bibinfo{pages}{042316}.

\bibitem[{\citenamefont{Dieks}(1982)}]{dieks-no-cloning}
\bibinfo{author}{\bibnamefont{Dieks}, \bibfnamefont{D.}}, \bibinfo{year}{1982},
  \bibinfo{journal}{Phys. Lett. A} \textbf{\bibinfo{volume}{92}},
  \bibinfo{pages}{271 }.

\bibitem[{\citenamefont{Dong} \emph{et~al.}(2008)\citenamefont{Dong, Zou, and
  Guo}}]{Dong2008}
\bibinfo{author}{\bibnamefont{Dong}, \bibfnamefont{Y.~L.}},
  \bibinfo{author}{\bibfnamefont{X.~B.} \bibnamefont{Zou}}, and
  \bibinfo{author}{\bibfnamefont{G.~C.} \bibnamefont{Guo}},
  \bibinfo{year}{2008}, \bibinfo{journal}{Phys. Rev. A}
  \textbf{\bibinfo{volume}{77}}, \bibinfo{pages}{034304}.

\bibitem[{\citenamefont{Du} \emph{et~al.}(2005)\citenamefont{Du, Durt, Zou, Li,
  Kwek, Lai, Oh, and Ekert}}]{PhysRevLett.94.040505}
\bibinfo{author}{\bibnamefont{Du}, \bibfnamefont{J.~F.}},
  \bibinfo{author}{\bibfnamefont{T.}~\bibnamefont{Durt}},
  \bibinfo{author}{\bibfnamefont{P.}~\bibnamefont{Zou}},
  \bibinfo{author}{\bibfnamefont{H.}~\bibnamefont{Li}},
  \bibinfo{author}{\bibfnamefont{L.~C.} \bibnamefont{Kwek}},
  \bibinfo{author}{\bibfnamefont{C.~H.} \bibnamefont{Lai}},
  \bibinfo{author}{\bibfnamefont{C.~H.} \bibnamefont{Oh}}, and
  \bibinfo{author}{\bibfnamefont{A.}~\bibnamefont{Ekert}},
  \bibinfo{year}{2005}, \bibinfo{journal}{Phys. Rev. Lett.}
  \textbf{\bibinfo{volume}{94}}, \bibinfo{pages}{040505}.

\bibitem[{\citenamefont{Duan and
  Guo}(1998{\natexlab{a}})}]{PhysRevLett.80.4999}
\bibinfo{author}{\bibnamefont{Duan}, \bibfnamefont{L.~M.}}, and
  \bibinfo{author}{\bibfnamefont{G.~C.} \bibnamefont{Guo}},
  \bibinfo{year}{1998}{\natexlab{a}}, \bibinfo{journal}{Phys. Rev. Lett.}
  \textbf{\bibinfo{volume}{80}}, \bibinfo{pages}{4999}.

\bibitem[{\citenamefont{Duan and Guo}(1998{\natexlab{b}})}]{Duan1998261}
\bibinfo{author}{\bibnamefont{Duan}, \bibfnamefont{L.~M.}}, and
  \bibinfo{author}{\bibfnamefont{G.~C.} \bibnamefont{Guo}},
  \bibinfo{year}{1998}{\natexlab{b}}, \bibinfo{journal}{Phys. Lett. A}
  \textbf{\bibinfo{volume}{243}}, \bibinfo{pages}{261 }.

\bibitem[{\citenamefont{Duan and Guo}(1999)}]{ISI:000080184300011}
\bibinfo{author}{\bibnamefont{Duan}, \bibfnamefont{L.~M.}}, and
  \bibinfo{author}{\bibfnamefont{G.~C.} \bibnamefont{Guo}},
  \bibinfo{year}{1999}, \bibinfo{journal}{Commun. Theor. Phys.}
  \textbf{\bibinfo{volume}{31}}, \bibinfo{pages}{223}.

\bibitem[{\citenamefont{D\"{u}r}(2001)}]{PhysRevA.63.020303}
\bibinfo{author}{\bibnamefont{D\"{u}r}, \bibfnamefont{W.}},
  \bibinfo{year}{2001}, \bibinfo{journal}{Phys. Rev. A}
  \textbf{\bibinfo{volume}{63}}, \bibinfo{pages}{020303}.

\bibitem[{\citenamefont{D\"{u}r and Cirac}(2000)}]{Dur2000}
\bibinfo{author}{\bibnamefont{D\"{u}r}, \bibfnamefont{W.}}, and
  \bibinfo{author}{\bibfnamefont{J.~I.} \bibnamefont{Cirac}},
  \bibinfo{year}{2000}, \bibinfo{journal}{J. Mod. Optic.}
  \textbf{\bibinfo{volume}{47}}, \bibinfo{pages}{247}.

\bibitem[{\citenamefont{Durt and Du}(2004)}]{PhysRevA.69.062316}
\bibinfo{author}{\bibnamefont{Durt}, \bibfnamefont{T.}}, and
  \bibinfo{author}{\bibfnamefont{J.~F.} \bibnamefont{Du}},
  \bibinfo{year}{2004}, \bibinfo{journal}{Phys. Rev. A}
  \textbf{\bibinfo{volume}{69}}, \bibinfo{pages}{062316}.

\bibitem[{\citenamefont{Durt} \emph{et~al.}(2005)\citenamefont{Durt,
  Fiur\'{a}\v{s}ek, and Cerf}}]{PhysRevA.72.052322}
\bibinfo{author}{\bibnamefont{Durt}, \bibfnamefont{T.}},
  \bibinfo{author}{\bibfnamefont{J.}~\bibnamefont{Fiur\'{a}\v{s}ek}}, and
  \bibinfo{author}{\bibfnamefont{N.~J.} \bibnamefont{Cerf}},
  \bibinfo{year}{2005}, \bibinfo{journal}{Phys. Rev. A}
  \textbf{\bibinfo{volume}{72}}, \bibinfo{pages}{052322}.

\bibitem[{\citenamefont{Durt and Nagler}(2003)}]{PhysRevA.68.042323}
\bibinfo{author}{\bibnamefont{Durt}, \bibfnamefont{T.}}, and
  \bibinfo{author}{\bibfnamefont{B.}~\bibnamefont{Nagler}},
  \bibinfo{year}{2003}, \bibinfo{journal}{Phys. Rev. A}
  \textbf{\bibinfo{volume}{68}}, \bibinfo{pages}{042323}.

\bibitem[{\citenamefont{Durt and Van~de Putte}(2011)}]{Durt2011}
\bibinfo{author}{\bibnamefont{Durt}, \bibfnamefont{T.}}, and
  \bibinfo{author}{\bibfnamefont{J.}~\bibnamefont{Van~de Putte}},
  \bibinfo{year}{2011}, \bibinfo{journal}{Int. J. Quant. Inf.}
  \textbf{\bibinfo{volume}{9}}, \bibinfo{pages}{915}.

\bibitem[{\citenamefont{Eisert} \emph{et~al.}(2000)\citenamefont{Eisert,
  Jacobs, Papadopoulos, and Plenio}}]{EisertJacobs}
\bibinfo{author}{\bibnamefont{Eisert}, \bibfnamefont{J.}},
  \bibinfo{author}{\bibfnamefont{K.}~\bibnamefont{Jacobs}},
  \bibinfo{author}{\bibfnamefont{P.}~\bibnamefont{Papadopoulos}}, and
  \bibinfo{author}{\bibfnamefont{M.~B.} \bibnamefont{Plenio}},
  \bibinfo{year}{2000}, \bibinfo{journal}{Phys. Rev. A}
  \textbf{\bibinfo{volume}{62}}, \bibinfo{pages}{052317}.

\bibitem[{\citenamefont{Ekert and Jozsa}(1996)}]{Ekert-JozsaRMP96}
\bibinfo{author}{\bibnamefont{Ekert}, \bibfnamefont{A.}}, and
  \bibinfo{author}{\bibfnamefont{R.}~\bibnamefont{Jozsa}},
  \bibinfo{year}{1996}, \bibinfo{journal}{Rev. Mod. Phys.}
  \textbf{\bibinfo{volume}{68}}, \bibinfo{pages}{733}.

\bibitem[{\citenamefont{Ekert and Renner}(2014)}]{Ekert-Renner-Privacy}
\bibinfo{author}{\bibnamefont{Ekert}, \bibfnamefont{A.}}, and
  \bibinfo{author}{\bibfnamefont{R.}~\bibnamefont{Renner}},
  \bibinfo{year}{2014}, \bibinfo{journal}{Nature}
  \textbf{\bibinfo{volume}{507}}, \bibinfo{pages}{443}.

\bibitem[{\citenamefont{Ekert}(1991)}]{PhysRevLett.67.661}
\bibinfo{author}{\bibnamefont{Ekert}, \bibfnamefont{A.~K.}},
  \bibinfo{year}{1991}, \bibinfo{journal}{Phys. Rev. Lett.}
  \textbf{\bibinfo{volume}{67}}, \bibinfo{pages}{661}.

\bibitem[{\citenamefont{Englert and Aharonov}(2001)}]{tale}
\bibinfo{author}{\bibnamefont{Englert}, \bibfnamefont{B.~G.}}, and
  \bibinfo{author}{\bibfnamefont{Y.}~\bibnamefont{Aharonov}},
  \bibinfo{year}{2001}, \bibinfo{journal}{Phys. Lett. A}
  \textbf{\bibinfo{volume}{284}}, \bibinfo{pages}{1}.

\bibitem[{\citenamefont{Fan}(2003)}]{Fan-mixed}
\bibinfo{author}{\bibnamefont{Fan}, \bibfnamefont{H.}}, \bibinfo{year}{2003},
  \bibinfo{journal}{Phys. Rev. A} \textbf{\bibinfo{volume}{68}},
  \bibinfo{pages}{054301}.

\bibitem[{\citenamefont{Fan}(2004)}]{Fan04}
\bibinfo{author}{\bibnamefont{Fan}, \bibfnamefont{H.}}, \bibinfo{year}{2004},
  \bibinfo{journal}{Phys. Rev. Lett.} \textbf{\bibinfo{volume}{92}},
  \bibinfo{pages}{177905}.

\bibitem[{\citenamefont{Fan} \emph{et~al.}(2003)\citenamefont{Fan, Imai,
  Matsumoto, and Wang}}]{PhysRevA.67.022317}
\bibinfo{author}{\bibnamefont{Fan}, \bibfnamefont{H.}},
  \bibinfo{author}{\bibfnamefont{H.}~\bibnamefont{Imai}},
  \bibinfo{author}{\bibfnamefont{K.}~\bibnamefont{Matsumoto}}, and
  \bibinfo{author}{\bibfnamefont{X.~B.} \bibnamefont{Wang}},
  \bibinfo{year}{2003}, \bibinfo{journal}{Phys. Rev. A}
  \textbf{\bibinfo{volume}{67}}, \bibinfo{pages}{022317}.

\bibitem[{\citenamefont{Fan} \emph{et~al.}(2004)\citenamefont{Fan, Korepin, and
  Roychowdhury}}]{VBSPRL}
\bibinfo{author}{\bibnamefont{Fan}, \bibfnamefont{H.}},
  \bibinfo{author}{\bibfnamefont{V.}~\bibnamefont{Korepin}}, and
  \bibinfo{author}{\bibfnamefont{V.}~\bibnamefont{Roychowdhury}},
  \bibinfo{year}{2004}, \bibinfo{journal}{Phys. Rev. Lett.}
  \textbf{\bibinfo{volume}{93}}, \bibinfo{pages}{227203}.

\bibitem[{\citenamefont{Fan} \emph{et~al.}(2007)\citenamefont{Fan, Liu, and
  Shi}}]{ISI:000248425500008}
\bibinfo{author}{\bibnamefont{Fan}, \bibfnamefont{H.}},
  \bibinfo{author}{\bibfnamefont{B.~Y.} \bibnamefont{Liu}}, and
  \bibinfo{author}{\bibfnamefont{K.~J.} \bibnamefont{Shi}},
  \bibinfo{year}{2007}, \bibinfo{journal}{Quant. Inf. Comput.}
  \textbf{\bibinfo{volume}{7}}, \bibinfo{pages}{551}.

\bibitem[{\citenamefont{Fan}
  \emph{et~al.}(2001{\natexlab{a}})\citenamefont{Fan, Matsumoto, and
  Wadati}}]{PhysRevA.64.064301}
\bibinfo{author}{\bibnamefont{Fan}, \bibfnamefont{H.}},
  \bibinfo{author}{\bibfnamefont{K.}~\bibnamefont{Matsumoto}}, and
  \bibinfo{author}{\bibfnamefont{M.}~\bibnamefont{Wadati}},
  \bibinfo{year}{2001}{\natexlab{a}}, \bibinfo{journal}{Phys. Rev. A}
  \textbf{\bibinfo{volume}{64}}, \bibinfo{pages}{064301}.

\bibitem[{\citenamefont{Fan}
  \emph{et~al.}(2001{\natexlab{b}})\citenamefont{Fan, Matsumoto, Wang, and
  Wadati}}]{PhysRevA.65.012304}
\bibinfo{author}{\bibnamefont{Fan}, \bibfnamefont{H.}},
  \bibinfo{author}{\bibfnamefont{K.}~\bibnamefont{Matsumoto}},
  \bibinfo{author}{\bibfnamefont{X.~B.} \bibnamefont{Wang}}, and
  \bibinfo{author}{\bibfnamefont{M.}~\bibnamefont{Wadati}},
  \bibinfo{year}{2001}{\natexlab{b}}, \bibinfo{journal}{Phys. Rev. A}
  \textbf{\bibinfo{volume}{65}}, \bibinfo{pages}{012304}.

\bibitem[{\citenamefont{Fan} \emph{et~al.}(2002)\citenamefont{Fan, Weihs,
  Matsumoto, and Imai}}]{ISI:000177872600134}
\bibinfo{author}{\bibnamefont{Fan}, \bibfnamefont{H.}},
  \bibinfo{author}{\bibfnamefont{G.}~\bibnamefont{Weihs}},
  \bibinfo{author}{\bibfnamefont{K.}~\bibnamefont{Matsumoto}}, and
  \bibinfo{author}{\bibfnamefont{H.}~\bibnamefont{Imai}}, \bibinfo{year}{2002},
  \bibinfo{journal}{Phys. Rev. A} \textbf{\bibinfo{volume}{66}},
  \bibinfo{pages}{024307}.

\bibitem[{\citenamefont{Fang} \emph{et~al.}(2011)\citenamefont{Fang, Song, and
  Ye}}]{Fang2011}
\bibinfo{author}{\bibnamefont{Fang}, \bibfnamefont{B.~L.}},
  \bibinfo{author}{\bibfnamefont{Q.~M.} \bibnamefont{Song}}, and
  \bibinfo{author}{\bibfnamefont{L.}~\bibnamefont{Ye}}, \bibinfo{year}{2011},
  \bibinfo{journal}{Phys. Rev. A} \textbf{\bibinfo{volume}{83}},
  \bibinfo{pages}{042309}.

\bibitem[{\citenamefont{Fang}
  \emph{et~al.}(2012{\natexlab{a}})\citenamefont{Fang, Wu, and Ye}}]{Fang2012}
\bibinfo{author}{\bibnamefont{Fang}, \bibfnamefont{B.~L.}},
  \bibinfo{author}{\bibfnamefont{T.}~\bibnamefont{Wu}}, and
  \bibinfo{author}{\bibfnamefont{L.}~\bibnamefont{Ye}},
  \bibinfo{year}{2012}{\natexlab{a}}, \bibinfo{journal}{Europhys. Lett.}
  \textbf{\bibinfo{volume}{97}}, \bibinfo{pages}{60002}.

\bibitem[{\citenamefont{Fang}
  \emph{et~al.}(2012{\natexlab{b}})\citenamefont{Fang, Wu, and Ye}}]{Fang2012a}
\bibinfo{author}{\bibnamefont{Fang}, \bibfnamefont{B.~L.}},
  \bibinfo{author}{\bibfnamefont{T.}~\bibnamefont{Wu}}, and
  \bibinfo{author}{\bibfnamefont{L.}~\bibnamefont{Ye}},
  \bibinfo{year}{2012}{\natexlab{b}}, \bibinfo{journal}{Quant. Inf. Comput.}
  \textbf{\bibinfo{volume}{12}}, \bibinfo{pages}{334}.

\bibitem[{\citenamefont{Fang and Ye}(2010)}]{Fang2010}
\bibinfo{author}{\bibnamefont{Fang}, \bibfnamefont{B.~L.}}, and
  \bibinfo{author}{\bibfnamefont{L.}~\bibnamefont{Ye}}, \bibinfo{year}{2010},
  \bibinfo{journal}{Phys. Lett. A} \textbf{\bibinfo{volume}{374}},
  \bibinfo{pages}{1966}.

\bibitem[{\citenamefont{Fang} \emph{et~al.}(2006)\citenamefont{Fang, Liu, Liu,
  Shi, and Zhang}}]{ISI:000242357000016}
\bibinfo{author}{\bibnamefont{Fang}, \bibfnamefont{M.}},
  \bibinfo{author}{\bibfnamefont{Y.~M.} \bibnamefont{Liu}},
  \bibinfo{author}{\bibfnamefont{J.}~\bibnamefont{Liu}},
  \bibinfo{author}{\bibfnamefont{S.~H.} \bibnamefont{Shi}}, and
  \bibinfo{author}{\bibfnamefont{Z.~J.} \bibnamefont{Zhang}},
  \bibinfo{year}{2006}, \bibinfo{journal}{Commun. Theor. Phys.}
  \textbf{\bibinfo{volume}{46}}, \bibinfo{pages}{849}.

\bibitem[{\citenamefont{Fasel} \emph{et~al.}(2002)\citenamefont{Fasel, Gisin,
  Ribordy, Scarani, and Zbinden}}]{prl89.107901}
\bibinfo{author}{\bibnamefont{Fasel}, \bibfnamefont{S.}},
  \bibinfo{author}{\bibfnamefont{N.}~\bibnamefont{Gisin}},
  \bibinfo{author}{\bibfnamefont{G.}~\bibnamefont{Ribordy}},
  \bibinfo{author}{\bibfnamefont{V.}~\bibnamefont{Scarani}}, and
  \bibinfo{author}{\bibfnamefont{H.}~\bibnamefont{Zbinden}},
  \bibinfo{year}{2002}, \bibinfo{journal}{Phys. Rev. Lett.}
  \textbf{\bibinfo{volume}{89}}, \bibinfo{pages}{107901}.

\bibitem[{\citenamefont{Feng} \emph{et~al.}(2005)\citenamefont{Feng, Duan, and
  Ji}}]{ISI:000230887300043}
\bibinfo{author}{\bibnamefont{Feng}, \bibfnamefont{Y.}},
  \bibinfo{author}{\bibfnamefont{R.~Y.} \bibnamefont{Duan}}, and
  \bibinfo{author}{\bibfnamefont{Z.~F.} \bibnamefont{Ji}},
  \bibinfo{year}{2005}, \bibinfo{journal}{Phys. Rev. A}
  \textbf{\bibinfo{volume}{72}}, \bibinfo{pages}{012313}.

\bibitem[{\citenamefont{Fenyes}(2012)}]{Fenyes2012}
\bibinfo{author}{\bibnamefont{Fenyes}, \bibfnamefont{A.}},
  \bibinfo{year}{2012}, \bibinfo{journal}{J. Math. Phys.}
  \textbf{\bibinfo{volume}{53}}, \bibinfo{pages}{012902}.

\bibitem[{\citenamefont{Filip}(2004{\natexlab{a}})}]{filip1}
\bibinfo{author}{\bibnamefont{Filip}, \bibfnamefont{R.}},
  \bibinfo{year}{2004}{\natexlab{a}}, \bibinfo{journal}{Phys. Rev. A}
  \textbf{\bibinfo{volume}{69}}, \bibinfo{pages}{032309}.

\bibitem[{\citenamefont{Filip}(2004{\natexlab{b}})}]{filip2}
\bibinfo{author}{\bibnamefont{Filip}, \bibfnamefont{R.}},
  \bibinfo{year}{2004}{\natexlab{b}}, \bibinfo{journal}{Phys. Rev. A}
  \textbf{\bibinfo{volume}{69}}, \bibinfo{pages}{052301}.

\bibitem[{\citenamefont{Filip} \emph{et~al.}(2004)\citenamefont{Filip,
  Fiur\'{a}\v{s}ek, and Marek}}]{filipreverse}
\bibinfo{author}{\bibnamefont{Filip}, \bibfnamefont{R.}},
  \bibinfo{author}{\bibfnamefont{J.}~\bibnamefont{Fiur\'{a}\v{s}ek}}, and
  \bibinfo{author}{\bibfnamefont{P.}~\bibnamefont{Marek}},
  \bibinfo{year}{2004}, \bibinfo{journal}{Phys. Rev. A}
  \textbf{\bibinfo{volume}{69}}, \bibinfo{pages}{012314}.

\bibitem[{\citenamefont{Fiurasek and Cerf}(2007)}]{Fiurasek2007}
\bibinfo{author}{\bibnamefont{Fiurasek}, \bibfnamefont{J.}}, and
  \bibinfo{author}{\bibfnamefont{N.~J.} \bibnamefont{Cerf}},
  \bibinfo{year}{2007}, \bibinfo{journal}{Phys. Rev. A}
  \textbf{\bibinfo{volume}{75}}, \bibinfo{pages}{052335}.

\bibitem[{\citenamefont{Fiurasek and Cerf}(2008)}]{Fiurasek2008}
\bibinfo{author}{\bibnamefont{Fiurasek}, \bibfnamefont{J.}}, and
  \bibinfo{author}{\bibfnamefont{N.~J.} \bibnamefont{Cerf}},
  \bibinfo{year}{2008}, \bibinfo{journal}{Phys. Rev. A}
  \textbf{\bibinfo{volume}{77}}, \bibinfo{pages}{052308}.

\bibitem[{\citenamefont{Fiur\'{a}\v{s}ek}(2001{\natexlab{a}})}]{PhysRevA.64.06%
2310}
\bibinfo{author}{\bibnamefont{Fiur\'{a}\v{s}ek}, \bibfnamefont{J.}},
  \bibinfo{year}{2001}{\natexlab{a}}, \bibinfo{journal}{Phys. Rev. A}
  \textbf{\bibinfo{volume}{64}}, \bibinfo{pages}{062310}.

\bibitem[{\citenamefont{Fiur\'{a}\v{s}ek}(2001{\natexlab{b}})}]{prl86.4942}
\bibinfo{author}{\bibnamefont{Fiur\'{a}\v{s}ek}, \bibfnamefont{J.}},
  \bibinfo{year}{2001}{\natexlab{b}}, \bibinfo{journal}{Phys. Rev. Lett.}
  \textbf{\bibinfo{volume}{86}}, \bibinfo{pages}{4942}.

\bibitem[{\citenamefont{Fiur\'{a}\v{s}ek}(2003)}]{fiurasek2003}
\bibinfo{author}{\bibnamefont{Fiur\'{a}\v{s}ek}, \bibfnamefont{J.}},
  \bibinfo{year}{2003}, \bibinfo{journal}{Phys. Rev. A}
  \textbf{\bibinfo{volume}{67}}, \bibinfo{pages}{052314}.

\bibitem[{\citenamefont{Fiur\'{a}\v{s}ek}(2004)}]{PhysRevA.70.032308}
\bibinfo{author}{\bibnamefont{Fiur\'{a}\v{s}ek}, \bibfnamefont{J.}},
  \bibinfo{year}{2004}, \bibinfo{journal}{Phys. Rev. A}
  \textbf{\bibinfo{volume}{70}}, \bibinfo{pages}{032308}.

\bibitem[{\citenamefont{Fiur\'{a}\v{s}ek}
  \emph{et~al.}(2004)\citenamefont{Fiur\'{a}\v{s}ek, Cerf, and
  Polzik}}]{FiurasekCerfPolzik}
\bibinfo{author}{\bibnamefont{Fiur\'{a}\v{s}ek}, \bibfnamefont{J.}},
  \bibinfo{author}{\bibfnamefont{N.~J.} \bibnamefont{Cerf}}, and
  \bibinfo{author}{\bibfnamefont{E.~S.} \bibnamefont{Polzik}},
  \bibinfo{year}{2004}, \bibinfo{journal}{Phys. Rev. Lett.}
  \textbf{\bibinfo{volume}{93}}, \bibinfo{pages}{180501}.

\bibitem[{\citenamefont{Fiur\'{a}\v{s}ek}
  \emph{et~al.}(2005)\citenamefont{Fiur\'{a}\v{s}ek, Filip, and
  Cerf}}]{ISI:000234463800006}
\bibinfo{author}{\bibnamefont{Fiur\'{a}\v{s}ek}, \bibfnamefont{J.}},
  \bibinfo{author}{\bibfnamefont{R.}~\bibnamefont{Filip}}, and
  \bibinfo{author}{\bibfnamefont{N.~J.} \bibnamefont{Cerf}},
  \bibinfo{year}{2005}, \bibinfo{journal}{Quant. Inf. Comput.}
  \textbf{\bibinfo{volume}{5}}, \bibinfo{pages}{583}.

\bibitem[{\citenamefont{Fiur\'{a}\v{s}ek}
  \emph{et~al.}(2002)\citenamefont{Fiur\'{a}\v{s}ek, Iblisdir, Massar, and
  Cerf}}]{cloning-orthogonalqubits}
\bibinfo{author}{\bibnamefont{Fiur\'{a}\v{s}ek}, \bibfnamefont{J.}},
  \bibinfo{author}{\bibfnamefont{S.}~\bibnamefont{Iblisdir}},
  \bibinfo{author}{\bibfnamefont{S.}~\bibnamefont{Massar}}, and
  \bibinfo{author}{\bibfnamefont{N.~J.} \bibnamefont{Cerf}},
  \bibinfo{year}{2002}, \bibinfo{journal}{Phys. Rev. A}
  \textbf{\bibinfo{volume}{65}}, \bibinfo{pages}{040302(R)}.

\bibitem[{\citenamefont{Fritz} \emph{et~al.}(2013)\citenamefont{Fritz, Sainz,
  Augusiak, Brask, Chaves, Leverrier, and Ac\'{i}n}}]{local-orthogonality}
\bibinfo{author}{\bibnamefont{Fritz}, \bibfnamefont{T.}},
  \bibinfo{author}{\bibfnamefont{A.~B.} \bibnamefont{Sainz}},
  \bibinfo{author}{\bibfnamefont{R.}~\bibnamefont{Augusiak}},
  \bibinfo{author}{\bibfnamefont{J.~B.} \bibnamefont{Brask}},
  \bibinfo{author}{\bibfnamefont{R.}~\bibnamefont{Chaves}},
  \bibinfo{author}{\bibfnamefont{A.}~\bibnamefont{Leverrier}}, and
  \bibinfo{author}{\bibfnamefont{A.}~\bibnamefont{Ac\'{i}n}},
  \bibinfo{year}{2013}, \bibinfo{journal}{Nature Commun.}
  \textbf{\bibinfo{volume}{4}}, \bibinfo{pages}{2263}.

\bibitem[{\citenamefont{Frowis and D\"{u}r}(2012)}]{macro-clone}
\bibinfo{author}{\bibnamefont{Frowis}, \bibfnamefont{F.}}, and
  \bibinfo{author}{\bibfnamefont{W.}~\bibnamefont{D\"{u}r}},
  \bibinfo{year}{2012}, \bibinfo{journal}{Phys. Rev. Lett.}
  \textbf{\bibinfo{volume}{109}}, \bibinfo{pages}{170401}.

\bibitem[{\citenamefont{Fuchs} \emph{et~al.}(1997)\citenamefont{Fuchs, Gisin,
  Griffiths, Niu, and Peres}}]{fuchs97}
\bibinfo{author}{\bibnamefont{Fuchs}, \bibfnamefont{C.~A.}},
  \bibinfo{author}{\bibfnamefont{N.}~\bibnamefont{Gisin}},
  \bibinfo{author}{\bibfnamefont{R.~B.} \bibnamefont{Griffiths}},
  \bibinfo{author}{\bibfnamefont{C.~S.} \bibnamefont{Niu}}, and
  \bibinfo{author}{\bibfnamefont{A.}~\bibnamefont{Peres}},
  \bibinfo{year}{1997}, \bibinfo{journal}{Phys. Rev. A}
  \textbf{\bibinfo{volume}{56}}, \bibinfo{pages}{1163}.

\bibitem[{\citenamefont{Fuchs and Peres}(1996)}]{PhysRevA.53.2038}
\bibinfo{author}{\bibnamefont{Fuchs}, \bibfnamefont{C.~A.}}, and
  \bibinfo{author}{\bibfnamefont{A.}~\bibnamefont{Peres}},
  \bibinfo{year}{1996}, \bibinfo{journal}{Phys. Rev. A}
  \textbf{\bibinfo{volume}{53}}, \bibinfo{pages}{2038}.

\bibitem[{\citenamefont{Galvao and Hardy}(2000)}]{GalvaoHardy}
\bibinfo{author}{\bibnamefont{Galvao}, \bibfnamefont{E.~F.}}, and
  \bibinfo{author}{\bibfnamefont{L.}~\bibnamefont{Hardy}},
  \bibinfo{year}{2000}, \bibinfo{journal}{Phys. Rev. A}
  \textbf{\bibinfo{volume}{62}}, \bibinfo{pages}{022301}.

\bibitem[{\citenamefont{Gedik and \c{C}akmak}(2012)}]{GedikPreprint}
\bibinfo{author}{\bibnamefont{Gedik}, \bibfnamefont{Z.}}, and
  \bibinfo{author}{\bibfnamefont{B.}~\bibnamefont{\c{C}akmak}},
  \bibinfo{year}{2012}, \bibinfo{journal}{arXiv:1203.3054} .

\bibitem[{\citenamefont{Ghiu}(2003)}]{PhysRevA.67.012323}
\bibinfo{author}{\bibnamefont{Ghiu}, \bibfnamefont{I.}}, \bibinfo{year}{2003},
  \bibinfo{journal}{Phys. Rev. A} \textbf{\bibinfo{volume}{67}},
  \bibinfo{pages}{012323}.

\bibitem[{\citenamefont{Ghiu and Karlsson}(2005)}]{ISI:000232228300059}
\bibinfo{author}{\bibnamefont{Ghiu}, \bibfnamefont{I.}}, and
  \bibinfo{author}{\bibfnamefont{A.}~\bibnamefont{Karlsson}},
  \bibinfo{year}{2005}, \bibinfo{journal}{Phys. Rev. A}
  \textbf{\bibinfo{volume}{72}}, \bibinfo{pages}{032331}.

\bibitem[{\citenamefont{Ghosh} \emph{et~al.}(1999)\citenamefont{Ghosh, Kar, and
  Roy}}]{Ghosh199917}
\bibinfo{author}{\bibnamefont{Ghosh}, \bibfnamefont{S.}},
  \bibinfo{author}{\bibfnamefont{G.}~\bibnamefont{Kar}}, and
  \bibinfo{author}{\bibfnamefont{A.}~\bibnamefont{Roy}}, \bibinfo{year}{1999},
  \bibinfo{journal}{Phys. Lett. A} \textbf{\bibinfo{volume}{261}},
  \bibinfo{pages}{17 }.

\bibitem[{\citenamefont{Ghosh} \emph{et~al.}(2004)\citenamefont{Ghosh, Kar, and
  Roy}}]{GKRlocal-cloning}
\bibinfo{author}{\bibnamefont{Ghosh}, \bibfnamefont{S.}},
  \bibinfo{author}{\bibfnamefont{G.}~\bibnamefont{Kar}}, and
  \bibinfo{author}{\bibfnamefont{A.}~\bibnamefont{Roy}}, \bibinfo{year}{2004},
  \bibinfo{journal}{Phys. Rev. A} \textbf{\bibinfo{volume}{69}},
  \bibinfo{pages}{052312}.

\bibitem[{\citenamefont{Ghosh} \emph{et~al.}(2001)\citenamefont{Ghosh, Kar,
  Roy, Sen(De), and Sen}}]{prl87.277902}
\bibinfo{author}{\bibnamefont{Ghosh}, \bibfnamefont{S.}},
  \bibinfo{author}{\bibfnamefont{G.}~\bibnamefont{Kar}},
  \bibinfo{author}{\bibfnamefont{A.}~\bibnamefont{Roy}},
  \bibinfo{author}{\bibfnamefont{A.}~\bibnamefont{Sen(De)}}, and
  \bibinfo{author}{\bibfnamefont{U.}~\bibnamefont{Sen}}, \bibinfo{year}{2001},
  \bibinfo{journal}{Phys. Rev. Lett.} \textbf{\bibinfo{volume}{87}},
  \bibinfo{pages}{277902}.

\bibitem[{\citenamefont{Gisin}(1998)}]{Gisin19981}
\bibinfo{author}{\bibnamefont{Gisin}, \bibfnamefont{N.}}, \bibinfo{year}{1998},
  \bibinfo{journal}{Phys. Lett. A} \textbf{\bibinfo{volume}{242}},
  \bibinfo{pages}{1 }.

\bibitem[{\citenamefont{Gisin and Huttner}(1997)}]{GisinHuttner}
\bibinfo{author}{\bibnamefont{Gisin}, \bibfnamefont{N.}}, and
  \bibinfo{author}{\bibfnamefont{B.}~\bibnamefont{Huttner}},
  \bibinfo{year}{1997}, \bibinfo{journal}{Phys. Lett. A}
  \textbf{\bibinfo{volume}{228}}, \bibinfo{pages}{13}.

\bibitem[{\citenamefont{Gisin and Massar}(1997)}]{PhysRevLett.79.2153}
\bibinfo{author}{\bibnamefont{Gisin}, \bibfnamefont{N.}}, and
  \bibinfo{author}{\bibfnamefont{S.}~\bibnamefont{Massar}},
  \bibinfo{year}{1997}, \bibinfo{journal}{Phys. Rev. Lett.}
  \textbf{\bibinfo{volume}{79}}, \bibinfo{pages}{2153}.

\bibitem[{\citenamefont{Gisin and Popescu}(1999)}]{PhysRevLett.83.432}
\bibinfo{author}{\bibnamefont{Gisin}, \bibfnamefont{N.}}, and
  \bibinfo{author}{\bibfnamefont{S.}~\bibnamefont{Popescu}},
  \bibinfo{year}{1999}, \bibinfo{journal}{Phys. Rev. Lett.}
  \textbf{\bibinfo{volume}{83}}, \bibinfo{pages}{432}.

\bibitem[{\citenamefont{Gisin} \emph{et~al.}(2002)\citenamefont{Gisin, Ribordy,
  Tittel, and Zbinden}}]{QKDreview}
\bibinfo{author}{\bibnamefont{Gisin}, \bibfnamefont{N.}},
  \bibinfo{author}{\bibfnamefont{G.}~\bibnamefont{Ribordy}},
  \bibinfo{author}{\bibfnamefont{W.}~\bibnamefont{Tittel}}, and
  \bibinfo{author}{\bibfnamefont{H.}~\bibnamefont{Zbinden}},
  \bibinfo{year}{2002}, \bibinfo{journal}{Rev. Mod. Phys.}
  \textbf{\bibinfo{volume}{74}}, \bibinfo{pages}{145}.

\bibitem[{\citenamefont{Goldenberg and Vaidman}(1996)}]{prl77.3265}
\bibinfo{author}{\bibnamefont{Goldenberg}, \bibfnamefont{L.}}, and
  \bibinfo{author}{\bibfnamefont{L.}~\bibnamefont{Vaidman}},
  \bibinfo{year}{1996}, \bibinfo{journal}{Phys. Rev. Lett.}
  \textbf{\bibinfo{volume}{77}}, \bibinfo{pages}{3265}.

\bibitem[{\citenamefont{Gottesman}(1998)}]{gottesman98}
\bibinfo{author}{\bibnamefont{Gottesman}, \bibfnamefont{D.}},
  \bibinfo{year}{1998}, \bibinfo{journal}{Phys. Rev. A}
  \textbf{\bibinfo{volume}{57}}, \bibinfo{pages}{127}.

\bibitem[{\citenamefont{Griffiths and Niu}(1997)}]{PhysRevA.56.1173}
\bibinfo{author}{\bibnamefont{Griffiths}, \bibfnamefont{R.~B.}}, and
  \bibinfo{author}{\bibfnamefont{C.~S.} \bibnamefont{Niu}},
  \bibinfo{year}{1997}, \bibinfo{journal}{Phys. Rev. A}
  \textbf{\bibinfo{volume}{56}}, \bibinfo{pages}{1173}.

\bibitem[{\citenamefont{Grosshans and Cerf}(2004)}]{prl92.047905}
\bibinfo{author}{\bibnamefont{Grosshans}, \bibfnamefont{F.}}, and
  \bibinfo{author}{\bibfnamefont{N.~J.} \bibnamefont{Cerf}},
  \bibinfo{year}{2004}, \bibinfo{journal}{Phys. Rev. Lett.}
  \textbf{\bibinfo{volume}{92}}, \bibinfo{pages}{047905}.

\bibitem[{\citenamefont{Grosshans and Grangier}(2001)}]{GrosshansGrangier}
\bibinfo{author}{\bibnamefont{Grosshans}, \bibfnamefont{F.}}, and
  \bibinfo{author}{\bibfnamefont{P.}~\bibnamefont{Grangier}},
  \bibinfo{year}{2001}, \bibinfo{journal}{Phys. Rev. A}
  \textbf{\bibinfo{volume}{64}}, \bibinfo{pages}{010301}.

\bibitem[{\citenamefont{Grosshans and Grangier}(2002)}]{prl88.057902}
\bibinfo{author}{\bibnamefont{Grosshans}, \bibfnamefont{F.}}, and
  \bibinfo{author}{\bibfnamefont{P.}~\bibnamefont{Grangier}},
  \bibinfo{year}{2002}, \bibinfo{journal}{Phys. Rev. Lett.}
  \textbf{\bibinfo{volume}{88}}, \bibinfo{pages}{057902}.

\bibitem[{\citenamefont{Guta and Matsumoto}(2006)}]{ISI:000241067100032}
\bibinfo{author}{\bibnamefont{Guta}, \bibfnamefont{M.}}, and
  \bibinfo{author}{\bibfnamefont{K.}~\bibnamefont{Matsumoto}},
  \bibinfo{year}{2006}, \bibinfo{journal}{Phys. Rev. A}
  \textbf{\bibinfo{volume}{74}}, \bibinfo{pages}{032305}.

\bibitem[{\citenamefont{Han} \emph{et~al.}(2010)\citenamefont{Han, Bae, Wang,
  and Hwang}}]{Xiangbin-Hwang}
\bibinfo{author}{\bibnamefont{Han}, \bibfnamefont{Y.~D.}},
  \bibinfo{author}{\bibfnamefont{J.}~\bibnamefont{Bae}},
  \bibinfo{author}{\bibfnamefont{X.~B.} \bibnamefont{Wang}}, and
  \bibinfo{author}{\bibfnamefont{W.~Y.} \bibnamefont{Hwang}},
  \bibinfo{year}{2010}, \bibinfo{journal}{Phys. Rev. A}
  \textbf{\bibinfo{volume}{82}}, \bibinfo{pages}{062318}.

\bibitem[{\citenamefont{Hardy and Song}(1999)}]{Hardy1999331}
\bibinfo{author}{\bibnamefont{Hardy}, \bibfnamefont{L.}}, and
  \bibinfo{author}{\bibfnamefont{D.~D.} \bibnamefont{Song}},
  \bibinfo{year}{1999}, \bibinfo{journal}{Phys. Lett. A}
  \textbf{\bibinfo{volume}{259}}, \bibinfo{pages}{331 }.

\bibitem[{\citenamefont{Hayashi} \emph{et~al.}(2005)\citenamefont{Hayashi,
  Horibe, and Hashimoto}}]{LatinSquare}
\bibinfo{author}{\bibnamefont{Hayashi}, \bibfnamefont{A.}},
  \bibinfo{author}{\bibfnamefont{M.}~\bibnamefont{Horibe}}, and
  \bibinfo{author}{\bibfnamefont{T.}~\bibnamefont{Hashimoto}},
  \bibinfo{year}{2005}, \bibinfo{journal}{Phys. Rev. A}
  \textbf{\bibinfo{volume}{71}}, \bibinfo{pages}{052331}.

\bibitem[{\citenamefont{Hayden} \emph{et~al.}(2004)\citenamefont{Hayden, Jozsa,
  Petz, and Winter}}]{HaydenCMP}
\bibinfo{author}{\bibnamefont{Hayden}, \bibfnamefont{P.}},
  \bibinfo{author}{\bibfnamefont{R.}~\bibnamefont{Jozsa}},
  \bibinfo{author}{\bibfnamefont{D.}~\bibnamefont{Petz}}, and
  \bibinfo{author}{\bibfnamefont{A.}~\bibnamefont{Winter}},
  \bibinfo{year}{2004}, \bibinfo{journal}{Commun. Math. Phys.}
  \textbf{\bibinfo{volume}{246}}, \bibinfo{pages}{359}.

\bibitem[{\citenamefont{Herbert}(1982)}]{Herbert1982}
\bibinfo{author}{\bibnamefont{Herbert}, \bibfnamefont{N.}},
  \bibinfo{year}{1982}, \bibinfo{journal}{Found. Phys.}
  \textbf{\bibinfo{volume}{12}}, \bibinfo{pages}{1171}.

\bibitem[{\citenamefont{Hetet} \emph{et~al.}(2008)\citenamefont{Hetet,
  Longdell, Alexander, Lam, and Sellars}}]{HetetLongdell}
\bibinfo{author}{\bibnamefont{Hetet}, \bibfnamefont{G.}},
  \bibinfo{author}{\bibfnamefont{J.~J.} \bibnamefont{Longdell}},
  \bibinfo{author}{\bibfnamefont{A.~L.} \bibnamefont{Alexander}},
  \bibinfo{author}{\bibfnamefont{P.~K.} \bibnamefont{Lam}}, and
  \bibinfo{author}{\bibfnamefont{M.~J.} \bibnamefont{Sellars}},
  \bibinfo{year}{2008}, \bibinfo{journal}{Phys. Rev. Lett.}
  \textbf{\bibinfo{volume}{100}}, \bibinfo{pages}{023601}.

\bibitem[{\citenamefont{Hillery and Bu\v{z}ek}(1997)}]{hillery-buzek}
\bibinfo{author}{\bibnamefont{Hillery}, \bibfnamefont{M.}}, and
  \bibinfo{author}{\bibfnamefont{V.}~\bibnamefont{Bu\v{z}ek}},
  \bibinfo{year}{1997}, \bibinfo{journal}{Phys. Rev. A}
  \textbf{\bibinfo{volume}{56}}, \bibinfo{pages}{1212}.

\bibitem[{\citenamefont{Holevo}(1982)}]{Holevo1982}
\bibinfo{author}{\bibnamefont{Holevo}, \bibfnamefont{A.~S.}},
  \bibinfo{year}{1982}, \emph{\bibinfo{title}{Probabilistic and Statistical
  Aspects of Quantum Theory}} (\bibinfo{publisher}{North-Holland, Amsterdam}).

\bibitem[{\citenamefont{Horodecki} \emph{et~al.}(1996)\citenamefont{Horodecki,
  Horodecki, and Horodecki}}]{Horodecki19961}
\bibinfo{author}{\bibnamefont{Horodecki}, \bibfnamefont{M.}},
  \bibinfo{author}{\bibfnamefont{P.}~\bibnamefont{Horodecki}}, and
  \bibinfo{author}{\bibfnamefont{R.}~\bibnamefont{Horodecki}},
  \bibinfo{year}{1996}, \bibinfo{journal}{Phys. Lett. A}
  \textbf{\bibinfo{volume}{223}}, \bibinfo{pages}{1 }.

\bibitem[{\citenamefont{Horodecki} \emph{et~al.}(2006)\citenamefont{Horodecki,
  Horodecki, Horodecki, and Piani}}]{ISI:000237090400009}
\bibinfo{author}{\bibnamefont{Horodecki}, \bibfnamefont{M.}},
  \bibinfo{author}{\bibfnamefont{P.}~\bibnamefont{Horodecki}},
  \bibinfo{author}{\bibfnamefont{R.}~\bibnamefont{Horodecki}}, and
  \bibinfo{author}{\bibfnamefont{M.}~\bibnamefont{Piani}},
  \bibinfo{year}{2006}, \bibinfo{journal}{Int. J. Quant. Inf.}
  \textbf{\bibinfo{volume}{4}}, \bibinfo{pages}{105}.

\bibitem[{\citenamefont{Horodecki and Horodecki}(1998)}]{ISI:000075192800004}
\bibinfo{author}{\bibnamefont{Horodecki}, \bibfnamefont{M.}}, and
  \bibinfo{author}{\bibfnamefont{R.}~\bibnamefont{Horodecki}},
  \bibinfo{year}{1998}, \bibinfo{journal}{Phys. Lett. A}
  \textbf{\bibinfo{volume}{244}}, \bibinfo{pages}{473}.

\bibitem[{\citenamefont{Horodecki} \emph{et~al.}(2009)\citenamefont{Horodecki,
  Horodecki, Horodecki, and Horodecki}}]{RMP-Horodecki}
\bibinfo{author}{\bibnamefont{Horodecki}, \bibfnamefont{R.}},
  \bibinfo{author}{\bibfnamefont{P.}~\bibnamefont{Horodecki}},
  \bibinfo{author}{\bibfnamefont{M.}~\bibnamefont{Horodecki}}, and
  \bibinfo{author}{\bibfnamefont{K.}~\bibnamefont{Horodecki}},
  \bibinfo{year}{2009}, \bibinfo{journal}{Rev. Mod. Phys.}
  \textbf{\bibinfo{volume}{81}}, \bibinfo{pages}{865}.

\bibitem[{\citenamefont{Hu} \emph{et~al.}(2010)\citenamefont{Hu, Zhang, and
  Ye}}]{Hu2010}
\bibinfo{author}{\bibnamefont{Hu}, \bibfnamefont{G.~Y.}},
  \bibinfo{author}{\bibfnamefont{W.~H.} \bibnamefont{Zhang}}, and
  \bibinfo{author}{\bibfnamefont{L.}~\bibnamefont{Ye}}, \bibinfo{year}{2010},
  \bibinfo{journal}{Opt. Commun.} \textbf{\bibinfo{volume}{283}},
  \bibinfo{pages}{200}.

\bibitem[{\citenamefont{Hu} \emph{et~al.}(2009)\citenamefont{Hu, Yu, and
  Wang}}]{yuzw}
\bibinfo{author}{\bibnamefont{Hu}, \bibfnamefont{J.~Z.}},
  \bibinfo{author}{\bibfnamefont{Z.~W.} \bibnamefont{Yu}}, and
  \bibinfo{author}{\bibfnamefont{X.~B.} \bibnamefont{Wang}},
  \bibinfo{year}{2009}, \bibinfo{journal}{Eur. Phys. J. D}
  \textbf{\bibinfo{volume}{51}}, \bibinfo{pages}{381}.

\bibitem[{\citenamefont{Huang} \emph{et~al.}(2001)\citenamefont{Huang, Li, Li,
  Zhang, Jiang, and Guo}}]{pra64.012315}
\bibinfo{author}{\bibnamefont{Huang}, \bibfnamefont{Y.~F.}},
  \bibinfo{author}{\bibfnamefont{W.~L.} \bibnamefont{Li}},
  \bibinfo{author}{\bibfnamefont{C.~F.} \bibnamefont{Li}},
  \bibinfo{author}{\bibfnamefont{Y.~S.} \bibnamefont{Zhang}},
  \bibinfo{author}{\bibfnamefont{Y.~K.} \bibnamefont{Jiang}}, and
  \bibinfo{author}{\bibfnamefont{G.~C.} \bibnamefont{Guo}},
  \bibinfo{year}{2001}, \bibinfo{journal}{Phys. Rev. A}
  \textbf{\bibinfo{volume}{64}}, \bibinfo{pages}{012315}.

\bibitem[{\citenamefont{Iblisdir}
  \emph{et~al.}(2005{\natexlab{a}})\citenamefont{Iblisdir, Ac\'{i}n, Cerf,
  Filip, Fiur\'{a}\v{s}ek, and Gisin}}]{PhysRevA.72.042328}
\bibinfo{author}{\bibnamefont{Iblisdir}, \bibfnamefont{S.}},
  \bibinfo{author}{\bibfnamefont{A.}~\bibnamefont{Ac\'{i}n}},
  \bibinfo{author}{\bibfnamefont{N.~J.} \bibnamefont{Cerf}},
  \bibinfo{author}{\bibfnamefont{R.}~\bibnamefont{Filip}},
  \bibinfo{author}{\bibfnamefont{J.}~\bibnamefont{Fiur\'{a}\v{s}ek}}, and
  \bibinfo{author}{\bibfnamefont{N.}~\bibnamefont{Gisin}},
  \bibinfo{year}{2005}{\natexlab{a}}, \bibinfo{journal}{Phys. Rev. A}
  \textbf{\bibinfo{volume}{72}}, \bibinfo{pages}{042328}.

\bibitem[{\citenamefont{Iblisdir}
  \emph{et~al.}(2005{\natexlab{b}})\citenamefont{Iblisdir, Ac\'{i}n, and
  Gisin}}]{Iblisdir2005}
\bibinfo{author}{\bibnamefont{Iblisdir}, \bibfnamefont{S.}},
  \bibinfo{author}{\bibfnamefont{A.}~\bibnamefont{Ac\'{i}n}}, and
  \bibinfo{author}{\bibfnamefont{N.}~\bibnamefont{Gisin}},
  \bibinfo{year}{2005}{\natexlab{b}},
  \bibinfo{journal}{arXiv:quant-ph/0505152v1} .

\bibitem[{\citenamefont{Irvine} \emph{et~al.}(2004)\citenamefont{Irvine,
  Lamas~Linares, de~Dood, and Bouwmeester}}]{PhysRevLett.92.047902}
\bibinfo{author}{\bibnamefont{Irvine}, \bibfnamefont{W.~T.~M.}},
  \bibinfo{author}{\bibfnamefont{A.}~\bibnamefont{Lamas~Linares}},
  \bibinfo{author}{\bibfnamefont{M.~J.~A.} \bibnamefont{de~Dood}}, and
  \bibinfo{author}{\bibfnamefont{D.}~\bibnamefont{Bouwmeester}},
  \bibinfo{year}{2004}, \bibinfo{journal}{Phys. Rev. Lett.}
  \textbf{\bibinfo{volume}{92}}, \bibinfo{pages}{047902}.

\bibitem[{\citenamefont{Ishizaka and Hiroshima}(2008)}]{Ishizaka-Hiroshima}
\bibinfo{author}{\bibnamefont{Ishizaka}, \bibfnamefont{S.}}, and
  \bibinfo{author}{\bibfnamefont{T.}~\bibnamefont{Hiroshima}},
  \bibinfo{year}{2008}, \bibinfo{journal}{Phys. Rev. Lett.}
  \textbf{\bibinfo{volume}{101}}, \bibinfo{pages}{240501}.

\bibitem[{\citenamefont{Janzing and Steudel}(2007)}]{ISI:000244532300030}
\bibinfo{author}{\bibnamefont{Janzing}, \bibfnamefont{D.}}, and
  \bibinfo{author}{\bibfnamefont{B.}~\bibnamefont{Steudel}},
  \bibinfo{year}{2007}, \bibinfo{journal}{Phys. Rev. A}
  \textbf{\bibinfo{volume}{75}}, \bibinfo{pages}{022309}.

\bibitem[{\citenamefont{Ji} \emph{et~al.}(2005)\citenamefont{Ji, Feng, and
  Ying}}]{yingms05}
\bibinfo{author}{\bibnamefont{Ji}, \bibfnamefont{Z.~F.}},
  \bibinfo{author}{\bibfnamefont{Y.}~\bibnamefont{Feng}}, and
  \bibinfo{author}{\bibfnamefont{M.~S.} \bibnamefont{Ying}},
  \bibinfo{year}{2005}, \bibinfo{journal}{Phys. Rev. A}
  \textbf{\bibinfo{volume}{72}}, \bibinfo{pages}{032324}.

\bibitem[{\citenamefont{Jiang and
  Yu}(2010{\natexlab{a}})}]{ISI:000278182800016}
\bibinfo{author}{\bibnamefont{Jiang}, \bibfnamefont{M.~M.}}, and
  \bibinfo{author}{\bibfnamefont{S.~X.} \bibnamefont{Yu}},
  \bibinfo{year}{2010}{\natexlab{a}}, \bibinfo{journal}{J. Math. Phys.}
  \textbf{\bibinfo{volume}{51}}, \bibinfo{pages}{052306}.

\bibitem[{\citenamefont{Jiang and
  Yu}(2010{\natexlab{b}})}]{ISI:000273639000005}
\bibinfo{author}{\bibnamefont{Jiang}, \bibfnamefont{M.~M.}}, and
  \bibinfo{author}{\bibfnamefont{S.~X.} \bibnamefont{Yu}},
  \bibinfo{year}{2010}{\natexlab{b}}, \bibinfo{journal}{Chin. Phys. Lett.}
  \textbf{\bibinfo{volume}{27}}, \bibinfo{pages}{010303}.

\bibitem[{\citenamefont{Jimenez}
  \emph{et~al.}(2010{\natexlab{a}})\citenamefont{Jimenez, Bergou, and
  Delgado}}]{Jimenez2010}
\bibinfo{author}{\bibnamefont{Jimenez}, \bibfnamefont{O.}},
  \bibinfo{author}{\bibfnamefont{J.}~\bibnamefont{Bergou}}, and
  \bibinfo{author}{\bibfnamefont{A.}~\bibnamefont{Delgado}},
  \bibinfo{year}{2010}{\natexlab{a}}, \bibinfo{journal}{Phys. Rev. A}
  \textbf{\bibinfo{volume}{82}}, \bibinfo{pages}{062307}.

\bibitem[{\citenamefont{Jimenez}
  \emph{et~al.}(2010{\natexlab{b}})\citenamefont{Jimenez, Roa, and
  Delgado}}]{Jimenez2010a}
\bibinfo{author}{\bibnamefont{Jimenez}, \bibfnamefont{O.}},
  \bibinfo{author}{\bibfnamefont{L.}~\bibnamefont{Roa}}, and
  \bibinfo{author}{\bibfnamefont{A.}~\bibnamefont{Delgado}},
  \bibinfo{year}{2010}{\natexlab{b}}, \bibinfo{journal}{Phys. Rev. A}
  \textbf{\bibinfo{volume}{82}}, \bibinfo{pages}{022328}.

\bibitem[{\citenamefont{Jing} \emph{et~al.}(2012)\citenamefont{Jing, Wang, Shi,
  Mu, and Fan}}]{Jing2013}
\bibinfo{author}{\bibnamefont{Jing}, \bibfnamefont{L.}},
  \bibinfo{author}{\bibfnamefont{Y.~N.} \bibnamefont{Wang}},
  \bibinfo{author}{\bibfnamefont{H.~D.} \bibnamefont{Shi}},
  \bibinfo{author}{\bibfnamefont{L.~Z.} \bibnamefont{Mu}}, and
  \bibinfo{author}{\bibfnamefont{H.}~\bibnamefont{Fan}}, \bibinfo{year}{2012},
  \bibinfo{journal}{Phys. Rev. A} \textbf{\bibinfo{volume}{86}},
  \bibinfo{pages}{062315}.

\bibitem[{\citenamefont{Jochym-O'Connor}
  \emph{et~al.}(2011)\citenamefont{Jochym-O'Connor, Bradler, and
  Wilde}}]{Jochym-OConnor2011}
\bibinfo{author}{\bibnamefont{Jochym-O'Connor}, \bibfnamefont{T.}},
  \bibinfo{author}{\bibfnamefont{K.}~\bibnamefont{Bradler}}, and
  \bibinfo{author}{\bibfnamefont{M.~M.} \bibnamefont{Wilde}},
  \bibinfo{year}{2011}, \bibinfo{journal}{J. Phys. A-Math. Theor.}
  \textbf{\bibinfo{volume}{44}}, \bibinfo{pages}{415306}.

\bibitem[{\citenamefont{Josza}(1994)}]{Josza1994}
\bibinfo{author}{\bibnamefont{Josza}, \bibfnamefont{R.}}, \bibinfo{year}{1994},
  \bibinfo{journal}{J. Mod. Optic.} \textbf{\bibinfo{volume}{41}},
  \bibinfo{pages}{2315}.

\bibitem[{\citenamefont{Kalev and Hen}(2008)}]{PhysRevLett.100.210502}
\bibinfo{author}{\bibnamefont{Kalev}, \bibfnamefont{A.}}, and
  \bibinfo{author}{\bibfnamefont{I.}~\bibnamefont{Hen}}, \bibinfo{year}{2008},
  \bibinfo{journal}{Phys. Rev. Lett.} \textbf{\bibinfo{volume}{100}},
  \bibinfo{pages}{210502}.

\bibitem[{\citenamefont{Karimipour}
  \emph{et~al.}(2002)\citenamefont{Karimipour, Bahraminasab, and
  Bagherinezhad}}]{Karimipour}
\bibinfo{author}{\bibnamefont{Karimipour}, \bibfnamefont{V.}},
  \bibinfo{author}{\bibfnamefont{A.}~\bibnamefont{Bahraminasab}}, and
  \bibinfo{author}{\bibfnamefont{S.}~\bibnamefont{Bagherinezhad}},
  \bibinfo{year}{2002}, \bibinfo{journal}{Phys. Rev. A}
  \textbf{\bibinfo{volume}{65}}, \bibinfo{pages}{052331}.

\bibitem[{\citenamefont{Karimipour and Rezakhani}(2002)}]{Karimipour-phase}
\bibinfo{author}{\bibnamefont{Karimipour}, \bibfnamefont{V.}}, and
  \bibinfo{author}{\bibfnamefont{A.~T.} \bibnamefont{Rezakhani}},
  \bibinfo{year}{2002}, \bibinfo{journal}{Phys. Rev. A}
  \textbf{\bibinfo{volume}{66}}, \bibinfo{pages}{052111}.

\bibitem[{\citenamefont{Karpov} \emph{et~al.}(2005)\citenamefont{Karpov, Navez,
  and Cerf}}]{c-entangle}
\bibinfo{author}{\bibnamefont{Karpov}, \bibfnamefont{E.}},
  \bibinfo{author}{\bibfnamefont{P.}~\bibnamefont{Navez}}, and
  \bibinfo{author}{\bibfnamefont{N.~J.} \bibnamefont{Cerf}},
  \bibinfo{year}{2005}, \bibinfo{journal}{Phys. Rev. A}
  \textbf{\bibinfo{volume}{72}}, \bibinfo{pages}{042314}.

\bibitem[{\citenamefont{Kay} \emph{et~al.}(2009)\citenamefont{Kay,
  Kaszlikowski, and Ramanathan}}]{Kay2009}
\bibinfo{author}{\bibnamefont{Kay}, \bibfnamefont{A.}},
  \bibinfo{author}{\bibfnamefont{D.}~\bibnamefont{Kaszlikowski}}, and
  \bibinfo{author}{\bibfnamefont{R.}~\bibnamefont{Ramanathan}},
  \bibinfo{year}{2009}, \bibinfo{journal}{Phys. Rev. Lett.}
  \textbf{\bibinfo{volume}{103}}, \bibinfo{pages}{050501}.

\bibitem[{\citenamefont{Kay} \emph{et~al.}(2012)\citenamefont{Kay, Ramanathan,
  and Kaszlikowski}}]{KayDagPreprint}
\bibinfo{author}{\bibnamefont{Kay}, \bibfnamefont{A.}},
  \bibinfo{author}{\bibfnamefont{R.}~\bibnamefont{Ramanathan}}, and
  \bibinfo{author}{\bibfnamefont{D.}~\bibnamefont{Kaszlikowski}},
  \bibinfo{year}{2012}, \bibinfo{journal}{arXiv:1208.5574} .

\bibitem[{\citenamefont{Kazakov}(2010)}]{ISI:000278194800002}
\bibinfo{author}{\bibnamefont{Kazakov}, \bibfnamefont{A.~Y.}},
  \bibinfo{year}{2010}, \bibinfo{journal}{Int. J. Quant. Inf.}
  \textbf{\bibinfo{volume}{8}}, \bibinfo{pages}{435}.

\bibitem[{\citenamefont{Kempe} \emph{et~al.}(2000)\citenamefont{Kempe, Simon,
  and Weihs}}]{Kempe00}
\bibinfo{author}{\bibnamefont{Kempe}, \bibfnamefont{J.}},
  \bibinfo{author}{\bibfnamefont{C.}~\bibnamefont{Simon}}, and
  \bibinfo{author}{\bibfnamefont{G.}~\bibnamefont{Weihs}},
  \bibinfo{year}{2000}, \bibinfo{journal}{Phys. Rev. A}
  \textbf{\bibinfo{volume}{62}}, \bibinfo{pages}{032302}.

\bibitem[{\citenamefont{Keyl and Werner}(1999)}]{ISI:000081019000005}
\bibinfo{author}{\bibnamefont{Keyl}, \bibfnamefont{M.}}, and
  \bibinfo{author}{\bibfnamefont{R.~F.} \bibnamefont{Werner}},
  \bibinfo{year}{1999}, \bibinfo{journal}{J. Math. Phys.}
  \textbf{\bibinfo{volume}{40}}, \bibinfo{pages}{3283}.

\bibitem[{\citenamefont{Khan and Howell}(2003)}]{Khan-Howell}
\bibinfo{author}{\bibnamefont{Khan}, \bibfnamefont{I.~A.}}, and
  \bibinfo{author}{\bibfnamefont{J.~C.} \bibnamefont{Howell}},
  \bibinfo{year}{2003}, \bibinfo{journal}{Phys. Rev. A}
  \textbf{\bibinfo{volume}{70}}, \bibinfo{pages}{010303(R)}.

\bibitem[{\citenamefont{Khan and Howell}(2004)}]{ISI:000221424900003}
\bibinfo{author}{\bibnamefont{Khan}, \bibfnamefont{I.~A.}}, and
  \bibinfo{author}{\bibfnamefont{J.~C.} \bibnamefont{Howell}},
  \bibinfo{year}{2004}, \bibinfo{journal}{Quant. Inf. Comput.}
  \textbf{\bibinfo{volume}{4}}, \bibinfo{pages}{114}.

\bibitem[{\citenamefont{Kimura} \emph{et~al.}(2006)\citenamefont{Kimura,
  Tanaka, and Ozawa}}]{OrthogonalArray}
\bibinfo{author}{\bibnamefont{Kimura}, \bibfnamefont{G.}},
  \bibinfo{author}{\bibfnamefont{H.}~\bibnamefont{Tanaka}}, and
  \bibinfo{author}{\bibfnamefont{M.}~\bibnamefont{Ozawa}},
  \bibinfo{year}{2006}, \bibinfo{journal}{Phys. Rev. A}
  \textbf{\bibinfo{volume}{73}}, \bibinfo{pages}{050301(R)}.

\bibitem[{\citenamefont{Kitaev}(2001)}]{Kitaev}
\bibinfo{author}{\bibnamefont{Kitaev}, \bibfnamefont{A.~Y.}},
  \bibinfo{year}{2001}, \bibinfo{journal}{Phys. USP.}
  \textbf{\bibinfo{volume}{44}}, \bibinfo{pages}{131}.

\bibitem[{\citenamefont{Koashi and Imoto}(1996)}]{koashi-crypto-mixed}
\bibinfo{author}{\bibnamefont{Koashi}, \bibfnamefont{M.}}, and
  \bibinfo{author}{\bibfnamefont{N.}~\bibnamefont{Imoto}},
  \bibinfo{year}{1996}, \bibinfo{journal}{Phys. Rev. Lett.}
  \textbf{\bibinfo{volume}{77}}, \bibinfo{pages}{2137}.

\bibitem[{\citenamefont{Koashi and Imoto}(1998)}]{prl81.4264}
\bibinfo{author}{\bibnamefont{Koashi}, \bibfnamefont{M.}}, and
  \bibinfo{author}{\bibfnamefont{N.}~\bibnamefont{Imoto}},
  \bibinfo{year}{1998}, \bibinfo{journal}{Phys. Rev. Lett.}
  \textbf{\bibinfo{volume}{81}}, \bibinfo{pages}{4264}.

\bibitem[{\citenamefont{Koashi and Imoto}(2002)}]{ISI:000177872600044}
\bibinfo{author}{\bibnamefont{Koashi}, \bibfnamefont{M.}}, and
  \bibinfo{author}{\bibfnamefont{N.}~\bibnamefont{Imoto}},
  \bibinfo{year}{2002}, \bibinfo{journal}{Phys. Rev. A}
  \textbf{\bibinfo{volume}{66}}, \bibinfo{pages}{022318}.

\bibitem[{\citenamefont{Kocsis} \emph{et~al.}(2013)\citenamefont{Kocsis, Xiang,
  Ralph, and Pryde}}]{XiangGuoYong2}
\bibinfo{author}{\bibnamefont{Kocsis}, \bibfnamefont{S.}},
  \bibinfo{author}{\bibfnamefont{G.~Y.} \bibnamefont{Xiang}},
  \bibinfo{author}{\bibfnamefont{T.~C.} \bibnamefont{Ralph}}, and
  \bibinfo{author}{\bibfnamefont{G.~J.} \bibnamefont{Pryde}},
  \bibinfo{year}{2013}, \bibinfo{journal}{Nat. Phys.}
  \textbf{\bibinfo{volume}{9}}, \bibinfo{pages}{23}.

\bibitem[{\citenamefont{Koike} \emph{et~al.}(2006)\citenamefont{Koike,
  Takahashi, Yonezawa, Takei, Braunstein, Aoki, and Furusawa}}]{KoikeTakahashi}
\bibinfo{author}{\bibnamefont{Koike}, \bibfnamefont{S.}},
  \bibinfo{author}{\bibfnamefont{H.}~\bibnamefont{Takahashi}},
  \bibinfo{author}{\bibfnamefont{H.}~\bibnamefont{Yonezawa}},
  \bibinfo{author}{\bibfnamefont{N.}~\bibnamefont{Takei}},
  \bibinfo{author}{\bibfnamefont{S.~L.} \bibnamefont{Braunstein}},
  \bibinfo{author}{\bibfnamefont{T.}~\bibnamefont{Aoki}}, and
  \bibinfo{author}{\bibfnamefont{A.}~\bibnamefont{Furusawa}},
  \bibinfo{year}{2006}, \bibinfo{journal}{Phys. Rev. Lett.}
  \textbf{\bibinfo{volume}{96}}, \bibinfo{pages}{060504}.

\bibitem[{\citenamefont{Korbicz} \emph{et~al.}(2014)\citenamefont{Korbicz,
  Horodecki, and Horodecki}}]{Horodecki-information-broadcasting}
\bibinfo{author}{\bibnamefont{Korbicz}, \bibfnamefont{J.~K.}},
  \bibinfo{author}{\bibfnamefont{P.}~\bibnamefont{Horodecki}}, and
  \bibinfo{author}{\bibfnamefont{R.}~\bibnamefont{Horodecki}},
  \bibinfo{year}{2014}, \bibinfo{journal}{Phys. Rev. Lett.}
  \textbf{\bibinfo{volume}{112}}, \bibinfo{pages}{120402}.

\bibitem[{\citenamefont{Kraus} \emph{et~al.}(2005)\citenamefont{Kraus, Gisin,
  and Renner}}]{prl95.080501}
\bibinfo{author}{\bibnamefont{Kraus}, \bibfnamefont{B.}},
  \bibinfo{author}{\bibfnamefont{N.}~\bibnamefont{Gisin}}, and
  \bibinfo{author}{\bibfnamefont{R.}~\bibnamefont{Renner}},
  \bibinfo{year}{2005}, \bibinfo{journal}{Phys. Rev. Lett.}
  \textbf{\bibinfo{volume}{95}}, \bibinfo{pages}{080501}.

\bibitem[{\citenamefont{Kretschmann}
  \emph{et~al.}(2008)\citenamefont{Kretschmann, Schlingemann, and
  Werner}}]{ISI:000254475300022}
\bibinfo{author}{\bibnamefont{Kretschmann}, \bibfnamefont{D.}},
  \bibinfo{author}{\bibfnamefont{D.}~\bibnamefont{Schlingemann}}, and
  \bibinfo{author}{\bibfnamefont{R.~F.} \bibnamefont{Werner}},
  \bibinfo{year}{2008}, \bibinfo{journal}{IEEE Trans. Inf. Theory}
  \textbf{\bibinfo{volume}{54}}, \bibinfo{pages}{1708}.

\bibitem[{\citenamefont{Kwek} \emph{et~al.}(2000)\citenamefont{Kwek, Oh, Wang,
  and Yeo}}]{ISI:000165200800033}
\bibinfo{author}{\bibnamefont{Kwek}, \bibfnamefont{L.~C.}},
  \bibinfo{author}{\bibfnamefont{C.~H.} \bibnamefont{Oh}},
  \bibinfo{author}{\bibfnamefont{X.~B.} \bibnamefont{Wang}}, and
  \bibinfo{author}{\bibfnamefont{Y.}~\bibnamefont{Yeo}}, \bibinfo{year}{2000},
  \bibinfo{journal}{Phys. Rev. A} \textbf{\bibinfo{volume}{62}},
  \bibinfo{pages}{052313}.

\bibitem[{\citenamefont{Lamas-Linares}
  \emph{et~al.}(2002)\citenamefont{Lamas-Linares, Simon, Howell, and
  Bouwmeester}}]{667603820020426}
\bibinfo{author}{\bibnamefont{Lamas-Linares}, \bibfnamefont{A.}},
  \bibinfo{author}{\bibfnamefont{C.}~\bibnamefont{Simon}},
  \bibinfo{author}{\bibfnamefont{J.~C.} \bibnamefont{Howell}}, and
  \bibinfo{author}{\bibfnamefont{D.}~\bibnamefont{Bouwmeester}},
  \bibinfo{year}{2002}, \bibinfo{journal}{Science}
  \textbf{\bibinfo{volume}{296}}, \bibinfo{pages}{712}.

\bibitem[{\citenamefont{Lamata} \emph{et~al.}(2008)\citenamefont{Lamata, Leon,
  Perez-Garcia, Salgado, and Solano}}]{lamatasequential}
\bibinfo{author}{\bibnamefont{Lamata}, \bibfnamefont{L.}},
  \bibinfo{author}{\bibfnamefont{J.}~\bibnamefont{Leon}},
  \bibinfo{author}{\bibfnamefont{D.}~\bibnamefont{Perez-Garcia}},
  \bibinfo{author}{\bibfnamefont{D.}~\bibnamefont{Salgado}}, and
  \bibinfo{author}{\bibfnamefont{E.}~\bibnamefont{Solano}},
  \bibinfo{year}{2008}, \bibinfo{journal}{Phys. Rev. Lett.}
  \textbf{\bibinfo{volume}{101}}, \bibinfo{pages}{180506}.

\bibitem[{\citenamefont{Lamoureux and Cerf}(2005)}]{ISI:000228015900003}
\bibinfo{author}{\bibnamefont{Lamoureux}, \bibfnamefont{L.~P.}}, and
  \bibinfo{author}{\bibfnamefont{N.~J.} \bibnamefont{Cerf}},
  \bibinfo{year}{2005}, \bibinfo{journal}{Quant. Inf. Comput.}
  \textbf{\bibinfo{volume}{5}}, \bibinfo{pages}{32}.

\bibitem[{\citenamefont{Lamoureux} \emph{et~al.}(2004)\citenamefont{Lamoureux,
  Navez, Fiur\'{a}\v{s}ek, and Cerf}}]{c-entagle0}
\bibinfo{author}{\bibnamefont{Lamoureux}, \bibfnamefont{L.~P.}},
  \bibinfo{author}{\bibfnamefont{P.}~\bibnamefont{Navez}},
  \bibinfo{author}{\bibfnamefont{J.}~\bibnamefont{Fiur\'{a}\v{s}ek}}, and
  \bibinfo{author}{\bibfnamefont{N.~J.} \bibnamefont{Cerf}},
  \bibinfo{year}{2004}, \bibinfo{journal}{Phys. Rev. A}
  \textbf{\bibinfo{volume}{69}}, \bibinfo{pages}{040301(R)}.

\bibitem[{\citenamefont{Leifer}(2006)}]{ISI:000241723100033}
\bibinfo{author}{\bibnamefont{Leifer}, \bibfnamefont{M.~S.}},
  \bibinfo{year}{2006}, \bibinfo{journal}{Phys. Rev. A}
  \textbf{\bibinfo{volume}{74}}, \bibinfo{pages}{042310}.

\bibitem[{\citenamefont{Lemr} \emph{et~al.}(2012)\citenamefont{Lemr,
  Bartkiewicz, Cernoch, Soubusta, and Miranowicz}}]{Lemr2012}
\bibinfo{author}{\bibnamefont{Lemr}, \bibfnamefont{K.}},
  \bibinfo{author}{\bibfnamefont{K.}~\bibnamefont{Bartkiewicz}},
  \bibinfo{author}{\bibfnamefont{A.}~\bibnamefont{Cernoch}},
  \bibinfo{author}{\bibfnamefont{J.}~\bibnamefont{Soubusta}}, and
  \bibinfo{author}{\bibfnamefont{A.}~\bibnamefont{Miranowicz}},
  \bibinfo{year}{2012}, \bibinfo{journal}{Phys. Rev. A}
  \textbf{\bibinfo{volume}{85}}, \bibinfo{pages}{050307(R)}.

\bibitem[{\citenamefont{Levente} \emph{et~al.}(2010)\citenamefont{Levente,
  Matyas, Peter, and Jozsef}}]{Levente2010}
\bibinfo{author}{\bibnamefont{Levente}, \bibfnamefont{S.}},
  \bibinfo{author}{\bibfnamefont{K.}~\bibnamefont{Matyas}},
  \bibinfo{author}{\bibfnamefont{A.}~\bibnamefont{Peter}}, and
  \bibinfo{author}{\bibfnamefont{J.}~\bibnamefont{Jozsef}},
  \bibinfo{year}{2010}, \bibinfo{journal}{Phys. Rev. A}
  \textbf{\bibinfo{volume}{81}}, \bibinfo{pages}{032323}.

\bibitem[{\citenamefont{Li and Shen}(2009)}]{ISI:000269858900005}
\bibinfo{author}{\bibnamefont{Li}, \bibfnamefont{D.~C.}}, and
  \bibinfo{author}{\bibfnamefont{Z.~Y.} \bibnamefont{Shen}},
  \bibinfo{year}{2009}, \bibinfo{journal}{Int. J. Theor. Phys.}
  \textbf{\bibinfo{volume}{48}}, \bibinfo{pages}{2777}.

\bibitem[{\citenamefont{Li} \emph{et~al.}(2005{\natexlab{a}})\citenamefont{Li,
  Li, Huang, and Li}}]{ISI:000232866900008}
\bibinfo{author}{\bibnamefont{Li}, \bibfnamefont{D.~F.}},
  \bibinfo{author}{\bibfnamefont{X.~R.} \bibnamefont{Li}},
  \bibinfo{author}{\bibfnamefont{H.~T.} \bibnamefont{Huang}}, and
  \bibinfo{author}{\bibfnamefont{X.~X.} \bibnamefont{Li}},
  \bibinfo{year}{2005}{\natexlab{a}}, \bibinfo{journal}{Int. J. Quant. Inf.}
  \textbf{\bibinfo{volume}{3}}, \bibinfo{pages}{551}.

\bibitem[{\citenamefont{Li} \emph{et~al.}(2005{\natexlab{b}})\citenamefont{Li,
  Li, Huang, and Li}}]{ISI:000231391900002}
\bibinfo{author}{\bibnamefont{Li}, \bibfnamefont{D.~F.}},
  \bibinfo{author}{\bibfnamefont{X.~R.} \bibnamefont{Li}},
  \bibinfo{author}{\bibfnamefont{H.~T.} \bibnamefont{Huang}}, and
  \bibinfo{author}{\bibfnamefont{X.~X.} \bibnamefont{Li}},
  \bibinfo{year}{2005}{\natexlab{b}}, \bibinfo{journal}{J. Math. Phys.}
  \textbf{\bibinfo{volume}{46}}, \bibinfo{pages}{082102}.

\bibitem[{\citenamefont{Li and Qiu}(2007)}]{ISI:000244768000011}
\bibinfo{author}{\bibnamefont{Li}, \bibfnamefont{L.~J.}}, and
  \bibinfo{author}{\bibfnamefont{D.~W.} \bibnamefont{Qiu}},
  \bibinfo{year}{2007}, \bibinfo{journal}{Phys. Lett. A}
  \textbf{\bibinfo{volume}{362}}, \bibinfo{pages}{143}.

\bibitem[{\citenamefont{Li} \emph{et~al.}(2009)\citenamefont{Li, Qiu, Li, Wu,
  and Zou}}]{ISI:000264951100014}
\bibinfo{author}{\bibnamefont{Li}, \bibfnamefont{L.~J.}},
  \bibinfo{author}{\bibfnamefont{D.~W.} \bibnamefont{Qiu}},
  \bibinfo{author}{\bibfnamefont{L.~Z.} \bibnamefont{Li}},
  \bibinfo{author}{\bibfnamefont{L.~H.} \bibnamefont{Wu}}, and
  \bibinfo{author}{\bibfnamefont{X.~F.} \bibnamefont{Zou}},
  \bibinfo{year}{2009}, \bibinfo{journal}{J. Phys. A-Math. Theor.}
  \textbf{\bibinfo{volume}{42}}, \bibinfo{pages}{175302}.

\bibitem[{\citenamefont{Li} \emph{et~al.}(2007)\citenamefont{Li, Feng, Liang,
  and Yu}}]{Li2007a}
\bibinfo{author}{\bibnamefont{Li}, \bibfnamefont{Y.~L.}},
  \bibinfo{author}{\bibfnamefont{J.}~\bibnamefont{Feng}},
  \bibinfo{author}{\bibfnamefont{B.~L.} \bibnamefont{Liang}}, and
  \bibinfo{author}{\bibfnamefont{Y.~F.} \bibnamefont{Yu}},
  \bibinfo{year}{2007}, \bibinfo{journal}{Int. J. Theor. Phys.}
  \textbf{\bibinfo{volume}{46}}, \bibinfo{pages}{2599}.

\bibitem[{\citenamefont{Li} \emph{et~al.}(2012)\citenamefont{Li, Zhao, Fei,
  Fan, and Liu}}]{Lizongguo}
\bibinfo{author}{\bibnamefont{Li}, \bibfnamefont{Z.~G.}},
  \bibinfo{author}{\bibfnamefont{M.~J.} \bibnamefont{Zhao}},
  \bibinfo{author}{\bibfnamefont{S.~M.} \bibnamefont{Fei}},
  \bibinfo{author}{\bibfnamefont{H.}~\bibnamefont{Fan}}, and
  \bibinfo{author}{\bibfnamefont{W.~M.} \bibnamefont{Liu}},
  \bibinfo{year}{2012}, \bibinfo{journal}{Quant. Inf. Comput.}
  \textbf{\bibinfo{volume}{12}}, \bibinfo{pages}{63}.

\bibitem[{\citenamefont{Lindblad}(1975)}]{Lindblad1875}
\bibinfo{author}{\bibnamefont{Lindblad}, \bibfnamefont{G.}},
  \bibinfo{year}{1975}, \bibinfo{journal}{Commun. Math. Phys.}
  \textbf{\bibinfo{volume}{40}}, \bibinfo{pages}{147}.

\bibitem[{\citenamefont{Lindblad}(1999)}]{ISI:000079331000009}
\bibinfo{author}{\bibnamefont{Lindblad}, \bibfnamefont{G.}},
  \bibinfo{year}{1999}, \bibinfo{journal}{Lett. Math. Phys.}
  \textbf{\bibinfo{volume}{47}}, \bibinfo{pages}{189}.

\bibitem[{\citenamefont{Lo and Chau}(1999)}]{LoChau99}
\bibinfo{author}{\bibnamefont{Lo}, \bibfnamefont{H.~K.}}, and
  \bibinfo{author}{\bibfnamefont{H.~F.} \bibnamefont{Chau}},
  \bibinfo{year}{1999}, \bibinfo{journal}{Science}
  \textbf{\bibinfo{volume}{283}}, \bibinfo{pages}{2050}.

\bibitem[{\citenamefont{van Loock and Braunstein}(2000)}]{prl84.3482}
\bibinfo{author}{\bibnamefont{van Loock}, \bibfnamefont{P.}}, and
  \bibinfo{author}{\bibfnamefont{S.~L.} \bibnamefont{Braunstein}},
  \bibinfo{year}{2000}, \bibinfo{journal}{Phys. Rev. Lett.}
  \textbf{\bibinfo{volume}{84}}, \bibinfo{pages}{3482}.

\bibitem[{\citenamefont{van Loock and
  Braunstein}(2001)}]{PhysRevLett.87.247901}
\bibinfo{author}{\bibnamefont{van Loock}, \bibfnamefont{P.}}, and
  \bibinfo{author}{\bibfnamefont{S.~L.} \bibnamefont{Braunstein}},
  \bibinfo{year}{2001}, \bibinfo{journal}{Phys. Rev. Lett.}
  \textbf{\bibinfo{volume}{87}}, \bibinfo{pages}{247901}.

\bibitem[{\citenamefont{Luo}(2010{\natexlab{a}})}]{Luo2010}
\bibinfo{author}{\bibnamefont{Luo}, \bibfnamefont{S.~L.}},
  \bibinfo{year}{2010}{\natexlab{a}}, \bibinfo{journal}{Phys. Lett. A}
  \textbf{\bibinfo{volume}{374}}, \bibinfo{pages}{1350}.

\bibitem[{\citenamefont{Luo}(2010{\natexlab{b}})}]{ISI:000277243800004}
\bibinfo{author}{\bibnamefont{Luo}, \bibfnamefont{S.~L.}},
  \bibinfo{year}{2010}{\natexlab{b}}, \bibinfo{journal}{Lett. Math. Phys.}
  \textbf{\bibinfo{volume}{92}}, \bibinfo{pages}{143}.

\bibitem[{\citenamefont{Luo} \emph{et~al.}(2009)\citenamefont{Luo, Li, and
  Cao}}]{ISI:000266500900237}
\bibinfo{author}{\bibnamefont{Luo}, \bibfnamefont{S.~L.}},
  \bibinfo{author}{\bibfnamefont{N.}~\bibnamefont{Li}}, and
  \bibinfo{author}{\bibfnamefont{X.~L.} \bibnamefont{Cao}},
  \bibinfo{year}{2009}, \bibinfo{journal}{Phys. Rev. A}
  \textbf{\bibinfo{volume}{79}}, \bibinfo{pages}{054305}.

\bibitem[{\citenamefont{Luo and Sun}(2010)}]{ISI:000280467400011}
\bibinfo{author}{\bibnamefont{Luo}, \bibfnamefont{S.~L.}}, and
  \bibinfo{author}{\bibfnamefont{W.}~\bibnamefont{Sun}}, \bibinfo{year}{2010},
  \bibinfo{journal}{Phys. Rev. A} \textbf{\bibinfo{volume}{82}},
  \bibinfo{pages}{012338}.

\bibitem[{\citenamefont{Ma} \emph{et~al.}(2011)\citenamefont{Ma, Wang, Sun, and
  Nori}}]{Majian}
\bibinfo{author}{\bibnamefont{Ma}, \bibfnamefont{J.}},
  \bibinfo{author}{\bibfnamefont{X.~G.} \bibnamefont{Wang}},
  \bibinfo{author}{\bibfnamefont{C.~P.} \bibnamefont{Sun}}, and
  \bibinfo{author}{\bibfnamefont{F.}~\bibnamefont{Nori}}, \bibinfo{year}{2011},
  \bibinfo{journal}{Phys. Rep.} \textbf{\bibinfo{volume}{509}},
  \bibinfo{pages}{89}.

\bibitem[{\citenamefont{Ma and Zhan}(2009)}]{ISI:000263307800011}
\bibinfo{author}{\bibnamefont{Ma}, \bibfnamefont{P.~C.}}, and
  \bibinfo{author}{\bibfnamefont{Y.~B.} \bibnamefont{Zhan}},
  \bibinfo{year}{2009}, \bibinfo{journal}{Commun. Theor. Phys.}
  \textbf{\bibinfo{volume}{51}}, \bibinfo{pages}{57}.

\bibitem[{\citenamefont{Ma} \emph{et~al.}(2009)\citenamefont{Ma, Zhan, and
  Zhang}}]{ISI:000267510900005}
\bibinfo{author}{\bibnamefont{Ma}, \bibfnamefont{P.~C.}},
  \bibinfo{author}{\bibfnamefont{Y.~B.} \bibnamefont{Zhan}}, and
  \bibinfo{author}{\bibfnamefont{L.~L.} \bibnamefont{Zhang}},
  \bibinfo{year}{2009}, \bibinfo{journal}{Int. J. Mod. Phys. B}
  \textbf{\bibinfo{volume}{23}}, \bibinfo{pages}{3231}.

\bibitem[{\citenamefont{Ma}(2007)}]{ma2007group}
\bibinfo{author}{\bibnamefont{Ma}, \bibfnamefont{Z.~Q.}}, \bibinfo{year}{2007},
  \emph{\bibinfo{title}{Group Theory for Physicists, p122-131, p353-364}}
  (\bibinfo{publisher}{World Scientific,Singapore}).

\bibitem[{\citenamefont{Macchiavello}(2003)}]{multiphase-esti}
\bibinfo{author}{\bibnamefont{Macchiavello}, \bibfnamefont{C.}},
  \bibinfo{year}{2003}, \bibinfo{journal}{Phys. Rev. A}
  \textbf{\bibinfo{volume}{67}}, \bibinfo{pages}{062302}.

\bibitem[{\citenamefont{Maccone}(2006)}]{ISI:000237147700049}
\bibinfo{author}{\bibnamefont{Maccone}, \bibfnamefont{L.}},
  \bibinfo{year}{2006}, \bibinfo{journal}{Phys. Rev. A}
  \textbf{\bibinfo{volume}{73}}, \bibinfo{pages}{042307}.

\bibitem[{\citenamefont{Martini} \emph{et~al.}(2004)\citenamefont{Martini,
  Pelliccia, and Sciarrino}}]{prl92.067901}
\bibinfo{author}{\bibnamefont{Martini}, \bibfnamefont{F.~D.}},
  \bibinfo{author}{\bibfnamefont{D.}~\bibnamefont{Pelliccia}}, and
  \bibinfo{author}{\bibfnamefont{F.}~\bibnamefont{Sciarrino}},
  \bibinfo{year}{2004}, \bibinfo{journal}{Phys. Rev. Lett.}
  \textbf{\bibinfo{volume}{92}}, \bibinfo{pages}{067901}.

\bibitem[{\citenamefont{Martini and Sciarrino}(2012)}]{REM-DeMartini}
\bibinfo{author}{\bibnamefont{Martini}, \bibfnamefont{F.~D.}}, and
  \bibinfo{author}{\bibfnamefont{F.}~\bibnamefont{Sciarrino}},
  \bibinfo{year}{2012}, \bibinfo{journal}{Rev. Mod. Phys.}
  \textbf{\bibinfo{volume}{84}}, \bibinfo{pages}{1765}.

\bibitem[{\citenamefont{Masanes} \emph{et~al.}(2006)\citenamefont{Masanes,
  Acin, and Gisin}}]{g-nonsignaling}
\bibinfo{author}{\bibnamefont{Masanes}, \bibfnamefont{L.}},
  \bibinfo{author}{\bibfnamefont{A.}~\bibnamefont{Acin}}, and
  \bibinfo{author}{\bibfnamefont{N.}~\bibnamefont{Gisin}},
  \bibinfo{year}{2006}, \bibinfo{journal}{Phys. Rev. A}
  \textbf{\bibinfo{volume}{73}}, \bibinfo{pages}{012112}.

\bibitem[{\citenamefont{Massar}(2000)}]{massar-antispin}
\bibinfo{author}{\bibnamefont{Massar}, \bibfnamefont{S.}},
  \bibinfo{year}{2000}, \bibinfo{journal}{Phys. Rev. A}
  \textbf{\bibinfo{volume}{62}}, \bibinfo{pages}{040101(R)}.

\bibitem[{\citenamefont{Massar and Popescu}(1995)}]{Massar1995}
\bibinfo{author}{\bibnamefont{Massar}, \bibfnamefont{S.}}, and
  \bibinfo{author}{\bibfnamefont{S.}~\bibnamefont{Popescu}},
  \bibinfo{year}{1995}, \bibinfo{journal}{Phys. Rev. Lett.}
  \textbf{\bibinfo{volume}{74}}, \bibinfo{pages}{1259}.

\bibitem[{\citenamefont{Mayer}(2001)}]{Mayers}
\bibinfo{author}{\bibnamefont{Mayer}, \bibfnamefont{D.}}, \bibinfo{year}{2001},
  \bibinfo{journal}{J. ACM} \textbf{\bibinfo{volume}{48}},
  \bibinfo{pages}{351}.

\bibitem[{\citenamefont{Mendonca} \emph{et~al.}(2008)\citenamefont{Mendonca,
  Gilchrist, and Doherty}}]{ISI:000258180300064}
\bibinfo{author}{\bibnamefont{Mendonca}, \bibfnamefont{P.~E. M.~F.}},
  \bibinfo{author}{\bibfnamefont{A.}~\bibnamefont{Gilchrist}}, and
  \bibinfo{author}{\bibfnamefont{A.~C.} \bibnamefont{Doherty}},
  \bibinfo{year}{2008}, \bibinfo{journal}{Phys. Rev. A}
  \textbf{\bibinfo{volume}{78}}, \bibinfo{pages}{012319}.

\bibitem[{\citenamefont{Meng and Zhu}(2009)}]{Meng2009}
\bibinfo{author}{\bibnamefont{Meng}, \bibfnamefont{F.~Y.}}, and
  \bibinfo{author}{\bibfnamefont{A.~D.} \bibnamefont{Zhu}},
  \bibinfo{year}{2009}, \bibinfo{journal}{J. Mod. Optic.}
  \textbf{\bibinfo{volume}{56}}, \bibinfo{pages}{1255}.

\bibitem[{\citenamefont{Milman} \emph{et~al.}(2003)\citenamefont{Milman,
  Ollivier, and Raimond}}]{qed-clone}
\bibinfo{author}{\bibnamefont{Milman}, \bibfnamefont{P.}},
  \bibinfo{author}{\bibfnamefont{H.}~\bibnamefont{Ollivier}}, and
  \bibinfo{author}{\bibfnamefont{J.~M.} \bibnamefont{Raimond}},
  \bibinfo{year}{2003}, \bibinfo{journal}{Phys. Rev. A}
  \textbf{\bibinfo{volume}{67}}, \bibinfo{pages}{012314}.

\bibitem[{\citenamefont{Mishra}(2012)}]{Mishra2012}
\bibinfo{author}{\bibnamefont{Mishra}, \bibfnamefont{D.~K.}},
  \bibinfo{year}{2012}, \bibinfo{journal}{Opt. Commun.}
  \textbf{\bibinfo{volume}{285}}, \bibinfo{pages}{1560}.

\bibitem[{\citenamefont{Miyadera and
  Imai}(2006{\natexlab{a}})}]{ISI:000243166700173}
\bibinfo{author}{\bibnamefont{Miyadera}, \bibfnamefont{T.}}, and
  \bibinfo{author}{\bibfnamefont{H.}~\bibnamefont{Imai}},
  \bibinfo{year}{2006}{\natexlab{a}}, \bibinfo{journal}{Phys. Rev. A}
  \textbf{\bibinfo{volume}{74}}, \bibinfo{pages}{064302}.

\bibitem[{\citenamefont{Miyadera and
  Imai}(2006{\natexlab{b}})}]{ISI:000240238300141}
\bibinfo{author}{\bibnamefont{Miyadera}, \bibfnamefont{T.}}, and
  \bibinfo{author}{\bibfnamefont{H.}~\bibnamefont{Imai}},
  \bibinfo{year}{2006}{\natexlab{b}}, \bibinfo{journal}{Phys. Rev. A}
  \textbf{\bibinfo{volume}{74}}, \bibinfo{pages}{024101}.

\bibitem[{\citenamefont{Modi} \emph{et~al.}(2012)\citenamefont{Modi, Brodutch,
  Cable, Paterek, and Vedral}}]{RMP-discord}
\bibinfo{author}{\bibnamefont{Modi}, \bibfnamefont{K.}},
  \bibinfo{author}{\bibfnamefont{A.}~\bibnamefont{Brodutch}},
  \bibinfo{author}{\bibfnamefont{H.}~\bibnamefont{Cable}},
  \bibinfo{author}{\bibfnamefont{T.}~\bibnamefont{Paterek}}, and
  \bibinfo{author}{\bibfnamefont{V.}~\bibnamefont{Vedral}},
  \bibinfo{year}{2012}, \bibinfo{journal}{Rev. Mod. Phys.}
  \textbf{\bibinfo{volume}{84}}, \bibinfo{pages}{1655}.

\bibitem[{\citenamefont{Mor}(1998)}]{Morprl1998}
\bibinfo{author}{\bibnamefont{Mor}, \bibfnamefont{T.}}, \bibinfo{year}{1998},
  \bibinfo{journal}{Phys. Rev. Lett.} \textbf{\bibinfo{volume}{80}},
  \bibinfo{pages}{3137}.

\bibitem[{\citenamefont{Mor and Terno}(1999)}]{ISI:000084148000024}
\bibinfo{author}{\bibnamefont{Mor}, \bibfnamefont{T.}}, and
  \bibinfo{author}{\bibfnamefont{D.~R.} \bibnamefont{Terno}},
  \bibinfo{year}{1999}, \bibinfo{journal}{Phys. Rev. A}
  \textbf{\bibinfo{volume}{60}}, \bibinfo{pages}{4341}.

\bibitem[{\citenamefont{Murao} \emph{et~al.}(1999)\citenamefont{Murao,
  Jonathan, Plenio, and Vedral}}]{PhysRevA.59.156}
\bibinfo{author}{\bibnamefont{Murao}, \bibfnamefont{M.}},
  \bibinfo{author}{\bibfnamefont{D.}~\bibnamefont{Jonathan}},
  \bibinfo{author}{\bibfnamefont{M.~B.} \bibnamefont{Plenio}}, and
  \bibinfo{author}{\bibfnamefont{V.}~\bibnamefont{Vedral}},
  \bibinfo{year}{1999}, \bibinfo{journal}{Phys. Rev. A}
  \textbf{\bibinfo{volume}{59}}, \bibinfo{pages}{156}.

\bibitem[{\citenamefont{Murao} \emph{et~al.}(2000)\citenamefont{Murao, Plenio,
  and Vedral}}]{PhysRevA.61.032311}
\bibinfo{author}{\bibnamefont{Murao}, \bibfnamefont{M.}},
  \bibinfo{author}{\bibfnamefont{M.~B.} \bibnamefont{Plenio}}, and
  \bibinfo{author}{\bibfnamefont{V.}~\bibnamefont{Vedral}},
  \bibinfo{year}{2000}, \bibinfo{journal}{Phys. Rev. A}
  \textbf{\bibinfo{volume}{61}}, \bibinfo{pages}{032311}.

\bibitem[{\citenamefont{Murao and Vedral}(2001)}]{muraovedral}
\bibinfo{author}{\bibnamefont{Murao}, \bibfnamefont{M.}}, and
  \bibinfo{author}{\bibfnamefont{V.}~\bibnamefont{Vedral}},
  \bibinfo{year}{2001}, \bibinfo{journal}{Phys. Rev. Lett.}
  \textbf{\bibinfo{volume}{86}}, \bibinfo{pages}{352}.

\bibitem[{\citenamefont{Nagali} \emph{et~al.}(2007)\citenamefont{Nagali,
  De~Angelis, Sciarrino, and De~Martini}}]{Nagali2007}
\bibinfo{author}{\bibnamefont{Nagali}, \bibfnamefont{E.}},
  \bibinfo{author}{\bibfnamefont{T.}~\bibnamefont{De~Angelis}},
  \bibinfo{author}{\bibfnamefont{F.}~\bibnamefont{Sciarrino}}, and
  \bibinfo{author}{\bibfnamefont{F.}~\bibnamefont{De~Martini}},
  \bibinfo{year}{2007}, \bibinfo{journal}{Phys. Rev. A}
  \textbf{\bibinfo{volume}{76}}, \bibinfo{pages}{042126}.

\bibitem[{\citenamefont{Nagali} \emph{et~al.}(2010)\citenamefont{Nagali,
  Giovannini, Marrucci, Slussarenko, Santamato, and
  Sciarrino}}]{nagaliexperiment}
\bibinfo{author}{\bibnamefont{Nagali}, \bibfnamefont{E.}},
  \bibinfo{author}{\bibfnamefont{D.}~\bibnamefont{Giovannini}},
  \bibinfo{author}{\bibfnamefont{L.}~\bibnamefont{Marrucci}},
  \bibinfo{author}{\bibfnamefont{S.}~\bibnamefont{Slussarenko}},
  \bibinfo{author}{\bibfnamefont{E.}~\bibnamefont{Santamato}}, and
  \bibinfo{author}{\bibfnamefont{F.}~\bibnamefont{Sciarrino}},
  \bibinfo{year}{2010}, \bibinfo{journal}{Phys. Rev. Lett.}
  \textbf{\bibinfo{volume}{105}}, \bibinfo{pages}{073602}.

\bibitem[{\citenamefont{Nagali} \emph{et~al.}(2009)\citenamefont{Nagali,
  Sansoni, Sciarrino, De~Martini, Marrucci, Piccirillo, Karimi, and
  Santamato}}]{nagali}
\bibinfo{author}{\bibnamefont{Nagali}, \bibfnamefont{E.}},
  \bibinfo{author}{\bibfnamefont{L.}~\bibnamefont{Sansoni}},
  \bibinfo{author}{\bibfnamefont{F.}~\bibnamefont{Sciarrino}},
  \bibinfo{author}{\bibfnamefont{F.}~\bibnamefont{De~Martini}},
  \bibinfo{author}{\bibfnamefont{L.}~\bibnamefont{Marrucci}},
  \bibinfo{author}{\bibfnamefont{B.}~\bibnamefont{Piccirillo}},
  \bibinfo{author}{\bibfnamefont{E.}~\bibnamefont{Karimi}}, and
  \bibinfo{author}{\bibfnamefont{E.}~\bibnamefont{Santamato}},
  \bibinfo{year}{2009}, \bibinfo{journal}{Nat. Photonics}
  \textbf{\bibinfo{volume}{3}}, \bibinfo{pages}{720}.

\bibitem[{\citenamefont{Nagy}(2009)}]{ISI:000268481100011}
\bibinfo{author}{\bibnamefont{Nagy}, \bibfnamefont{G.}}, \bibinfo{year}{2009},
  \bibinfo{journal}{Rep. Math. Phys.} \textbf{\bibinfo{volume}{63}},
  \bibinfo{pages}{447}.

\bibitem[{\citenamefont{Namiki}(2011)}]{Namiki2011}
\bibinfo{author}{\bibnamefont{Namiki}, \bibfnamefont{R.}},
  \bibinfo{year}{2011}, \bibinfo{journal}{Phys. Rev. A}
  \textbf{\bibinfo{volume}{83}}, \bibinfo{pages}{040302(R)}.

\bibitem[{\citenamefont{Navez and Cerf}(2003)}]{PhysRevA.68.032313}
\bibinfo{author}{\bibnamefont{Navez}, \bibfnamefont{P.}}, and
  \bibinfo{author}{\bibfnamefont{N.~J.} \bibnamefont{Cerf}},
  \bibinfo{year}{2003}, \bibinfo{journal}{Phys. Rev. A}
  \textbf{\bibinfo{volume}{68}}, \bibinfo{pages}{032313}.

\bibitem[{\citenamefont{Nha and Carmichael}(2005)}]{ISI:000228632100051}
\bibinfo{author}{\bibnamefont{Nha}, \bibfnamefont{H.}}, and
  \bibinfo{author}{\bibfnamefont{H.~J.} \bibnamefont{Carmichael}},
  \bibinfo{year}{2005}, \bibinfo{journal}{Phys. Rev. A}
  \textbf{\bibinfo{volume}{71}}, \bibinfo{pages}{032336}.

\bibitem[{\citenamefont{Niederberger}
  \emph{et~al.}(2005)\citenamefont{Niederberger, Scarani, and
  Gisin}}]{pns-clone}
\bibinfo{author}{\bibnamefont{Niederberger}, \bibfnamefont{A.}},
  \bibinfo{author}{\bibfnamefont{V.}~\bibnamefont{Scarani}}, and
  \bibinfo{author}{\bibfnamefont{N.}~\bibnamefont{Gisin}},
  \bibinfo{year}{2005}, \bibinfo{journal}{Phys. Rev. A}
  \textbf{\bibinfo{volume}{71}}, \bibinfo{pages}{042316}.

\bibitem[{\citenamefont{Nielsen and Chuang}(2000)}]{Nielsen2000}
\bibinfo{author}{\bibnamefont{Nielsen}, \bibfnamefont{M.~A.}}, and
  \bibinfo{author}{\bibfnamefont{I.~C.} \bibnamefont{Chuang}},
  \bibinfo{year}{2000}, \emph{\bibinfo{title}{Quantum Computation and Quantum
  Information}} (\bibinfo{publisher}{Cambridge University Press, Cambridge}).

\bibitem[{\citenamefont{Nikolopoulos and Alber}(2005)}]{Nikolopoulos2005}
\bibinfo{author}{\bibnamefont{Nikolopoulos}, \bibfnamefont{G.~M.}}, and
  \bibinfo{author}{\bibfnamefont{G.}~\bibnamefont{Alber}},
  \bibinfo{year}{2005}, \bibinfo{journal}{Phys. Rev. A}
  \textbf{\bibinfo{volume}{72}}, \bibinfo{pages}{032320}.

\bibitem[{\citenamefont{Nikolopoulos}
  \emph{et~al.}(2006)\citenamefont{Nikolopoulos, Ranade, and
  Alber}}]{Nikolopoulos2006}
\bibinfo{author}{\bibnamefont{Nikolopoulos}, \bibfnamefont{G.~M.}},
  \bibinfo{author}{\bibfnamefont{K.~S.} \bibnamefont{Ranade}}, and
  \bibinfo{author}{\bibfnamefont{G.}~\bibnamefont{Alber}},
  \bibinfo{year}{2006}, \bibinfo{journal}{Phys. Rev. A}
  \textbf{\bibinfo{volume}{73}}, \bibinfo{pages}{032325}.

\bibitem[{\citenamefont{Niu and Griffiths}(1998)}]{PhysRevA.58.4377}
\bibinfo{author}{\bibnamefont{Niu}, \bibfnamefont{C.~S.}}, and
  \bibinfo{author}{\bibfnamefont{R.~B.} \bibnamefont{Griffiths}},
  \bibinfo{year}{1998}, \bibinfo{journal}{Phys. Rev. A}
  \textbf{\bibinfo{volume}{58}}, \bibinfo{pages}{4377}.

\bibitem[{\citenamefont{Niu and Griffiths}(1999)}]{PhysRevA.60.2764}
\bibinfo{author}{\bibnamefont{Niu}, \bibfnamefont{C.~S.}}, and
  \bibinfo{author}{\bibfnamefont{R.~B.} \bibnamefont{Griffiths}},
  \bibinfo{year}{1999}, \bibinfo{journal}{Phys. Rev. A}
  \textbf{\bibinfo{volume}{60}}, \bibinfo{pages}{2764}.

\bibitem[{\citenamefont{Niu}(2009)}]{ISI:000269486400010}
\bibinfo{author}{\bibnamefont{Niu}, \bibfnamefont{X.~F.}},
  \bibinfo{year}{2009}, \bibinfo{journal}{Int. J. Theor. Phys.}
  \textbf{\bibinfo{volume}{48}}, \bibinfo{pages}{2599}.

\bibitem[{\citenamefont{Nuida} \emph{et~al.}(2010)\citenamefont{Nuida, Kimura,
  and Miyadera}}]{ISI:000282433200038}
\bibinfo{author}{\bibnamefont{Nuida}, \bibfnamefont{K.}},
  \bibinfo{author}{\bibfnamefont{G.}~\bibnamefont{Kimura}}, and
  \bibinfo{author}{\bibfnamefont{T.}~\bibnamefont{Miyadera}},
  \bibinfo{year}{2010}, \bibinfo{journal}{J. Math. Phys.}
  \textbf{\bibinfo{volume}{51}}, \bibinfo{pages}{093505}.

\bibitem[{\citenamefont{Olivares and Paris}(2008)}]{Olivares2008}
\bibinfo{author}{\bibnamefont{Olivares}, \bibfnamefont{S.}}, and
  \bibinfo{author}{\bibfnamefont{M.~G.~A.} \bibnamefont{Paris}},
  \bibinfo{year}{2008}, \bibinfo{journal}{Eur. Phys. J.-Spec. Top.}
  \textbf{\bibinfo{volume}{160}}, \bibinfo{pages}{319}.

\bibitem[{\citenamefont{Olivares} \emph{et~al.}(2006)\citenamefont{Olivares,
  Paris, and Andersen}}]{Olivares2006}
\bibinfo{author}{\bibnamefont{Olivares}, \bibfnamefont{S.}},
  \bibinfo{author}{\bibfnamefont{M.~G.~A.} \bibnamefont{Paris}}, and
  \bibinfo{author}{\bibfnamefont{U.~L.} \bibnamefont{Andersen}},
  \bibinfo{year}{2006}, \bibinfo{journal}{Phys. Rev. A}
  \textbf{\bibinfo{volume}{73}}, \bibinfo{pages}{062330}.

\bibitem[{\citenamefont{Osborne and Verstraete}(2006)}]{Osborne2006}
\bibinfo{author}{\bibnamefont{Osborne}, \bibfnamefont{T.~J.}}, and
  \bibinfo{author}{\bibfnamefont{F.}~\bibnamefont{Verstraete}},
  \bibinfo{year}{2006}, \bibinfo{journal}{Phys. Rev. Lett.}
  \textbf{\bibinfo{volume}{96}}, \bibinfo{pages}{220503}.

\bibitem[{\citenamefont{Ou and Fan}(2007)}]{Ouyc2007}
\bibinfo{author}{\bibnamefont{Ou}, \bibfnamefont{Y.~C.}}, and
  \bibinfo{author}{\bibfnamefont{H.}~\bibnamefont{Fan}}, \bibinfo{year}{2007},
  \bibinfo{journal}{Phys. Rev. A} \textbf{\bibinfo{volume}{75}},
  \bibinfo{pages}{062308}.

\bibitem[{\citenamefont{Ou} \emph{et~al.}(2008)\citenamefont{Ou, Fan, and
  Fei}}]{Ouyc2008}
\bibinfo{author}{\bibnamefont{Ou}, \bibfnamefont{Y.~C.}},
  \bibinfo{author}{\bibfnamefont{H.}~\bibnamefont{Fan}}, and
  \bibinfo{author}{\bibfnamefont{S.~M.} \bibnamefont{Fei}},
  \bibinfo{year}{2008}, \bibinfo{journal}{Phys. Rev. A}
  \textbf{\bibinfo{volume}{78}}, \bibinfo{pages}{012311}.

\bibitem[{\citenamefont{Owari and Hayashi}(2006)}]{OH06}
\bibinfo{author}{\bibnamefont{Owari}, \bibfnamefont{M.}}, and
  \bibinfo{author}{\bibfnamefont{M.}~\bibnamefont{Hayashi}},
  \bibinfo{year}{2006}, \bibinfo{journal}{Phys. Rev. A}
  \textbf{\bibinfo{volume}{74}}, \bibinfo{pages}{032108}.

\bibitem[{\citenamefont{Ozdemir} \emph{et~al.}(2007)\citenamefont{Ozdemir,
  Bartkiewicz, Liu, and Miranowicz}}]{Ozdemir2007}
\bibinfo{author}{\bibnamefont{Ozdemir}, \bibfnamefont{S.~K.}},
  \bibinfo{author}{\bibfnamefont{K.}~\bibnamefont{Bartkiewicz}},
  \bibinfo{author}{\bibfnamefont{Y.~X.} \bibnamefont{Liu}}, and
  \bibinfo{author}{\bibfnamefont{A.}~\bibnamefont{Miranowicz}},
  \bibinfo{year}{2007}, \bibinfo{journal}{Phys. Rev. A}
  \textbf{\bibinfo{volume}{76}}, \bibinfo{pages}{042325}.

\bibitem[{\citenamefont{Pan} \emph{et~al.}(2012)\citenamefont{Pan, Chen, Lu,
  Weinfurter, Zeilinger, and Zukowski}}]{REM-Pan}
\bibinfo{author}{\bibnamefont{Pan}, \bibfnamefont{J.~W.}},
  \bibinfo{author}{\bibfnamefont{Z.~B.} \bibnamefont{Chen}},
  \bibinfo{author}{\bibfnamefont{C.~Y.} \bibnamefont{Lu}},
  \bibinfo{author}{\bibfnamefont{H.}~\bibnamefont{Weinfurter}},
  \bibinfo{author}{\bibfnamefont{A.}~\bibnamefont{Zeilinger}}, and
  \bibinfo{author}{\bibfnamefont{M.}~\bibnamefont{Zukowski}},
  \bibinfo{year}{2012}, \bibinfo{journal}{Rev. Mod. Phys.}
  \textbf{\bibinfo{volume}{84}}, \bibinfo{pages}{777}.

\bibitem[{\citenamefont{Pan} \emph{et~al.}(2011)\citenamefont{Pan, Liu, Yang,
  and Fan}}]{Pan2011}
\bibinfo{author}{\bibnamefont{Pan}, \bibfnamefont{X.~Y.}},
  \bibinfo{author}{\bibfnamefont{G.~Q.} \bibnamefont{Liu}},
  \bibinfo{author}{\bibfnamefont{L.~L.} \bibnamefont{Yang}}, and
  \bibinfo{author}{\bibfnamefont{H.}~\bibnamefont{Fan}}, \bibinfo{year}{2011},
  \bibinfo{journal}{Appl. Phys. Lett.} \textbf{\bibinfo{volume}{99}},
  \bibinfo{pages}{051113}.

\bibitem[{\citenamefont{Pang} \emph{et~al.}(2011)\citenamefont{Pang, Wu, and
  Chen}}]{ISI:000298380700008}
\bibinfo{author}{\bibnamefont{Pang}, \bibfnamefont{S.~S.}},
  \bibinfo{author}{\bibfnamefont{S.~J.} \bibnamefont{Wu}}, and
  \bibinfo{author}{\bibfnamefont{Z.~B.} \bibnamefont{Chen}},
  \bibinfo{year}{2011}, \bibinfo{journal}{Phys. Rev. A}
  \textbf{\bibinfo{volume}{84}}, \bibinfo{pages}{062313}.

\bibitem[{\citenamefont{Pati}(1999)}]{PhysRevLett.83.2849}
\bibinfo{author}{\bibnamefont{Pati}, \bibfnamefont{A.~K.}},
  \bibinfo{year}{1999}, \bibinfo{journal}{Phys. Rev. Lett.}
  \textbf{\bibinfo{volume}{83}}, \bibinfo{pages}{2849}.

\bibitem[{\citenamefont{Pati}(2000)}]{ISI:000085336900037}
\bibinfo{author}{\bibnamefont{Pati}, \bibfnamefont{A.~K.}},
  \bibinfo{year}{2000}, \bibinfo{journal}{Phys. Rev. A}
  \textbf{\bibinfo{volume}{61}}, \bibinfo{pages}{022308}.

\bibitem[{\citenamefont{Pati and Braunstein}(2000)}]{no-deleting}
\bibinfo{author}{\bibnamefont{Pati}, \bibfnamefont{A.~K.}}, and
  \bibinfo{author}{\bibfnamefont{S.~L.} \bibnamefont{Braunstein}},
  \bibinfo{year}{2000}, \bibinfo{journal}{Nature}
  \textbf{\bibinfo{volume}{404}}, \bibinfo{pages}{164 }.

\bibitem[{\citenamefont{Pawlowski and Brukner}(2009)}]{bell-non-signaling}
\bibinfo{author}{\bibnamefont{Pawlowski}, \bibfnamefont{M.}}, and
  \bibinfo{author}{\bibfnamefont{C.}~\bibnamefont{Brukner}},
  \bibinfo{year}{2009}, \bibinfo{journal}{Phys. Rev. Lett.}
  \textbf{\bibinfo{volume}{102}}, \bibinfo{pages}{030403}.

\bibitem[{\citenamefont{Pelliccia} \emph{et~al.}(2003)\citenamefont{Pelliccia,
  Schettini, Sciarrino, Sias, and De~Martini}}]{PhysRevA.68.042306}
\bibinfo{author}{\bibnamefont{Pelliccia}, \bibfnamefont{D.}},
  \bibinfo{author}{\bibfnamefont{V.}~\bibnamefont{Schettini}},
  \bibinfo{author}{\bibfnamefont{F.}~\bibnamefont{Sciarrino}},
  \bibinfo{author}{\bibfnamefont{C.}~\bibnamefont{Sias}}, and
  \bibinfo{author}{\bibfnamefont{F.}~\bibnamefont{De~Martini}},
  \bibinfo{year}{2003}, \bibinfo{journal}{Phys. Rev. A}
  \textbf{\bibinfo{volume}{68}}, \bibinfo{pages}{042306}.

\bibitem[{\citenamefont{Peres}(1995)}]{Peres95}
\bibinfo{author}{\bibnamefont{Peres}, \bibfnamefont{A.}}, \bibinfo{year}{1995},
  \emph{\bibinfo{title}{Quantum Theory: Concepts and Methods}}
  (\bibinfo{publisher}{Springer}).

\bibitem[{\citenamefont{Peres}(1996{\natexlab{a}})}]{prl77.3264}
\bibinfo{author}{\bibnamefont{Peres}, \bibfnamefont{A.}},
  \bibinfo{year}{1996}{\natexlab{a}}, \bibinfo{journal}{Phys. Rev. Lett.}
  \textbf{\bibinfo{volume}{77}}, \bibinfo{pages}{3264}.

\bibitem[{\citenamefont{Peres}(1996{\natexlab{b}})}]{Peres96}
\bibinfo{author}{\bibnamefont{Peres}, \bibfnamefont{A.}},
  \bibinfo{year}{1996}{\natexlab{b}}, \bibinfo{journal}{Phys. Rev. Lett.}
  \textbf{\bibinfo{volume}{77}}, \bibinfo{pages}{1413}.

\bibitem[{\citenamefont{Peres and Wootters}(1991)}]{PeresWootters}
\bibinfo{author}{\bibnamefont{Peres}, \bibfnamefont{A.}}, and
  \bibinfo{author}{\bibfnamefont{W.~K.} \bibnamefont{Wootters}},
  \bibinfo{year}{1991}, \bibinfo{journal}{Phys. Rev. Lett.}
  \textbf{\bibinfo{volume}{66}}, \bibinfo{pages}{1119}.

\bibitem[{\citenamefont{Piani} \emph{et~al.}(2008)\citenamefont{Piani,
  Horodecki, and Horodecki}}]{ISI:000253764500008}
\bibinfo{author}{\bibnamefont{Piani}, \bibfnamefont{M.}},
  \bibinfo{author}{\bibfnamefont{P.}~\bibnamefont{Horodecki}}, and
  \bibinfo{author}{\bibfnamefont{R.}~\bibnamefont{Horodecki}},
  \bibinfo{year}{2008}, \bibinfo{journal}{Phys. Rev. Lett.}
  \textbf{\bibinfo{volume}{100}}, \bibinfo{pages}{090502}.

\bibitem[{\citenamefont{Popescu}(2014)}]{popescu-non-locality}
\bibinfo{author}{\bibnamefont{Popescu}, \bibfnamefont{S.}},
  \bibinfo{year}{2014}, \bibinfo{journal}{Nature Phys.}
  \textbf{\bibinfo{volume}{10}}, \bibinfo{pages}{264}.

\bibitem[{\citenamefont{Prevedel} \emph{et~al.}(2009)\citenamefont{Prevedel,
  Cronenberg, Tame, Paternostro, Walther, Kim, and
  Zeilinger}}]{dickestate-experiment}
\bibinfo{author}{\bibnamefont{Prevedel}, \bibfnamefont{R.}},
  \bibinfo{author}{\bibfnamefont{G.}~\bibnamefont{Cronenberg}},
  \bibinfo{author}{\bibfnamefont{M.~S.} \bibnamefont{Tame}},
  \bibinfo{author}{\bibfnamefont{M.}~\bibnamefont{Paternostro}},
  \bibinfo{author}{\bibfnamefont{P.}~\bibnamefont{Walther}},
  \bibinfo{author}{\bibfnamefont{M.~S.} \bibnamefont{Kim}}, and
  \bibinfo{author}{\bibfnamefont{A.}~\bibnamefont{Zeilinger}},
  \bibinfo{year}{2009}, \bibinfo{journal}{Phys. Rev. Lett.}
  \textbf{\bibinfo{volume}{103}}, \bibinfo{pages}{020503}.

\bibitem[{\citenamefont{Qiu}(2002)}]{ISI:000175743800069}
\bibinfo{author}{\bibnamefont{Qiu}, \bibfnamefont{D.~W.}},
  \bibinfo{year}{2002}, \bibinfo{journal}{Phys. Rev. A}
  \textbf{\bibinfo{volume}{65}}, \bibinfo{pages}{052329}.

\bibitem[{\citenamefont{Qiu}(2006)}]{ISI:000238380400032}
\bibinfo{author}{\bibnamefont{Qiu}, \bibfnamefont{D.~W.}},
  \bibinfo{year}{2006}, \bibinfo{journal}{J. Phys. A-Math. Gen.}
  \textbf{\bibinfo{volume}{39}}, \bibinfo{pages}{5135}.

\bibitem[{\citenamefont{Qiu}(2008)}]{ISI:000252862000060}
\bibinfo{author}{\bibnamefont{Qiu}, \bibfnamefont{D.~W.}},
  \bibinfo{year}{2008}, \bibinfo{journal}{Phys. Rev. A}
  \textbf{\bibinfo{volume}{77}}, \bibinfo{pages}{012328}.

\bibitem[{\citenamefont{Raeisi} \emph{et~al.}(2012)\citenamefont{Raeisi,
  Tittel, and Simon}}]{revsecloning}
\bibinfo{author}{\bibnamefont{Raeisi}, \bibfnamefont{S.}},
  \bibinfo{author}{\bibfnamefont{W.}~\bibnamefont{Tittel}}, and
  \bibinfo{author}{\bibfnamefont{C.}~\bibnamefont{Simon}},
  \bibinfo{year}{2012}, \bibinfo{journal}{Phys. Rev. Lett.}
  \textbf{\bibinfo{volume}{108}}, \bibinfo{pages}{120404}.

\bibitem[{\citenamefont{Rastegin}(2002)}]{ISI:000179502200033}
\bibinfo{author}{\bibnamefont{Rastegin}, \bibfnamefont{A.~E.}},
  \bibinfo{year}{2002}, \bibinfo{journal}{Phys. Rev. A}
  \textbf{\bibinfo{volume}{66}}, \bibinfo{pages}{042304}.

\bibitem[{\citenamefont{Rastegin}(2003{\natexlab{a}})}]{ISI:000185716700021}
\bibinfo{author}{\bibnamefont{Rastegin}, \bibfnamefont{A.~E.}},
  \bibinfo{year}{2003}{\natexlab{a}}, \bibinfo{journal}{Phys. Rev. A}
  \textbf{\bibinfo{volume}{68}}, \bibinfo{pages}{032303}.

\bibitem[{\citenamefont{Rastegin}(2003{\natexlab{b}})}]{ISI:000180804600031}
\bibinfo{author}{\bibnamefont{Rastegin}, \bibfnamefont{A.~E.}},
  \bibinfo{year}{2003}{\natexlab{b}}, \bibinfo{journal}{Phys. Rev. A}
  \textbf{\bibinfo{volume}{67}}, \bibinfo{pages}{012305}.

\bibitem[{\citenamefont{Rastegin}(2010)}]{ISI:000282921000007}
\bibinfo{author}{\bibnamefont{Rastegin}, \bibfnamefont{A.~E.}},
  \bibinfo{year}{2010}, \bibinfo{journal}{Quant. Inf. Comput.}
  \textbf{\bibinfo{volume}{10}}, \bibinfo{pages}{971}.

\bibitem[{\citenamefont{Raynal and Lutkenhaus}(2005)}]{ISI:000231564200074}
\bibinfo{author}{\bibnamefont{Raynal}, \bibfnamefont{P.}}, and
  \bibinfo{author}{\bibfnamefont{N.}~\bibnamefont{Lutkenhaus}},
  \bibinfo{year}{2005}, \bibinfo{journal}{Phys. Rev. A}
  \textbf{\bibinfo{volume}{72}}, \bibinfo{pages}{022342}.

\bibitem[{\citenamefont{Reck} \emph{et~al.}(1994)\citenamefont{Reck, Zeilinger,
  Bernstein, and Bertani}}]{Reck1994}
\bibinfo{author}{\bibnamefont{Reck}, \bibfnamefont{M.}},
  \bibinfo{author}{\bibfnamefont{A.}~\bibnamefont{Zeilinger}},
  \bibinfo{author}{\bibfnamefont{H.~J.} \bibnamefont{Bernstein}}, and
  \bibinfo{author}{\bibfnamefont{P.}~\bibnamefont{Bertani}},
  \bibinfo{year}{1994}, \bibinfo{journal}{Phys. Rev. Lett.}
  \textbf{\bibinfo{volume}{73}}, \bibinfo{pages}{58}.

\bibitem[{\citenamefont{Reimpell and Werner}(2007)}]{Biased}
\bibinfo{author}{\bibnamefont{Reimpell}, \bibfnamefont{M.}}, and
  \bibinfo{author}{\bibfnamefont{R.~F.} \bibnamefont{Werner}},
  \bibinfo{year}{2007}, \bibinfo{journal}{Phys. Rev. A}
  \textbf{\bibinfo{volume}{75}}, \bibinfo{pages}{062334}.

\bibitem[{\citenamefont{Ren} \emph{et~al.}(2011)\citenamefont{Ren, Xiang, and
  Fan}}]{Renxj2011}
\bibinfo{author}{\bibnamefont{Ren}, \bibfnamefont{X.}},
  \bibinfo{author}{\bibfnamefont{Y.}~\bibnamefont{Xiang}}, and
  \bibinfo{author}{\bibfnamefont{H.}~\bibnamefont{Fan}}, \bibinfo{year}{2011},
  \bibinfo{journal}{Eur. Phys. J. D} \textbf{\bibinfo{volume}{65}},
  \bibinfo{pages}{621}.

\bibitem[{\citenamefont{Renes}(2004)}]{Renes2004}
\bibinfo{author}{\bibnamefont{Renes}, \bibfnamefont{J.~M.}},
  \bibinfo{year}{2004}, \bibinfo{journal}{Phys. Rev. A}
  \textbf{\bibinfo{volume}{70}}, \bibinfo{pages}{052314}.

\bibitem[{\citenamefont{Rezakhani} \emph{et~al.}(2005)\citenamefont{Rezakhani,
  Siadatnejad, and Ghaderi}}]{Rezakhani2005278}
\bibinfo{author}{\bibnamefont{Rezakhani}, \bibfnamefont{A.}},
  \bibinfo{author}{\bibfnamefont{S.}~\bibnamefont{Siadatnejad}}, and
  \bibinfo{author}{\bibfnamefont{A.}~\bibnamefont{Ghaderi}},
  \bibinfo{year}{2005}, \bibinfo{journal}{Phys. Lett. A}
  \textbf{\bibinfo{volume}{336}}, \bibinfo{pages}{278 }.

\bibitem[{\citenamefont{Ricci} \emph{et~al.}(2005)\citenamefont{Ricci,
  Sciarrino, Cerf, Filip, Fiur\'{a}\v{s}ek, and De~Martini}}]{Ricci2005}
\bibinfo{author}{\bibnamefont{Ricci}, \bibfnamefont{M.}},
  \bibinfo{author}{\bibfnamefont{F.}~\bibnamefont{Sciarrino}},
  \bibinfo{author}{\bibfnamefont{N.~J.} \bibnamefont{Cerf}},
  \bibinfo{author}{\bibfnamefont{R.}~\bibnamefont{Filip}},
  \bibinfo{author}{\bibfnamefont{J.}~\bibnamefont{Fiur\'{a}\v{s}ek}}, and
  \bibinfo{author}{\bibfnamefont{F.}~\bibnamefont{De~Martini}},
  \bibinfo{year}{2005}, \bibinfo{journal}{Phys. Rev. Lett.}
  \textbf{\bibinfo{volume}{95}}, \bibinfo{pages}{090504}.

\bibitem[{\citenamefont{Ricci} \emph{et~al.}(2004)\citenamefont{Ricci,
  Sciarrino, Sias, and De~Martini}}]{PhysRevLett.92.047901}
\bibinfo{author}{\bibnamefont{Ricci}, \bibfnamefont{M.}},
  \bibinfo{author}{\bibfnamefont{F.}~\bibnamefont{Sciarrino}},
  \bibinfo{author}{\bibfnamefont{C.}~\bibnamefont{Sias}}, and
  \bibinfo{author}{\bibfnamefont{F.}~\bibnamefont{De~Martini}},
  \bibinfo{year}{2004}, \bibinfo{journal}{Phys. Rev. Lett.}
  \textbf{\bibinfo{volume}{92}}, \bibinfo{pages}{047901}.

\bibitem[{\citenamefont{Rivest} \emph{et~al.}(1978)\citenamefont{Rivest,
  Shamir, and Adleman}}]{RSA78}
\bibinfo{author}{\bibnamefont{Rivest}, \bibfnamefont{R.~L.}},
  \bibinfo{author}{\bibfnamefont{A.}~\bibnamefont{Shamir}}, and
  \bibinfo{author}{\bibfnamefont{L.~M.} \bibnamefont{Adleman}},
  \bibinfo{year}{1978}, \bibinfo{journal}{Comm. ACM}
  \textbf{\bibinfo{volume}{21}}, \bibinfo{pages}{120}.

\bibitem[{\citenamefont{Rodriguez-Rosario}
  \emph{et~al.}(2010)\citenamefont{Rodriguez-Rosario, Modi, and
  Aspuru-Guzik}}]{ISI:000274001500049}
\bibinfo{author}{\bibnamefont{Rodriguez-Rosario}, \bibfnamefont{C.~A.}},
  \bibinfo{author}{\bibfnamefont{K.}~\bibnamefont{Modi}}, and
  \bibinfo{author}{\bibfnamefont{A.}~\bibnamefont{Aspuru-Guzik}},
  \bibinfo{year}{2010}, \bibinfo{journal}{Phys. Rev. A}
  \textbf{\bibinfo{volume}{81}}, \bibinfo{pages}{012313}.

\bibitem[{\citenamefont{Roubert and Braun}(2008)}]{ISI:000260574100041}
\bibinfo{author}{\bibnamefont{Roubert}, \bibfnamefont{B.}}, and
  \bibinfo{author}{\bibfnamefont{D.}~\bibnamefont{Braun}},
  \bibinfo{year}{2008}, \bibinfo{journal}{Phys. Rev. A}
  \textbf{\bibinfo{volume}{78}}, \bibinfo{pages}{042311}.

\bibitem[{\citenamefont{Rybar and Ziman}(2008)}]{Rybar2008}
\bibinfo{author}{\bibnamefont{Rybar}, \bibfnamefont{T.}}, and
  \bibinfo{author}{\bibfnamefont{M.}~\bibnamefont{Ziman}},
  \bibinfo{year}{2008}, \bibinfo{journal}{Phys. Rev. A}
  \textbf{\bibinfo{volume}{78}}, \bibinfo{pages}{052114}.

\bibitem[{\citenamefont{Saberi and Mardoukhi}(2012)}]{Saberi2012}
\bibinfo{author}{\bibnamefont{Saberi}, \bibfnamefont{H.}}, and
  \bibinfo{author}{\bibfnamefont{Y.}~\bibnamefont{Mardoukhi}},
  \bibinfo{year}{2012}, \bibinfo{journal}{Phys. Rev. A}
  \textbf{\bibinfo{volume}{85}}, \bibinfo{pages}{052323}.

\bibitem[{\citenamefont{Sabuncu} \emph{et~al.}(2007)\citenamefont{Sabuncu,
  Andersen, and Leuchs}}]{experiementalcvcloning}
\bibinfo{author}{\bibnamefont{Sabuncu}, \bibfnamefont{M.}},
  \bibinfo{author}{\bibfnamefont{U.~L.} \bibnamefont{Andersen}}, and
  \bibinfo{author}{\bibfnamefont{G.}~\bibnamefont{Leuchs}},
  \bibinfo{year}{2007}, \bibinfo{journal}{Phys. Rev. Lett.}
  \textbf{\bibinfo{volume}{98}}, \bibinfo{pages}{170503}.

\bibitem[{\citenamefont{Sagawa and Ueda}(2008)}]{ISI:000252862000045}
\bibinfo{author}{\bibnamefont{Sagawa}, \bibfnamefont{T.}}, and
  \bibinfo{author}{\bibfnamefont{M.}~\bibnamefont{Ueda}}, \bibinfo{year}{2008},
  \bibinfo{journal}{Phys. Rev. A} \textbf{\bibinfo{volume}{77}},
  \bibinfo{pages}{012313}.

\bibitem[{\citenamefont{Sanguinetti}
  \emph{et~al.}(2010)\citenamefont{Sanguinetti, Pomarico, Sekatski, Zbinden,
  and Gisin}}]{clone-radiometry}
\bibinfo{author}{\bibnamefont{Sanguinetti}, \bibfnamefont{B.}},
  \bibinfo{author}{\bibfnamefont{E.}~\bibnamefont{Pomarico}},
  \bibinfo{author}{\bibfnamefont{P.}~\bibnamefont{Sekatski}},
  \bibinfo{author}{\bibfnamefont{H.}~\bibnamefont{Zbinden}}, and
  \bibinfo{author}{\bibfnamefont{N.}~\bibnamefont{Gisin}},
  \bibinfo{year}{2010}, \bibinfo{journal}{Phys. Rev. Lett.}
  \textbf{\bibinfo{volume}{105}}, \bibinfo{pages}{080503}.

\bibitem[{\citenamefont{Scarani and Gisin}(2001)}]{scarani2001}
\bibinfo{author}{\bibnamefont{Scarani}, \bibfnamefont{V.}}, and
  \bibinfo{author}{\bibfnamefont{N.}~\bibnamefont{Gisin}},
  \bibinfo{year}{2001}, \bibinfo{journal}{Phys. Rev. A}
  \textbf{\bibinfo{volume}{65}}, \bibinfo{pages}{012311}.

\bibitem[{\citenamefont{Scarani} \emph{et~al.}(2005)\citenamefont{Scarani,
  Iblisdir, Gisin, and Ac\'{i}n}}]{RevModPhys.77.1225}
\bibinfo{author}{\bibnamefont{Scarani}, \bibfnamefont{V.}},
  \bibinfo{author}{\bibfnamefont{S.}~\bibnamefont{Iblisdir}},
  \bibinfo{author}{\bibfnamefont{N.}~\bibnamefont{Gisin}}, and
  \bibinfo{author}{\bibfnamefont{A.}~\bibnamefont{Ac\'{i}n}},
  \bibinfo{year}{2005}, \bibinfo{journal}{Rev. Mod. Phys.}
  \textbf{\bibinfo{volume}{77}}, \bibinfo{pages}{1225}.

\bibitem[{\citenamefont{Schon} \emph{et~al.}(2005)\citenamefont{Schon, Solano,
  Verstraete, Cirac, and Wolf}}]{PhysRevLett.95.110503}
\bibinfo{author}{\bibnamefont{Schon}, \bibfnamefont{C.}},
  \bibinfo{author}{\bibfnamefont{E.}~\bibnamefont{Solano}},
  \bibinfo{author}{\bibfnamefont{F.}~\bibnamefont{Verstraete}},
  \bibinfo{author}{\bibfnamefont{J.~I.} \bibnamefont{Cirac}}, and
  \bibinfo{author}{\bibfnamefont{M.~M.} \bibnamefont{Wolf}},
  \bibinfo{year}{2005}, \bibinfo{journal}{Phys. Rev. Lett.}
  \textbf{\bibinfo{volume}{95}}, \bibinfo{pages}{110503}.

\bibitem[{\citenamefont{Sciarrino and De~Martini}(2005)}]{1to3phasecloning}
\bibinfo{author}{\bibnamefont{Sciarrino}, \bibfnamefont{F.}}, and
  \bibinfo{author}{\bibfnamefont{F.}~\bibnamefont{De~Martini}},
  \bibinfo{year}{2005}, \bibinfo{journal}{Phys. Rev. A}
  \textbf{\bibinfo{volume}{72}}, \bibinfo{pages}{062313}.

\bibitem[{\citenamefont{Sciarrino and De~Martini}(2007)}]{Sciarrino2007}
\bibinfo{author}{\bibnamefont{Sciarrino}, \bibfnamefont{F.}}, and
  \bibinfo{author}{\bibfnamefont{F.}~\bibnamefont{De~Martini}},
  \bibinfo{year}{2007}, \bibinfo{journal}{Phys. Rev. A}
  \textbf{\bibinfo{volume}{76}}, \bibinfo{pages}{012330}.

\bibitem[{\citenamefont{Sciarrino}
  \emph{et~al.}(2004{\natexlab{a}})\citenamefont{Sciarrino, Sias, Ricci, and
  De~Martini}}]{Sciarrino200434}
\bibinfo{author}{\bibnamefont{Sciarrino}, \bibfnamefont{F.}},
  \bibinfo{author}{\bibfnamefont{C.}~\bibnamefont{Sias}},
  \bibinfo{author}{\bibfnamefont{M.}~\bibnamefont{Ricci}}, and
  \bibinfo{author}{\bibfnamefont{F.}~\bibnamefont{De~Martini}},
  \bibinfo{year}{2004}{\natexlab{a}}, \bibinfo{journal}{Phys. Lett. A}
  \textbf{\bibinfo{volume}{323}}, \bibinfo{pages}{34 }.

\bibitem[{\citenamefont{Sciarrino}
  \emph{et~al.}(2004{\natexlab{b}})\citenamefont{Sciarrino, Sias, Ricci, and
  De~Martini}}]{stochastic}
\bibinfo{author}{\bibnamefont{Sciarrino}, \bibfnamefont{F.}},
  \bibinfo{author}{\bibfnamefont{C.}~\bibnamefont{Sias}},
  \bibinfo{author}{\bibfnamefont{M.}~\bibnamefont{Ricci}}, and
  \bibinfo{author}{\bibfnamefont{F.}~\bibnamefont{De~Martini}},
  \bibinfo{year}{2004}{\natexlab{b}}, \bibinfo{journal}{Phys. Rev. A}
  \textbf{\bibinfo{volume}{70}}, \bibinfo{pages}{052305}.

\bibitem[{\citenamefont{Sekatski} \emph{et~al.}(2010)\citenamefont{Sekatski,
  Sanguinetti, Pomarico, Gisin, and Simon}}]{Sekatski2010}
\bibinfo{author}{\bibnamefont{Sekatski}, \bibfnamefont{P.}},
  \bibinfo{author}{\bibfnamefont{B.}~\bibnamefont{Sanguinetti}},
  \bibinfo{author}{\bibfnamefont{E.}~\bibnamefont{Pomarico}},
  \bibinfo{author}{\bibfnamefont{N.}~\bibnamefont{Gisin}}, and
  \bibinfo{author}{\bibfnamefont{C.}~\bibnamefont{Simon}},
  \bibinfo{year}{2010}, \bibinfo{journal}{Phys. Rev. A}
  \textbf{\bibinfo{volume}{82}}, \bibinfo{pages}{053814}.

\bibitem[{\citenamefont{Shen} \emph{et~al.}(2011)\citenamefont{Shen, Hao, and
  Long}}]{Shen2011}
\bibinfo{author}{\bibnamefont{Shen}, \bibfnamefont{Y.}},
  \bibinfo{author}{\bibfnamefont{L.~A.} \bibnamefont{Hao}}, and
  \bibinfo{author}{\bibfnamefont{G.~L.} \bibnamefont{Long}},
  \bibinfo{year}{2011}, \bibinfo{journal}{Chin. Phys. Lett.}
  \textbf{\bibinfo{volume}{28}}, \bibinfo{pages}{010306}.

\bibitem[{\citenamefont{Shor}(1994)}]{Shor94}
\bibinfo{author}{\bibnamefont{Shor}, \bibfnamefont{P.~W.}},
  \bibinfo{year}{1994}, in \emph{\bibinfo{booktitle}{35-th Annual Symposium on
  Foundations of Computer Science}} (\bibinfo{publisher}{IEEE, Los Alamitos,
  CA}), pp. \bibinfo{pages}{56--65}.

\bibitem[{\citenamefont{Shor and Preskill}(2000)}]{Shor2000}
\bibinfo{author}{\bibnamefont{Shor}, \bibfnamefont{P.~W.}}, and
  \bibinfo{author}{\bibfnamefont{J.}~\bibnamefont{Preskill}},
  \bibinfo{year}{2000}, \bibinfo{journal}{Phys. Rev. Lett.}
  \textbf{\bibinfo{volume}{85}}, \bibinfo{pages}{441}.

\bibitem[{\citenamefont{Simon} \emph{et~al.}(2001)\citenamefont{Simon,
  Bu\v{z}ek, and Gisin}}]{PhysRevLett.87.170405}
\bibinfo{author}{\bibnamefont{Simon}, \bibfnamefont{C.}},
  \bibinfo{author}{\bibfnamefont{V.}~\bibnamefont{Bu\v{z}ek}}, and
  \bibinfo{author}{\bibfnamefont{N.}~\bibnamefont{Gisin}},
  \bibinfo{year}{2001}, \bibinfo{journal}{Phys. Rev. Lett.}
  \textbf{\bibinfo{volume}{87}}, \bibinfo{pages}{170405}.

\bibitem[{\citenamefont{Simon} \emph{et~al.}(2000)\citenamefont{Simon, Weihs,
  and Zeilinger}}]{PhysRevLett.84.2993}
\bibinfo{author}{\bibnamefont{Simon}, \bibfnamefont{C.}},
  \bibinfo{author}{\bibfnamefont{G.}~\bibnamefont{Weihs}}, and
  \bibinfo{author}{\bibfnamefont{A.}~\bibnamefont{Zeilinger}},
  \bibinfo{year}{2000}, \bibinfo{journal}{Phys. Rev. Lett.}
  \textbf{\bibinfo{volume}{84}}, \bibinfo{pages}{2993}.

\bibitem[{\citenamefont{Siomau and Fritzsche}(2010{\natexlab{a}})}]{Siomau2010}
\bibinfo{author}{\bibnamefont{Siomau}, \bibfnamefont{M.}}, and
  \bibinfo{author}{\bibfnamefont{S.}~\bibnamefont{Fritzsche}},
  \bibinfo{year}{2010}{\natexlab{a}}, \bibinfo{journal}{Eur. Phys. J. D}
  \textbf{\bibinfo{volume}{57}}, \bibinfo{pages}{293}.

\bibitem[{\citenamefont{Siomau and
  Fritzsche}(2010{\natexlab{b}})}]{Siomau2010a}
\bibinfo{author}{\bibnamefont{Siomau}, \bibfnamefont{M.}}, and
  \bibinfo{author}{\bibfnamefont{S.}~\bibnamefont{Fritzsche}},
  \bibinfo{year}{2010}{\natexlab{b}}, \bibinfo{journal}{Eur. Phys. J. D}
  \textbf{\bibinfo{volume}{60}}, \bibinfo{pages}{417}.

\bibitem[{\citenamefont{Song and Qin}(2008)}]{Song2008}
\bibinfo{author}{\bibnamefont{Song}, \bibfnamefont{W.}}, and
  \bibinfo{author}{\bibfnamefont{T.}~\bibnamefont{Qin}}, \bibinfo{year}{2008},
  \bibinfo{journal}{Commun. Theor. Phys.} \textbf{\bibinfo{volume}{49}},
  \bibinfo{pages}{1515}.

\bibitem[{\citenamefont{Soubusta} \emph{et~al.}(2008)\citenamefont{Soubusta,
  Bartuskova, Cernoch, Dusek, and Fiurasek}}]{Soubusta2008}
\bibinfo{author}{\bibnamefont{Soubusta}, \bibfnamefont{J.}},
  \bibinfo{author}{\bibfnamefont{L.}~\bibnamefont{Bartuskova}},
  \bibinfo{author}{\bibfnamefont{A.}~\bibnamefont{Cernoch}},
  \bibinfo{author}{\bibfnamefont{M.}~\bibnamefont{Dusek}}, and
  \bibinfo{author}{\bibfnamefont{J.}~\bibnamefont{Fiurasek}},
  \bibinfo{year}{2008}, \bibinfo{journal}{Phys. Rev. A}
  \textbf{\bibinfo{volume}{78}}, \bibinfo{pages}{052323}.

\bibitem[{\citenamefont{Tamaki} \emph{et~al.}(2003)\citenamefont{Tamaki,
  Koashi, and Imoto}}]{TKI03}
\bibinfo{author}{\bibnamefont{Tamaki}, \bibfnamefont{K.}},
  \bibinfo{author}{\bibfnamefont{M.}~\bibnamefont{Koashi}}, and
  \bibinfo{author}{\bibfnamefont{N.}~\bibnamefont{Imoto}},
  \bibinfo{year}{2003}, \bibinfo{journal}{Phys. Rev. Lett.}
  \textbf{\bibinfo{volume}{90}}, \bibinfo{pages}{167904}.

\bibitem[{\citenamefont{Umegaki and K\"odai}(1962)}]{Umegaki1962}
\bibinfo{author}{\bibnamefont{Umegaki}, \bibfnamefont{U.~H.}}, and
  \bibinfo{author}{\bibnamefont{K\"odai}}, \bibinfo{year}{1962},
  \bibinfo{journal}{Math. Sem. Rep.} \textbf{\bibinfo{volume}{14}},
  \bibinfo{pages}{59}.

\bibitem[{\citenamefont{Usami} \emph{et~al.}(2003)\citenamefont{Usami, Nambu,
  Tsuda, Matsumoto, and Nakamura}}]{ISI:000185192100032}
\bibinfo{author}{\bibnamefont{Usami}, \bibfnamefont{K.}},
  \bibinfo{author}{\bibfnamefont{Y.}~\bibnamefont{Nambu}},
  \bibinfo{author}{\bibfnamefont{Y.}~\bibnamefont{Tsuda}},
  \bibinfo{author}{\bibfnamefont{K.}~\bibnamefont{Matsumoto}}, and
  \bibinfo{author}{\bibfnamefont{K.}~\bibnamefont{Nakamura}},
  \bibinfo{year}{2003}, \bibinfo{journal}{Phys. Rev. A}
  \textbf{\bibinfo{volume}{68}}, \bibinfo{pages}{022314}.

\bibitem[{\citenamefont{Vaidman} \emph{et~al.}(1987)\citenamefont{Vaidman,
  Aharonov, and Albert}}]{meanking}
\bibinfo{author}{\bibnamefont{Vaidman}, \bibfnamefont{L.}},
  \bibinfo{author}{\bibfnamefont{Y.}~\bibnamefont{Aharonov}}, and
  \bibinfo{author}{\bibfnamefont{D.~Z.} \bibnamefont{Albert}},
  \bibinfo{year}{1987}, \bibinfo{journal}{Phys. Rev. Lett.}
  \textbf{\bibinfo{volume}{58}}, \bibinfo{pages}{14}.

\bibitem[{\citenamefont{Valente} \emph{et~al.}(2012)\citenamefont{Valente, Li,
  Poizat, G\'{e}rard, Kwek, Santos, and Auff\`{e}ves}}]{ValenteLi}
\bibinfo{author}{\bibnamefont{Valente}, \bibfnamefont{D.}},
  \bibinfo{author}{\bibfnamefont{Y.}~\bibnamefont{Li}},
  \bibinfo{author}{\bibfnamefont{J.~P.} \bibnamefont{Poizat}},
  \bibinfo{author}{\bibfnamefont{L.~C.} \bibnamefont{G\'{e}rard}},
  \bibinfo{author}{\bibfnamefont{L.~C.} \bibnamefont{Kwek}},
  \bibinfo{author}{\bibfnamefont{M.~F.} \bibnamefont{Santos}}, and
  \bibinfo{author}{\bibfnamefont{A.}~\bibnamefont{Auff\`{e}ves}},
  \bibinfo{year}{2012}, \bibinfo{journal}{Phys. Rev. A}
  \textbf{\bibinfo{volume}{86}}, \bibinfo{pages}{022333}.

\bibitem[{\citenamefont{Vedral}(2002)}]{vedralRMP}
\bibinfo{author}{\bibnamefont{Vedral}, \bibfnamefont{V.}},
  \bibinfo{year}{2002}, \bibinfo{journal}{Rev. Mod. Phys.}
  \textbf{\bibinfo{volume}{74}}, \bibinfo{pages}{197}.

\bibitem[{\citenamefont{Vidal}(2003)}]{PhysRevLett.91.147902}
\bibinfo{author}{\bibnamefont{Vidal}, \bibfnamefont{G.}}, \bibinfo{year}{2003},
  \bibinfo{journal}{Phys. Rev. Lett.} \textbf{\bibinfo{volume}{91}},
  \bibinfo{pages}{147902}.

\bibitem[{\citenamefont{Vidal} \emph{et~al.}(1999)\citenamefont{Vidal, Latorre,
  Pascual, and Tarrach}}]{ISI:000081247000022}
\bibinfo{author}{\bibnamefont{Vidal}, \bibfnamefont{G.}},
  \bibinfo{author}{\bibfnamefont{J.~I.} \bibnamefont{Latorre}},
  \bibinfo{author}{\bibfnamefont{P.}~\bibnamefont{Pascual}}, and
  \bibinfo{author}{\bibfnamefont{R.}~\bibnamefont{Tarrach}},
  \bibinfo{year}{1999}, \bibinfo{journal}{Phys. Rev. A}
  \textbf{\bibinfo{volume}{60}}, \bibinfo{pages}{126}.

\bibitem[{\citenamefont{Walgate and Hardy}(2002)}]{walgate}
\bibinfo{author}{\bibnamefont{Walgate}, \bibfnamefont{J.}}, and
  \bibinfo{author}{\bibfnamefont{L.}~\bibnamefont{Hardy}},
  \bibinfo{year}{2002}, \bibinfo{journal}{Phys. Rev. Lett.}
  \textbf{\bibinfo{volume}{89}}, \bibinfo{pages}{147901}.

\bibitem[{\citenamefont{Walgate} \emph{et~al.}(2000)\citenamefont{Walgate,
  Short, Hardy, and Vedral}}]{walgateshort}
\bibinfo{author}{\bibnamefont{Walgate}, \bibfnamefont{J.}},
  \bibinfo{author}{\bibfnamefont{A.~J.} \bibnamefont{Short}},
  \bibinfo{author}{\bibfnamefont{L.}~\bibnamefont{Hardy}}, and
  \bibinfo{author}{\bibfnamefont{V.}~\bibnamefont{Vedral}},
  \bibinfo{year}{2000}, \bibinfo{journal}{Phys. Rev. Lett.}
  \textbf{\bibinfo{volume}{85}}, \bibinfo{pages}{4972}.

\bibitem[{\citenamefont{Walker and Braunstein}(2007)}]{broadcastingclassic}
\bibinfo{author}{\bibnamefont{Walker}, \bibfnamefont{T.~A.}}, and
  \bibinfo{author}{\bibfnamefont{S.~L.} \bibnamefont{Braunstein}},
  \bibinfo{year}{2007}, \bibinfo{journal}{Phys. Rev. Lett.}
  \textbf{\bibinfo{volume}{98}}, \bibinfo{pages}{080501}.

\bibitem[{\citenamefont{Wang} \emph{et~al.}(2007)\citenamefont{Wang, Hiroshima,
  Tomita, and Hayashi}}]{Wangphysreport}
\bibinfo{author}{\bibnamefont{Wang}, \bibfnamefont{X.~B.}},
  \bibinfo{author}{\bibfnamefont{T.}~\bibnamefont{Hiroshima}},
  \bibinfo{author}{\bibfnamefont{A.}~\bibnamefont{Tomita}}, and
  \bibinfo{author}{\bibfnamefont{M.}~\bibnamefont{Hayashi}},
  \bibinfo{year}{2007}, \bibinfo{journal}{Phys. Rep.}
  \textbf{\bibinfo{volume}{448}}, \bibinfo{pages}{1}.

\bibitem[{\citenamefont{Wang and Yang}(2009{\natexlab{a}})}]{Wang2009}
\bibinfo{author}{\bibnamefont{Wang}, \bibfnamefont{X.~W.}}, and
  \bibinfo{author}{\bibfnamefont{G.~J.} \bibnamefont{Yang}},
  \bibinfo{year}{2009}{\natexlab{a}}, \bibinfo{journal}{Phys. Rev. A}
  \textbf{\bibinfo{volume}{79}}, \bibinfo{pages}{062315}.

\bibitem[{\citenamefont{Wang and Yang}(2009{\natexlab{b}})}]{Wang2009a}
\bibinfo{author}{\bibnamefont{Wang}, \bibfnamefont{X.~W.}}, and
  \bibinfo{author}{\bibfnamefont{G.~J.} \bibnamefont{Yang}},
  \bibinfo{year}{2009}{\natexlab{b}}, \bibinfo{journal}{Phys. Rev. A}
  \textbf{\bibinfo{volume}{79}}, \bibinfo{pages}{064306}.

\bibitem[{\citenamefont{Wang}
  \emph{et~al.}(2011{\natexlab{a}})\citenamefont{Wang, Zhang, Yang, Tang, and
  Xie}}]{Wang2011}
\bibinfo{author}{\bibnamefont{Wang}, \bibfnamefont{X.~W.}},
  \bibinfo{author}{\bibfnamefont{D.~Y.} \bibnamefont{Zhang}},
  \bibinfo{author}{\bibfnamefont{G.~J.} \bibnamefont{Yang}},
  \bibinfo{author}{\bibfnamefont{S.~Q.} \bibnamefont{Tang}}, and
  \bibinfo{author}{\bibfnamefont{L.~J.} \bibnamefont{Xie}},
  \bibinfo{year}{2011}{\natexlab{a}}, \bibinfo{journal}{Phys. Rev. A}
  \textbf{\bibinfo{volume}{84}}, \bibinfo{pages}{042310}.

\bibitem[{\citenamefont{Wang} \emph{et~al.}(2012)\citenamefont{Wang, Shi, Jing,
  Xiong, Lei, Mu, and Fan}}]{1751-8121-45-2-025304}
\bibinfo{author}{\bibnamefont{Wang}, \bibfnamefont{Y.~N.}},
  \bibinfo{author}{\bibfnamefont{H.~D.} \bibnamefont{Shi}},
  \bibinfo{author}{\bibfnamefont{L.}~\bibnamefont{Jing}},
  \bibinfo{author}{\bibfnamefont{Z.~X.} \bibnamefont{Xiong}},
  \bibinfo{author}{\bibfnamefont{J.}~\bibnamefont{Lei}},
  \bibinfo{author}{\bibfnamefont{L.~Z.} \bibnamefont{Mu}}, and
  \bibinfo{author}{\bibfnamefont{H.}~\bibnamefont{Fan}}, \bibinfo{year}{2012},
  \bibinfo{journal}{J. Phys. A-Math. Theor.} \textbf{\bibinfo{volume}{45}},
  \bibinfo{pages}{025304}.

\bibitem[{\citenamefont{Wang}
  \emph{et~al.}(2011{\natexlab{b}})\citenamefont{Wang, Shi, Xiong, Jing, Ren,
  Mu, and Fan}}]{PhysRevA.84.034302}
\bibinfo{author}{\bibnamefont{Wang}, \bibfnamefont{Y.~N.}},
  \bibinfo{author}{\bibfnamefont{H.~D.} \bibnamefont{Shi}},
  \bibinfo{author}{\bibfnamefont{Z.~X.} \bibnamefont{Xiong}},
  \bibinfo{author}{\bibfnamefont{L.}~\bibnamefont{Jing}},
  \bibinfo{author}{\bibfnamefont{X.~J.} \bibnamefont{Ren}},
  \bibinfo{author}{\bibfnamefont{L.~Z.} \bibnamefont{Mu}}, and
  \bibinfo{author}{\bibfnamefont{H.}~\bibnamefont{Fan}},
  \bibinfo{year}{2011}{\natexlab{b}}, \bibinfo{journal}{Phys. Rev. A}
  \textbf{\bibinfo{volume}{84}}, \bibinfo{pages}{034302}.

\bibitem[{\citenamefont{Weedbrook} \emph{et~al.}(2008)\citenamefont{Weedbrook,
  Grosse, Symul, Lam, and Ralph}}]{Weedbrook2008}
\bibinfo{author}{\bibnamefont{Weedbrook}, \bibfnamefont{C.}},
  \bibinfo{author}{\bibfnamefont{N.~B.} \bibnamefont{Grosse}},
  \bibinfo{author}{\bibfnamefont{T.}~\bibnamefont{Symul}},
  \bibinfo{author}{\bibfnamefont{P.~K.} \bibnamefont{Lam}}, and
  \bibinfo{author}{\bibfnamefont{T.~C.} \bibnamefont{Ralph}},
  \bibinfo{year}{2008}, \bibinfo{journal}{Phys. Rev. A}
  \textbf{\bibinfo{volume}{77}}, \bibinfo{pages}{052313}.

\bibitem[{\citenamefont{Weedbrook} \emph{et~al.}(2012)\citenamefont{Weedbrook,
  Pirandola, Garcia-Patron, Cerf, Ralph, Shapiro, and Lloyd}}]{RMP-gaussian}
\bibinfo{author}{\bibnamefont{Weedbrook}, \bibfnamefont{C.}},
  \bibinfo{author}{\bibfnamefont{S.}~\bibnamefont{Pirandola}},
  \bibinfo{author}{\bibfnamefont{R.}~\bibnamefont{Garcia-Patron}},
  \bibinfo{author}{\bibfnamefont{N.~J.} \bibnamefont{Cerf}},
  \bibinfo{author}{\bibfnamefont{T.~C.} \bibnamefont{Ralph}},
  \bibinfo{author}{\bibfnamefont{J.~H.} \bibnamefont{Shapiro}}, and
  \bibinfo{author}{\bibfnamefont{S.}~\bibnamefont{Lloyd}},
  \bibinfo{year}{2012}, \bibinfo{journal}{Rev. Mod. Phys.}
  \textbf{\bibinfo{volume}{84}}, \bibinfo{pages}{621}.

\bibitem[{\citenamefont{Wehrl}(1978)}]{RevModPhys.50.221}
\bibinfo{author}{\bibnamefont{Wehrl}, \bibfnamefont{A.}}, \bibinfo{year}{1978},
  \bibinfo{journal}{Rev. Mod. Phys.} \textbf{\bibinfo{volume}{50}},
  \bibinfo{pages}{221}.

\bibitem[{\citenamefont{Werner} \emph{et~al.}(2009)\citenamefont{Werner, Franz,
  and Werner}}]{Werner}
\bibinfo{author}{\bibnamefont{Werner}, \bibfnamefont{A.~H.}},
  \bibinfo{author}{\bibfnamefont{T.}~\bibnamefont{Franz}}, and
  \bibinfo{author}{\bibfnamefont{R.~F.} \bibnamefont{Werner}},
  \bibinfo{year}{2009}, \bibinfo{journal}{Phys. Rev. Lett.}
  \textbf{\bibinfo{volume}{103}}, \bibinfo{pages}{220504}.

\bibitem[{\citenamefont{Werner}(1998)}]{PhysRevA.58.1827}
\bibinfo{author}{\bibnamefont{Werner}, \bibfnamefont{R.~F.}},
  \bibinfo{year}{1998}, \bibinfo{journal}{Phys. Rev. A}
  \textbf{\bibinfo{volume}{58}}, \bibinfo{pages}{1827}.

\bibitem[{\citenamefont{Wootters and Fields}(1989)}]{Wootters1989}
\bibinfo{author}{\bibnamefont{Wootters}, \bibfnamefont{W.~K.}}, and
  \bibinfo{author}{\bibfnamefont{B.~D.} \bibnamefont{Fields}},
  \bibinfo{year}{1989}, \bibinfo{journal}{Ann. Phys.-New York}
  \textbf{\bibinfo{volume}{191}}, \bibinfo{pages}{363}.

\bibitem[{\citenamefont{Wootters and Zurek}(1982)}]{Wootters1982}
\bibinfo{author}{\bibnamefont{Wootters}, \bibfnamefont{W.~K.}}, and
  \bibinfo{author}{\bibfnamefont{W.~H.} \bibnamefont{Zurek}},
  \bibinfo{year}{1982}, \bibinfo{journal}{Nature}
  \textbf{\bibinfo{volume}{299}}, \bibinfo{pages}{802}.

\bibitem[{\citenamefont{Wootters and Zurek}(2009)}]{Wootters2009}
\bibinfo{author}{\bibnamefont{Wootters}, \bibfnamefont{W.~K.}}, and
  \bibinfo{author}{\bibfnamefont{W.~H.} \bibnamefont{Zurek}},
  \bibinfo{year}{2009}, \bibinfo{journal}{Phys. Today}
  \textbf{\bibinfo{volume}{62}}, \bibinfo{pages}{76}.

\bibitem[{\citenamefont{Wu and Wu}(2012)}]{Wu2012}
\bibinfo{author}{\bibnamefont{Wu}, \bibfnamefont{F.}}, and
  \bibinfo{author}{\bibfnamefont{X.~H.} \bibnamefont{Wu}},
  \bibinfo{year}{2012}, \bibinfo{journal}{Quant. Inf. Process.}
  \textbf{\bibinfo{volume}{11}}, \bibinfo{pages}{1}.

\bibitem[{\citenamefont{Wu and Guo}(2011)}]{ISI:000291254100001}
\bibinfo{author}{\bibnamefont{Wu}, \bibfnamefont{Y.~C.}}, and
  \bibinfo{author}{\bibfnamefont{G.~C.} \bibnamefont{Guo}},
  \bibinfo{year}{2011}, \bibinfo{journal}{Phys. Rev. A}
  \textbf{\bibinfo{volume}{83}}, \bibinfo{pages}{062301}.

\bibitem[{\citenamefont{Xiang} \emph{et~al.}(2010)\citenamefont{Xiang, Ralph,
  Lund, Walk, and Pryde}}]{XiangGuoYong1}
\bibinfo{author}{\bibnamefont{Xiang}, \bibfnamefont{G.~Y.}},
  \bibinfo{author}{\bibfnamefont{T.~C.} \bibnamefont{Ralph}},
  \bibinfo{author}{\bibfnamefont{A.~P.} \bibnamefont{Lund}},
  \bibinfo{author}{\bibfnamefont{N.}~\bibnamefont{Walk}}, and
  \bibinfo{author}{\bibfnamefont{G.~J.} \bibnamefont{Pryde}},
  \bibinfo{year}{2010}, \bibinfo{journal}{Nat. Photonics}
  \textbf{\bibinfo{volume}{4}}, \bibinfo{pages}{316}.

\bibitem[{\citenamefont{Xiong} \emph{et~al.}(2012)\citenamefont{Xiong, Shi,
  Wang, Jing, Lei, Mu, and Fan}}]{PhysRevA.85.012334}
\bibinfo{author}{\bibnamefont{Xiong}, \bibfnamefont{Z.~X.}},
  \bibinfo{author}{\bibfnamefont{H.~D.} \bibnamefont{Shi}},
  \bibinfo{author}{\bibfnamefont{Y.~N.} \bibnamefont{Wang}},
  \bibinfo{author}{\bibfnamefont{L.}~\bibnamefont{Jing}},
  \bibinfo{author}{\bibfnamefont{J.}~\bibnamefont{Lei}},
  \bibinfo{author}{\bibfnamefont{L.~Z.} \bibnamefont{Mu}}, and
  \bibinfo{author}{\bibfnamefont{H.}~\bibnamefont{Fan}}, \bibinfo{year}{2012},
  \bibinfo{journal}{Phys. Rev. A} \textbf{\bibinfo{volume}{85}},
  \bibinfo{pages}{012334}.

\bibitem[{\citenamefont{Xu} \emph{et~al.}(2008)\citenamefont{Xu, Li, Chen, Zou,
  and Guo}}]{XuLiChen}
\bibinfo{author}{\bibnamefont{Xu}, \bibfnamefont{J.~S.}},
  \bibinfo{author}{\bibfnamefont{C.~F.} \bibnamefont{Li}},
  \bibinfo{author}{\bibfnamefont{L.}~\bibnamefont{Chen}},
  \bibinfo{author}{\bibfnamefont{X.~B.} \bibnamefont{Zou}}, and
  \bibinfo{author}{\bibfnamefont{G.~C.} \bibnamefont{Guo}},
  \bibinfo{year}{2008}, \bibinfo{journal}{Phys. Rev. A}
  \textbf{\bibinfo{volume}{78}}, \bibinfo{pages}{032322}.

\bibitem[{\citenamefont{Yan} \emph{et~al.}(2009)\citenamefont{Yan, Gao, and
  Zhao}}]{Yan2009}
\bibinfo{author}{\bibnamefont{Yan}, \bibfnamefont{L.~H.}},
  \bibinfo{author}{\bibfnamefont{Y.~F.} \bibnamefont{Gao}}, and
  \bibinfo{author}{\bibfnamefont{J.~G.} \bibnamefont{Zhao}},
  \bibinfo{year}{2009}, \bibinfo{journal}{Int. J. Theor. Phys.}
  \textbf{\bibinfo{volume}{48}}, \bibinfo{pages}{2445}.

\bibitem[{\citenamefont{Yang} \emph{et~al.}(2008)\citenamefont{Yang, Yu, Zhang,
  and Liu}}]{ISI:000254541100254}
\bibinfo{author}{\bibnamefont{Yang}, \bibfnamefont{J.}},
  \bibinfo{author}{\bibfnamefont{Y.~F.} \bibnamefont{Yu}},
  \bibinfo{author}{\bibfnamefont{Z.~M.} \bibnamefont{Zhang}}, and
  \bibinfo{author}{\bibfnamefont{S.~H.} \bibnamefont{Liu}},
  \bibinfo{year}{2008}, \bibinfo{journal}{Phys. Rev. A}
  \textbf{\bibinfo{volume}{77}}, \bibinfo{pages}{034302}.

\bibitem[{\citenamefont{Yang} \emph{et~al.}(2007)\citenamefont{Yang, Zhao, Liu,
  and Chen}}]{Yang2007}
\bibinfo{author}{\bibnamefont{Yang}, \bibfnamefont{S.}},
  \bibinfo{author}{\bibfnamefont{M.~S.} \bibnamefont{Zhao}},
  \bibinfo{author}{\bibfnamefont{N.~L.} \bibnamefont{Liu}}, and
  \bibinfo{author}{\bibfnamefont{Z.~B.} \bibnamefont{Chen}},
  \bibinfo{year}{2007}, \bibinfo{journal}{Chin. Phys. Lett.}
  \textbf{\bibinfo{volume}{24}}, \bibinfo{pages}{3048}.

\bibitem[{\citenamefont{Yoshida} \emph{et~al.}(2010)\citenamefont{Yoshida,
  Miyadera, and Imai}}]{Yoshida2010}
\bibinfo{author}{\bibnamefont{Yoshida}, \bibfnamefont{M.}},
  \bibinfo{author}{\bibfnamefont{T.}~\bibnamefont{Miyadera}}, and
  \bibinfo{author}{\bibfnamefont{H.}~\bibnamefont{Imai}}, \bibinfo{year}{2010},
  \bibinfo{journal}{Int. Sym. on ISITA} , \bibinfo{pages}{917}.

\bibitem[{\citenamefont{Yu} \emph{et~al.}(2007)\citenamefont{Yu, Zhang, and
  Ye}}]{PhysRevA.76.034303}
\bibinfo{author}{\bibnamefont{Yu}, \bibfnamefont{L.~B.}},
  \bibinfo{author}{\bibfnamefont{W.~H.} \bibnamefont{Zhang}}, and
  \bibinfo{author}{\bibfnamefont{L.}~\bibnamefont{Ye}}, \bibinfo{year}{2007},
  \bibinfo{journal}{Phys. Rev. A} \textbf{\bibinfo{volume}{76}},
  \bibinfo{pages}{034303}.

\bibitem[{\citenamefont{Yu} \emph{et~al.}(2004)\citenamefont{Yu, Feng, Zhou,
  and Zhan}}]{ISI:000221638800002}
\bibinfo{author}{\bibnamefont{Yu}, \bibfnamefont{Y.~F.}},
  \bibinfo{author}{\bibfnamefont{J.}~\bibnamefont{Feng}},
  \bibinfo{author}{\bibfnamefont{X.~Q.} \bibnamefont{Zhou}}, and
  \bibinfo{author}{\bibfnamefont{M.~S.} \bibnamefont{Zhan}},
  \bibinfo{year}{2004}, \bibinfo{journal}{Int. J. Theor. Phys.}
  \textbf{\bibinfo{volume}{43}}, \bibinfo{pages}{21}.

\bibitem[{\citenamefont{Yu}(2009)}]{Yu2009}
\bibinfo{author}{\bibnamefont{Yu}, \bibfnamefont{Z.~W.}}, \bibinfo{year}{2009},
  \bibinfo{journal}{Chin. Phys. Lett.} \textbf{\bibinfo{volume}{26}},
  \bibinfo{pages}{080304}.

\bibitem[{\citenamefont{Yuen}(1986)}]{yuen-no-cloning}
\bibinfo{author}{\bibnamefont{Yuen}, \bibfnamefont{H.~P.}},
  \bibinfo{year}{1986}, \bibinfo{journal}{Phys. Lett. A}
  \textbf{\bibinfo{volume}{113}}, \bibinfo{pages}{405}.

\bibitem[{\citenamefont{Zanardi}(1998)}]{zanardi}
\bibinfo{author}{\bibnamefont{Zanardi}, \bibfnamefont{P.}},
  \bibinfo{year}{1998}, \bibinfo{journal}{Phys. Rev. A}
  \textbf{\bibinfo{volume}{58}}, \bibinfo{pages}{3484}.

\bibitem[{\citenamefont{Zavatta} \emph{et~al.}(2011)\citenamefont{Zavatta,
  Fiurasek, and Bellini}}]{naturephoton2011}
\bibinfo{author}{\bibnamefont{Zavatta}, \bibfnamefont{A.}},
  \bibinfo{author}{\bibfnamefont{J.}~\bibnamefont{Fiurasek}}, and
  \bibinfo{author}{\bibfnamefont{M.}~\bibnamefont{Bellini}},
  \bibinfo{year}{2011}, \bibinfo{journal}{Nat. Photonics}
  \textbf{\bibinfo{volume}{5}}, \bibinfo{pages}{52}.

\bibitem[{\citenamefont{Zhan} \emph{et~al.}(2009)\citenamefont{Zhan, Zhang,
  Gao, Li, and Zeng}}]{Zhan2009}
\bibinfo{author}{\bibnamefont{Zhan}, \bibfnamefont{X.~G.}},
  \bibinfo{author}{\bibfnamefont{D.~Y.} \bibnamefont{Zhang}},
  \bibinfo{author}{\bibfnamefont{F.}~\bibnamefont{Gao}},
  \bibinfo{author}{\bibfnamefont{Y.~L.} \bibnamefont{Li}}, and
  \bibinfo{author}{\bibfnamefont{H.~S.} \bibnamefont{Zeng}},
  \bibinfo{year}{2009}, \bibinfo{journal}{Commun. Theor. Phys.}
  \textbf{\bibinfo{volume}{51}}, \bibinfo{pages}{1023}.

\bibitem[{\citenamefont{Zhan}(2005)}]{ISI:000227361300007}
\bibinfo{author}{\bibnamefont{Zhan}, \bibfnamefont{Y.~B.}},
  \bibinfo{year}{2005}, \bibinfo{journal}{Phys. Lett. A}
  \textbf{\bibinfo{volume}{336}}, \bibinfo{pages}{317}.

\bibitem[{\citenamefont{Zhang}
  \emph{et~al.}(2000{\natexlab{a}})\citenamefont{Zhang, Li, Wang, and
  Guo}}]{ZhangLiWangGuo}
\bibinfo{author}{\bibnamefont{Zhang}, \bibfnamefont{C.~W.}},
  \bibinfo{author}{\bibfnamefont{C.~F.} \bibnamefont{Li}},
  \bibinfo{author}{\bibfnamefont{Z.~Y.} \bibnamefont{Wang}}, and
  \bibinfo{author}{\bibfnamefont{G.~C.} \bibnamefont{Guo}},
  \bibinfo{year}{2000}{\natexlab{a}}, \bibinfo{journal}{Phys. Rev. A}
  \textbf{\bibinfo{volume}{62}}, \bibinfo{pages}{042302}.

\bibitem[{\citenamefont{Zhang}
  \emph{et~al.}(2000{\natexlab{b}})\citenamefont{Zhang, Wang, Li, and
  Guo}}]{ISI:000087567900029}
\bibinfo{author}{\bibnamefont{Zhang}, \bibfnamefont{C.~W.}},
  \bibinfo{author}{\bibfnamefont{Z.~Y.} \bibnamefont{Wang}},
  \bibinfo{author}{\bibfnamefont{C.~F.} \bibnamefont{Li}}, and
  \bibinfo{author}{\bibfnamefont{G.~C.} \bibnamefont{Guo}},
  \bibinfo{year}{2000}{\natexlab{b}}, \bibinfo{journal}{Phys. Rev. A}
  \textbf{\bibinfo{volume}{61}}, \bibinfo{pages}{062310}.

\bibitem[{\citenamefont{Zhang} \emph{et~al.}(2008)\citenamefont{Zhang, Xie, and
  van Loock}}]{ZhangJing08PRA}
\bibinfo{author}{\bibnamefont{Zhang}, \bibfnamefont{J.}},
  \bibinfo{author}{\bibfnamefont{C.~D.} \bibnamefont{Xie}}, and
  \bibinfo{author}{\bibfnamefont{P.}~\bibnamefont{van Loock}},
  \bibinfo{year}{2008}, \bibinfo{journal}{Phys. Rev. A}
  \textbf{\bibinfo{volume}{77}}, \bibinfo{pages}{022316}.

\bibitem[{\citenamefont{Zhang} \emph{et~al.}(2005)\citenamefont{Zhang, Xie, and
  Peng}}]{ZhangJing05PRL}
\bibinfo{author}{\bibnamefont{Zhang}, \bibfnamefont{J.}},
  \bibinfo{author}{\bibfnamefont{C.~D.} \bibnamefont{Xie}}, and
  \bibinfo{author}{\bibfnamefont{K.~C.} \bibnamefont{Peng}},
  \bibinfo{year}{2005}, \bibinfo{journal}{Phys. Rev. Lett.}
  \textbf{\bibinfo{volume}{95}}, \bibinfo{pages}{170501}.

\bibitem[{\citenamefont{Zhang} \emph{et~al.}(2006)\citenamefont{Zhang, Xie, and
  Peng}}]{PhysRevA.73.042315}
\bibinfo{author}{\bibnamefont{Zhang}, \bibfnamefont{J.}},
  \bibinfo{author}{\bibfnamefont{C.~D.} \bibnamefont{Xie}}, and
  \bibinfo{author}{\bibfnamefont{K.~C.} \bibnamefont{Peng}},
  \bibinfo{year}{2006}, \bibinfo{journal}{Phys. Rev. A}
  \textbf{\bibinfo{volume}{73}}, \bibinfo{pages}{042315}.

\bibitem[{\citenamefont{Zhang}
  \emph{et~al.}(2010{\natexlab{a}})\citenamefont{Zhang, Dai, Cao, and
  Yang}}]{Zhang2010a}
\bibinfo{author}{\bibnamefont{Zhang}, \bibfnamefont{W.~H.}},
  \bibinfo{author}{\bibfnamefont{J.~L.} \bibnamefont{Dai}},
  \bibinfo{author}{\bibfnamefont{Z.~L.} \bibnamefont{Cao}}, and
  \bibinfo{author}{\bibfnamefont{M.}~\bibnamefont{Yang}},
  \bibinfo{year}{2010}{\natexlab{a}}, \bibinfo{journal}{Opt. Commun.}
  \textbf{\bibinfo{volume}{283}}, \bibinfo{pages}{1956}.

\bibitem[{\citenamefont{Zhang}
  \emph{et~al.}(2010{\natexlab{b}})\citenamefont{Zhang, Dai, Cao, and
  Yang}}]{Zhang2010b}
\bibinfo{author}{\bibnamefont{Zhang}, \bibfnamefont{W.~H.}},
  \bibinfo{author}{\bibfnamefont{J.~L.} \bibnamefont{Dai}},
  \bibinfo{author}{\bibfnamefont{Z.~L.} \bibnamefont{Cao}}, and
  \bibinfo{author}{\bibfnamefont{M.}~\bibnamefont{Yang}},
  \bibinfo{year}{2010}{\natexlab{b}}, \bibinfo{journal}{Opt. Commun.}
  \textbf{\bibinfo{volume}{283}}, \bibinfo{pages}{3818}.

\bibitem[{\citenamefont{Zhang and Ye}(2009)}]{Zhang2009}
\bibinfo{author}{\bibnamefont{Zhang}, \bibfnamefont{W.~H.}}, and
  \bibinfo{author}{\bibfnamefont{L.}~\bibnamefont{Ye}}, \bibinfo{year}{2009},
  \bibinfo{journal}{Chin. Phys. B} \textbf{\bibinfo{volume}{18}},
  \bibinfo{pages}{3702}.

\bibitem[{\citenamefont{Zhang} \emph{et~al.}(2007)\citenamefont{Zhang, Yu, Ye,
  and Dai}}]{Zhang2007}
\bibinfo{author}{\bibnamefont{Zhang}, \bibfnamefont{W.~H.}},
  \bibinfo{author}{\bibfnamefont{L.~B.} \bibnamefont{Yu}},
  \bibinfo{author}{\bibfnamefont{L.}~\bibnamefont{Ye}}, and
  \bibinfo{author}{\bibfnamefont{J.~L.} \bibnamefont{Dai}},
  \bibinfo{year}{2007}, \bibinfo{journal}{Phys. Lett. A}
  \textbf{\bibinfo{volume}{360}}, \bibinfo{pages}{726}.

\bibitem[{\citenamefont{Zhang}
  \emph{et~al.}(2013{\natexlab{a}})\citenamefont{Zhang, Wang, Xiang, Jing, Mu,
  Korepin, and Fan}}]{Zhangnew}
\bibinfo{author}{\bibnamefont{Zhang}, \bibfnamefont{Y.~L.}},
  \bibinfo{author}{\bibfnamefont{Y.~N.} \bibnamefont{Wang}},
  \bibinfo{author}{\bibfnamefont{X.~R.} \bibnamefont{Xiang}},
  \bibinfo{author}{\bibfnamefont{L.}~\bibnamefont{Jing}},
  \bibinfo{author}{\bibfnamefont{L.~Z.} \bibnamefont{Mu}},
  \bibinfo{author}{\bibfnamefont{V.}~\bibnamefont{Korepin}}, and
  \bibinfo{author}{\bibfnamefont{H.}~\bibnamefont{Fan}},
  \bibinfo{year}{2013}{\natexlab{a}}, \bibinfo{journal}{Phys. Rev. A}
  \textbf{\bibinfo{volume}{87}}, \bibinfo{pages}{022302}.

\bibitem[{\citenamefont{Zhang}
  \emph{et~al.}(2013{\natexlab{b}})\citenamefont{Zhang, Zhang, Mu, and
  Fan}}]{zhangnew2}
\bibinfo{author}{\bibnamefont{Zhang}, \bibfnamefont{Y.~L.}},
  \bibinfo{author}{\bibfnamefont{Y.~R.} \bibnamefont{Zhang}},
  \bibinfo{author}{\bibfnamefont{L.~Z.} \bibnamefont{Mu}}, and
  \bibinfo{author}{\bibfnamefont{H.}~\bibnamefont{Fan}},
  \bibinfo{year}{2013}{\natexlab{b}}, \bibinfo{journal}{Phys. Rev. A}
  \textbf{\bibinfo{volume}{88}}, \bibinfo{pages}{052314}.

\bibitem[{\citenamefont{Zhang} \emph{et~al.}(2012)\citenamefont{Zhang, Tang,
  Xie, and Wang}}]{Zhang2012}
\bibinfo{author}{\bibnamefont{Zhang}, \bibfnamefont{Z.~Y.}},
  \bibinfo{author}{\bibfnamefont{S.~Q.} \bibnamefont{Tang}},
  \bibinfo{author}{\bibfnamefont{L.~J.} \bibnamefont{Xie}}, and
  \bibinfo{author}{\bibfnamefont{X.~W.} \bibnamefont{Wang}},
  \bibinfo{year}{2012}, \bibinfo{journal}{Int. J. Theor. Phys.}
  \textbf{\bibinfo{volume}{51}}, \bibinfo{pages}{805}.

\bibitem[{\citenamefont{Zhao} \emph{et~al.}(2005)\citenamefont{Zhao, Zhang,
  Zhou, Chen, Lu, Karlsson, and Pan}}]{Zhaozhi05}
\bibinfo{author}{\bibnamefont{Zhao}, \bibfnamefont{Z.}},
  \bibinfo{author}{\bibfnamefont{A.~N.} \bibnamefont{Zhang}},
  \bibinfo{author}{\bibfnamefont{X.~Q.} \bibnamefont{Zhou}},
  \bibinfo{author}{\bibfnamefont{Y.~A.} \bibnamefont{Chen}},
  \bibinfo{author}{\bibfnamefont{C.~Y.} \bibnamefont{Lu}},
  \bibinfo{author}{\bibfnamefont{A.}~\bibnamefont{Karlsson}}, and
  \bibinfo{author}{\bibfnamefont{J.~W.} \bibnamefont{Pan}},
  \bibinfo{year}{2005}, \bibinfo{journal}{Phys. Rev. Lett.}
  \textbf{\bibinfo{volume}{95}}, \bibinfo{pages}{030502}.

\bibitem[{\citenamefont{Zheng}(2004)}]{ISI:000223715500003}
\bibinfo{author}{\bibnamefont{Zheng}, \bibfnamefont{S.~B.}},
  \bibinfo{year}{2004}, \bibinfo{journal}{Chin. Phys. Lett.}
  \textbf{\bibinfo{volume}{21}}, \bibinfo{pages}{1689}.

\bibitem[{\citenamefont{Zheng}(2005)}]{ISI:000230374500016}
\bibinfo{author}{\bibnamefont{Zheng}, \bibfnamefont{S.~B.}},
  \bibinfo{year}{2005}, \bibinfo{journal}{J. Phys. B-Mol. Opt.}
  \textbf{\bibinfo{volume}{38}}, \bibinfo{pages}{1499}.

\bibitem[{\citenamefont{Zhou} \emph{et~al.}(2006)\citenamefont{Zhou, Zeng, and
  You}}]{ISI:000235859600008}
\bibinfo{author}{\bibnamefont{Zhou}, \bibfnamefont{D.~L.}},
  \bibinfo{author}{\bibfnamefont{B.}~\bibnamefont{Zeng}}, and
  \bibinfo{author}{\bibfnamefont{L.}~\bibnamefont{You}}, \bibinfo{year}{2006},
  \bibinfo{journal}{Phys. Lett. A} \textbf{\bibinfo{volume}{352}},
  \bibinfo{pages}{41}.

\bibitem[{\citenamefont{Zhou}(2011)}]{Zhou2011}
\bibinfo{author}{\bibnamefont{Zhou}, \bibfnamefont{Y.~H.}},
  \bibinfo{year}{2011}, \bibinfo{journal}{Chin. Phys. B}
  \textbf{\bibinfo{volume}{20}}, \bibinfo{pages}{080305}.

\bibitem[{\citenamefont{Zhou and Guo}(2000)}]{ISI:000085836300022}
\bibinfo{author}{\bibnamefont{Zhou}, \bibfnamefont{Z.~W.}}, and
  \bibinfo{author}{\bibfnamefont{G.~C.} \bibnamefont{Guo}},
  \bibinfo{year}{2000}, \bibinfo{journal}{Phys. Rev. A}
  \textbf{\bibinfo{volume}{61}}, \bibinfo{pages}{032108}.

\bibitem[{\citenamefont{Zou} \emph{et~al.}(2006)\citenamefont{Zou, Dong, and
  Guo}}]{Zou2006}
\bibinfo{author}{\bibnamefont{Zou}, \bibfnamefont{X.~B.}},
  \bibinfo{author}{\bibfnamefont{Y.~L.} \bibnamefont{Dong}}, and
  \bibinfo{author}{\bibfnamefont{G.~C.} \bibnamefont{Guo}},
  \bibinfo{year}{2006}, \bibinfo{journal}{Phys. Lett. A}
  \textbf{\bibinfo{volume}{360}}, \bibinfo{pages}{44}.

\bibitem[{\citenamefont{Zou and Mathis}(2005)}]{Zou2005-M}
\bibinfo{author}{\bibnamefont{Zou}, \bibfnamefont{X.~B.}}, and
  \bibinfo{author}{\bibfnamefont{W.}~\bibnamefont{Mathis}},
  \bibinfo{year}{2005}, \bibinfo{journal}{Phys. Rev. A}
  \textbf{\bibinfo{volume}{72}}, \bibinfo{pages}{022306}.

\bibitem[{\citenamefont{Zou} \emph{et~al.}(2003)\citenamefont{Zou, Pahlke, and
  Mathis}}]{Zou2003}
\bibinfo{author}{\bibnamefont{Zou}, \bibfnamefont{X.~B.}},
  \bibinfo{author}{\bibfnamefont{K.}~\bibnamefont{Pahlke}}, and
  \bibinfo{author}{\bibfnamefont{W.}~\bibnamefont{Mathis}},
  \bibinfo{year}{2003}, \bibinfo{journal}{Phys. Rev. A}
  \textbf{\bibinfo{volume}{67}}, \bibinfo{pages}{024304}.

\end{thebibliography}

\end{document}